\documentclass[numbers = noenddot, twoside=true, open = right, a4wide, 11pt , parskip=half, 
liststotocnumbered,headings=optiontohead, captions=tableheading]{scrreprt}

\PassOptionsToPackage{hyphens}{url}\usepackage{hyperref}
\usepackage[utf8]{inputenc}
\usepackage[T1]{fontenc}
\usepackage{lmodern}
\usepackage[english]{babel}
\usepackage{graphicx}
\usepackage{nicefrac}
\usepackage{a4wide}
\usepackage{amsmath}
\usepackage{parskip}
\usepackage{amssymb}
\usepackage{lscape}
\usepackage{fancyhdr}
\usepackage{appendix}
\usepackage{booktabs}
\usepackage{multirow,arydshln,tabu}
\usepackage{setspace}
\usepackage[small,hang,bf]{caption}
\usepackage{subcaption,booktabs} 
\usepackage{array}
\usepackage{caption}
\usepackage{url}
\usepackage{natbib}
\usepackage{listings}
\usepackage{ae,aecompl}
\usepackage[table]{xcolor}
\usepackage{enumitem} 
\usepackage{pdflscape, afterpage} 

\newcommand\degr{\hbox{$^\circ$}}
\newcommand\farcs{\hbox{$.\!\!^{\prime\prime}$}}
\newcommand\arcsec{\hbox{$^{\prime\prime}$}}

\newcommand\aap{A\&A}                
\newcommand\aaps{A\&AS}              
\newcommand\aj{AJ}                   
\newcommand\apj{ApJ}                 
\newcommand\apjl{ApJ}                
\newcommand\apjs{ApJS}               
\newcommand\mnras{MNRAS}             
\newcommand\pasa{Publ. Astron. Soc. Australia}  

\begin{document}
\raggedbottom

\pagestyle{empty}

\begin{titlepage}
	\begin{center}
		\Huge \textsc{}\\
		\textsc{\huge Galaxy And Mass Assembly (GAMA): Bulge-disk decomposition of KiDS and VIKING data in the nearby universe} \\
		\vspace{3.5cm}
		\textsc{\Large Dissertation \\}
		\vspace{0.5cm}
		\textsc{\large zur Erlangung des Doktorgrades \\ an der Fakult\"at für  Mathematik, Informatik und Naturwissenschaften\\}
		\textsc{\large Fachbereich Physik\\}
		\textsc{\large der Universit\"at Hamburg\\}
		\normalsize
		\vspace{3cm}
		\textsc{\large vorgelegt von\\}
	  \textsc{\LARGE Sarah Casura\\}
	\vspace{3cm}
	\textsc{\large Hamburg\\2022}
	\normalsize
	\end{center}
\pagenumbering{gobble}

\end{titlepage}

 \cleardoublepage

\pagenumbering{gooble}

\pagestyle{plain}


\begin{flushright}
 \begin{table}[!b]
 \setlength{\tabcolsep}{0pt}
  \begin{tabular}{l r}
   Gutachter der Dissertation: & Prof. Dr. Jochen Liske \\
				     & A/Prof. Dr. Aaron S. G. Robotham \\
				     & \\
   Zusammensetzung der Pr\"ufungskommission: & Prof. Dr. Jochen Liske \\  
& Prof. Dr. Robi Banerjee \\
& Prof. Dr. Peter Hauschildt \\
& Prof. Dr. Dieter Horns \\
& Prof. Dr. Johannes Haller \\
& \\
Vorsitzender der Pr\"ufungskommission: & Prof. Dr. Peter Hauschildt \\ 
& \\
Datum der Disputation: & 17.06.2022  \\ 
& \\
Vorsitzender des Fach-Promotionsausschusses PHYSIK: & Prof. Dr. Wolfgang J. Parak \\
& \\
Leiter des Fachbereichs PHYSIK: & Prof. Dr. G\"unter H. W. Sigl\\
& \\
Dekan der Fakultät MIN: & Prof. Dr. Heinrich Graener
  \end{tabular}

 \end{table}

\end{flushright}

\pagenumbering{Roman}
\chapter*{Eidesstattliche Versicherung}
\bigskip
 Hiermit versichere ich an Eides statt, die vorliegende Dissertationsschrift selbst verfasst und keine anderen als die angegebenen Hilfsmittel und Quellen benutzt zu haben. \\
 
Hamburg, den 27.04.2022
\bigskip
\bigskip

{\begin{flushright}
  (Sarah Casura) \\
 \end{flushright}
 }
 
\newpage 
 
\chapter*{Zusammenfassung}
Quantitative Strukturanalysen von Galaxien sind unerlässlich um theoretische Modelle und Simulationen einzugrenzen, welche die Entstehung und Entwicklung von Galaxien sowie des gesamten Universums beschreiben. Entsprechende Daten werden von Himmelsdurchmusterungen in hoher Qualität geliefert. Um diese vollumfänglich nutzen zu können, sind automatisierte Methoden notwendig.  
Diese Promotionsarbeit präsentiert Strukturparameter für die Komponenten einer großen Anzahl von GAMA Galaxien im nahen Universum und verbessert die Bildanalyse, die Modellierung von Galaxien sowie die Nachbearbeitung solcher Modelle zur Qualitätssicherung, im Kontext von automatisierten Galaxiezerlegungsstudien. 
Die Probe umfasst 13096 Galaxien in den GAMA II Equatorialregionen bei Rotverschiebungen $z$\,<\,0.08. Die Bilddaten stammen von den KiDS und VIKING Himmelsdurchmusterungen, welche den optischen und nah-infraroten Wellenlängenbereich abdecken ($ugriZYJHK_s$-Filter). Wir fitten die Oberflächenhelligkeitsverteilung jeder Galaxie in jedem Filter mit drei Modellen: einer S\'ersicfunktion, einer S\'ersic- plus Exponentialfunktion sowie einer Punktquelle plus Exponentialfunktion. Die Modellanpassung erfolgt mit einer vollautomatisierten Markov-Chain-Monte-Carlo-Analyse mit dem bayesschen zweidimensionalen Profilanpassungs-Code \texttt{ProFit}. Die Vorarbeit, einschlie{\ss}lich der Bildsegmentierung, Hintergrundsubtraktion, Modellierung der Punktspreizfunktion und Schätzung von Anfangswerten, wird mit dem Bildanalysen-Code \texttt{ProFound} ausgeführt. Nach der Modellierung wird eine Modellauswahl getroffen und Galaxien für welche keines der Modelle angemessen ist - z.B. asymmetrische und verschmelzende Systeme - markiert. Die Modellierungsqualität wird durch visuelle Inspektion, Vergleiche mit bestehenden Veröffentlichungen sowie zwischen unabhängigen Ergebnissen für mehrfach-Aufnahmen derselben Galaxie und mit speziell auf diese Arbeit zugeschnittenen Simulationen überprüft. Diese werden auch für eine Untersuchung von systematischen Fehlerquellen verwendet. Die Modellierungsresultate erweisen sich als verlässlich für eine große Spannbreite an verschiedenen Galaxietypen und Bildqualitäten. Es liegen nur minimale systematische Messabweichungen vor. Die systematischen Fehler sind einen Faktor von 2-3 gr\"o{\ss}er als zufällige Fehler. Die automatisierte Modellauswahl stimmt mit einer Genauigkeit von >\,90\,\% mit der visuellen Inspektion einer Teilprobe von Galaxien überein. Eine Untersuchung der $g-r$-Farben von Galaxiekomponenten und des zugehörigen Farbe-Helligkeits-Diagramms ergibt, dass diese im Einklang mit bestehenden Veröffentlichungen sind, trotz der größeren Flexibilität der Modelle dieser Arbeit. Alle Resultate sind auf der GAMA-Datenbank integriert. 

\newpage

\chapter*{Abstract}
Quantitative measurements of the structural components of the galaxy population are crucial to constrain theory and simulations, which - in turn - constrain the formation and evolution of galaxies and the universe as a whole. This requires analysing large and diverse samples of galaxies at multiple wavelengths. Corresponding high-quality data is delivered by current and future large galaxy imaging surveys. To fully exploit these, automated analysis methods need to be developed further.  
In this thesis, we derive a catalogue of robust structural parameters for the components of a large sample of nearby GAMA galaxies while at the same time contributing to the advancement of image analysis, surface brightness fitting and post-processing routines for quality assurance in the context of automated large-scale bulge-disk decomposition studies. 
The sample consists of 13096 galaxies at redshifts $z$\,<\,0.08 in the GAMA II equatorial survey regions with imaging data from the Kilo-Degree Survey (KiDS) and the VISTA Kilo-Degree INfrared Galaxy (VIKING) survey spanning the optical and near-infrared ($u, g, r, i, Z, Y, J, H$ and $K_s$ bands). We fit three models to the surface brightness distribution of each galaxy in each band individually: a single S\'ersic model, a S\'ersic plus exponential and a point source plus exponential. The fitting is performed with a fully automated Markov-chain Monte Carlo (MCMC) analysis using the Bayesian two-dimensional profile fitting code \texttt{ProFit}. All preparatory work, including image segmentation, background subtraction, point spread function estimation, and obtaining initial guesses, is carried out using the complementary image analysis package \texttt{ProFound}. After fitting the galaxies, we perform model selection and flag galaxies for which none of our models are appropriate, mainly mergers and irregular galaxies. The fit quality is assessed by visual inspections, comparison to previous works, comparison of independent fits of galaxies in the overlap regions between KiDS tiles and bespoke simulations. The latter two are also used for a detailed investigation of systematic error sources. We find that our fit results are robust across various galaxy types and image qualities with minimal biases. Errors given by the MCMC underestimate the true errors typically by factors 2-3. Automated model selection criteria are accurate to >\,90\,\% as calibrated by visual inspection of a subsample of galaxies. We also present $g-r$ component colours and the corresponding colour-magnitude diagram, consistent with previous works despite our increased fit flexibility. All results are integrated into the GAMA database.

\newpage 

\chapter*{Previous publications}
The following publications make use of results obtained in this thesis:
\begin{itemize}
\item Casura~S., Liske~J., Robotham~A.~S.~G., Brough~S., Driver~S.~P., Graham~A.~W., H{\"a}u{\ss}ler~B., et al., ``Galaxy And Mass Assembly (GAMA): Bulge-disk decomposition of KiDS data in the nearby universe", MNRAS (submitted)
\item Oh S., Colless M., D'Eugenio F., Croom S.~M., Cortese L., Groves B., Kewley L.~J., et al., ``The SAMI Galaxy Survey: the difference between ionized gas and stellar velocity dispersions", 2022, MNRAS, 512, 1765. doi:10.1093/mnras/stac509
\item H{\"a}u{\ss}ler B., Vika M., Bamford S.~P., Johnston E.~J., Brough S., Casura S., Holwerda B.~W., et al., ``Galapagos-2/Galfitm/GAMA -- multi-wavelength measurement of galaxy structure: separating the properties of spheroid and disk components in modern surveys", 2022, arXiv, arXiv:2204.05907
\item Oh S., Colless M., Barsanti S., Casura S., Cortese L., van de Sande J., Owers M.~S., et al., ``The SAMI Galaxy Survey: decomposed stellar kinematics of galaxy bulges and disks", 2020, MNRAS, 495, 4638. doi:10.1093/mnras/staa1330
\item Cluver M.~E., Jarrett T.~H., Taylor E.~N., Hopkins A.~M., Brough S., Casura S., Holwerda B.~W., et al., “Galaxy and Mass Assembly (GAMA): Demonstrating the Power of WISE in the Study of Galaxy Groups to z\,<\,0.1”, 2020, ApJ, 898, 20. doi:10.3847/1538-4357/ab9cb8
\item Robotham A.~S.~G., Davies L.~J.~M., Driver S.~P., Koushan S., Taranu D.~S., Casura S., Liske J., “ProFound: Source Extraction and Application to Modern Survey Data”, 2018, MNRAS, 476, 3137. doi:10.1093/mnras/sty440
\end{itemize}

\cleardoublepage

\pagestyle{fancy}


\fancyhf{}
\fancyhead[RE]{\color[gray]{0}\leftmark}
\fancyhead[LO]{\color[gray]{0}\rightmark}
\fancyfoot[RO,LE]{\thepage}

\doublespacing
\tableofcontents
\onehalfspacing

\newpage
\thispagestyle{plain}

\chapter{Introduction}
\label{chap:intro}
\pagenumbering{arabic} 

Most of the information we obtain about the universe is in the form of electromagnetic radiation. The majority of electromagnetic radiation is emitted by galaxies, which are one of the most important constituents of the universe. Studying the galaxy population, their structure and evolution is therefore one of the main pathways to understand our universe \citep[e.g.][]{ZeilikandGregory, Hammerbook, Kembhavi2020, Haeussler2022}. 

Galaxies are complex objects consisting mainly of stars, gas, dust, dark matter and a supermassive black hole, sometimes including an active galactic nucleus (AGN). They differ greatly in their physical properties including their size, mass and its distribution in different components, their age, composition (relative amounts of gas and dust, types of stars contained, abundances and distribution of different elements, dark matter component), star formation activity, assembly and evolutionary history. Most of those properties can only be inferred indirectly and suffer from many uncertainties \citep[e.g.][see also Section~\ref{sec:galaxystructure} for an overview over different galaxy structures]{ZeilikandGregory, Vika2013, Hammerbook}.

The picture becomes even more complex when considering the interactions between the different constituents of galaxies as well as their evolution over time. For example, stars can form from gas via collapse and then feed back into the environment from which they formed via radiation and mass-loss processes \citep[e.g.][and references therein]{Kelvin2012, Driver2016}. A significant fraction of the radiation emitted by stars is absorbed by dust and re-emitted at longer wavelengths \citep{Driver2007, Popescu2011}. The visual appearance of a galaxy at any given point in time is therefore not only determined by the relative amounts of stars, gas and dust it contains, but also their relative distributions and the galaxy geometry and orientation with respect to the line of sight. Further, most galaxies are not isolated systems but exhibit a range of interactions with their environments. Exactly how all of these aspects are related and lead to the formation and evolution of a variety of galaxies is still not fully understood, with much debate on the relative importance of different processes such as mergers, feedback or secular processes \citep{Driver2009, Lange2016, Hammerbook, Nedkova2021}. 

To constrain the formation and evolution of galaxies, observations are key. The two main branches of observations of galaxies are photometry and spectroscopy. While the latter has the advantage of providing the intensity as a function of wavelength, the former offers larger samples at higher spatial resolution and can also include at least some colour information via multi-wavelength observations in different broad-band filters \citep{ZeilikandGregory, Vika2014, Haeussler2022}. For both methods, the optical part of the electromagnetic spectrum is of special importance. In part, this is because optical telescopes (and the analysis methods of corresponding data) have reached greater maturity than their counterparts at other wavelengths due to their much longer history. However, there is also physical motivation for studying galaxies at (near-)visible wavelengths since this is the range in which most stars emit most of their radiation \citep{ZeilikandGregory, Driver2012, Vika2013}.

Along with technological advances, observations of galaxies have become more sophisticated over time \citep{Driver2009, Haeussler2022}. To fully exploit the increased data quality and quantity of modern galaxy surveys and maximise their science return, corresponding advances in analysis methods are needed \citep{Hammerbook, Robotham2017, Robotham2022, Haeussler2022}. Quantitative measurements of the sizes and structures of the galaxy population are particularly important to constrain theory and simulations, which - in turn - constrain the formation and evolution of galaxies and the universe as a whole \citep[e.g.][and see also Section~\ref{sec:SBmodelling} for a more comprehensive overview of galaxy surveys and corresponding analysis methods]{Driver2009, Simard2011, Kelvin2012, Nedkova2021}.

This thesis presents the analysis of the photometric images of 13096 galaxies from the Galaxy and Mass Assembly \citep[GAMA;][]{Driver2009} survey in nine broad-band optical and near-infrared (NIR) filters. We fit several models to the surface brightness distribution of each galaxy in each band, in order to obtain information about their internal stellar structure. The resulting catalogue can be used to analyse the properties of galaxies as a function of wavelength, morphology or other parameters and serves as a basis for the comparison to theory and simulations. At the same time, we have contributed to the advancement of software, methodologies and analysis techniques in the field of large automated bulge-disk decomposition studies of galaxies.

We provide more details on the aims and achievements of this thesis at the end of Section~\ref{sec:backgroundandaims}, after introducing the more general context. The remainder of this chapter then presents the data (Section~\ref{sec:dataandsample}) as well as the analysis methods and code packages used (Section~\ref{sec:methodsandcode}), including a discussion of the distinguishing features of this study compared to previous work. Chapter~\ref{chap:pipeline} details the pipeline we developed for the bulge-disk decomposition of our sample of galaxies, including preparatory work and post-processing; and the evolution of all steps over time. The main results of this pipeline are then shown in Chapter~\ref{chap:results}. Chapter~\ref{chap:QC} focuses on the quality control of the fits by comparison to previous work and a detailed investigation into systematic uncertainties and biases. We conclude with a summary and outlook in Chapter~\ref{chap:conclusion}. We assume a standard cosmology of $H_0$\,=\,70\,km\,s$^{-1}$\,Mpc$^{-1}$, $\Omega_m$\,=\,0.3 and $\Omega_{\lambda}$\,=\,0.7 throughout. 

Parts of this work have already been submitted for publication in \citet{Casura2022}. In particular, Section~\ref{sec:SBmodelling}, parts of Section~\ref{sec:scienceaims}, most of Sections~\ref{sec:gama}, \ref{sec:kids}, \ref{sec:sampleselection}, \ref{sec:profit} and \ref{sec:profound}, Section~\ref{sec:pipelineoverview}, parts of Section~\ref{sec:statsoverview}, Sections~\ref{sec:v04parameterdistributions}, \ref{sec:colours} and~\ref{sec:cataloguelimitations}, Chapter~\ref{chap:QC} and the majority of Chapter~\ref{chap:conclusion} are either heavily based on or directly taken from \citet{Casura2022} with only minor modifications. This includes Figures~\ref{fig:exampleseg}, \ref{fig:examplefit}, \ref{fig:tightseg}, \ref{fig:examplepsfoutput} and~\ref{fig:examplefit-2}, the top panel of Figure~\ref{fig:ncompstats}, Figures~\ref{fig:resultshists}, \ref{fig:BThist}, \ref{fig:magrecovery}, \ref{fig:colourplots}, \ref{fig:examplefithighbt}, \ref{fig:colourvskennedy}, \ref{fig:sizemass} and \ref{fig:comparelee}, the right panels of Figures~\ref{fig:compareoverlap} and~\ref{fig:comparesimulations}, Figures~\ref{fig:examplebadsim}, \ref{fig:exampleoutliersim}, \ref{fig:exampledoublesim} and~\ref{fig:differrnorm} as well as Tables~\ref{tab:fittingparameters}, \ref{tab:singlefitlimits}, \ref{tab:modelselconfusionr}, \ref{tab:modelselconfusiongijoint}, \ref{tab:results}, \ref{tab:stats} and \ref{tab:errorunderestimate}. The remaining work of this thesis has not been presented elsewhere to date. Where we have taken large amounts of content from \citet{Casura2022}, we also indicate this at the beginning of the corresponding chapter or section. 

The written work presented here is accompanied by several data products which are bundled in the \texttt{BDDecomp} data management unit (DMU) on the GAMA team database\footnote{\url{http://www.gama-survey.org/db/}}. It includes several catalogues containing the results of the preparatory work, the galaxy fitting and the post-processing as well as a detailed description of all processing steps and the columns contained in each catalogue. Various diagnostic plots for each galaxy fit and all inputs used for the fitting (i.e. the outputs of the preparatory work) are stored on the GAMA file server. To date, there have been four releases of the \texttt{BDDecomp} DMU (\texttt{v01} to \texttt{v04}), with the fifth (\texttt{v05}) to be published alongside this thesis. More information on the \texttt{BDDecomp} DMU and its different releases is given in Section~\ref{sec:bddecompdmu}. In addition, all of the results stored on the GAMA database as well as many test runs and the corresponding code can be found on the local machines at Hamburg observatory, specifically on \texttt{/hs/fs11/data/gama/profit/} and the corresponding directories on the \texttt{fs12}, \texttt{fs13} and \texttt{fs14} machines. The ``readme"-files in those directories provide orientation. The (publicly available) KiDS and (preprocessed) VIKING data are also on the \texttt{fs11} machine, in the \texttt{/hs/fs11/data/gama/data/imaging/} directory.

\section{Background and aims}
\label{sec:backgroundandaims}

In this section, we explain the context and motivation for the current work. We begin with a more detailed description of the different types of galaxies and their structure (Section~\ref{sec:galaxystructure}), followed by a brief overview on galaxy surface brightness modelling including available tools and common practices in the field (Section~\ref{sec:SBmodelling}). This provides the background for the aims of this thesis, which are described in Section~\ref{sec:scienceaims}. Most of Section~\ref{sec:SBmodelling} and parts of Section~\ref{sec:scienceaims} are taken from the introduction of \citet{Casura2022}.

\subsection{The structure of galaxies}
\label{sec:galaxystructure}

Here, we give a brief overview over the different types of present-day (low $z$) galaxies and their main properties, mostly following the discussion in \citet{ZeilikandGregory}. We note that the picture drawn here is somewhat simplified and galaxies are more complex in detail, with many exceptions to the rule. 

\begin{figure}
  \begin{center}
	\includegraphics[width=0.8\textwidth]{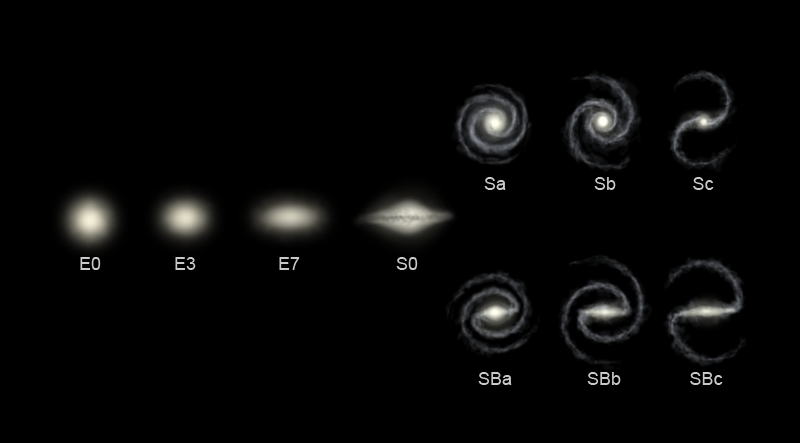}
    \caption{The Hubble sequence. Elliptical galaxies are labelled $E$ and shown in the left part of the diagram, with increasing numbers referring to increasing ellipticities. Spiral galaxies are shown on the right, split into ``ordinary" $S$ (top) and barred $SB$ (bottom) spirals. Letters $a$ to $c$ represent the transition from ``early-types" to ``late-types", characterised by an increased dominance of the disk relative to the bulge combined with less tightly wound spiral arms. Image credit: Ville Koistinen, Wikimedia Commons [\url{https://commons.wikimedia.org/wiki/File:Hubble_sequence_photo.png}].} 
    \label{fig:hubblesequence}
  \end{center}
\end{figure}

Traditionally, three main types of galaxies are distinguished according to their visual appearance: elliptical galaxies, spiral galaxies and irregular galaxies. These are represented in the Hubble sequence introduced by \citet{Hubble1926} with a schematic shown in Figure~\ref{fig:hubblesequence}. 

Elliptical galaxies have a smooth appearance with near to no substructure, gas or dust. Their stellar population is relatively old with little ongoing star formation such that they appear reddish in colour. They are pressure-supported with random stellar motions and typically follow a \citet{deVaucouleurs1948} radial profile, i.e. the intensity $I$ drops approximately as $10^{r^{-1/4}}$ with $r$ being the radius from the centre of the galaxy. 

Spiral galaxies show more structure than elliptical galaxies. They classically have two main stellar components: a bulge and a disk. The bulge sits at the centre and is similar to an elliptical galaxy in its properties, due to which bulges and ellipticals are often classed together and named spheroids. The disk is flat and rotationally supported, contains gas and dust and has a younger stellar population with ongoing star formation and bluer colours. It typically follows an exponential decrease in intensity from the centre, although with local variations due to substructure like spiral arms. 

As can be seen in Figure~\ref{fig:hubblesequence}, spiral galaxies are subdivided into two branches according to whether or not they have a central bar. The bar is disk-like in its stellar composition, but, as the name suggests, has an elongated shape. There are also spiral galaxies that host a pseudo-bulge instead of the classical bulge described above. Pseudo-bulges, like classical bulges, appear approximately spherical and are located at the centre of the galaxy. However, pseudo-bulges are younger and bluer in colour than classical bulges and show a less pronounced increase in intensity towards the centre \citep[e.g.][]{Kennedy2016, Lange2016, Haeussler2022}. 

The distinct differences between spheroids and disks can most easily be explained by different formation channels \citep[e.g.][]{Driver2013, Lange2016}. In the two-phase model, spheroids form in a so-called ``hot" mode, whereby they rapidly assemble their mass through collapse and mergers at early times. This leads to (now) old stellar populations with a high concentration at the nucleus and a generally spheroidal, pressure-supported shape. Disks are formed later and more slowly in the ``cold" mode via the accretion of gas regulated by feedback processes. Therefore, disk stellar populations are younger than spheroids, flattened due to the conservation of angular momentum and rotationally supported. Bars and pseudo-bulges can form from disk instabilities. 

In between spiral and elliptical galaxies lie S0 or lenticular galaxies. They have both a dominant bulge and disk, but no spiral arms or other prominent substructure. Their colour and composition is intermediate in between that of spiral galaxies and elliptical galaxies; their formation channel is still being debated \citep[e.g.][]{Barsanti2021}. 

The last of the three main categories of galaxies are irregulars, which show no clear symmetry. They tend to be bluer in colour than spiral galaxies and can also contain substructure, gas and dust. They are thought to be the result of interactions and mergers between galaxies, which can disturb the regular morphology and trigger star formation. 

In addition to these categories of giant galaxies, there are dwarf galaxies. Dwarfs can be elliptical or irregular in shape. They are the most numerous type, but contribute only a small fraction to the total stellar mass and light in the universe. They are also difficult to detect due to their faintness.

The relative fractions of galaxies that fall into each category depend strongly on the observational limits. For magnitude-limited surveys, spirals are usually dominant in number, followed by ellipticals since these two classes of galaxies are brightest. In the local volume, however, most galaxies are small irregular galaxies with spirals only contributing around 33\,\% and ellipticals 13\,\% \citep{ZeilikandGregory}. 

For the GAMA survey, \citet{Driver2022} have derived morphological classifications for all galaxies at a redshift of $z$\,<\,0.08. They find that approximately 10\,\% of their sample are elliptical galaxies, 45\,\% bulge-disk systems (i.e. spirals and S0) and another 45\,\% late-type systems with only a single component, including asymmetric systems (irregulars and disk-only systems with no discernible bulge). Since this sample of galaxies closely matches our sample selection (see Section~\ref{sec:sampleselection}), we might expect a very similar distribution of galaxy types for our catalogue.

\subsection{Galaxy surface brightness modelling}
\label{sec:SBmodelling}

To obtain physical quantities from the observations of galaxies provided by surveys, they need to be modelled. Due to its importance in understanding the history of the universe, the quantitative modelling of galaxy surface brightness distributions has a long history dating back to \citet{deVaucouleurs1948}, \citet{Sersic1963} and even earlier works; see \citet{Graham2013b} for a review of the development of light profile models. While the early works focused on azimuthally averaged galaxy profiling with a single functional form \citep[e.g.][]{Kormendy1977}, modern codes allow users to decompose galaxies into several distinct components (e.g. bulges, disks) and to take into account the full two-dimensional information. To this end, there are many different techniques, methods and code packages, all of which have become increasingly sophisticated as the quality and quantity of available astronomical data have grown. 

Broadly, they can be divided into parametric and non-parametric modelling, as well as one-dimensional and two-dimensional methods. Which of these is most appropriate to use depends on the science case and the available data. This work falls into the regime of large-scale automated analyses of galaxies with often barely resolved components, for which we want to obtain structural parameters that are easily comparable between galaxies. Hence, two-dimensional parametric analysis is most appropriate (see also the discussion in \citealt{Robotham2017} and references therein). 

Examples of such two-dimensional, parametric fitting tools used for large-scale automated analyses include \texttt{GIM2D} \citep{Simard2002}, \texttt{BUDDA} \citep{deSouza2004}, \texttt{GALFIT3} \citep{Peng2010}, \texttt{GALFITM} \citep{Vika2013}, \texttt{IMFIT} \citep{Erwin2015}, \texttt{ProFit} \citep{Robotham2017} and \texttt{PHI} \citep{Argyle2018}. 
Each of these tools comes with its own advantages and disadvantages, which goes to show how difficult the problem of galaxy modelling is, especially when automated for large samples of the very diverse galaxy population. Usually, some form of post-processing is needed to assess the influence of systematic uncertainties, judge the convergence, exclude bad fits and identify the most appropriate model to use for each galaxy. This can be achieved via visual inspection (for small enough samples), logical filters, frequentist statistics such as the $F$-test, Bayesian inference, or similar methods \citep[see, e.g.,][]{Allen2006, Gadotti2009, Simard2011, Vika2014, Meert2015, Lange2016, Mendez-Abreu2017}. 

Despite the associated difficulties (e.g. convergence and quality of fit metrics), many authors have performed two-dimensional surface brightness profile fitting for large numbers of galaxies, modelling the radial light profile as a simple functional form, most often a S\'ersic function \citep[][to name just a few]{Blanton2003, Blanton2005, Barden2005, Trujillo2006, Hyde2009, LaBarbera2010, Kelvin2012, vanderWel2012, Haeussler2013, Shibuya2015, Sanchez-Janssen2016}. 
The results of such analyses have been used to derive a number of key relations between different galaxy properties, their formation and evolutionary history, and interactions with the environment. 
For example, many works have studied the distribution of, and relation between, size and mass or luminosity for different galaxy types (split by e.g. S\'ersic index or colour), sometimes including morphology, surface brightness, internal velocity, environment, wavelength, colour, or redshift effects \citep[e.g.][]{Shen2003, Barden2005, Blanton2005, Trujillo2006, Hyde2009, LaBarbera2010, Kelvin2014, vanderWel2014, Lange2015, Shibuya2015, Nedkova2021}.

With improving data quality of surveys, the fitting of more than one component - i.e. decomposing galaxies - has become more common. While some authors, such as \citet{Gadotti2009}, \citet{Salo2015} or \citet{Gao2017} also account for bars, central point sources, spiral arms or other additional morphological features, most works focus on the bulge and disk. The focus on only two components is especially true when running automated analyses of large samples, since in many cases the data quality is not sufficient to meaningfully constrain more than one or two components, or it would require extensive manual tuning based on visual inspection. From a more physical point of view, the majority of the stellar mass in the local universe resides in ellipticals, disks and classical bulges, with pseudo-bulges and bars only contributing a few percent (\citealt{Gadotti2009}; and see also Section~\ref{sec:galaxystructure}). Hence, for automated analyses it is common practice to fit only two components, where the term ``bulge" is used to describe the central component, irrespective of whether it is a classical bulge, pseudo-bulge, bar, lens, AGN, or a mixture thereof, while ``disk" refers to a more extended component with typically lower surface brightness and potential additional structure such as spiral arms, breaks, flares or rings. 

Examples of large bulge-disk decomposition studies include \citet{Simard2002, Simard2011, Allen2006, Benson2007, Gadotti2009, Lackner2012, Fernandez-Lorenzo2014, Head2014, Mendel2014, Vika2014, Meert2015, Meert2016, Kennedy2016, Kim2016, Lange2016, Dimauro2018, Bottrell2019, Cook2019, Barsanti2021, DominguezSanchez2022, Haeussler2022, Hashemizadeh2022}; and \citet{Robotham2022}. 
Such catalogues can then be used to determine the relative numbers of different galaxy components as well as their luminosity or stellar mass functions, size-mass or size-luminosity relations, including their redshift evolution and dependence on other properties of the galaxy and its environment (similar to the studies of entire galaxies mentioned earlier). For example, this has been done by \citet{Driver2007b, Dutton2011, Tasca2014, Kennedy2016, Lange2016, Moffett2016} and \citet{Dimauro2019}.

In addition, quantitative measures for the components of galaxies aid the comparison of observational data to theory and simulations. Bulges and disks are often decisively different not only in their visual appearance but also in their structure, dynamics, stellar populations, gas and dust content and are thought to have different formation pathways (Section~\ref{sec:galaxystructure} and \citealt{Cole2000, Cook2009, Driver2013, Lange2016, Dimauro2018, Lagos2018, Oh2020}). Consequently, bulge-disk decomposition studies provide stringent constraints on the formation and evolutionary histories of galaxies and their physical properties that are not easily measured directly such as the dark matter halo, the build-up of stellar mass (in different components) over time, or merger histories 
\citep[examples include][]{Driver2013, Bottrell2017b, Bluck2019, Rodriguez-Gomez2019, deGraaff2022}. 
Hence, consistently measuring the structure of the stellar components is essential to make full use of current and future large-scale observational surveys such as the Kilo-Degree Survey \citep[KiDS;][]{deJong2013} and the VISTA Kilo-Degree INfrared Galaxy \citep[VIKING;][]{Edge2013} Survey or the Legacy Survey of Space and Time \citep[LSST;][]{Ivezic2019}, 
and of cosmological hydrodynamical simulations such as Illustris \citep{Vogelsberger2014} and IllustrisTNG \citep[The Next Generation;][]{Pillepich2018} or Evolution and Assembly of GaLaxies and their Environments \citep[EAGLE;][]{Schaye2015}.

\subsection{Science aims}
\label{sec:scienceaims}

This work is the first in a series of planned contributions to the field of galaxy structure and evolution, with the final aim of the series being to constrain the nature and distribution of dust in galaxy disks. \citet{Driver2007} have shown that in the $B$-band of the Millennium Galaxy Catalogue (MGC), an average of 37\,\% of photons produced in the disk and 71\,\% of photons produced in the bulge are absorbed before even leaving their galaxy of origin. Accounting for this effect of internal dust attenuation is therefore vital to avoid biases in structure formation and galaxy evolution studies, especially since the dust properties evolve as well. Due to the geometry of the system (see also Section~\ref{sec:galaxystructure}), bulges and disks are affected differently and the severity of the attenuation depends strongly on the inclination angle of the galaxy relative to our line of sight (see figure 11 in \citealt{Driver2007}). Bulges and disks therefore need to be studied separately. 

A sufficiently large sample of robust structural parameters for these two components can be used to constrain the nature and distribution of dust. This can be achieved by comparing the distribution of bulges and disks in the luminosity-size plane as a function of inclination to dust radiative transfer models such as those presented in \citet{Popescu2011} and preceding papers of this series. The analysis is particularly powerful if observations in multiple filters are available since dust attenuation also varies strongly as a function of wavelength \citep{Popescu2011}. 

This thesis presents the first step towards achieving this final goal: we obtain single S\'ersic fits and bulge-disk decompositions for 13096 GAMA galaxies in the KiDS $u$, $g$, $r$ and $i$ bands and the VIKING $Z$, $Y$, $J$, $H$ and $K_s$ bands. We choose \texttt{ProFit} (see Section~\ref{sec:profit}) as our modelling software due to its Bayesian nature (allowing full MCMC treatment including more realistic error estimates), its suitability to large-scale automated analyses and its ability, in combination with \texttt{ProFound} (Section~\ref{sec:profound}), to serve as a fully self-contained package covering all steps of the analysis from image segmentation through to model fitting. We supplement this functionality with our own routines for the rejection of unsuitable fits, model selection, and a characterisation of systematic uncertainties. The aim of our future work is to use those structural parameters for the bulges and disks to significantly expand the analysis of \citet{Driver2007}, exploiting our more and better data at several wavelengths. 

With these science aims in mind, we are most interested in obtaining structural parameters that are directly comparable amongst each other, i.e. consistent within the dataset; and correctly represent the statistical properties of the entire sample, with less emphasis placed on capturing all aspects of the detailed structure of individual galaxies. Consequently, we choose to model a maximum of two components for each galaxy and use the terms ``bulge" and ``disk" in their widest senses, in line with previous automated decompositions of large samples. In particular the ``bulges" we obtain are often mixtures of classical or pseudo-bulges, bars, lenses and AGN. Similarly, we place more emphasis on the central, high surface brightness regions of galaxies by modelling only a relatively tight region around each galaxy of interest. While most of the fits we obtain are not perfect (because galaxies are more complex than two simple components), they do achieve the aims specified above and are comparable to similar studies.

While much of our procedure was focused towards obtaining suitable fits for our final goal of studying dust properties, there are many other analyses that can be built on our results. Some of the most obvious of these include deriving the stellar mass functions of bulges and disks, studying component colours and investigating the trends of structural parameters such as bulge or disk size with wavelength, all of which belong to our plans for future work. Additionally, (an earlier version of) the resulting catalogue has been used to aid the kinematic bulge-disk decomposition of a sample of galaxies in the Sydney-AAO Multi-object Integral-field spectroscopy (SAMI) Galaxy Survey \citep{Oh2020}, to examine the properties of galaxy groups \citep{Cluver2020}, to investigate the difference between ionised gas and stellar velocity dispersions \citep{Oh2022}, to cross-check the results of other multi-wavelength bulge-disk decomposition studies \citep{Haeussler2022, Robotham2022}, to study the alignment of galaxy spin axes with filaments of the cosmic web as a function of different galaxy (component) properties (Barsanti et al., in prep.) and to validate a cosmological galaxy simulation against observations using an unsupervised machine learning technique (Turner et al., in prep.). Several student projects, including Bachelor's and Master's theses, have also made use of our results. For example, \citet{Roschlaub2022} tested the usage of a new deconvolution algorithm presented in \citet{Nammour2021} in the context of galaxy fitting, Targaczewski (in prep.) is working on estimating the supermassive black hole mass from the bulge S\'ersic index for a sample of AGN in our catalogue, Ehbrecht (in prep.) is investigating under which circumstances (seeing and noise) a given S\'ersic profile can be reasonably constrained, Porter-Temple (in prep.) is going to look at the difference the number of spiral arms make on the bulge-to-disk flux ratio and Porter (in prep.) will compare the bulge-to-disk flux ratios of void galaxies to those of the remaining GAMA sample. 

Apart from these scientific insights enabled by the final product of our pipeline (the catalogue), advancements on the technical side should not be neglected. As briefly mentioned before, improvements in methodology are crucial to make full use of current and future datasets of large-scale galaxy surveys; and we have contributed to this aspect in several ways. First of all, \texttt{ProFit}, and even more so \texttt{ProFound}, were under active development during the time of our own pipeline development, with this thesis project being one of the first large-scale automated applications of both packages. Consequently, our work has contributed to improving both packages by discovering bugs, suggesting additional features and serving as inspiration for the implementation of various automated procedures. In addition, we have added routines for the swapping of bulge and disk components (see Section~\ref{sec:galaxyfitting}) and post-processing of fits (mainly model selection and flagging of bad fits, Section~\ref{sec:postprocessing}); and have performed numerous tests with varying combinations of \texttt{ProFound} and \texttt{ProFit} routines and their tuning parameters for the preparatory work (image segmentation, background subtraction, point spread function estimation, obtaining initial guesses) as well as the galaxy fitting. The resulting procedures, as well as the numerous alternatives that we found to be less optimal are described in detail in Chapter~\ref{chap:pipeline} and can serve as guides for similar bulge-disk decomposition works on current and future large-scale galaxy surveys. Our detailed analysis of systematic uncertainties based on the overlap sample and bespoke simulations (Section~\ref{sec:systematics}) can also give orientation towards constraining and quantifying the dominant sources of error in other analyses. 

The goal of this thesis therefore is not only to present our final catalogue of robust structural parameters for the components of galaxies, but also the large amounts of technical work surrounding the pipeline development.

\section{Data and sample}
\label{sec:dataandsample} 

After the general introduction of the context and aims of this thesis, we now present the data products that we use from GAMA (Section~\ref{sec:gama}), KiDS (Section~\ref{sec:kids}) and VIKING (Section~\ref{sec:viking}) followed by the selection of our sample of galaxies (Section~\ref{sec:sampleselection}). Sections~\ref{sec:gama}, \ref{sec:kids} and~\ref{sec:sampleselection} are largely based on \citet[their section 2]{Casura2022}. 

\subsection{GAMA}
\label{sec:gama}

The Galaxy and Mass Assembly (GAMA)\footnote{\label{foot:gama}\url{http://www.gama-survey.org}} survey is a large low-redshift spectroscopic survey covering $\sim$\,238\,000 galaxies in 286\,deg$^2$ of sky (split into 5 survey regions) out to a redshift of approximately 0.6 and a depth of $r$\,<\,19.8\,mag. The observations were taken using the AAOmega spectrograph on the Anglo-Australian Telescope and were completed in 2014. The survey strategy and spectroscopic data reduction are described in detail in \citet{Driver2009, Baldry2010, Robotham2010, Driver2011, Hopkins2013, Baldry2014} and \citet{Liske2015}. 

In addition to the spectroscopic data, the GAMA team collected imaging data on the same galaxies from a number of independent surveys in more than 20 bands with wavelengths between 150\,nm and 500\,$\mu$m. Details of the imaging surveys and the photometric data reduction are given in \citet{Liske2015, Driver2016, Driver2022}; and relevant publications of the corresponding independent surveys. The combined spectroscopic and multiwavelength photometric data at this depth, resolution and completeness provide a unique opportunity to study a variety of properties of the low-redshift galaxy population.

In this work, we focus on the KiDS and VIKING imaging data in the nine optical and near-infrared filters $u, g, r, i, Z, Y, J, H, K_s$ (see Sections~\ref{sec:kids} and~\ref{sec:viking}) in the GAMA II equatorial survey regions, which are 3 regions of size 12\degr\,$\times$\,5\degr\ located along the equator at 9, 12 and 14.5 hours in right ascencion (the G09, G12 and G15 regions). For our sample selection, we make use of the equatorial input catalogue\footnote{For the sake of reproducibility, we always give the exact designation of a catalogue on the GAMA database in parentheses: the data management unit (DMU) that produced the catalogue (e.g. \texttt{EqInputCat}) followed by the catalogue name (e.g. \texttt{TilingCat}) and the version used (e.g. \texttt{v46}).} \citep[\texttt{EqInputCat:TilingCatv46},][]{Baldry2010} and the most recent version of the redshifts originally described by \citet{Baldry2012} (\texttt{LocalFlowCorrection:DistancesFramesv14}), see details in Section~\ref{sec:sampleselection}. For the stellar mass-size relation (Section~\ref{sec:sizemass}), we also use the Data Release (DR) 3 version of the stellar mass catalogue first presented in \citet{Taylor2011} (\texttt{StellarMasses:StellarMassesv19}); for the comparison to previous work (Section~\ref{sec:comparelee}) we use the single S\'ersic fits of \citet{Kelvin2012} (\texttt{SersicPhotometry:SersicCatSDSSv09}); and in order to correct galaxy colours for Galactic extinction, we use the corresponding table provided along with the equatorial input catalogue (\texttt{EqInputCat:GalacticExtinctionv03}). All of these catalogues can be obtained from the GAMA database.$^{\ref{foot:gama}}$ 

\subsection{KiDS}
\label{sec:kids}

The Kilo-Degree Survey \citep[KiDS,][]{deJong2013} is a wide-field imaging survey in the Southern sky using the VLT Survey Telescope (VST) at the ESO Paranal Observatory. 1350\,deg$^2$ are mapped in the optical broad-band filters $u, g, r, i$; while the VIKING Survey (\citealt{Edge2013}, Section~\ref{sec:viking}) provides the corresponding near-infrared data in the $Z, Y, J, H, K_s$ bands. The GAMA II equatorial survey regions have been covered as of DR3.0.  

KiDS provides $\sim$\,1\degr\,$\times$\,1\degr\ science tiles calibrated to absolute values of flux with associated weight maps (inverse variance) and binary masks. The science tiles are composed of 5 dithers (4 in $u$) totalling 1000, 900, 1800 and 1200\,s exposure time in $u,g,r,i$, with all dithers aligned in the right ascenscion and declination axes (i.e. no rotational dithers) and taken in immediate succession. The $r$-band observations were performed during the best seeing conditions in dark time; while $g$, $u$ and $i$ have progressively worse seeing and $i$ was additionally taken during grey time or bright moon. During co-addition, the dithers across all four bands were re-gridded onto a common pixel scale of $0\farcs$2. The magnitude zeropoint of the science tiles is close to zero with small corrections given in the image headers for DR4.0. The $r$-band point spread function (PSF) size is typically $0\farcs$7 and the limiting magnitudes in $u,g,r,i$ are $\sim$\,24.2, 25.1, 25.0, 23.7\,mag respectively (5$\sigma$ in a 2\arcsec aperture). This high image quality, depth, survey size and wide wavelength coverage in combination with VIKING make KiDS data unique. For details, see \citet{Kuijken2019}. 

For this work, we use the the $u$, $g$, $r$ and $i$ band science tiles, weight maps and masks from KiDS DR4.0 \citep{Kuijken2019}, which are publicly available\footnote{\url{http://kids.strw.leidenuniv.nl/DR4/index.php}} for our selected sample of galaxies (Section~\ref{sec:sampleselection}). Our primary band, used for most analyses during pipeline development, is the $r$-band since it is the deepest and was taken during the best seeing conditions. Observations in $g$ and $i$ are of comparable quality, such that these three bands together form our core bands where we obtain the best and most comparable fits (these are also what \citealt{Casura2022} is based on). The $u$-band is of considerable worse data quality especially in terms of depth and therefore we place less emphasis on its analysis. More details are given in Section~\ref{sec:pipelinedevelopment}, also explaining our choice of which of the KiDS data products to use (Section~\ref{sec:otherprepworkchoices}).

\subsection{VIKING}
\label{sec:viking}

The VISTA Kilo-degree INfrared Galaxy (VIKING) survey is a wide-field, intermediate-depth near-infrared imaging survey using the Visible and Infrared Survey Telescope for Astronomy (VISTA) at the ESO Paranal Observatory. 1350\,deg$^2$ were mapped in the broad-band filters $Z, Y, J, H, K_s$ over two areas of sky matched to the KiDS footprint. The observations are complete and the fourth and final public data release is described in \citet{Edge2020}. 

VIKING provides astrometrically and photometrically calibrated tiles of size $\sim$\,1.5\degr\ in right ascension (RA) and $\sim$\,1\degr\ in declination (Dec) composed of 6 pawprints each, arranged in a fashion to cover the gaps between detector chips. Each tile is observed twice, with observations sometimes years apart: first in the $Z, Y, J$ filters with total exposure times of 480, 400 and 200\,s; and then again in the $J, H, K_s$ filters with exposure times of 200, 300 and 480\,s. Combining the two $J$-band observations results in median magnitude limits of 21.4, 20.6, 20.1, 19.0 and 18.6\,mag in the $Z, Y, J, H$ and $K_s$ bands respectively, although some fields can be shallower by up to 0.3\,mag which is the quality threshold applied by the VIKING team. In addition to the stacked tiles, the pawprints are publicly available. Both pawprints and stacks are approximately aligned in RA and Dec, have a pixel size close to $0\farcs$34 and various Vega magnitude zeropoints around 30, with the exact values given in the image headers. They also both have associated confidence maps which give the per-pixel exposure time and exclude the ``bad patches" of two detectors that have flat-fielding issues. More details are given in \citet{Edge2020}.

For this work, we use the $Z, Y, J, H$ and $K_s$ individual detector images from the pawprints. In particular, we use the preprocessed versions from \citet{Wright2019} that were kindly provided to us by Angus Wright and the KiDS team. These have the advantage that they are specifically processed in order to allow a consistent analysis in combination with KiDS data: they have been rotated slightly to be exactly aligned in RA and Dec, corrected for atmospheric extinction, re-calibrated to remove the exposure time from the image units and re-scaled onto a common AB magnitude zeropoint of 30. In addition, \citet{Wright2019} also perform background subtraction, produce weight maps and perform a quality control of each chip. More details and the reasons for our choice to use these preprocessed individual chips instead of the VIKING stacked tiles are given in Section~\ref{sec:vikingdataproducts}. 

\subsection{Sample selection}
\label{sec:sampleselection}

Our main sample consists of all GAMA II equatorial region main survey targets with a reliable redshift in the range 0.005\,<\,$z$\,<\,0.08, which are a total of 12958 objects.\footnote{In detail, we select all targets with NQ\,$\geq$\,3, SURVEY\textunderscore CLASS\,$\geq$\,4 and 0.005\,<\,Z\textunderscore CMB\,<\,0.08 from \texttt{EqInputCat:TilingCatv46} joined to \texttt{LocalFlowCorrection:DistancesFramesv14} on CATAID.} In addition, we include all 2404 targets of the ``GAMA sample" of the SAMI Galaxy Survey\footnote{Taking the CATIDs listed in sami\textunderscore sel\textunderscore 20140413\textunderscore v1.9\textunderscore publiclist from \url{https://sami-survey.org/data/target_catalogue}} \citep{Bryant2015}, the majority of which are already in our main sample. The combination of both results in the full sample of 13096 unique physical objects, which were imaged a total of 14966 times in each of the KiDS $g$, $r$ and $i$ bands due to small overlap regions between the tiles. 11301, 1742, 31 and 22 objects were imaged once, twice, three and four times respectively. For versions of our bulge-disk decomposition pipeline up to \texttt{BDDecomp} \texttt{v04} \citep{Casura2022}, we keep these multiple data matches to the same physical object separate during all processing steps to serve as an internal consistency check (Chapter~\ref{chap:QC}). For \texttt{v05}, where we add the KiDS $u$ and the five VIKING bands to the analysis, we then changed this to instead fit all data matches in the same band jointly. This is necessary due to our decision to work at the individual detector chip level for VIKING (see Section~\ref{sec:viking}), resulting in many data matches to the same physical objects (more than 20 for some objects). See Section~\ref{sec:pipelineupdates} for details.

\section{Methods and code}
\label{sec:methodsandcode}

With the galaxy sample and input data defined, we now turn towards the methods used in the analysis of those. After a brief introduction into the theoretical background of Bayesian analysis and S\'ersic modelling in Sections~\ref{sec:bayesiananalysis} and~\ref{sec:sersicmodels}, we present the two main code packages used for the galaxy modelling and preparatory work respectively, \texttt{ProFit} and \texttt{ProFound} (Sections~\ref{sec:profit} and~\ref{sec:profound}; taken from section~2 of \citet{Casura2022}). 

\subsection{Bayesian analysis}
\label{sec:bayesiananalysis}

Bayesian probability theory, originally introduced by Reverend Tho\-mas Bayes in the 18$^\mathrm{th}$ century and further developed by Laplace in the early 19$^\mathrm{th}$ century, has recently experienced a ``re-discovery" with rapidly increasing popularity. In the alternative (for a long time more popular) frequentist approach, probability is defined as the ``long-run relative frequency" of an outcome and hence requires many repetitions of an experiment. In contrast to this, Bayesian probability is defined as a ``degree of belief", which has proven to be a powerful approach especially in fields such as astronomy, where the repeatability of experiments is often very limited \citep{Sivia2006, Gregory2005}. We briefly review its most basic concepts here following \citet{Sivia2006}. We refer the reader to this and similar works for a more detailed treatment. 

Bayesian analysis can be used both for parameter estimation and model selection. Its most basic building block is Bayes' theorem, which can be expressed as
\begin{equation}
\label{eq:bayes}
\mathrm{p}(B \vert A, C) = \frac{\mathrm{p}(A \vert B, C) \mathrm{p}(B \vert C)}{\mathrm{p}(A \vert C)}. 
\end{equation}
Here, $\mathrm{p}(B \vert A, C)$ denotes the probability p of statement B given information A and C. 

In the context of parameter estimation, this becomes
\begin{equation}
\label{eq:bayesparameter}
\mathrm{p}(\theta \vert data, M) = \frac{\mathrm{p}(data \vert \theta, M) \mathrm{p}(\theta \vert M)}{\mathrm{p}(data \vert M)}, 
\end{equation}
where $\theta$ denotes one or several parameters of model M. It gives the posterior probability of the parameter(s) [$\mathrm{p}(\theta \vert data, M)$, the quantity that is desired in parameter estimation], as the product of the probability of the data given the parameter values and the model [$\mathrm{p}(data \vert \theta, M)$, which is easily computed] with the prior of the parameter [$\mathrm{p}(\theta \vert M)$, its probability distribution before taking any data] and a normalisation constant that depends only on the data and the model, not the parameter [$\mathrm{p}(data \vert M)$, called evidence]. Note that if the model contains several parameters, the posterior probability is a joint probability for all of those parameters and one needs to marginalise (integrate) over all other parameters to obtain the correct one-dimensional posterior probability of any given parameter of choice. In practice, the most efficient way to achieve this usually is sampling the likelihood space via Markov Chain Monte Carlo (MCMC) or similar methods. Provided convergence is achieved, the MCMC chain points are representative samples from the joint posterior, allowing easy projection of the distribution onto any axis (parameter) of interest. 

For model selection, Bayes' theorem (Equation~\ref{eq:bayes}) takes the following form: 
\begin{equation}
\label{eq:bayesmodel}
\mathrm{p}(M_k \vert data, I) = \frac{\mathrm{p}(data \vert M_k, I) \mathrm{p}(M_k \vert I)}{\mathrm{p}(data \vert I)}. 
\end{equation}
This now expresses the posterior probability of Model k as the product of the model likelihood with the model prior, normalised by the probability of the data given any relevant background information $I$. The ratio of two model posteriors gives the odds ratio between two models: 
\begin{equation}
\label{eq:oddsratio}
O_{1:2} = \frac{\mathrm{p}(M_1 \vert data, I)}{\mathrm{p}(M_2 \vert data, I)} = \frac{\mathrm{p}(data \vert M_1, I)}{\mathrm{p}(data \vert M_2, I)}\frac{ \mathrm{p}(M_1 \vert I)}{\mathrm{p}(M_2 \vert I)},
\end{equation}
where the normalisation constant cancelled out since it is the same for both models. This gives the relative probability of the two models (for any parameter values) given the same set of data and background information. The last term on the right hand side is the prior odds ratio, i.e. the relative probability of the two models before taking any data. Unless there is a strong reason to prefer one model over the other (e.g. previous data contained in $I$), this ratio should be set to 1 to allow a fair comparison of models. 

The most relevant term is hence the first term on the right hand side, called the Bayes factor. It consists of the likelihoods for each model, which is the probability of the data given the model, marginalised (integrated) over all possible parameter values: 
\begin{equation}
\label{eq:marglike}
\begin{split}
\mathrm{p}(data \vert M_k, I) &= \int \mathrm{p}(data, \theta \vert M_k, I) d\theta\\
&= \int \mathrm{p}(data \vert \theta, M_k, I) \mathrm{p}(\theta \vert M_k, I) d\theta. 
\end{split}
\end{equation}
Due to this integral over all parameters, the marginalised likelihood will generally become smaller when adding parameters that do not improve the fit significantly (since the likelihood $\mathrm{p}(data \vert \theta, M_k, I)$ is then approximately constant while the integral over the prior probabi\-lities for the parameters $\mathrm{p}(\theta \vert M_k, I)$ decreases due to the larger prior range). Equation~\ref{eq:oddsratio} hence naturally implements Ockham's factor (also called Ockham's razor) which states that if two models can explain the data equally well, then the simpler one should be preferred (i.e. to avoid overfitting). 
Note that the model likelihood in Equation~\ref{eq:marglike} is also the same as the evidence in parameter estimation (the denominator in Equation~\ref{eq:bayesparameter}) and for this reason the Bayes factor can also be called the ratio of evidences or the ratio of marginalised likelihoods. Computing the model likelihood is often non-trivial especially for models with many para\-meters, since it requires high-dimensional integrals. We return to this in Section~\ref{sec:postprocessing}.

\subsection{S\'ersic models}
\label{sec:sersicmodels}

The most important component for both parameter estimation and model selection - apart from the data - is the model $M$ with its parameters $\theta$. If the model is wrong in the sense that it cannot represent the data adequately, then any estimated parameter values will have limited validity. Comparing two inadequate models consequently also produces an odds ratio with little meaning. We hence carefully explain our choice of models (and parameters) here, including its implications.

The most common radial profile to fit the surface brightness distribution of galaxies with is the \citet{Sersic1963} function. It gives the intensity $I$ as a function of radius $r$:
\begin{equation}
    I(r)=I_e \exp\left[-b_n \left(\left(\frac{r}{R_e}\right)^{1/n}-1\right)\right].
	\label{eq:sersic}
\end{equation}
Here, $n$ is the S\'ersic index (the main shape parameter), $R_e$ is the effective radius where half of the total flux is included (the size), $I_e$ is the intensity at $R_e$ (the overall normalisation) and $b_n$ is a normalisation constant which can be calculated from $n$. The S\'ersic function becomes a Gaussian for $n$\,=\,0.5, exponential for $n$\,=\,1 and a \citet{deVaucouleurs1948} profile for $n$\,=\,4 and can be adjusted to account for ellipticity and/or boxyness in two dimensions \citep{Robotham2017}. Due to this flexibility, the S\'ersic function can fit a wide variety of galaxy (component) shapes and types such as exponential disks or classical de Vaucouleurs bulges (Section~\ref{sec:galaxystructure}) and correspondingly is extremely popular in the galaxy fitting community \citep[and references therein]{Graham2005}. It is, however, a purely empirically derived function with no profound physical meaning. 

Adding several S\'ersic profiles together allows to fit multiple galaxy components. Common examples include a de Vaucouleurs bulge plus an exponential disk (i.e. S\'ersic functions with $n$ fixed to 4 and 1 respectively), a (free $n$) S\'ersic bulge plus exponential disk or a double S\'ersic profile. In all cases, the bulge may or may not be forced to be round (i.e. circular instead of elliptical in two dimensions) and there can additionally be further constraints on the parameters such that for example the centres of the bulge and disk components must align. Further morphological features, such as bars, can also be accounted for by adding more (S\'ersic or other) profiles. 

However, there are also problems associated with increasing numbers of components and fitting parameters: parameter degeneracies become more common, convergence is more difficult to achieve and when the fitting algorithm has converged it needs to be ensured that the solution is also physical, i.e. that the different model components do indeed represent dif\-fe\-rent physical components of the galaxy. As briefly mentioned in Section~\ref{sec:SBmodelling}, this problem is particularly pronounced for automated analyses of large samples of galaxies with varying properties, where manual intervention and control of the fits is limited. Depending on the image quality, physical properties and redshift of the galaxy, it is possible that only one component can reasonably be constrained by the data even though the galaxy physically consists of several. Others may need three of more components to adequately represent all morphological features. Conversely, there are also a number of galaxies (e.g. ellipticals) that physically contain just a single component, leaving any additional model components unconstrained. While in theory these should be easily recognized during model selection, it is often non-trivial in practice due to overlapping point sources, neighbouring objects, imperfect PSF estimates or sky subtraction, image artifacts and similar issues. 

For these reasons, we decided to follow most other works in the field of large automated ana\-lyses of galaxy surface brightness fitting (Section~\ref{sec:SBmodelling}) and fit a maximum of two components to each galaxy. For the disk component, we use an exponential profile that is elliptical in two dimensions. This ensures that we can correctly capture disks with differing brightnesses, sizes and inclination angles. We do not allow the S\'ersic index to vary (i.e. we fit exponential disks instead of S\'ersic disks) as most disks do follow an exponential profile on average \citep{ZeilikandGregory}, with spiral arms only contributing local perturbations. Other deviations from the exponential profile, such as disk breaks and flares are most common in the disk outskirts, where we place less emphasis by considering only a relatively tight region around each galaxy for fitting. For the bulge profile, however, we leave the S\'ersic index free in order to not only capture classical de Vaucouleurs bulges (with S\'ersic indices around 4) but also pseudo-bulges (with S\'ersic indices usually below 2) and bars (also low S\'ersic indices and additionally elongated shapes; see Section~\ref{sec:galaxystructure}). For the same reason we also do not constrain the bulge - or, more precisely, the central component - to be round. However, we do constrain the bulge and disk to lie precisely on top of each other to avoid one of the components wandering off to fit overlapping point sources. 

In addition to this two-component model, we fit two simpler ones: a single S\'ersic model to represent those galaxies that either physically have just one component or where the data quality is not sufficient to constrain more than one component; and a ``1.5-component" point source bulge plus exponential disk model. While the former is also routinely done in the literature since it is a wise decision in general to start with a simple model before adding complexity (also for Bayesian model selection), the latter is less common. The reason for us adding this model is that after fitting, we identified a population of bulges that were clearly present, but had ill-constrained parameters due to their faintness and small sizes. To obtain reliable magnitudes for these bulges, we considered it best to limit the freedom of the fitting parameters to a minimum and fit the point source model (consisting of only a magnitude and a position) instead of the S\'ersic bulge model.

In summary, we fit three models to each galaxy: a single S\'ersic, a S\'ersic plus exponential and a point source plus exponential. More details are given in Sections~\ref{sec:galaxyfitting} and~\ref{sec:modellingdecisions}. We note that according to the morphological classification performed by \citet{Driver2022} (see Section~\ref{sec:galaxystructure}), the majority of galaxies in our sample can be approximately represented by one of these three models. However, in a statistical sense, our models are not an appropriate representation of the detailed structure of most galaxies. We therefore expect highly correlated residuals caused by - for example - spiral arms, rings, nuclear lenses, bars, AGN and bulges (especially if several of these features are present and the ``bulge" S\'ersic function is forced to compromise between fitting them) as well as asymmetries and irregular features of all kinds. These model shortcomings need to be considered when assessing the fit quality and for model selection, see Sections~\ref{sec:postprocessing} and \ref{sec:swappingandoutliers}.

\subsection{ProFit}
\label{sec:profit}

Once the data are obtained and the model and its parameters are defined, the decision comes to the software to use for fitting. We opt to use \texttt{ProFit}\footnote{\url{https://github.com/ICRAR/ProFit}} (v1.3.2) which is a free and open-source, fully Bayesian two-dimensional profile fitting code specifically developed to fit the surface brightness distributions of galaxies \citep{Robotham2017}. \texttt{ProFit} offers great flexibility: there are several built-in profiles to choose from, it is easy to add several components of the same or different profiles, there is a choice of likelihood calculations and optimisation algorithms that can be used (various downhill gradient options, genetic algorithms, over 60 variants of MCMC methods), parameters can be fitted in linear or logarithmic space, it is possible to add complex priors for each, as well as constraints relating several parameters; and much more. The pixel integrations are performed using a standalone C++ library (\texttt{libprofit}), making it both faster and more accurate than other commonly used algorithms such as \texttt{GALFIT} (\citealt{Peng2010}; see detailed comparison in \citealt{Robotham2017}). This allows us to fit galaxies with the computationally more expensive MCMC algorithms, overcoming the main problems of downhill gradient based optimisers: their susceptibility to initial guesses and their inability to easily derive realistic error estimates \citep[e.g.][]{Lange2016}. This makes \texttt{ProFit} highly suitable for the decomposition of large sets of galaxies with little user intervention.

\subsection{ProFound}
\label{sec:profound} 

\texttt{ProFit} (Section~\ref{sec:profit}) requires a number of inputs apart from the (sky-subtracted) science image and the chosen model to fit, most importantly initial parameter guesses, a segmentation map specifying which pixels to fit, a sigma (error) map and a PSF image. To provide these inputs in a robust and consistent manner, the sister package \texttt{ProFound}\footnote{\url{https://github.com/asgr/ProFound/}} \citep{Robotham2018} was developed, which also serves as a stand-alone source finding and image analysis tool. The main novelties of \texttt{ProFound} compared to other commonly used free and open-source packages such as \texttt{Source Extractor} \citep{Bertin1996} are that, rather than elliptical apertures, \texttt{ProFound} uses dilated ``segments" (collections of pixels of arbitrary shape) with watershed de-blending across saddle-points in flux. This means that the flux from each pixel is attributed to exactly one source (or the background) and apertures are never overlapping or nested. It also allows for extracting more complex object shapes than ellipses while still capturing the total flux due to the segment dilation (expansion) process. This makes it less prone to catastrophic segmentation failures (such as fragmentation of bright sources or blending of several sources into one aperture), reducing the need for manual intervention or multiple runs with ``hot" and ``cold" deblending settings, making \texttt{ProFound} particularly suitable for large-scale automated analysis of deep extragalactic surveys \citep{Robotham2018, Davies2018, Bellstedt2020}.

Apart from the segmentation map, the main function of the package, \texttt{profoundProFound}, also returns estimated sky and sky-RMS maps (if not given as inputs) and a wealth of ancillary data including a list of segments and their properties such as their size, ellipticity and the flux contained. The latter is particularly useful to obtain reasonable initial parameter guesses for galaxy fitting; or for identifying certain types of sources (e.g. stars for PSF estimation). The package also contains many additional functions for further image analysis and processing, all within the same framework. In addition, combining \texttt{ProFound} with \texttt{ProFit} allows the user to estimate a PSF (see Section~\ref{sec:preparatorysteps}), entirely removing any dependence on external tools. Finally, both packages come with comprehensive documentation and many extended examples and vignettes which serve as great resources for newcomers to the fields of source extraction and galaxy fitting. 

We use \texttt{ProFound} (v1.9.2) along with \texttt{ProFit} (v1.3.2) for all preparatory steps (image segmentation/source identification, sky subtraction, initial parameter estimates and PSF determination; see Section~\ref{sec:preparatorysteps} for details) producing the inputs needed for the galaxy fitting with \texttt{ProFit}. 

\clearpage
\newpage
\chapter{Bulge-disk decomposition pipeline}
\label{chap:pipeline}

In this chapter, we present the bulge-disk decomposition pipeline we built around the \texttt{ProFound} and \texttt{ProFit} software packages. It consists of three main steps: preparatory work, galaxy surface brightness fitting and post-processing. The main inputs to the pipeline are a list of galaxies (GAMA CATAIDs) and the path to a directory containing (KiDS or VIKING) imaging files with associated weight maps and masks. The fully automated pipeline then performs all necessary steps to proceed from these basic inputs to a final catalogue of properties of the galaxies and their components, including numerous quality assessment parameters.

We start with an overview of all steps of the pipeline in Section~\ref{sec:pipelineoverview}. This is based on section~3 in \citet{Casura2022} and describes the status quo at the time of submission, with \texttt{v04} being the corresponding \texttt{BDDecomp} DMU version on the GAMA database. It is limited to the processing of the KiDS $g$, $r$ and $i$ bands. Section~\ref{sec:pipelinedevelopment} details how we have arrived at all of the procedures and the evolution of the pipeline from \texttt{v01} through to \texttt{v04}. Most of these studies focused on the KiDS $r$-band only. Finally, in Section~\ref{sec:pipelineupdates} we summarise the changes made to the pipeline after the submission of \citet{Casura2022}. These updates were mainly performed in order to process VIKING data and resulted in \texttt{v05} of the \texttt{BDDecomp} DMU, which encompasses all 9 KiDS and VIKING bands ($u, g, r, i, Z, Y, J, H, K_s$).
 
We use the free and open-source programming language \texttt{R} \citep{Rv3.6} for all scripting.

\section{Pipeline overview}
\label{sec:pipelineoverview}

This section is directly taken from section~3 in \citet{Casura2022} with very minor adjustments and gives an overview of the entire bulge-disk decomposition pipeline, including all preparatory steps, the actual galaxy fitting and the post-processing. It reflects the status of the data processing at the time of submission, meaning it refers to the processing of the KiDS $g$, $r$ and $i$ bands and \texttt{v04} of the corresponding \texttt{BDDecomp} DMU on the GAMA database. The adjustments relative to \citet{Casura2022} were mainly to fit the style of this thesis and to point to relevant further tests, examples and diagnostics in Section~\ref{sec:pipelinedevelopment} that are not contained in \citet{Casura2022}. 

\subsection{Preparatory steps} 
\label{sec:preparatorysteps}
The first part of our bulge-disk decomposition pipeline consists of several image processing steps that we group under the term ``preparatory work". 

\subsubsection{Cutouts and masking}

KiDS imaging tiles are registered to the same pixel grid across all four bands (with matching weight maps and masks), such that a joint analysis of the bands is straightforward. They are also aligned such that the $x$-axis corresponds to right ascencion (RA) and the $y$-axis to declination (Dec). Hence, we obtain a 400\arcsec\,$\times$\,400\arcsec\ cutout of the KiDS tile, associated weight map and mask for each object in our sample, and for each of the three KiDS bands we used ($g,r,i$). The masks of all three bands are then combined and all pixels which have a value greater than 0 in any of the masks are excluded from analysis. This results in approximately 20\,\% of all pixels being masked out. This large fraction of masking is primarily due to the reflection halos of bright stars that are also clearly visible in the data (see \citealt{deJong2015} for details). We combine the masks in this way to ensure that the pixels used for analysis are exactly the same in all bands and so the results are most directly comparable. Objects for which the central pixel is masked ($\sim$\,20\,\% of all galaxies) are skipped in the galaxy fitting. More details on the choice of KiDS data products and the cutout size to use are given in Section~\ref{sec:otherprepworkchoices}. 

\subsubsection{Image segmentation}

We perform image segmentation in order to determine which pixels to fit for each of our objects, identify other nearby sources, improve the background subtraction and obtain reasonable initial guesses for the galaxy parameters. This is performed on the joint $g, r, i$ cutouts with \texttt{ProFound} in several steps.  

First, we add the cutouts in the $g$, $r$, and $i$ bands using inverse variance weighting and compute the joint weight map. We then estimate the (joint $gri$) sky by running the stacked image through \texttt{profoundProFound} passing in the correct magnitude zeropoint, mask and weight, but leaving \texttt{skycut} on its default of 1. This means that all pixels with a flux at least 1$\sigma$ above the median are progressively assigned to segments (collections of pixels belonging to an object) using an iterative process: starting with the brightest pixel in the image, segments are grown by adding neighbouring pixels with lower flux; new segments are started when a pixel shows more flux than its neighbours (within some tolerance) or when all neighbouring pixels above the \texttt{skycut} value have been assigned. Once all pixels above \texttt{skycut} have been assigned, the resulting segments are additionally expanded until flux convergence is reached. For more details, see \citet{Robotham2018}. 

\begin{figure}[ht!]
  \begin{center}
	\includegraphics[width=0.8\textwidth]{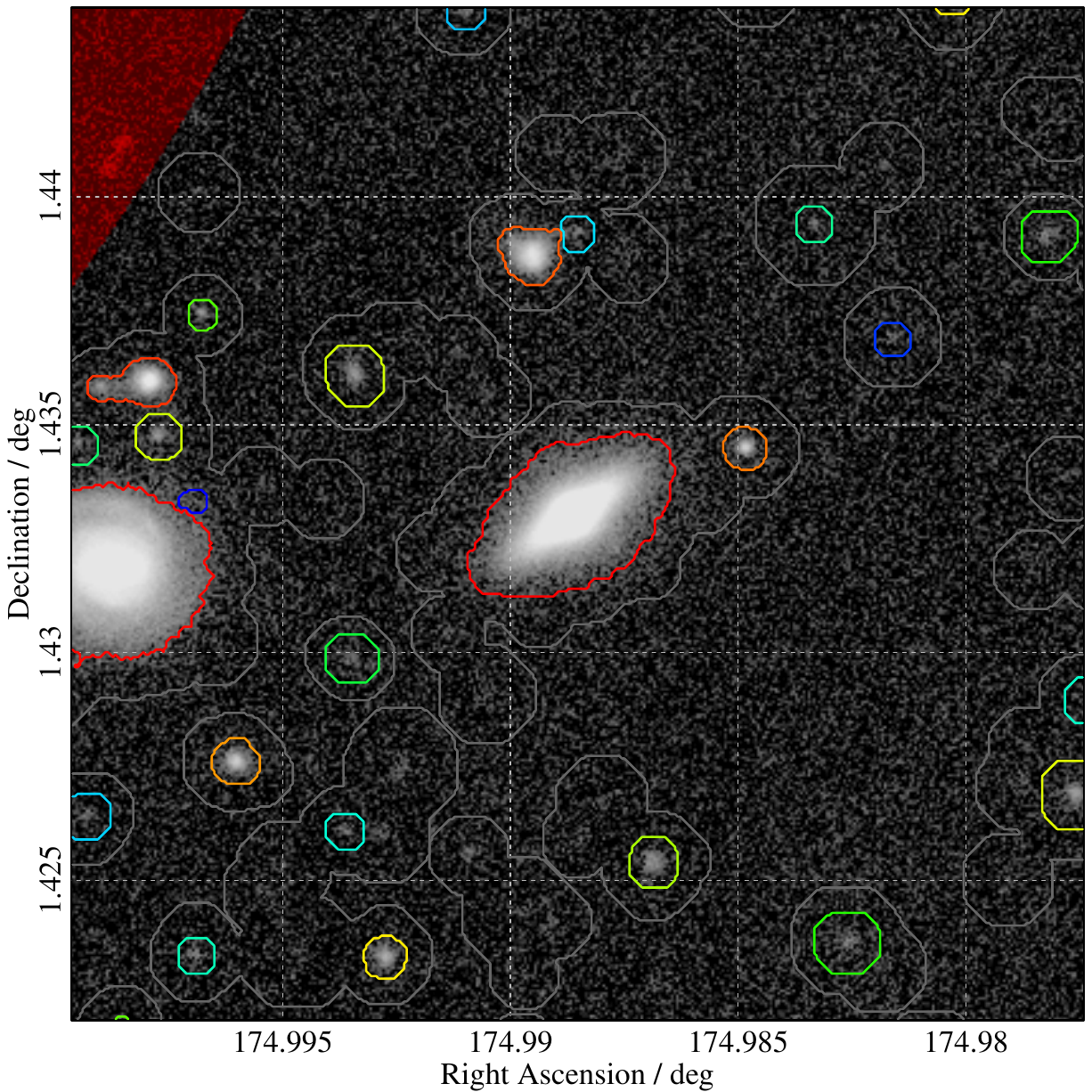}
    \caption{The \texttt{ProFound} segmentation map obtained for the GAMA galaxy 396740 overlaid on the KiDS $r$-band image. Note this is only a cutout of the full segmentation map showing the central 100\arcsec\,$\times$\,100\arcsec. Identified objects (segments) are shown with contours, coloured from red to blue according to the flux contained. Grey contours indicate the more dilated segmentation map used for the background subtraction. Masked areas are shaded red.} 
    \label{fig:exampleseg}
  \end{center}
\end{figure}

Along the way, \texttt{ProFound} estimates the sky background several times since object detection relies on accurate background subtraction and vice versa.\footnote{The sky variance can also be estimated, but in our case this is already provided as the KiDS weight maps and given to \texttt{ProFound} as an input.} For the final sky estimate, the already-dilated segments are expanded even further to ensure that no object flux will bias the background determination. This very aggressive object mask is indicated with grey contours in Figure \ref{fig:exampleseg}. We use it for the joint-$gri$ sky estimate here and also for the band-specific background determination detailed below (performed in the same way).  

For the galaxy fitting, however, we decided to use tighter segments that do not push that deeply into the sky. Besides speeding up the fit, this results in the best possible fit to the inner, high signal-to-noise regions of the galaxy that we are most interested in and reduces the sensitivity to background subtraction problems, flux from the wings of other objects and features that cannot be captured by our models such as disk breaks and flares, and edge-on disks requiring the inclined disk model \citep{vanderKruit1981}. Note, however, that this choice comes with some trade-offs, most notably that the fit frequently overpredicts the flux outside the segment boundary. We address this in more detail in Section~\ref{sec:postprocessing}. 

To obtain these tighter segmentation maps, we run \texttt{profoundProFound} again with the sky now fixed and a higher \texttt{skycut} value of 2. This means that only pixels with a flux at least 2$\sigma$ above the background level are considered in the segmentation, which ensures that fewer noise fluctuations are ``detected" and segment borders are smooth. In order to capture all flux of the galaxy wings, the segment for the object of interest (only) is then expanded further (using \texttt{profoundMakeSegimDilate}) such that its area increases by typically around 30\,\%. This last step also ensures unbiased smooth borders of the segment since it is entirely independent of noise fluctuations. The resulting segmentation map is indicated with coloured contours in Figure \ref{fig:exampleseg} and is used for galaxy fitting in all bands, so that exactly the same pixels are fitted in each band (the segmentation statistics are re-calculated in each band). 

Section~\ref{sec:segchoices} contains more details on the procedures for defining the segmentation map, reasons behind our decision to use different segmentation maps for background subtraction and galaxy fitting and the effect that different segment sizes have on galaxy fitting.
 
\subsubsection{Background subtraction}

KiDS tiles are background-subtracted already, however we opt to use the sky estimated by \texttt{ProFound} to even out inhomogeneities on smaller scales. For this, we split our 400\arcsec\,$\times$\,400\arcsec\ cutout into 16 square boxes and mask out all objects using the aggressively dilated segmentation map indicated with grey contours in Figure \ref{fig:exampleseg}. The sky is then estimated as the median of the remaining (background) pixels in each box independently; and the solutions between the boxes interpolated with a bicubic spline.\footnote{This is done by \texttt{profoundProFound} internally; with the box size and the order of the interpolation spline being some of the variables we set.} This is done for each band independently, however the segmentation maps used to mask out objects are the same in all bands.

This procedure for the background subtraction was chosen after extensive testing during pipeline development which is summarised in Section~\ref{sec:backgroundstudies}. In short, we found that the \texttt{ProFound} sky adopted here does not subtract object wings while still homogenising the background well enough to avoid having to fit it along with the object of interest (introducing possible parameter degeneracies). It also decreases the sensitivity of the fit to the chosen segment size. 

\subsubsection{Sigma maps}

Once the image segmentation and background subtraction is completed, we also calculate the sigma (error) map for each cutout (independently in each band). This is a combination of the KiDS weight map (where $\sigma$\,=\,1\,/\,$\sqrt{weight}$) and the object shot noise. The latter is estimated as $\sqrt{N}$, where $N$ is the number of photons per pixel (using positive-valued pixels only). This, in turn, is obtained by converting the image into counts using the gain provided in the meta-data associated with each KiDS tile.

\subsubsection{PSF estimation}

PSF fitting is performed on the background-subtracted 400\arcsec\,$\times$\,400\arcsec\ cutouts with corresponding masks and sigma maps in each band. The segmentation statistics returned by \texttt{ProFound} are used to identify isolated stars (round, bright, small and highly concentrated objects with few nearby segments). More details on the star candidate selection are given in Section~\ref{sec:psfdetails}. These objects are then fit with a \citet{Moffat1969} function using \texttt{ProFit}; fitting all parameters except boxyness, i.e. the position, magnitude, full width at half maximum (FWHM), concentration index, axial ratio and position angle. Scale parameters are fitted in logarithmic space, a Normal likelihood function is used, initial guesses are taken from the segmentation statistics and we use the \texttt{BFGS} algorithm from \texttt{optim} \citep{Rv3.6}, which is a fast downhill gradient optimisation using a quasi-Newton method published simultaneously by \citet{Broyden1970, Fletcher1970, Goldfarb1970, Shanno1970}. 

Some of the objects fitted above may not actually be suitable for PSF estimation as they can be too faint or bright (close to saturation), have irregular features, bad pixels or additional small objects included in the fitting segment. Unsuitable objects are excluded by a combination of hard cuts in reduced chi-squared ($\chi^2_\nu$), position and magnitude relative to the \texttt{ProFound} estimates and an iterative 2$\sigma$-clipping in FWHM, concentration index, angle and axial ratio. Again, more details can be found in Section~\ref{sec:psfdetails}. Finally, we take the median of the Moffat parameters of a maximum of 8 suitable stars (the closest 2 from each quadrant where possible to ensure an even distribution around the position of interest) and use these Moffat parameters to create a model PSF image. The size of the PSF image is adjusted to include at least 99\,\% of the total flux; or to a maximum of the median segment size within which the stars were fitted, with pixels in the corners of the image set to zero to avoid having a rectangular PSF. 

More details of the PSF fitting procedure and example diagnostic plots are given in Section~\ref{sec:psfdetails}, which also includes a description of other options we explored for obtaining PSFs, a summary of the PSF quality control and an overview of the PSF effects on galaxy fitting. 

\subsubsection{Outputs}

For the fitting, we are only interested in the central galaxy and the closest neighbouring sources (for potential simultaneous fitting and to gain a better overview during visual inspection). Hence, we do not save the entire 2000\,$\times$\,2000\,pix$^2$ cutouts used in the preparatory work as that would unnecessarily waste storage space and computational time used on reading and writing files. Instead, the image, corresponding mask, segmentation map, sigma map and sky image are cut down to the smallest possible size that includes the object of interest (centred) and its neighbouring (touching) segments before saving. These 5 files, the model PSF image and some ancillary information such as the segment statistics are the main outputs of the preparatory work pipeline and serve as inputs for the galaxy fitting, which we describe in Section~\ref{sec:galaxyfitting}.

\subsection{Galaxy fitting}
\label{sec:galaxyfitting}
The second major step in the pipeline is the model fitting, based on the outputs of the preparatory work. 

\subsubsection{Inputs and models}

We use the Bayesian code \texttt{ProFit} \citep{Robotham2017} to perform 2-dimensional multi-component surface brightness modelling in each band independently, assuming elliptical geometry and a (combination of) S\'ersic function(s) as the radial profile. The \citet{Sersic1963} function is given in Equation~\ref{eq:sersic} and described by three main parameters: the S\'ersic index $n$ giving the overall shape (with special cases $n$\,=\,0.5: Gaussian; $n$\,=\,1: exponential and $n$\,=\,4: \citet{deVaucouleurs1948} profile), the effective radius $R_e$ including half of the total flux and the overall normalisation which we specify as total magnitude $m$. In addition, in 2 dimensions the axial ratio $b/a$ gives the ratio of the minor to the major axis of the elliptical model and the position angle PA its orientation, while $x$ and $y$ are used to define the position in RA and Dec. Throughout this thesis, $R_e$ refers to the effective radius along the major axis of the elliptical model. The S\'ersic model is detailed in \citet{Graham2005}; and see also Section~\ref{sec:sersicmodels}. 

The data inputs for \texttt{ProFit} are a background-subtracted image, corresponding mask, segmentation map, sigma map and PSF. All of these are obtained during the preparatory steps (Section~\ref{sec:preparatorysteps}). On the modelling side, the main choices are the profile(s) to fit with initial parameter guesses and priors, the likelihood function to use, the fitting algorithm and convergence criteria; which are detailed below. In short, we choose to fit each object with 3 different models in a four-step procedure: 
\begin{enumerate}[label=(\roman*)]
\item{Single component S\'ersic fits with initial guesses from segmentation statistics.}
\item{Double component S\'ersic bulge plus exponential disk fits with initial guesses from single component fits.}
\item{Double component re-fits for a subset of galaxies which seemed to have the bulge and disk components swapped in step (ii), see below.}
\item{``1.5-component" point source bulge plus exponential disk fits with initial guesses from double component fits.}
\end{enumerate} 
Further details on this choice of models are given in Sections~\ref{sec:sersicmodels} and~\ref{sec:modellingdecisions}. Note that, for brevity, we will call the central component ``bulge" throughout this work, even if it may not be a classical bulge. In particular, we do not distinguish classical bulges from pseudo-bulges, bars, AGNs, nuclear disks, combinations thereof or anything else that may emit light near the centre of a galaxy. Hence, we also use the term ``bulge" for 1.5-component fits where the central component is unresolved and for double component central components with low S\'ersic index and/or low axial ratios. 

\begin{table}[h]
	\centering
	\caption{The fitting parameters for each of our three models.}
	\label{tab:fittingparameters}
	\begin{tabular}{l l c l cc l cc} 
		\hline
		 && single && \multicolumn{2}{c}{double} && \multicolumn{2}{c}{1.5-comp.} \\
		 parameter && && bulge & disk && bulge & disk \\
		 \hline
		$x$-centre && free && \multicolumn{2}{c}{free} && \multicolumn{2}{c}{free} \\
		$y$-centre && free &&  \multicolumn{2}{c}{free} && \multicolumn{2}{c}{free}\\
		$m$ && free && free & free && free & free\\
		$\log_{10}(R_e)$&& free && free & free && N/A & free\\
		$\log_{10}(n)$ && free && free & 1 && N/A & 1 \\
		$\log_{10}(b/a)$&& free && free & free && N/A & free\\
		PA && free && free & free && N/A & free\\
		boxyness && 0 && 0 & 0 && N/A & 0\\
		\hline
	\end{tabular}
\end{table}

To implement our three models, we make use of two of the many models built into \texttt{ProFit}, namely the S\'ersic and point source models. We fit all parameters except \texttt{boxyness} (i.e. we do not allow deviations of components from an elliptical shape) and, for the double and 1.5-component models, tying the positions of the two components together. Exponential disks are implemented using a S\'ersic profile with the S\'ersic index fixed to 1. This leaves 7 free parameters for our single S\'ersic and 1.5-component models and 11 free parameters of the double component fits, which are summarised in Table~\ref{tab:fittingparameters}; and see also Section~\ref{sec:modellingdecisions}. Scale parameters (S\'ersic index, effective radius and axial ratio) are treated in logarithmic space throughout, i.e. the actual fitting parameters are $\log_{10}(X)$ for scale parameters $X$ (Section~\ref{sec:fittingspecifics}). 

The 1.5-component model is needed for around 15-30\,\% of our double component systems where the bulge is too small relative to the image resolution to meaningfully constrain its S\'ersic parameters (the exact number depends on the band due to the different PSF sizes). With the point source profile, at least we can determine the existence of a second component and constrain its magnitude and hence the bulge-to-total (or AGN-to-total, bar-to-total, etc.) flux ratio. An example is given in Figure~\ref{fig:examplefit1.5} (Section~\ref{sec:modellingdecisions}). 

If the centre of an object is masked or the PSF estimation failed (which happens if large fractions of the surrounding area are masked), then the object is skipped and no fits are obtained. This affects approximately 20\,\% of the galaxies. All other objects are fitted with all three models; and the best model is selected subsequently (see Section~\ref{sec:postprocessing} on details of the model selection and Section~\ref{sec:statistics} for the corresponding statistics).

\subsubsection{Initial guesses}

Since we use MCMC algorithms, our fits do not strongly depend on the initial guesses. However, reasonable starting parameters are still required for convergence within finite computing times. 

The initial guesses for the single component S\'ersic parameters are obtained directly from the segmentation statistics output by \texttt{profoundProFound} (Section~\ref{sec:profound}) where we use the position, magnitude, effective radius (\texttt{R50}), axial ratio and angle as given; and the inverse of the concentration (1/\texttt{con}) for the S\'ersic index. 

For the double component fits, we convert the single component fits into initial guesses as follows: the position is taken unchanged, the magnitude of the single component fit is split equally between the two components, the bulge and disk effective radii are taken as 1/2 and 1 times the single component effective radius respectively, the S\'ersic index of the bulge is set to 4 and its axial ratio to 1 (round), the disk axial ratio is set to the axial ratio of the single component fit and the position angles of both components are taken as that of the single component fit. See also Section~\ref{sec:modellingdecisions}. 

Initial guesses for the 1.5-component fits are taken from the double component fits (after making sure the components are not swapped, see below), where the bulge magnitude is used as the point source magnitude and the disk parameters are taken unchanged. 

\subsubsection{Priors, intervals, constraints}

All parameters are limited to fixed intervals. In addition, there can be constraints between parameters (such that, e.g., the bulge and disk positions can be tied together). If a (trial) parameter is outside the bounds of its interval or constraint during any step of the fitting process, \texttt{ProFit} moves it back onto the limit before the likelihood is evaluated.

The limits for single-component fits are given in Table~\ref{tab:singlefitlimits}. In addition, the position angle is constrained such that if it leaves its interval, it is not just moved back onto the limit but jumps back 180\degr\ (which is the same angle, just more in the centre of the fitting interval). 
\begin{table}
	\centering
	\caption{The fitting limits for single-component fits.}
	\label{tab:singlefitlimits}
	\begin{tabular}{lll} 
		\hline
		parameter(s) & lower limit & upper limit \\
		\hline
		$x$- and $y$-centre & 0 & cutout side length\\
		magnitude & 10 & 25\\ 
		effective radius & 0.5 pixels & $\sqrt{2}$ cutout side length\\
		S\'ersic index & 0.1 & 20\\
		axial ratio & 0.05 & 1\\
		position angle & -90\degr\ & 270\degr\ \\
		\hline
	\end{tabular}
\end{table}

There are no additional priors or constraints for single component  fits. This means that in effect, we use unnormalised uniform priors which are 1 everywhere in the respective interval and zero otherwise. For scale  parameters (which are fitted in logarithmic space) the priors are uniform in logarithmic space, corresponding to \citet{Jeffreys1946}, i.e. uninformative, priors. 

The limits and constraints for double and 1.5-component fits are the same as for the single component fits (for both bulges and disks), except for the magnitude where the individual component magnitudes have infinity as their upper limit and instead the \emph{total} magnitude is constrained to be within  the magnitude limits. This is most consistent and also allows the fitting procedure to discard one of the two components for systems which can equally well be fitted with a single S\'ersic function (we then take this into account in the model selection). Further considerations on additional constraints between parameters (that we did not enforce in the end) are summarised in Section~\ref{sec:modellingdecisions}. 

Note that the above procedure results in unnormalised posteriors. The lack of normalization does not impede our analysis because the only time when we compare posteriors is during model selection, where we effectively fold the normalisation into the calibration during visual inspection (Section~\ref{sec:postprocessing}). 

\subsubsection{Likelihood function}

We use a Normal likelihood function for all fits. We have tested a t-distribution likelihood function which is less sensitive to outliers/unfittable regions; but found that the Normal likelihood function is better suited to our needs for several reasons. 

First of all, the t-distribution fits often preferred to use the freedom of the bulge parameters to fit disk features instead (e.g. rings, bumps, flares, etc. that cannot be captured by the exponential model), treating the bulge as an outlier since the t-distribution prefers a few strong outliers (the bulge pixels) over many weak ones. 

Second, the t-distribution fits fail for galaxies which are perfectly fitted by the model since then the errors truly are distributed Normally. This is a relatively common occurrence.

Hence some galaxies ($\sim$\,20\,\%) need to be fitted with a Normal distribution anyway, which, third, makes model selection much harder since the likelihood values obtained with different likelihood functions cannot easily be compared to each other. 

See Section~\ref{sec:fittingspecifics} for more details and examples.

\subsubsection{Fit and convergence}

All fits are performed on the sky-subtracted image within the galaxy segment only using the \texttt{convergeFit} function from the \texttt{AllStarFit} package \citep{AllStarFit}. 
This function uses a combination of different downhill gradient algorithms available in the \texttt{nloptr} package \citep{nloptr} followed by several MCMC fits with \texttt{LaplacesDemon} \citep{LaplacesDemon} until convergence is reached. The exact procedure is described in Section~\ref{sec:fittingspecifics}. 

The downhill gradient algorithms are used first to improve the initial guesses. The MCMC chain is not very sensitive to the initial guesses, but converges much faster if starting closer to the peak of the likelihood. Once the MCMC chains have converged, 2000 further likelihood points are collected to ensure a stationary sample for the subsequent analysis of the galaxy. We test this in Section~\ref{sec:fittingspecifics}. 

We only fit the primary object of interest. While simultaneously fitting neighbouring sources is possible in \texttt{ProFit} and might have improved the fit on a few objects, the effects are generally small since the galaxies we study are not in highly crowded fields and the segmentation process usually excludes the vast majority of the flux from other sources. This is especially true since we use tight fitting segments within which the galaxy flux is dominant (cf. Section~\ref{sec:preparatorysteps}); and considering that the watershed algorithm of \texttt{ProFound} cleanly separates even overlapping sources, so neighbours are automatically masked (Section~\ref{sec:profound}). Hence we opted for the simpler and computationally cheaper option of only fitting the main objects. We confirm that this does not lead to major biases in Section~\ref{sec:parameterrecovery}).

\begin{figure}
	\includegraphics[width=\textwidth]{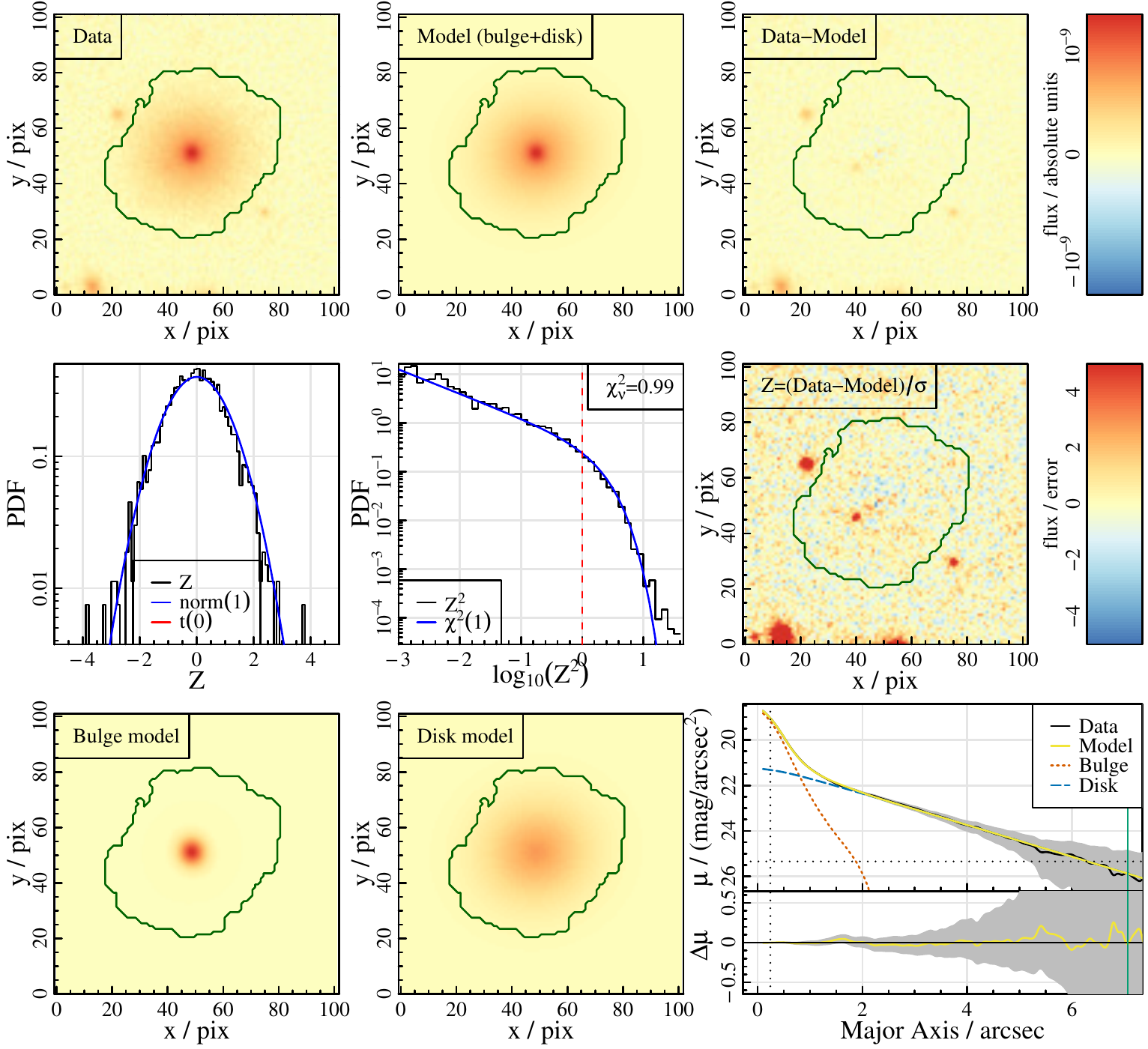}
    \caption{The result of the 2-component (S\'ersic bulge plus exponential disk) fit for the galaxy 611298 in the KiDS $r$-band. \textbf{Top row:} the data, 2-component model and residual between them shown in absolute values of flux given by the colour bar on the right. The green contour indicates the segment used for fitting. Note that the flux scaling here is non-linear and optimised to increase visibility of galaxy features, but is the same in all 3 panels. \textbf{Middle row:} goodness of fit statistics. The right panel is the normalised residual $Z$ (colour bar on the right) capped at $\pm$\,5$\sigma$. The left panel is the one-dimensional distribution (measured probability density function, PDF) of $Z$ within the segment; with blue and red curves showing a Normal distribution with a standard deviation of 1 and a Students-t distribution with the relevant degrees of freedom for comparison. The middle panel shows the measured PDF of $Z^2$ compared to a $\chi^2$-distribution of 1 degree of freedom (blue). The reduced chi-squared, $\chi^2_\nu$ (sum over $Z^2$ divided by the degrees of freedom of the fit) is given in the top right corner. \textbf{Bottom row:} The bulge and disk models in 2 dimensions on the same flux scale as the top row; and the bulge, disk, and total model compared against the data in one-dimensional form (azimuthally averaged over elliptical annuli). The FWHM of the PSF and the approximate 1$\sigma$ surface brightness limits are indicated by vertical and horizontal dotted lines for orientation. The vertical solid green line indicates the segment radius beyond which our model is extrapolated. The pixel scale is $0\farcs$2 for KiDS data, i.e. 1\arcsec\ corresponds to 5\,pix.} 
    \label{fig:examplefit}
\end{figure}

An example fit for an object which is well-represented by our 2-component model is shown in Figure~\ref{fig:examplefit}. 

\subsubsection{Component swapping}

Approximately 20-30\,\% of the double component fits have their bulge and disk components swapped, i.e. the exponential component fitting the central region and the S\'ersic component fitting the wings (this is a common problem in galaxy fitting, first pointed out by \citealt{Allen2006}). In particular, the freedom of the S\'ersic component is often used to fit disks that do not follow pure exponential profiles while at the same time being the dominant component in terms of flux (which is the case for most galaxies). To solve this problem, we devised an empirical swapping procedure guided by the visual inspection of a subsample of our galaxies. 

First, we select the galaxies that are most likely to have swapped components based on a cut in the plane of the ratio of S\'ersic indices and the ratio of effective radii for the single component fits and the bulge of the double component fits. The reasoning for choosing this parameter space to calibrate the cut was that we would generally expect bulges alone to be more concentrated (i.e. smaller effective radius and higher S\'ersic index) than when mixed with their respective disks in the single S\'ersic fits. This results in approximately 30\,\% of our sample being flagged as possibly swapped, which we then re-fit in a second step.

The re-fit is performed in exactly the same way as the original fit, except that we now use the results of the previous double-component fit as initial guesses, swapping around the bulge and disk components (except for the bulge S\'ersic index for which we use a value of 4). While the MCMC chain is less sensitive to initial guesses than a downhill gradient algorithm, it will still show some dependency for finite run-times. In particular, in our double component model the two components are \emph{nearly} interchangeable with the only difference being the S\'ersic index (fixed to 1 for the disk, free for the bulge). Hence there will always be 2 high maxima in likelihood space, which are far apart in the 11-dimensional parameter space. Moving from one to the other would require changing 9 parameters (all except position) at once in the right direction and hence is statistically unlikely. Therefore, we assist the code in finding the other maximum by manually swapping the initial guesses. 

In approximately 5\,\% of all re-fits, the code still converges on the same fit as before the swapping, but in most cases we find another likelihood maximum which corresponds to the bulge and disk components being reversed. As a third step we then select between the old and the new fit to obtain the physically more appropriate one. For this we first check whether either of the fits has a bulge S\'ersic index smaller than 2 \emph{and} a bulge effective radius at least 10\,\% larger than the disk effective radius \emph{and} a bulge-to-total ratio above 0.7 (i.e. the ``bulge" component is close to exponential, larger than the disk and contains the majority of the flux). If this is the case for only one of the fits, we choose the other one. If it is true for both or neither of the fits, then we apply our main criterion, which is that we choose the fit with the higher absolute value of bulge flux in the central pixel. These selection criteria are again based on visual inspection guided by the notion that we expect the bulge to be smaller and steeper than the disk and have proven to work very well. Note that the fit we select in this way is the one that is physically better motivated (i.e. with the bulge at the centre), and not necessarily the one which is statistically better.

After this procedure, the number of galaxies which still have the bulge and disk components swapped (and are classified as double component fits in model selection) is reduced to $\sim$\,1-2\,\%. The corresponding diagnostic plot based on the visual inspection as well as an example are shown in Section~\ref{sec:swappingandoutliers} (Figures~\ref{fig:swapdiagnostic}, \ref{fig:examplefitswap1} and~\ref{fig:examplefitswap2}). 

\subsection{Post-processing}
\label{sec:postprocessing}
To assess and improve the quality of the fits, we perform a number of post-processing steps, namely the flagging of bad fits, model selection, and truncating fits to segment radii. 

\subsubsection{Flagging of bad fits}

After all three models have been fitted to all objects, we run them through our outlier flagging process (separately in each band). Each model is treated separately first; they are then combined during the model selection (see below). 

The criteria for flagging bad fits (outliers) are: a very irregular fitting segment, an extreme bulge-to-total flux ratio, numerical integration problems, a parameter hitting its fit limits, poor $\chi^2$ statistics, a large distance between the input and fitted positions and a small fraction of model flux within the fitting segment. Additionally, there are some cautionary flags that identify fits which should be treated with extra care. All criteria are derived from and calibrated against visual inspection and described in more detail below and in Section~\ref{sec:swappingandoutliers} (including diagnostic plots). For orientation, we give the percentage of affected $r$-band fits in parentheses for each criterion below. The corresponding percentages for the $g$ and $i$ band fits are given in Table~\ref{tab:v04outlierstats} in Section~\ref{sec:outlierstats}. Note, however, that bad fits tend to fall into multiple of these categories, so the total number of bad fits is smaller than the sum of flagged objects in each category. Overall, approximately 9\,\%, 11\,\% and 10\,\% of all non-skipped fits are flagged in the $g$, $r$ and $i$ bands, respectively (after model selection).

\begin{description}[font=\normalfont]
\item[Very irregular segment (5.4\,\%):]{we calculate the difference between the magnitude of the model contained within the segment and the magnitude contained within the ``segment radius", which is defined as the maximum distance between the centre of the fit and the edge of segment. Objects where this magnitude difference is larger than 0.3 are flagged, as this is an indication for irregular segments (shredded, partly masked or cut off by another object for example). Note this criterion often shows overlap with the criterion on the fraction of model flux contained within the segment (see description below).}

\item[Extreme bulge-to-total ratio (0.1\,\%):]{we flag double component and 1.5-component fits with a bulge-to-total ratio smaller than 0.001 or larger than 0.999 because in these cases the second component has negligible flux and a single component fit is better suited.}

\item[Numerical integration problems (0.2\,\%):]{\texttt{ProFit} includes an oversampling scheme for accurate pixel flux integration where pixels containing steep flux gradients are recursively oversampled up to an oversampling factor of 4096; in the central pixel even up to $\sim$\,$10^9$ \citep[for more details see][]{Robotham2017}. However, for very extreme model parameters, even this procedure may not be accurate enough anymore, leading to significant errors in the pixel flux calculations. This could be improved by changing the default oversampling values to achieve higher accuracy (at the cost of increased computational time), however we opted for simply excluding those cases since usually this only happens for unresolved bulges which are better represented by the 1.5-component fits anyway.}

\item[Parameter hitting limit (5.8\,\%):]{we flag objects where the magnitude, effective radius or S\'ersic index hit either of their limits (cf. Section~\ref{sec:galaxyfitting}); or the axial ratio hit its lower limit (for double component fits this applies to both components individually). The axial ratio upper limit is not flagged because fits are allowed to be exactly round, but there is a cautionary flag for all objects which hit any of its parameter limits (6.5\,\%). We also add a cautionary flag for suspiciously small or large errors on any parameter, where ``suspicious" is defined as being an outlier in the respective distribution of errors (2.1\,\%).}

\item[Poor $\chi^2$ statistics (0.1\,\%):]{we flag fits with a $\chi_\nu^2$ larger than 80; or where the $\chi^2$ in the central pixel is more than 1000 times larger than the average $\chi^2$ per pixel since that is an indication that the bulge was not fitted.}

\item[Large distance between input and output position (0.3\,\%):]{we flag fits with a distance between the input and output position of more than 2\arcsec\ (10\,pix), which are usually highly asymmetrical objects, mergers, objects with very nearby other objects (especially small objects embedded in the wings of much larger objects), or objects in regions of the image with unmasked instrumental effects. Often the fitted object then is not the one that we intended to fit. There is also a cautionary flag for offsets above 1\arcsec\ (1.3\,\%).}

\item[Small fraction of model flux within fitting segment (1.4\,\%):]{we flag fits where the amount of model flux (of any component) that falls within the fitting segment is less than 20\,\%. With so little flux to work on \texttt{ProFit} cannot constrain the parameters well anymore and these are often objects which are cut off by a masked region (e.g. a bright star) or other nearby objects. There is a cautionary flag for objects where the fraction of model flux (of any component) that falls within the segment is less than 50\,\% (9.3\,\%).}
\end{description}

\subsubsection{Model selection}

As detailed in Section~\ref{sec:bayesiananalysis}, model selection in Bayesian analysis is performed by computing the posterior odds ratio, which in turn depends on the marginalised likelihoods for the two models in question. Since this is often difficult to compute in practice, many information criteria tests have been developed which are based on the (non-marginalised) likelihood (or $\chi^2$) combined with some penalty term depending on the number of model parameters. This penalty term serves to judge whether a more complicated model is justified and takes the role of Ockham's factor. Commonly used tests include the Akaike information criterion \citep[AIC,][]{Akaike1974}, the Bayesian information criterion \citep[BIC,][]{Schwarz1978}, or the deviance information criterion \citep[DIC,][]{Spiegelhalter2002}. We choose to use the deviance information criterion, which is usually recommended over the AIC or BIC in Bayesian analysis \citep{Hilbe2017} and straightforward to compute from an MCMC output. Brief tests using the BIC or the estimated log marginal likelihood output by \texttt{LaplacesDemon} showed similar results.

The DIC is a direct output of the \texttt{LaplacesDemon} function (see Section~\ref{sec:galaxyfitting}) and is defined as: 
\begin{equation}
\label{eq:dic}
\mathrm{DIC} = Dev + pD = Dev + \mathrm{var}(Dev)/2,
\end{equation}
where $pD$ is a measure of the number of free parameters in the model and $Dev$\,=\,$-2$\,$\times$\,log-likelihood is the deviance. In theory, then, if the DIC difference $\Delta$DIC between two models is negative, the first model is preferred and if it is positive, then the second model is preferred; with differences larger than approximately 4 being considered meaningful \citep{Hilbe2017}. However, for the case of galaxy fitting where many features are present that cannot be captured by the model (bars, spiral arms, disk breaks or flares, tidal tails, mergers, foreground objects, etc.), we want to choose the model that we consider physically more appropriate rather than better in a strictly statistical sense. This requires visual classification, logical filters, detailed simulations or a manual calibration of the $\Delta$DIC cut (or whichever other chosen diagnostic) by visual inspection of a representative sub-sample \citep[e.g.][]{Allen2006, Simard2011, Vika2014, Argyle2018, Kruk2018, Cook2019, Robotham2022}. We choose the latter approach, which has the added advantages that we do not need to worry about normalising our likelihoods (cf. Section~\ref{sec:galaxyfitting}), hence circumventing dependencies of the results on prior widths; nor the fact that our pixel values are correlated (due to the PSF) -- these effects are simply folded into the visual calibration. 

We use a random sample of $\sim$\,700 non-skipped objects per band (i.e. $\sim$\,2000 objects in total) for the calibration; and a further 1000 $r$-band objects that were previously inspected for cross-checking the results. In addition, our model selection procedure takes into account some of the outlier flagging. 
For each of the $\sim$\,700 objects in each band, we visually inspected the fits of all three models and classified the object into one of the categories: ``single component", ``1.5-component", ``double component", ``not sure if 1.5- or double component", ``not sure at all", ``unfittable" (outlier). We then calculate the DIC differences between all three models (i.e. $\Delta$DIC$_{1-1.5}$, $\Delta$DIC$_{1-2}$ and $\Delta$DIC$_{1.5-2}$) and calibrate them for model selection in two steps: first, we select between single component fit or not; of the ones that are not single component fits we then select between double component or 1.5-component fits.

For the first step of model selection calibration, the $\Delta$DIC$_{1-1.5}$ and $\Delta$DIC$_{1-2}$ cuts are optimised such that the minimum number of fits is classified wrongly. ``Wrong" in this case means a fit was manually classified as ``single" but is now a double/1.5; or a fit was manually classified as ``1.5", ``double", or ``not sure if 1.5 or double" but is now a single. ``Unfittable" and ``not sure at all" cases are ignored. For the second step of model selection calibration, the $\Delta$DIC$_{1.5-2}$ cut is optimised in the same way; where ``wrong" now means that the fit was manually classified as ``1.5" but is now a double or vice versa, with all other categories being ignored. For the two steps of the calibration, we bootstrap the manual sample 1000 and 500 times respectively and repeat the optimisation to get an estimate of the error on the chosen DIC cuts. These errors are chosen as the 1$\sigma$ quantiles (i.e. they contain the central 68\,\% of DIC cut distributions). Hence, all our calibrated DIC cuts have a median, a lower limit and an upper limit. Any object within these limits is flagged as unsure in the model selection, i.e. the DIC differences are not conclusive for this object. 

To perform the actual model selection, the calibrated DIC cuts in each band are then applied to the entire sample, again in a two-step procedure: the single component fit is selected if neither of the 1.5- or double component fits are significantly better (as indicated by the DIC differences). Double component fits need to be significantly better than 1.5-component fits, too. In all cases, if the DIC difference is very clear, we do the model selection first; then flag objects as outliers if needed.\footnote{This means that it is possible (and not uncommon) that a galaxy which is classified as an outlier has a non-flagged fit in another model (but the fit that was chosen was significantly better than the other one, despite it being an outlier).} In the unsure region of the DIC difference, we choose the model that is not flagged as outlier; if neither is flagged, the DIC cut is applied.  

Compared against visual inspection (keeping in mind that visual classification is not free of errors either), roughly 7\,\%, 9\,\% and 6\,\% of the galaxies end up in the wrong category in total in the $g$, $r$ and $i$ bands respectively (in both steps of model selection combined, ignoring cases which were visually classified as ``unsure"). Table~\ref{tab:modelselconfusionr} gives the detailed confusion matrix for the $r$-band. Note that we do not consider the success of the outlier flagging here, so for outliers we show what the galaxy would have been classified as if it were not flagged (absolute value of the \texttt{NCOMP} column in our catalogue). We highlight those galaxies that are correctly classified in bold and show those that were ignored during the model selection calibration process in grey font. The remaining (black) numbers add up to the 9\,\% quoted above. Corresponding confusion matrices for the $g$ and $i$ bands are given in Section~\ref{sec:swappingandoutliers} (Tables~\ref{tab:modelselconfusiong} and~\ref{tab:modelselconfusioni}).
Note that since we minimise the \emph{total} number of fits classed wrongly, there is a slight bias against the rarer categories in the automated model selection. For example, the relative fraction of true 1.5-component objects (as per the visual inspection) that is classified wrongly by the automated selection is higher simply because 1.5-component objects are much rarer than single or double component objects. 

\begin{table}
	\centering
	\caption{The confusion matrix for our model selection based on a DIC difference cut compared against visual inspection for the $r$-band. All values are in percent of the total number of visually inspected $r$-band galaxies. Bold font highlights galaxies classified correctly, while grey shows those that were ignored during the calibration.}
	\label{tab:modelselconfusionr}
	\begin{tabu}{lcrrrc} 
		\hline
		 & \multicolumn{5}{r}{number of components} \\
		visual classification & & 1 & 1.5 & 2 &\\
		\hline
		``single" && \textbf{41.6} & 0 & 2.7 &\\
		``1.5" && 2.2 & \textbf{2.4} & 0.9 &\\
		``double" && 3.1 & 0.1 & \textbf{9.2} &\\
		``1.5 or double" && 0.3 & \textbf{0.6} & \textbf{3.0} &\\
		\rowfont{\color{gray}}
		``unsure" && 16.1 & 0.4 & 13.1 &\\
		\rowfont{\color{gray}}
		``unfittable" && 0.9 & 0.6 & 2.7 & \\
        \hline
	\end{tabu}
\end{table}

In addition to this band-specific model selection, we perform a joint model selection for all three bands. For this, we sum the DIC values of all three bands for each model before computing the DIC differences. Then we perform the same optimisation procedure as for the single bands (using all $\sim$\,2000 visually classified objects across the three bands) to obtain cuts in DIC difference which we subsequently apply for the model selection. Note that the model selected in this way is by necessity a compromise between the different bands, which have different depth and seeing. In this procedure, approximately 9\,\% of fits are classified wrongly across all bands compared to visual classification. The corresponding confusion matrix is shown in Table~\ref{tab:modelselconfusionjoint} in Section~\ref{sec:swappingandoutliers}.

The accuracy of the model selection is also confirmed using simulations, to the extent to which our simulations allow us to do so (see Section~\ref{sec:simulationsmodelselection} for details).

\subsubsection{Truncating to segment radii}

As detailed in Section~\ref{sec:preparatorysteps}, we produce segmentation maps that define the fitting region, meaning that only pixels within the fitting segment are considered during the evaluation of the likelihood of the model (equivalent to giving all pixels outside the segment zero weight in the fit). We choose tight fitting segments (cf. Section~\ref{sec:preparatorysteps}) in order to obtain the best possible fit in the inner, high signal-to-noise ratio regions of the galaxies and be less sensitive to disk breaks, flares, nearby other objects, sky subtraction problems and similar. The disadvantage of this approach is that profiles are not necessarily forced to zero for large radii, i.e. our S\'ersic fits often show unphysically large effective radii combined with high S\'ersic indices. 

To mitigate this effect, we define a ``segment radius" for each galaxy segment, which is simply the maximum distance between the fitted galaxy centre and the edge of the segment and can be understood as the upper limit to within which our model is valid. We then calculate the ``segment magnitude", $m_{seg}$, which is the magnitude of the (intrinsic, not PSF-convolved) profile integrated to the segment radius (rather than infinity); and the ``segment effective radius", $R_{e, seg}$, which is the radius containing half of the flux defined by the segment magnitude. These values (and quantities derived from them, such as segment bulge-to-total flux ratios) are provided in the catalogue (labelled \texttt{*\_SEGRAD}) and we strongly recommend using these instead of the S\'ersic values integrated to infinity whenever they are available. For a direct parameter comparison to other works, the values in those catalogues should also be appropriately truncated.

In the following, we explain this recommendation in more detail; with further points to note in Sections~\ref{sec:segchoices}, \ref{sec:comparelee} and \ref{sec:systematics}.  

\begin{figure}
    \includegraphics[width=0.5\textwidth]{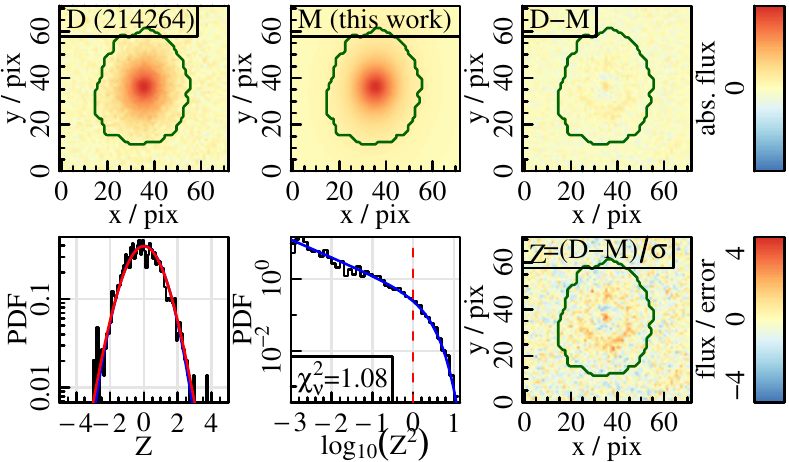}
    \includegraphics[width=0.5\textwidth]{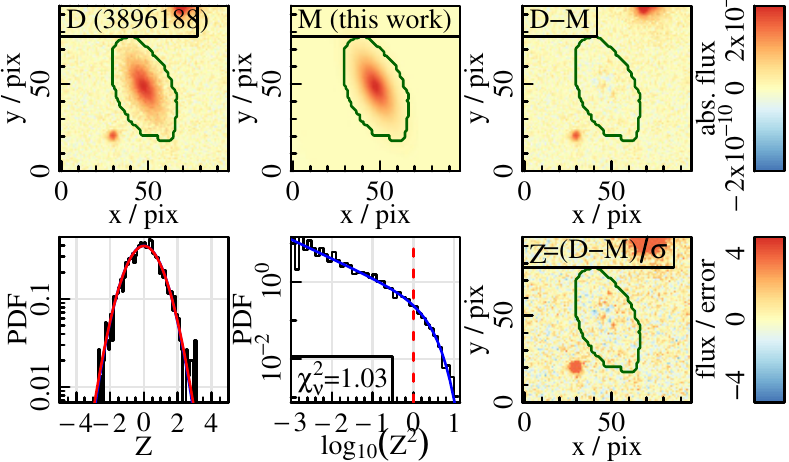}
    \includegraphics[width=0.5\textwidth]{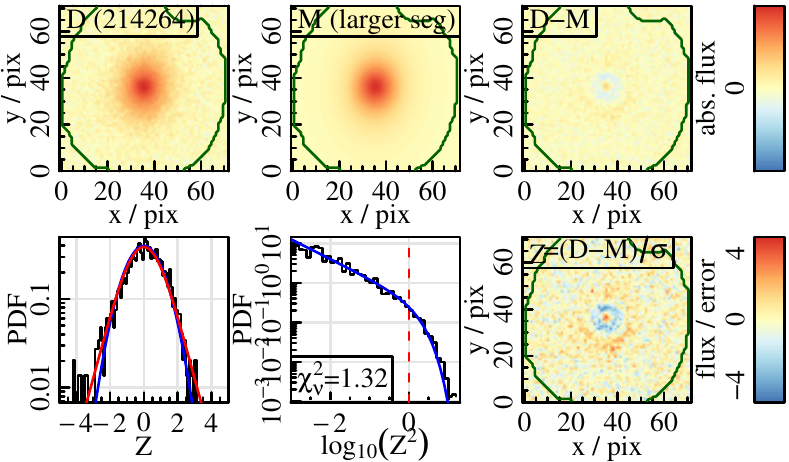}
    \includegraphics[width=0.5\textwidth]{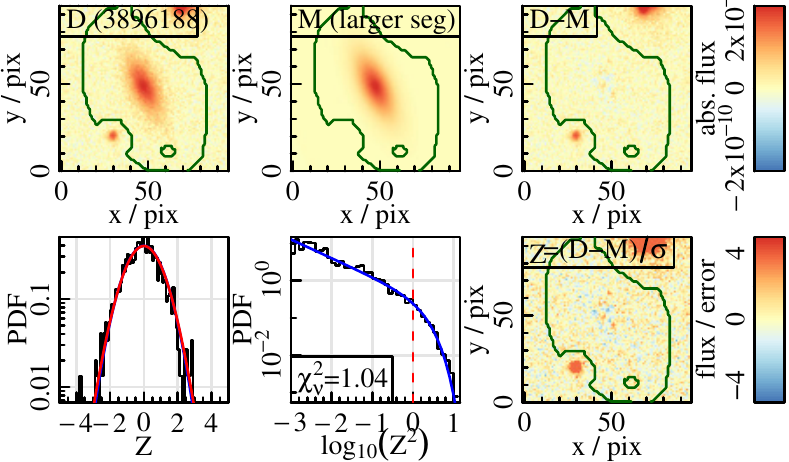}
    \includegraphics[width=0.5\textwidth]{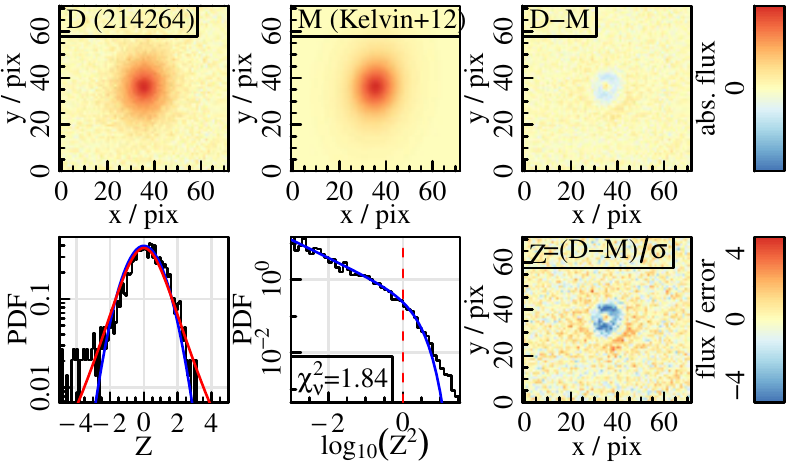}
    \includegraphics[width=0.5\textwidth]{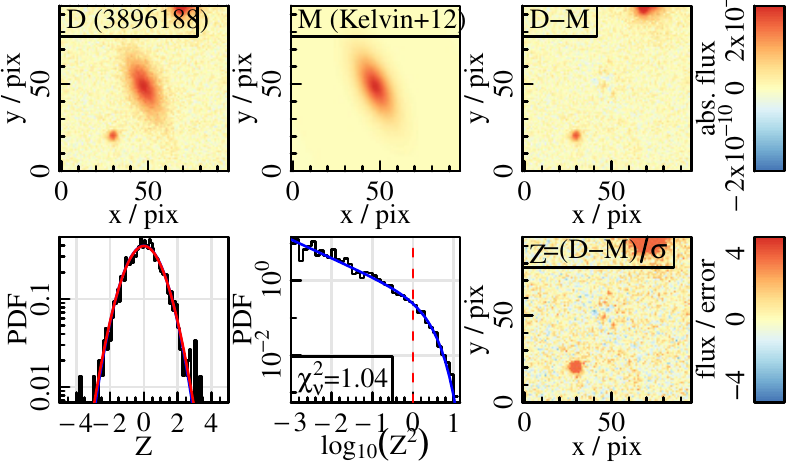}
    \includegraphics[width=0.5\textwidth]{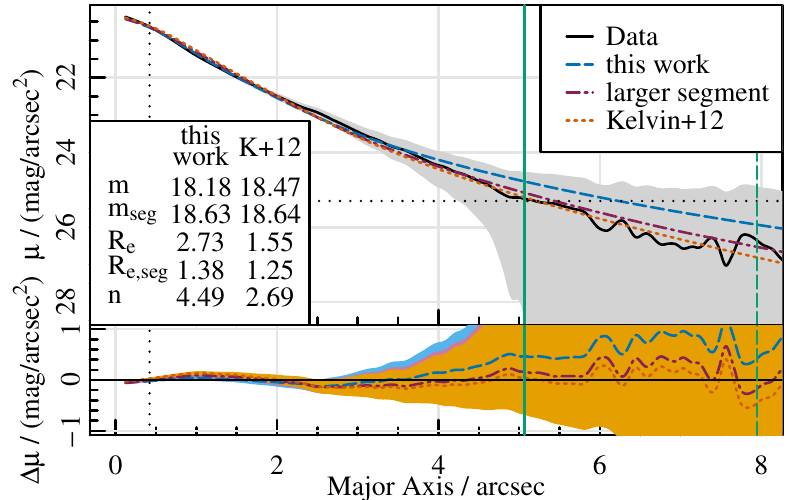}
    \includegraphics[width=0.5\textwidth]{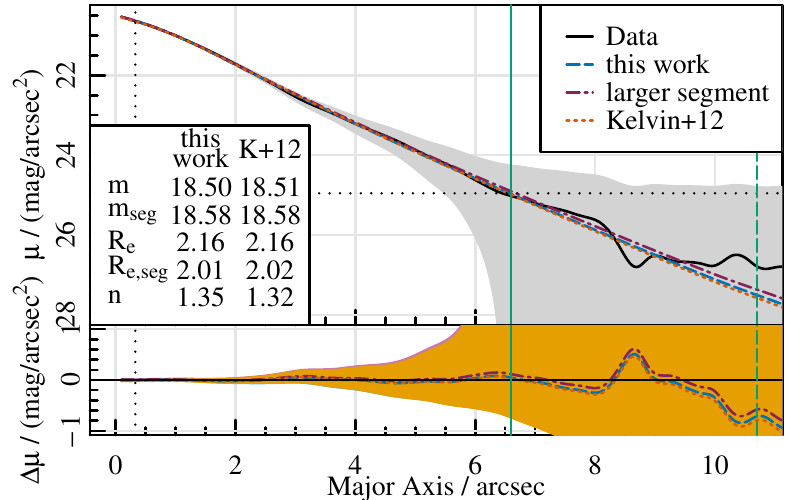}
    \caption{\textbf{Left:} detailed comparison of our single-S\'ersic fit, our fit using a larger segment, and the \citet{Kelvin2012} fit to the galaxy 214264, which in reality is a 1.5-component system. \textbf{Right:} the same for galaxy 3896188, which is well-described by a single S\'ersic component. \textbf{Top two rows:} our fit to the KiDS $r$-band data with panels the same as those in Figure~\ref{fig:examplefit}. \textbf{Rows three and four:} the fit we obtained by using a larger fitting segment as indicated. \textbf{Rows five and six:} the \citet{Kelvin2012} fits (originally performed on SDSS $r$-band data) evaluated on the KiDS $r$-band data, which use a fitting region larger than the cutout shown. \textbf{Bottom panels:} direct comparison of the one-dimensional profiles, see text for details.
    }
    \label{fig:tightseg}
\end{figure}

Figure~\ref{fig:tightseg} illustrates the effects produced by our tight fitting segments, how to mitigate those by truncating the magnitude and effective radius appropriately; and the circumstances under which this correction is necessary. For two example galaxies - 214264 and 3896188 - we show a detailed comparison of our single S\'ersic fit to both a fit using a larger segment (from \texttt{v05} of our pipeline, see Section~\ref{sec:pipelineupdates}) and to the fit obtained in \citet{Kelvin2012}. We present a more general (statistical) comparison of our fit results to those of \citet{Kelvin2012} in Section~\ref{sec:comparelee}, where we also give more details on how they derived their fits. For the purposes of the analysis in this section it suffices to say that \citet{Kelvin2012} used much larger fitting regions than we do, while the remaining analysis is in many ways analogous to ours (although they use different data, code and procedures in detail). 

Focusing on the left half of Figure~\ref{fig:tightseg}, the top two rows show the KiDS $r$-band data, our single S\'ersic model, the residual and various goodness of fit statistics as described in the caption of Figure~\ref{fig:examplefit}. Rows three and four show the same for a larger fitting segment as indicated by the green contour. Rows five and six again show the same for the \citet{Kelvin2012} fit, where we note that this was originally performed on $r$-band Sloan Digital Sky Survey (SDSS, \citealt{York2000}) data but is now evaluated on the $r$-band KiDS data. The \citet{Kelvin2012} fits were performed on cutouts larger than the size shown here, i.e. they include all visible pixels (and more) in the fit. Note that the reduced chi-squared value quoted in the bottom middle panel of each set of plots always is evaluated within the smallest segment so that they can be directly compared. 

Finally, the bottom panels show a direct comparison of the one-dimensional profiles of all three fits, which we will now study in detail. We show the surface brightness (azimuthally averaged over elliptical annuli) against the projected major axis for the data (solid black line with grey uncertainty region), our model fit for the fiducial segment (dashed blue line) and the larger segment (dash-dotted pink line) and the \citet{Kelvin2012} model fit (dotted orange line). The vertical green solid and dashed lines indicate the segment radii (for the two segment sizes respectively) beyond which our model is an extrapolation. The vertical dotted line shows the half width at half maximum (HWHM) of the PSF and the horizontal dotted line is the 1$\sigma$ surface brightness limit of the data. The inset in the bottom left of this plot shows the fitted magnitude $m$, effective radius $R_e$ in arcseconds and S\'ersic index $n$ values for our and the \citet{Kelvin2012} fits; and the corresponding segment-radius-truncated values for $m$ and $R_e$. Below that, we show the difference between all three models and the data (with errors): our fiducial fit in blue with a dashed line, the fit in the larger segment in pink with a dash-dotted line and the \citet{Kelvin2012} fit in orange with a dotted line.

Our model is a better fit to the inner regions of the galaxy than the \citet{Kelvin2012} fit (out to about 2\arcsec, also evident from the two-dimensional plots and from the reduced $\chi^2$-value within the segment decreasing from 1.84 to 1.08), owing to the higher S\'ersic index which better represents the steep bulge at the centre. However, it has a large effective radius and considerable amounts of model flux at large radii which are not observed in the data. In particular in the region beyond the segment radius, where our model is merely extrapolated, it is clearly oversubtracting the data (also visible in the 2-dimensional plots). Correnspondingly, the truncated segment quantities differ substantially from the fitted S\'ersic values. The \citet{Kelvin2012} fits, instead, use a larger fitting region and hence follow the data out to larger radii, which results in a worse fit of the central regions but does not contain such large amounts of excess flux beyond the surface brightness limit. Hence, truncating to segment radii has a smaller effect on the parameter values. The truncated values for both models are then in reasonable agreement with each other, except for the S\'ersic index, for which no truncated version exists as it would be unclear how to define such a value. Our fit in the larger segment is in between the two others in all respects, since it has a fitting region intermediate to the other two.

Note that the differences only come about when the model is not (in a formal statistical sense) a good representation of the data, i.e. when there is a need to compromise between fitting different regions. In the case of the left side of Figure~\ref{fig:tightseg}, the galaxy shown is better described by a 1.5-component model ($m^B$\,=\,20.47, $m^D$\,=\,18.79, $R_e^D$\,=\,1.89\arcsec), although in general there are many objects in our sample for which even a two-component model cannot capture all aspects of the data. For comparison, in the right half of Figure~\ref{fig:tightseg}, we show a galaxy that is well-described by a single S\'ersic model: here, both our and the \citet{Kelvin2012} fits arrive at virtually the same solution despite the different fitting regions. In fact, all three models and the data are nearly indistinguishable all the way down to the 1$\sigma$ surface brightness limit. 

In short, there is no perfect way to fit a S\'ersic function to an object which intrinsically does \emph{not} have a pure S\'ersic profile. For such objects, which unfortunately comprise the majority of our sample, the fitted parameters will always depend on the exact fitting region used as well as the quality of the data (its depth in particular). Most previous works, including \citet{Kelvin2012}, opted to use large fitting regions in order to include enough sky pixels to ensure that the profiles are constrained to approach zero flux at large radii (although a S\'ersic function technically never reaches zero exactly). Here, we choose a different approach by using smaller fitting segments. This means that the profiles are not constrained to approach zero flux at large radii. Instead more emphasis is placed on adequately representing the inner regions of the galaxies. We choose this approach since it is most appropriate for our science case, where we are primarily interested in comparing the high signal-to-noise regions of galaxies from the same data set amongst each other. 
In addition, it decreases the sensitivity of our fits to deviations from a S\'ersic profile in the low surface brightness wings of objects (arguably no galaxy truly follows a S\'ersic profile to infinity) as well as nearby other objects and inaccuracies in the sky subtraction. 
We stress that this means that our parameters are not directly comparable to other works using larger fitting segments. In particular, our S\'ersic indices tend to be systematically higher (see Section~\ref{sec:comparelee}) since high S\'ersic indices result in high amounts of flux at large radii and are hence suppressed when constraining the models to zero flux at large radii. Magnitudes and effective radii can be compared to those of other studies by truncating to segment radii.

\section{Pipeline development}
\label{sec:pipelinedevelopment}
This section contains technical details on the development of the bulge-disk decomposition pipeline and the numerous tests performed at various stages during that process with a par\-ti\-cu\-lar focus on the preparatory work. Building on the brief overview over the final \texttt{v04} pipeline in Section~\ref{sec:pipelineoverview} \citep[reproduced from][their section~3; a summary of which can also be found in the description accompanying the \texttt{BDDecomp} DMU on the GAMA database]{Casura2022}, we provide additional information on each step. This includes supplementary tests and diagnostic plots as well as explanations of how we have arrived at each of the decisions listed in Section~\ref{sec:pipelineoverview} and what other options have been explored over the course of the pipeline development. Further, we detail the pipeline evolution from \texttt{v01} through to \texttt{v04}; while the changes made for \texttt{v05} are given in Section~\ref{sec:pipelineupdates}. A summary of the key changes between all five different versions is provided in Section~\ref{sec:bddecompdmu}. 

Note, however, that many of the procedures presented here have been developed in an iterative way since most choices are interconnected and influence or depend on each other (both within the preparatory work pipeline and also for the galaxy fitting). \texttt{ProFit} and even more so \texttt{ProFound} were also actively developed during the time of pipeline development with frequent changes in the procedures and default values, necessitating corresponding adaptations to our routines. Consequently, the decisions presented here were not done in any strict chronolocical order and instead many of the test runs varying all of those parameters were performed and repeated numerous times at different points during pipeline development. We do not describe all of those iterations here, but instead try to summarise the most important findings and give evidence of when each decision was made (be it originally or finally). Unless stated otherwise, the tests in this section have been carried out on the KiDS $r$-band data.

\subsection{Data inputs and setup}
\label{sec:otherprepworkchoices}
We begin with a more detailed description on the general setup of the preparatory work pipeline and the data products it uses as inputs. 

\subsubsection{KiDS data products}

Before starting the data processing, we needed to decide which of the KiDS data products to use. The basic data unit for KiDS photometric data is a science tile. Each tile is approximately 1\,deg$^2$ or 18500\,$\times$\,19500\,pix$^2$ in size, is astrometrically and photometrically calibrated with a uniform pixel size of $0\farcs2$ and a magnitude zeropoint of zero. They are stacked images (coadds) composed of 5 (4 in $u$) slightly offset (dithered) frames taken in direct succession and arranged to close the gaps between the charge-coupled device (CCD) chips (see details in \citealt{deJong2015, deJong2017, Kuijken2019} and cf. also Section~\ref{sec:kids}). 

Since both re-gridding and stacking of individual exposures can result in problems for accurate photometry (e.g. correlations between pixels, abrupt changes in the background levels or PSF distortions), we briefly considered working with individual dithers instead of the science tiles. However, the dithers are not publicly available which means that they did not benefit from the same analysis and data reduction procedures as the full tiles. In particular, parts of the processing and quality control were performed on the full tiles and not at the individual dither level, see details in \citet{deJong2015, deJong2017, Kuijken2019}, leaving the dithers at a lower data quality. The science tiles, on the other hand, show a very high data quality despite the three main problems of re-gridded and stacked images listed above: pixel correlations are present, but limited to very small scales (see Section~\ref{sec:backgroundstudies}). Abrupt changes in the background levels at the edges of individual CCDs are very rare due to the careful background subtraction performed by the KiDS team before stacking. PSF distortions are minimal since all dithers were taken in direct temporal succession, so the seeing did not vary much between individual exposures. 

We concluded that it is more advisable to use the publicly available science tiles for KiDS, which the team themselves also use to create their photometric catalogues. This also significantly facilitated the data processing and galaxy fitting, especially since \texttt{ProFit} did not yet support multi-frame fitting at that stage. Note that for VIKING data, our decision was different and we decided to use the individual frames instead of the coadds (see Section~\ref{sec:vikingdataproducts}).

Each science tile has an associated weight map and flag image (mask). The weight images give the inverse variance for each pixel in the same flux units as the science tiles. They include information from the dithering pattern, flat fields, dark frames and the first step of the masking: defects that affect individual frames, such as cosmic rays, hot and cold pixels, saturated pixels or satellite tracks are masked in individual frames when adding them up to tiles, with the weight of that pixel (in the final co-add) being reduced accordingly \citep{deJong2015}.

We noted at various points during our analysis (see, e.g., Section~\ref{sec:backgroundstudies}), that the KiDS weight maps are conservatively estimated, i.e. the errors resulting from $1/\sqrt{weight}$ are slightly larger than the typical standard deviation of sky pixels (after our background subtraction and object masking routines). However, in view of the wealth of information included in the weight maps and in particular the strongly and abruptly varying weights across the tiling pattern, we decided it is still better to use those conservative weights than to not use them at all or try to estimate them using \texttt{ProFound}. Due to this, perfect fits (both for PSF estimation and galaxy fitting) usually have $\chi^2_\nu$ values around 0.7 to 0.9 instead of the nominal expectation value of unity. 

In addition to the weight maps, each KiDS tile has an associated mask image. These are mostly produced by an automated procedure developed specifically for the purpose of masking critical areas related to bright stars: saturated pixels, readout spikes due to saturated pixels, spikes caused by diffraction by the mirror supports and up to three reflection halos produced by the optics components \citep{deJong2015}. In addition, for DR1.0 and DR2.0, defects not related to bright stars were manually masked. In DR3.0 the manual masks were not included \citep{deJong2017}. For DR4.0 a semi-automatic procedure was then developed which includes the majority of these remaining issues with minimal manual intervention \citep{Kuijken2019}. 

The masks are binary flags where each of the affected areas has a different value. It is therefore possible to exclude only certain types of areas, such as saturated pixels, while still using those in others, e.g. weak tertiary reflection halos. However, we have decided against such an approach and instead mask all pixels with a flag value greater than zero. This results in approximately 20\,\% of pixels being masked out in $gri$ and correspondingly, $\sim$\,20\,\% of galaxies in our sample are skipped since their centre is masked. This is a large fraction, but it is not a limiting factor to our analysis. Our emphasis is on obtaining the most directly comparable fits to a statistically large sample of galaxies, for which it is optimal to use this highest data quality and treat all bands consistently. The sample size could instead be increased by including more galaxies at higher redshift if desired. 

If there is a particular interest in individual masked galaxies, it is still possible to re-run those ignoring some or all of the mask values. We have done this in the context of a Master's project focusing on AGN, of which there are only few in our sample (Targaczewski in prep.). However, this in turn required careful visual control of the resulting fits. For our own large scale automated analysis, it is hence preferable to use all flag values as masks. Importantly, the masks will randomly affect all types of galaxies and therefore not introduce any bias in our statistical analysis. 

In summary, we opt to use the KiDS science tiles with associated weight maps and masks, including all binary flag values. For \texttt{v01} and \texttt{v02} of the \texttt{BDDecomp} DMU we used KiDS DR1.0, DR2.0 and DR3.0, which were incremental data releases and together cover all of the GAMA equatorial regions. From \texttt{v03} onwards, we moved to KiDS DR4.0, which became newly available then. This did not add any tiles in our region of interest which was complete already in DR3.0. However, in contrast to the previous incremental data releases, DR4.0 was a complete re-release of all tiles with a number of processing changes. Most notably, the re-processing included a photometric homogenisation across all tiles (see \citealt{Kuijken2019}). DR4.0 therefore supersedes all previous KiDS data releases. 

\subsubsection{Pipeline setup}

Given the list of galaxies in our sample and the list of KiDS tiles with associated weight maps and masks, the first decision to make is whether one should work on a ``per-galaxy-basis", i.e. taking cutouts of the corresponding tile(s) for each galaxy or a ``per-tile-basis", i.e. treating all galaxies within a tile at the same time. We opted to work on the galaxy level, mostly due to computational memory limits since the tiles are rather large ($\sim$\,1.5\,GB each). Hence for each galaxy in our sample, we find the corresponding RA and Dec position from the GAMA database, identify the tile(s) that this position falls into and take a cutout around the galaxy position, with corresponding cutouts taken from the weight and mask images. 

This approach also has the advantage that the galaxies, which are the more physically meaningful quantity, are taken as the basis instead of the tiles, which can vary in different datasets. This allowed a relatively easy expansion from KiDS to VIKING data; and - in that context - a switch from treating several data matches to the same galaxy individually to treating them jointly (see Section~\ref{sec:pipelineupdates}). It also facilitates multi-band fitting, which we will attempt in future work. 

The relatively large cutout size of 400\arcsec\,$\times$\,400\arcsec\ was chosen to allow for a reasonable number of stars for PSF estimation (see Section~\ref{sec:psfdetails}); but is small enough to be handled computationally without problems. All preparatory work (image segmentation, background subtraction, PSF estimation) is carried out on these cutouts. We then further reduce the cutout size to only the region of interest around the galaxy itself before storage and subsequent galaxy fitting (see Section~\ref{sec:preparatorysteps}). 

Note that we have decided to split the pipeline into three main steps: the preparatory work, the galaxy fitting and the post-processing (Section~\ref{sec:pipelineoverview}). After each of these steps, the results are stored to disk. This has the advantages that the galaxies can be fitted with different models or procedures using the same preparatory work; and that the galaxy fitting is better reproducible since all inputs are stored in the preparatory work directory. Also, the post-processing is more flexible in the sense that e.g. model selection and outlier flagging can be re-calibrated without having to re-fit any galaxies. This is particularly important since the galaxy fitting takes the vast majority of computational time, with the preparatory work contributing around 2-3\,\% of the total run-time (6-7\,\% for VIKING) and the post-processing being negligible. Since each galaxy is treated independently of all other galaxies, the pipeline runs in parallel on many cores.

\subsection{Background subtraction choices}
\label{sec:backgroundstudies}
After the trivial step of taking cutouts, the first main task of the preparatory work pipeline is to perform image segmentation to identify the pixels that belong to the object of interest and mask out neighbouring sources. This, however, is intimately linked to the background subtraction, since a reliable source identification needs a well-known background estimate and vice versa. Hence, the main function of \texttt{ProFound}, \texttt{profoundProFound}, iteratively performs image segmentation and background subtraction simultaneously by default. However, since KiDS image tiles are already background-subtracted, the question arose as to whether we should use this \texttt{ProFound}-estimated sky (e.g. to even out smaller-scale background inhomogeneities and to make the treatment of KiDS data more comparable to the treatment of VIKING data) or set it to zero a priori. 

In addition, \texttt{ProFit} has the option to fit a constant background along with the source(s) of interest, which could also be used for local sky estimates instead of or in addition to the \texttt{ProFound}-estimated sky. We will refer to the former as (\texttt{ProFit}) background fitting and the latter as (\texttt{ProFound}) sky subtraction. Note that the two methods are not equivalent and describe slightly different ``types" of sky: \texttt{ProFound} explicitly attempts to exclude faint undetected background sources and extended low-surface brightness wings of objects, while \texttt{ProFit} includes all undetected light in the background estimate (see details below).

Hence, the main choices to make with regards to background subtraction are whether or not to use the \texttt{ProFound}-estimated sky and/or fit a constant background with \texttt{ProFit} along with the source (in addition to the background subtraction already performed by the KiDS team); and if so, which algorithms and options to use for the background estimation. Interconnected choices are how deep to go in the object detection (most strongly influenced by the value of \texttt{skycut} in \texttt{profoundProFound}) and how large to make the segments for fitting objects. 

The decisions made with regards to these choices are summarised in Table~\ref{tab:bgdecisions}. In the following, we give more details on the metrics that we used to inform these choices. Note that these studies were only performed on KiDS $r$-band data. The final pipeline uses a joint treatment of the $g$, $r$ and $i$ bands as described in Section~\ref{sec:preparatorysteps} and has been updated in many other respects, too. However, the choices made at this stage remained. We also note that while we focus on the background subtraction in this section, we inevitably touch upon image segmentation, too. In particular, we discuss the choices made for object masking during background estimation, while the details of the segmentation maps used for the galaxy fitting are de-coupled (one of the choices made during the background studies) and explained in more detail in Section~\ref{sec:segchoices}. 

\begin{table}
\begin{center}
\caption{Sky subtraction/background fitting choices}
\label{tab:bgdecisions}
\begin{tabular}{|l|l|}
\hline
Choice to make & Decision made\\
\hline
Perform \texttt{ProFound} sky subtraction & Yes \\
Sky box size & 1/4 of large cutout side length, i.e. 500\,pix or 100\arcsec\\ 
Sky grid size & Equal to sky box size (default)\\
Interpolation type & Bilinear (default) \\
Clipping & Yes (default)\\
Type of sky estimate & Median (default)\\
\hline
Improve sky with FFT & Yes up to \texttt{BDDecomp v02}, then no\\
Object masking (for FFT) & Aggressive \\
\hline
Perform \texttt{ProFit} background fitting & No \\
Size of cutout to fit & Star segment only\\
Skycut value (see Section~\ref{sec:segchoices}) & 1 for sky subtraction, 2 for object fitting\\
\hline
\end{tabular}
\end{center}
\end{table}

\subsubsection{\texttt{ProFound} sky subtraction}

As a first metric to judge the effects of different sky subtraction options, we use the distribution of \texttt{ProFit} backgrounds fitted to stars used during PSF estimation around a test sample of $\sim$\,200 galaxies (one in each KiDS tile, approximately uniformly distributed across the galaxy magnitude range). The stars have the advantage that they show much less variation than galaxies and can usually be (near-)perfectly fitted with \citet{Moffat1969} functions. They also do not suffer from PSF uncertainties since they do not need PSFs; they are more numerous than our galaxies and fitting them is much faster since simple downhill gradient algorithms suffice for such simple systems. In the following, we give a brief overview of the main results of these detailed investigations. 

\begin{figure}
\begin{center}
	\includegraphics[width=0.8\textwidth]{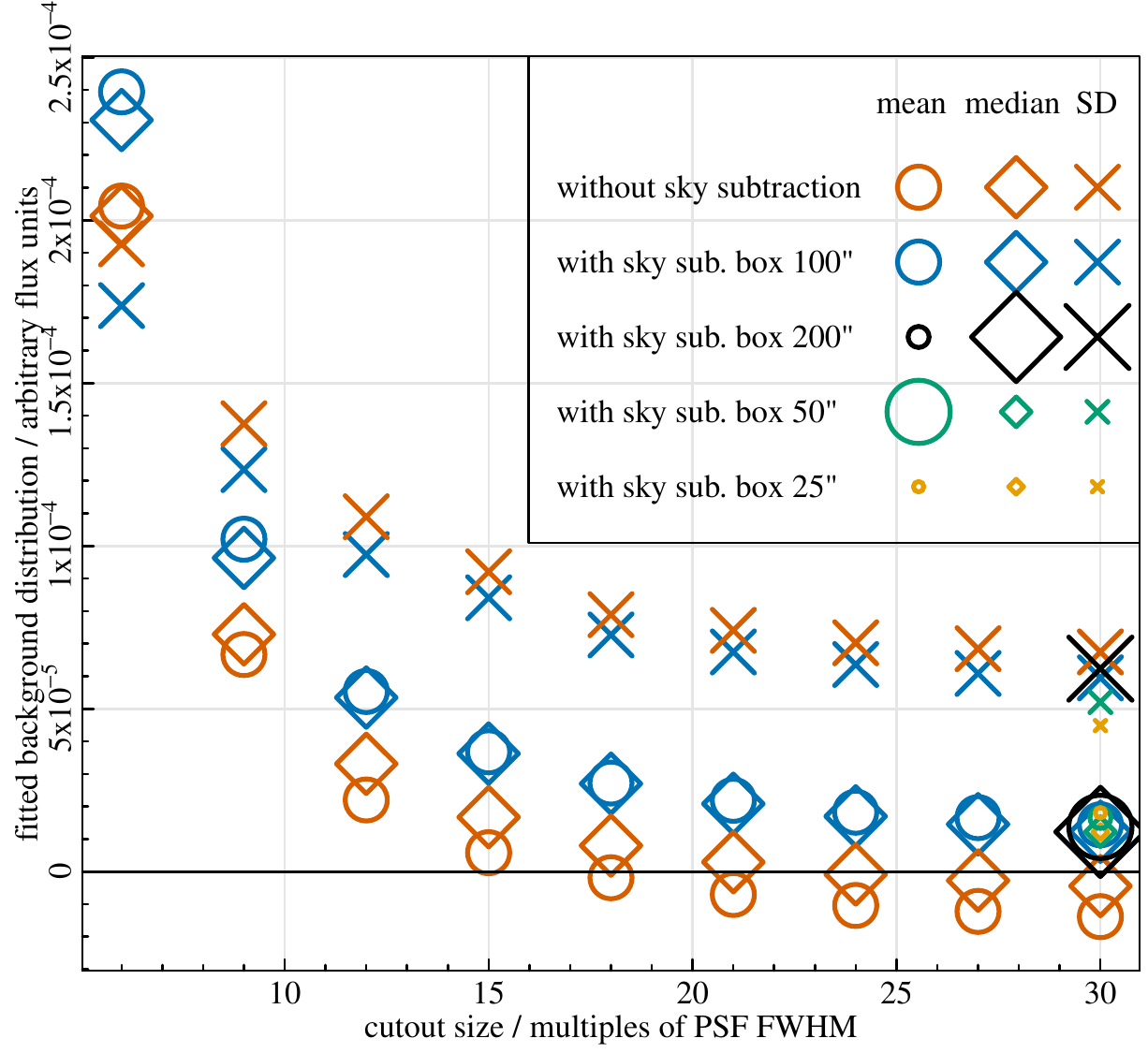}
	\includegraphics[width=0.8\textwidth]{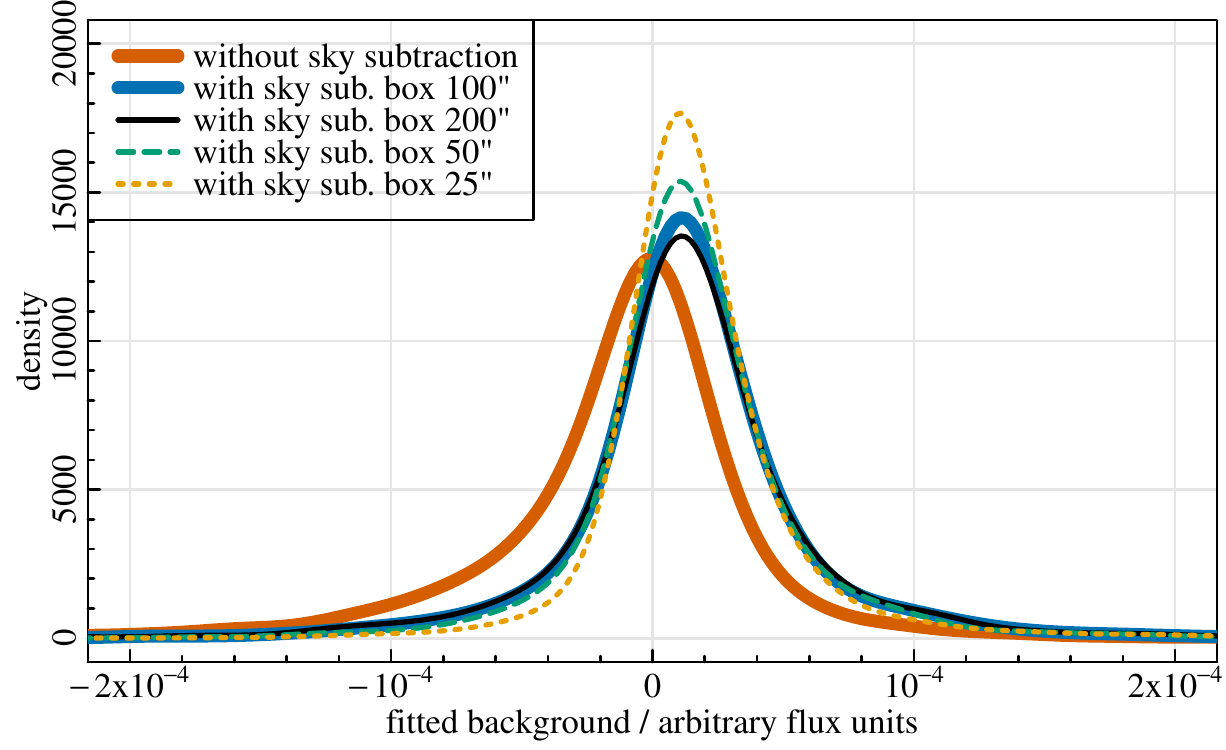}
    \caption{\textbf{Top panel:} the mean (circles), median (diamonds) and standard deviation (crosses) of the distribution of fitted \texttt{ProFit} backgrounds as a function of the fitting cutout size and with (blue) or without (red) previous \texttt{ProFound} sky subtraction. For the largest cutout size, we additionally show the effect of the chosen box size for sky subtraction (black, green, yellow). \textbf{Bottom panel:} the full distribution of backgrounds for the largest cutout size (corresponding to the rightmost set of symbols in the top panel).}
    \label{fig:bgdistribution}
\end{center}
\end{figure}

For Figure~\ref{fig:bgdistribution}, we fitted a constant background in addition to the Moffat parameters to a few thousand stars around a random test sample of galaxies using \texttt{ProFit}. We used all pixels within a square cutout around the star for fitting, except those that are identified as belonging to other objects during the segmentation procedure. The cutout side length ranges from 6 to 30 times the FWHM of the PSF given in the header of the corresponding KiDS tile, which is typically around 0.6-0.8\arcsec. The top panel shows the mean, median and standard deviation of the distribution of fitted backgrounds (in arbitrary units of flux since the cutouts were normalised before fitting) for different cutout sizes and for the two cases of also performing a \texttt{ProFound} sky subtraction before fitting (blue symbols) or not (red symbols). 
For the largest cutout size, we also show results for different sky box sizes, which is set to the fiducial value of 100\arcsec\ for all other runs: black symbols denote an increase in the box size by a factor of 2, while green and yellow symbols represent a decrease by a factor of 2 and 4 respectively. 
The bottom panel shows the full distribution of fitted backgrounds for the largest cutout size, corresponding to the rightmost set of symbols in the top panel. 

For small cutout sizes, the distributions are clearly biased positive (top panel of Figure~\ref{fig:bgdistribution}), i.e. the background fit is dominated by the wings of the star. For larger cutout sizes, the distributions then converge onto a slightly positive or negative value (with and without \texttt{ProFound} sky subtraction respectively). In general, the distributions with \texttt{ProFound} sky subtraction are slightly narrower and more symmetric than those without (i.e. the standard deviation is smaller and the mean and median are closer together). Due to these reasons (width, asymmetry and convergence onto negative values for non-\texttt{ProFound}-sky-subtracted images), we concluded that the \texttt{ProFound} sky subtraction is useful to even out smaller-scale inhomogeneities even though KiDS images are already background subtracted. In addition, it will make the comparison to VIKING data (which have potentially different backgrounds to KiDS) more consistent. 

The fact that the mean and median of the fitted distributions with previous sky subtraction converge onto a positive value instead of zero is expected due to the different ``types" of sky that are estimated by the two options: \texttt{ProFit} will estimate a slightly higher sky than \texttt{ProFound} because the former simply uses all non-object pixels while the latter explicitly tries to avoid being biased by undetected (wings of) objects by clipping outliers and using the part of the flux distribution smaller than the mode only to estimate the median and standard deviation of the background in each box. 
 
The box size should be chosen as small as possible to be able to detect small-scale inhomogeneities in the background, yet larger than the largest (complexes of) objects so as to not subtract their wings. For our purposes, we found a box size of 100\arcsec\ to be suitable. Increasing the box size by a factor of two only increases the standard deviation of the distribution, but does not change the mean or median. Decreasing the box size by a factor of two decreases the standard deviation but also shifts the mean away from the median, making the distribution more skewed. Decreasing the box size by a factor of four (i.e. to 25\arcsec) makes the distribution significantly more asymmetric and - upon visual inspection of a random sample of fitted backgrounds - also reveals clear correlations between the fitted background and the location of objects in the images. Hence the box size of 100\arcsec\ seems to be optimal in that it is independent of the object distribution yet still detects some small-scale background variations. 

We also varied other parameters to test their influence on the sky estimate (grid size on which to place the boxes, bilinear or bicubic interpolation between boxes, clipping outliers or not, using the mean or median for the sky estimate, dilating the object masks more or less) but have found either little influence or adverse effects on the sky estimate. Hence we decided to use the \texttt{profoundProFound}-defaults for these parameters, which were optimised for typical optical and NIR survey data (such as KiDS) during the development of \texttt{ProFound} \citep{Robotham2018}.

Finally, we should mention that we have also carried out tests where we divide each of the \texttt{ProFit} background fits by its respective error and investigate the resulting distributions. In theory, this should result in a Gaussian with mean zero and standard deviation one. A small positive shift away from the mean is to be expected according to the previous analysis, although much smaller than one standard deviation (Figure~\ref{fig:bgdistribution}). A standard deviation larger than one would indicate that the background offset (i.e. the positive shift) is not constant for different cutouts. This analysis, however, relies on realistically estimating the uncertainty of the fitted background value. We have tried this in two different ways: either during the downhill gradient fitting with \texttt{optim} (see Section~\ref{sec:preparatorysteps}) or directly from the KiDS weight map. Both methods have unresolved problems, so we did not use these results further and only add a short description here for completeness. 

In the first case, we use \texttt{optim}'s option to return the Hessian along with the best fit value of all parameters. The diagonal elements of the inverse of the Hessian matrix are the variances of the parameters and hence the entry corresponding to the background fit can be used to estimate its uncertainty (ignoring covariances and systematic uncertainties). For unknown reasons, however, the variance of the background was often negative, resulting in an imaginary standard deviation. Taking the absolute value before taking the square root gives errors of the correct order of magnitude, such that the normalised background distributions then have a standard deviation of approximately one and a slight positive offset.

For the second approach, we use the fact that each background pixel is drawn from a normal distribution with a standard deviation given by $\sigma$\,=\,1\,/\,$\sqrt{weight}$. \texttt{ProFit} essentially finds the mean of this background in each cutout weighted according to the errors ($\sigma$). The error on the mean (uncertainty coming from randomness that cannot be avoided) is $\sigma_{bg}$\,=\,$\overline{\sigma}/\sqrt{N}$ where $\overline{\sigma}$ is the mean of $\sigma$ and $N$ the number of pixels used to fit the background. This ignores any possible correlations between pixels or fitting parameters and assumes that the weight map is an accurate representation of the true errors on pixels. The problem with this estimate is that it results in background uncertainties that are approximately a factor of five smaller than those estimated from the Hessian; and the background fit distributions are correspondingly broader. 

It is unclear why the two methods provide such different results for the background uncertainty. There are correlations between pixels on small scales (see below) that will additionally contribute to the uncertainty estimated from the weight map, but this is unlikely to explain a factor of five. Correlations between parameters are ignored in both cases and hence cannot explain the difference. The weight maps tend to be on the conservative side compared to the true pixel errors (see Section~\ref{sec:otherprepworkchoices}), but this also does not explain a factor of five and moreover would tend to make the difference even larger if corrected for. Given these open issues, we decided to focus on the distribution of fitted backgrounds in absolute values as presented above; and did not consider the normalised distributions further.

\subsubsection{\texttt{ProFit} background fitting}

\begin{figure}[ht!]
	\includegraphics[width=\textwidth]{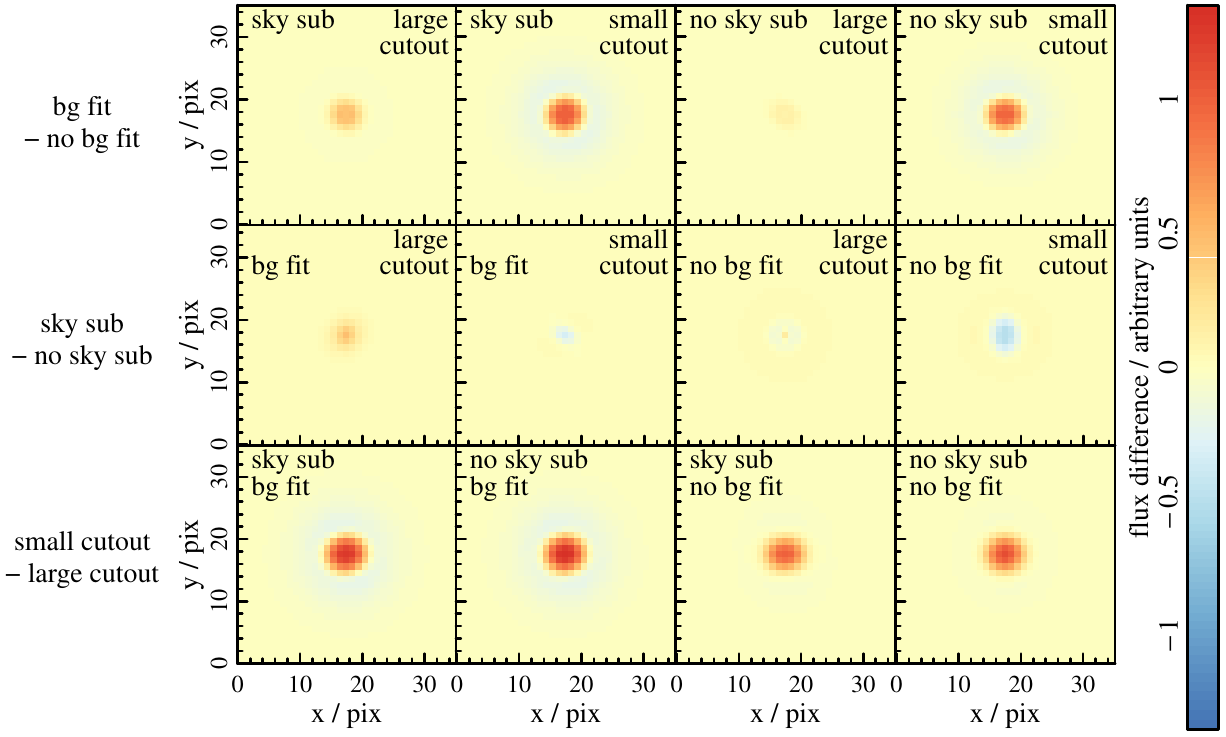}
    \caption{Stacked differences between model PSFs obtained from test runs with different combinations of \texttt{ProFound} sky subtraction (sky sub), \texttt{ProFit} background fitting (bg fit) and cutout sizes as labelled. \textbf{Top row:} the difference between model PSFs obtained by fitting or not fitting a \texttt{ProFit} background for four different combinations of \texttt{ProFound} sky subtraction and cutout size as labelled in the panels. ``Small" cutouts refer to a side length of 6 times the PSF FWHM, ``large" means 18 times the FWHM. \textbf{Middle row:} the difference between model PSFs obtained with \texttt{ProFound} sky subtraction vs. those obtained without, for four different combinations of \texttt{ProFit} background fitting and cutout sizes, as indicated. \textbf{Bottom row:} the difference in model PSFs obtained from fitting large or small cutouts for four different combinations of \texttt{ProFound} sky subtraction and \texttt{ProFit} background fitting, as labelled.}
    \label{fig:psfstack}
\end{figure}

Since the above analysis relies on the distribution of fitted \texttt{ProFit} backgrounds as a metric, it does not allow to decide whether or not the fitted background should be used in addition to the \texttt{ProFound} estimated sky. Therefore, as a further test, Figure~\ref{fig:psfstack} shows the systematic differences that occur in the PSF estimates as a function of the different background fitting and cutout size options. This now uses the model PSFs estimated for the 223 test galaxies (one in each KiDS tile within our region) rather than the fits to individual stars. For any two runs, the differences between the 223 pairs of model PSFs are calculated and then added together (stacked) to show systematic trends. Ideally, PSF estimates should be robust against changes in the processing details (e.g. the cutout sizes used for fitting them), such that the differences should be small.

The first row shows the difference between fitting a \texttt{ProFit} background or not for four different combinations of cutout size and \texttt{ProFound} sky subtraction as labelled in the panels. The larger cutouts (first and third panel) show less of a difference than the smaller cutouts. Note, for comparison to Figure~\ref{fig:bgdistribution}, that ``small" cutouts always mean 6 times the PSF FWHM given in the header, while ``large" cutouts refer to 18 times the PSF FWHM (we unfortunately did not perform these test runs with larger cutout sizes than this due to the order in which the decisions were originally taken). For both cutout sizes, there is also less of a difference in the fitted PSFs without \texttt{ProFound} sky subtraction (third and fourth panel) than with. This means that performing a \texttt{ProFit} background fit on the already-\texttt{ProFound}-sky-subtracted image systematically changes the PSF estimate - likely because there is no background left to fit and so the additional degrees of freedom are used to change (improve) the PSF fit instead. Note all residuals here are positive at the centre, indicating that fitting a \texttt{ProFit} background generally makes the PSFs more peaked at the centre.

The second row shows the difference between subtracting a \texttt{ProFound} sky or not for four different combinations of cutout size and \texttt{ProFit} background fitting. The second and fourth panels which show differences for small cutouts show negative residuals while the first and third panels (large cutouts) show positive residuals. In general, all residuals are small, which confirms that the PSF estimate is generally robust against performing a \texttt{ProFound} sky subtraction or not (i.e. the sky subtraction does not subtract the wings of the PSF or similar). The best results (i.e. smallest residuals) are obtained using the large cutouts and no \texttt{ProFit} background fitting (third panel). When using smaller cutouts there seem to be less residuals with \texttt{ProFit} background fitting (second panel) than without (fourth panel). This can be interpreted as follows: for small cutouts, the star dominates the cutout region, so the \texttt{ProFound} sky subtraction does not influence the fit of the star. This is even clearer when we additionally fit a background within the small cutout region. For larger cutouts, the PSF estimate is also robust against performing a \texttt{ProFound} sky subtraction or not, provided no additional \texttt{ProFit} background is fitted. An additional background fit makes the PSF estimate dependent on the sky subtraction (first panel), most likely because now the fit has too many degrees of freedom. 

The third row of Figure~\ref{fig:psfstack} shows the difference between a large and a small cutout for four different combinations of \texttt{ProFound} sky subtraction and \texttt{ProFit} background fitting. As observed before, the most robust results are obtained with \texttt{ProFound} sky subtraction only (third panel). Both results using \texttt{ProFit} background fitting (first two panels) are more dependent on the cutout size used.

As a result of these studies, we saw our decision to use the \texttt{ProFound} sky subtraction reinforced and decided to not use additional \texttt{ProFit} background fitting: after \texttt{ProFound} sky subtraction, the background is homogeneous enough that an additional \texttt{ProFit} background fitting changes the PSF rather than fitting the actual background. Also, as evident from Figure~\ref{fig:bgdistribution}, we need very large cutouts for robust \texttt{ProFit} background fits. This has problems on its own as it will necessarily lead to more contamination from the wings of other objects or undetected faint objects (masking is never perfect), increases the chance of hitting an image edge or masked area, and increases computational time. In addition, there is a degeneracy between the fitted background and the concentration index of the Moffat parameter in that it can never be unambiguously determined whether the background is used to fit the wings of the star or the wings of a low-concentration Moffat function are used to fit the background (or wings of other objects).\footnote{Similar arguments apply to the S\'ersic index of a galaxy.} For very large cutout sizes, the star may even become insignificant relative to the background so that the fitting does not focus on the main object of interest anymore (and instead the Moffat function is used primarily to even out background inhomogeneities). Fitting galaxies introduces a wealth of additional problems if their outskirts do not follow S\'ersic functions anymore, e.g. due to disk breaks, flares, rings and similar (cf. Section~\ref{sec:postprocessing}). 

In summary, we opted to use the \texttt{ProFound} estimated sky with default parameters except for the box size of 100\arcsec; but not to perform an additional \texttt{ProFit} background fit. This in turn allows to use tight segmentation maps for fitting. These choices were made before the first DMU release and hence did not change between \texttt{v01} and \texttt{v04}.

\subsubsection{Pixel correlations}

To detect problems in the sky subtraction and object masking, an analysis of correlations between pixels can be useful. \texttt{ProFound} offers a set of functions to achieve this via a Fast Fourier Transform (FFT) of the normalised image (each pixel divided by its error) after sky subtraction and object masking. This FFT image gives information on the pixel correlations on different spatial scales. In the ideal case, this should result in pure uncorrelated white Gaussian noise. The idea then is that power on small scales represents undetected sources, while on larger scales it captures complex features of the sky background. 

For KiDS images, we expect some residual correlation on small spatial scales. This is mainly because before stacking, all KiDS frames have been re-gridded onto the same pixel scale of $0\farcs$2 from their native scales of $\sim$\,0$\farcs$21. This ensures that the tiles are pixel-matched across all four bands with the axes aligned in RA and Dec (see Sections~\ref{sec:kids}, \ref{sec:preparatorysteps} and \citealt{Kuijken2019}). However, it will also introduce correlations on very small scales. In addition, there can be intrinsic instrumental correlations, typically on the scale of a few pixels. Investigating these in detail goes beyond the scope of this work. The PSF, although it is several pixels across, is not expected to play a significant role for seeing-limited data such as KiDS since most noise is ``added" at a later stage. For perfect sky subtraction and object masking, we hence expect correlations to reach zero beyond the scale of a few pixels. In practice, we might expect a small positive excess resulting from very faint, individually undetectable objects. 

\begin{figure}
	\includegraphics[width=0.5\textwidth]{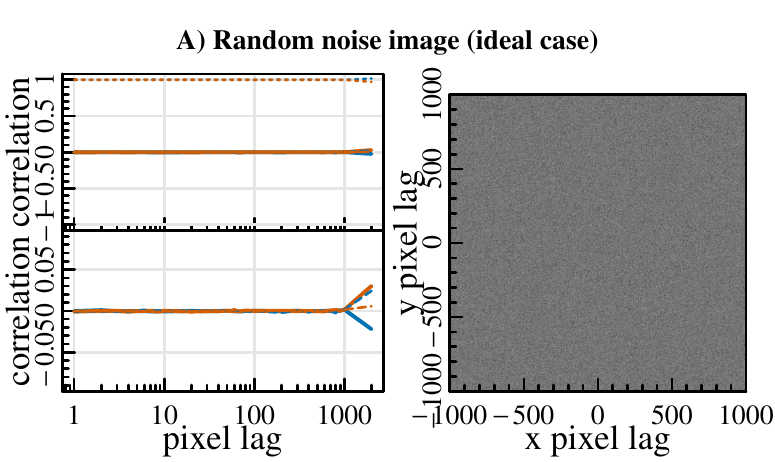}
	\includegraphics[width=0.5\textwidth]{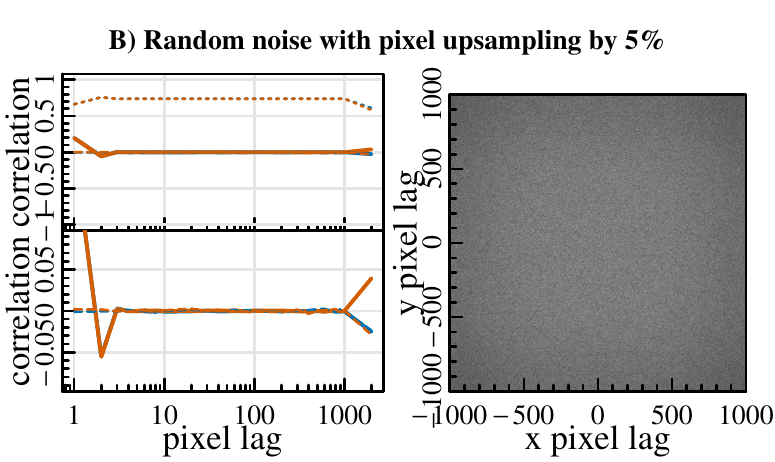}
	\includegraphics[width=0.5\textwidth]{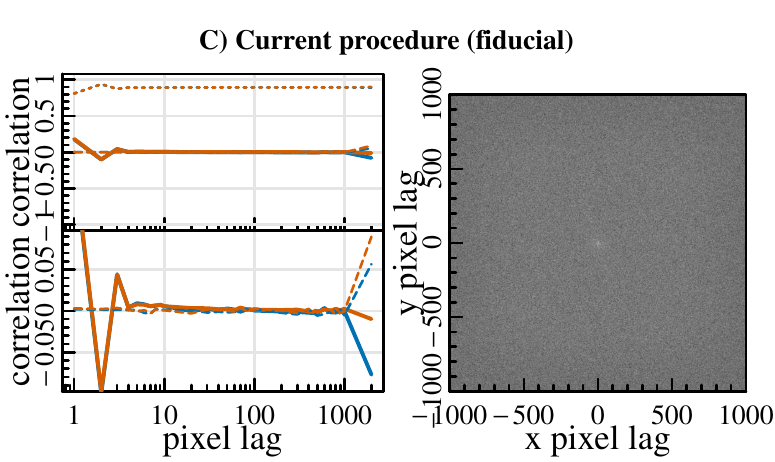}
	\includegraphics[width=0.5\textwidth]{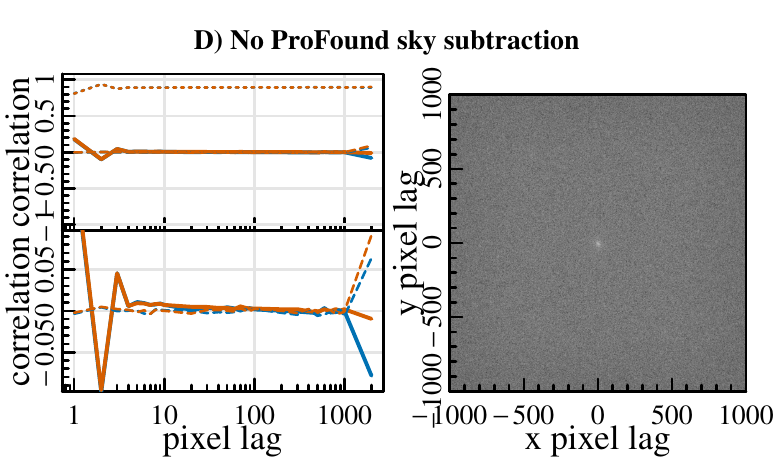}
	\includegraphics[width=0.5\textwidth]{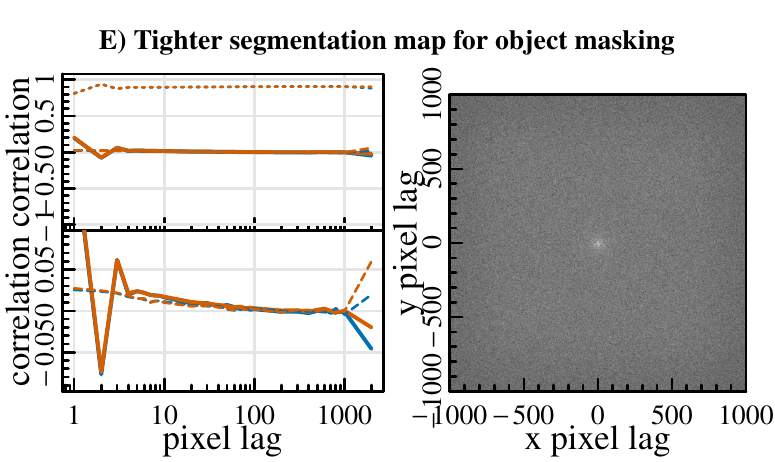}
	\includegraphics[width=0.5\textwidth]{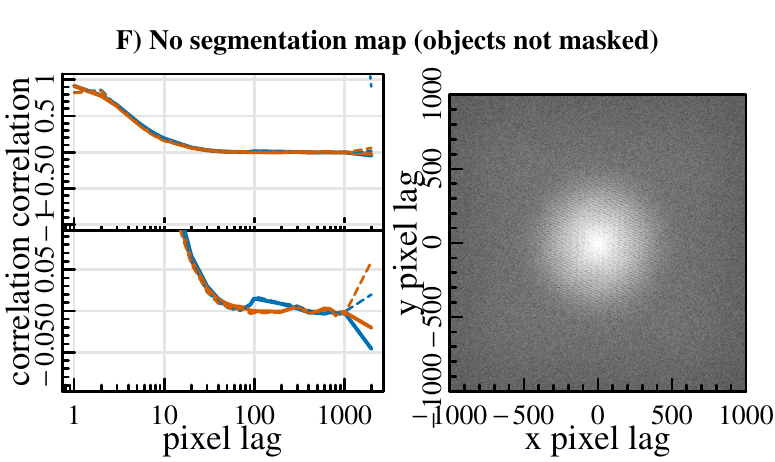}
	\includegraphics[width=\textwidth]{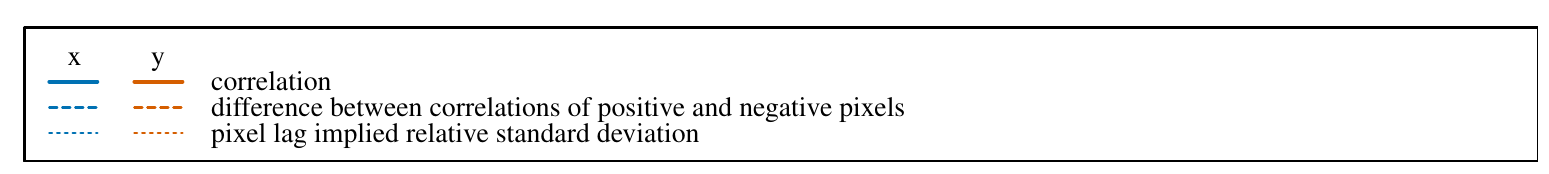}
    \caption{Correlations between pixels for \textbf{A)} a random noise image, \textbf{B)} a random noise image with pixel upsampling by 5\,\%, \textbf{C)} a real image around the example galaxy 396740 (the same as in Figure~\ref{fig:exampleseg} for comparison) treated as in the final pipeline, \textbf{D)} the same real image as C), but leaving out the \texttt{ProFound} sky subtraction during treatment, \textbf{E)} the same real image as C) but using tighter segmentation maps for object masking (coloured instead of grey contours in Figure~\ref{fig:exampleseg}) and \textbf{F)} the same real image as C) without any object masking. The left panels always show the correlogram, i.e. pixel correlation versus pixel lag in $x$ and $y$, with the meaning of the different lines indicated by the legend at the bottom of the plot. The bottom panel is a zoom (on the $y$-axis) into the top panel. The images to the right of the correlograms show the corresponding FFT image, scaled such that excess power is white (phase ignored, asinh scaling) and small scales are at the centre of the image.}
    \label{fig:corplots}
\end{figure}

Figure~\ref{fig:corplots} shows the main diagnostic outputs produced by the \texttt{profoundPixelCorrelation} function for six different input images (A-F). In the left panel of each set of plots, we show a so-called correlogram: pixel correlation as a function of $x$ (RA) and $y$ (Dec) pixel lag (solid blue and red lines respectively). In addition, the dashed lines show the difference between the correlations of positive pixels and those of negative pixels (also in $x$ and $y$); while the dotted lines indicate the pixel lag implied relative standard deviation. All correlograms are shown once with a $y$-axis range of -1 to 1 (top of the left panels) and once as a zoom into the range -0.1 to 0.1 (bottom). The $x$-axis is the same for all correlograms, ranging from a lag of 1 pixel to 2000 (the image size) on a logarithmic scale. Generally, a positive excess in this plot indicates that object detection or sky subtraction are not aggressive enough, while systematically negative correlations suggest that the image contains large pools of negative pixels, hinting towards problems in the sky subtraction. To the right of each set of correlograms, we show the corresponding two-dimensional FFT image, where white colour means a relative excess power (phase ignored) and small scales are centrally located. Note the linear axes scales compared to the logarithmic $x$-axis of the correlogram; while conversely the colour scale now uses an asinh stretching instead of the linear $y$-axis of the correlogram. While the quantitative interpretation of these images is less straight-forward, they can be used to detect correlations at different scales and in different directions at least qualitatively. In both cases, masked pixels (including those identified as belonging to objects) are excluded from the analysis by \texttt{ProFound}. 

Starting from the top left, the first set of panels A), show the result obtained for a generated image of pure Gaussian noise with a mean of zero and standard deviation of one. As expected, the pixel correlations are zero on all scales, with the standard deviation being one. The exception is the last data point (at 2000 pixels), where the lines diverge due to unrealiable low number statistics once we reach the image size. We hence only consider values up to 1000\,pix on the $x$-axis for all correlograms. Correspondingly, the FFT image contains no structure (showing that the random number generator is working well). This respresents the ideal case which we would like to achieve for our real images (after sky subtraction and object masking), but are likely to never fully reach given the limitations explained above.

Panels B) show the first step of complications that is known to exist in real (KiDS) images: pixel correlations due to the resampling of the raw images before stacking. For this analysis, we have created a random noise image of slightly smaller dimension (1905\,$\times$\,1905\,pix$^2$), which we have then upsampled onto the usual 2001\,$\times$\,2001\,pix$^2$ grid and subsequently passed through the FFT analysis. This roughly corresponds to the upsampling performed by the KiDS team from a native instrumental pixel scale of $\sim$\,0$\farcs$21 to 0$\farcs$2 in the tiles. This introduces correlations on small scales (up to $\sim$\,3\,pix), which are clearly visible in the correlograms. Since positive and negative pixels are equally affected by the upsampling, the difference between positive and negative correlations (dashed lines) stay zero. Note, however, that since we have not corrected the weight map (which is uniformly one for a random image of standard deviation one), the pixel lag implied standard deviation is now below one (i.e. the true standard deviation of the upsampled pixels is smaller than one). 

Panels C) show the results obtained for one of our real images after sky subtraction and object masking with our final pipeline. We choose the same example galaxy as that shown in Figure~\ref{fig:exampleseg}, to allow a direct comparison with the segmentation maps. The most striking feature here is the characteristic pattern of correlations at very small scales, which - judging from panels B) - we mostly attribute to the upsampling (and possibly intrinsic instrumental effects on small scales, which we did not investigate). These features consistently persist for all images (although we only show one example here). Beyond a scale of $\sim$\,3\,pix, the correlation drops to zero rapidly, although a small excess can be seen at scales up to 10\,pix, maybe even up to 100\,pix. These correlations are also visible as a faint white dot at the centre of the FFT image. However, there seem to be no asymmetries in our analysis (no structure in the FFT, $x$ and $y$ correlations always the same, negative and positive pixels affected equally). Note that the standard deviation is also slightly below one here, implying that the true standard deviation of background pixels (after sky subtraction and object masking) is slightly smaller than that implied by the weight maps provided by KiDS. We discuss this further in Section~\ref{sec:otherprepworkchoices}.

Panels D) show the same as panels C), only that in this case, we do not perform the \texttt{ProFound} sky subtraction. The plots are nearly indistinguishable from those in panels C), demonstrating that pixel correlations are neither introduced nor removed by our sky subtraction procedure and instead correlations are primarily caused by remaining undetected objects. Since the box size of the sky substraction is 100\,\arcsec\ or 500\,pix (Section~\ref{sec:preparatorysteps}), we only expect effects on large scales here. 

We test the effect of undetected (wings of) objects in panels E), where we use a tighter segmentation map for object masking. Instead of the segmentation map generated for the background estimation, we use that produced for fitting the galaxies (cf. Section~\ref{sec:preparatorysteps} and Figure~\ref{fig:exampleseg}: here, we use the coloured contours instead of the grey ones for object masking). This goes less deep in both object detection and segment dilation. In other words, we mask less of the faintest objects; and the regions masked around each (bright or faint) object are smaller. This results in a clear positive excess at scales up to $\sim$\,100\,pix. The dashed lines also deviate from zero at these scales, indicating that positive pixels are more correlated on average than negative pixels (since the undetected sources of light are all positive). The FFT shows a corresponding white dot at its centre, but no additional two-dimensional structure. 

Finally, panels F) show the pathological case of ignoring the segmentation map alltogether, i.e. no objects are masked (except for those that are saturated and hence included in the KiDS masks already at an earlier stage). The correlation of pixels reaches values near unity for small scales and is substantial up to scales of 100\,pix or even beyond. The dashed lines closely follow the solid lines, indicating that this is entirely caused by correlations between positive pixels (i.e. there are no negative flux objects in the image). The asymmetry between $x$ and $y$ correlations at scales of $100-200$\,pix is unique to this particular image and not generally visible for other example galaxies. The corresponding FFT image also shows a large power excess at scales up to several hundred pixels. 

In conclusion, our procedure for background subtraction and object masking seems to remove the vast majority of correlations between pixels, with only the characteristic features of up-sampled pixels remaining at very small scales; and a slight excess caused by faint, individually undetectable objects at slightly larger scales. It is, however, important to use the very agressive, deep and highly dilated object mask that we generate for accurate background subtraction; rather than the tighter fitting segmentation map (for details on the latter, see Section~\ref{sec:segchoices}). For this reason, we eventually decided to use two separate segmentation maps for these two processes, to better accommodate the different aims that we try to achieve with them (cf. Section~\ref{sec:preparatorysteps}). This change came into effect with \texttt{v03} of the \texttt{BDDecomp} DMU. 

In addition to the above, it should be mentioned that \texttt{ProFound} offers a function to improve an already-estimated sky on the basis of an FFT analysis (\texttt{profoundSkySplitFFT}). The idea is that, on large scales, excess power captures complex features of the sky background that cannot be described by the bilinear or bicubic interpolation between boxes that is performed by \texttt{profoundProFound}. Hence, this part of the FFT image can be used to refine the sky estimate; the ``new" (refined) sky is returned by the function. In early versions of the pipeline (up to the release of \texttt{BDDecomp v02}), we have used this refined sky. However, starting from \texttt{v03} we reverted to using the ``original" sky estimated by \texttt{profoundProFound} since it turned out to be more robust especially in crowded fields and/or in the presence of nearby masked objects, image edges and artifacts. On the other hand, for ``well-behaved" fields there is very little structure to be removed on large scales (see Figure~\ref{fig:corplots}). Hence for most images, this additional step of sky refinement had very little effect, while for some fields it had adverse effects (also confirmed by simulations). For our large automated analysis with varying image properties and intrinsically relatively flat sky, we therefore found \texttt{profoundSkySplitFFT} to be unsuitable. 

\subsection{PSF estimation details}
\label{sec:psfdetails}
The last major step in the preparatory work pipeline is the PSF estimation. An accurate and precise estimate of the PSF is crucial to the success of the galaxy fitting. Any deviations of the model PSF from the truth will introduce systematic uncertainties in the galaxy parameters that are not easily accounted for. Unfortunately, however, the PSF estimation is not straight-forward as there are many uncertainties involved and choices to make: how to identify and select stars suitable for PSF estimation, which function and algorithm to fit them with (or whether to fit them in the first place), how many stars to use and how to combine the different estimates into a final model PSF for each galaxy of interest.

In view of these difficulties, we first contacted the KiDS team to ask whether they would be able to provide PSFs even though they are not publicly released. The team were willing to provide PSFs in the form of shapelet functions for each tile. However, these are not the PSFs that the team use for their own weak lensing analyses, since those are performed on individual pawprints rather than stacked tiles. They did therefore not undergo a detailed quality control. A brief comparison with a selection of stars in a number of test tiles showed that the shapelets did not fully capture the extended wings of some PSFs. In addition, we already anticipated the addition of VIKING data, for which there are no such shapelets available. We therefore decided that it would be best to devise our own method of PSF estimation to obtain consistent estimates in all bands and have full control over their quality. 

Since PSF estimation using \texttt{ProFound} and \texttt{ProFit} has not previously been performed, there were no established procedures for us to follow. Consequently, our treatment was developed as a mixture between theoretical considerations, procedures that have proven to work well with other software and a considerable fraction of trial and error combined with visual inspection of the results. A very brief overview of the resulting process is given in Section~\ref{sec:preparatorysteps}. Here, we describe the treatment in much more detail, explaining some of the reasons for our choices and other possibilities that we explored. We finish with an overview of the PSF quality control and its effects on the galaxy fitting.

Many of the details changed during the pipeline evolution. For example, the star candidate selection process depends on certain cuts that need to be adjusted when making significant changes to the way segments are defined (i.e. between \texttt{v02} and \texttt{v03} of the \texttt{BDDecomp} DMU). Since these cuts were mostly tuned by visual inspection and also changed frequently between test runs, we do not give detailed justifications for each pipeline version and instead focus on the status quo in \texttt{v04} (which is also what \citealt{Casura2022} is based on). We note that, while the detailed numbers may have changed several times, the overall procedure remained the same.

\subsubsection{Star candidate selection} 

The first decision to make is whether to estimate PSFs separately for each individual object of interest, or fit all stars in a large region (e.g. an entire KiDS tile) at once and then interpolate the solution for each position of interest. Both approaches have their advantages and disadvantages. We took the first approach, whereby for each position of interest we independently estimate a PSF. This has the benefits that local PSF variations are captured without worrying about interpolation uncertainties. It also allows to flexibly change or improve individual PSFs without having to re-fit all stars in the field; and it allows to use the independent PSF estimates of neighbouring galaxies for quality control (see below). In addition, it is straightforward to handle computationally since it can just be done along with the local background subtraction and other preparatory work that is run for each galaxy. 

Disadvantages of this approach are that each individual PSF estimate is based on a relatively low number of stars (to keep it local), potentially introducing more noise; and that there will be a number of galaxies with missing PSFs if no suitable nearby stars are available (for example because large areas in the vicinity of the galaxy are masked). However, the latter only affects less than 1\,\% of galaxies (see Table~\ref{tab:results} in Section~\ref{sec:statistics}) the majority of which tend to be in regions of decreased data quality (near bright, saturated stars, image edges or artifacts). To decrease the noise on our PSF estimates, we introduce a fitting step with \texttt{ProFit} instead of - as is sometimes done in other software - just averaging a large number of star cutouts. We find the fitted Moffat parameters for stars near each other to be very consistent (see below). 

In summary, we perform PSF estimation on the single-band 400\arcsec\,$\times$\,400\arcsec\ cutouts with associated sigma (error) maps and masks, after background subtraction and image segmentation. For \texttt{v04} of the pipeline, we use the segments from the stacked $gri$ images, but re-calculate the segment statistics for each individual band. For the test runs which this section is based on, we only used $r$-band images. 

From the segmentation statistics returned by \texttt{profoundProFound}, we select relatively round and isolated objects as follows: 
\begin{itemize}
\item objects that do not touch other segments, masked regions or image edges (edge fraction\,=\,1)
\item objects with a regular boundary geometry (edge excess <\,1) 
\item objects with an axial ratio (minor/major axis) larger than 0.5
\item objects which were not flagged as possibly spurious
\end{itemize}
These are trivial selections based on theoretical considerations to remove the segments with least reliable statistics; with the exception of the axial ratio. For the latter, we originally used an axial ratio cut based on the PSF ellipticity given in the header of KiDS science tiles. The header value refers to the average PSF ellipticity across the entire tile, which is typically very low (corresponding to axial ratios $\gtrsim$\,0.9). However, we found that the PSF can vary strongly locally across the tile and is often elongated significantly especially towards the tile corners (see Figure~\ref{fig:tileoverview}), with axial ratios as low as $\sim$\,0.6. Due to these local variations, a cut based on the average PSF ellipticity proved to be unsuitable and we instead decided to simply use a fixed cut of 0.5 on the axial ratio to remove very elongated objects from further analysis.

\begin{figure}[ht!]
\begin{center}
	\includegraphics[width=0.8\textwidth]{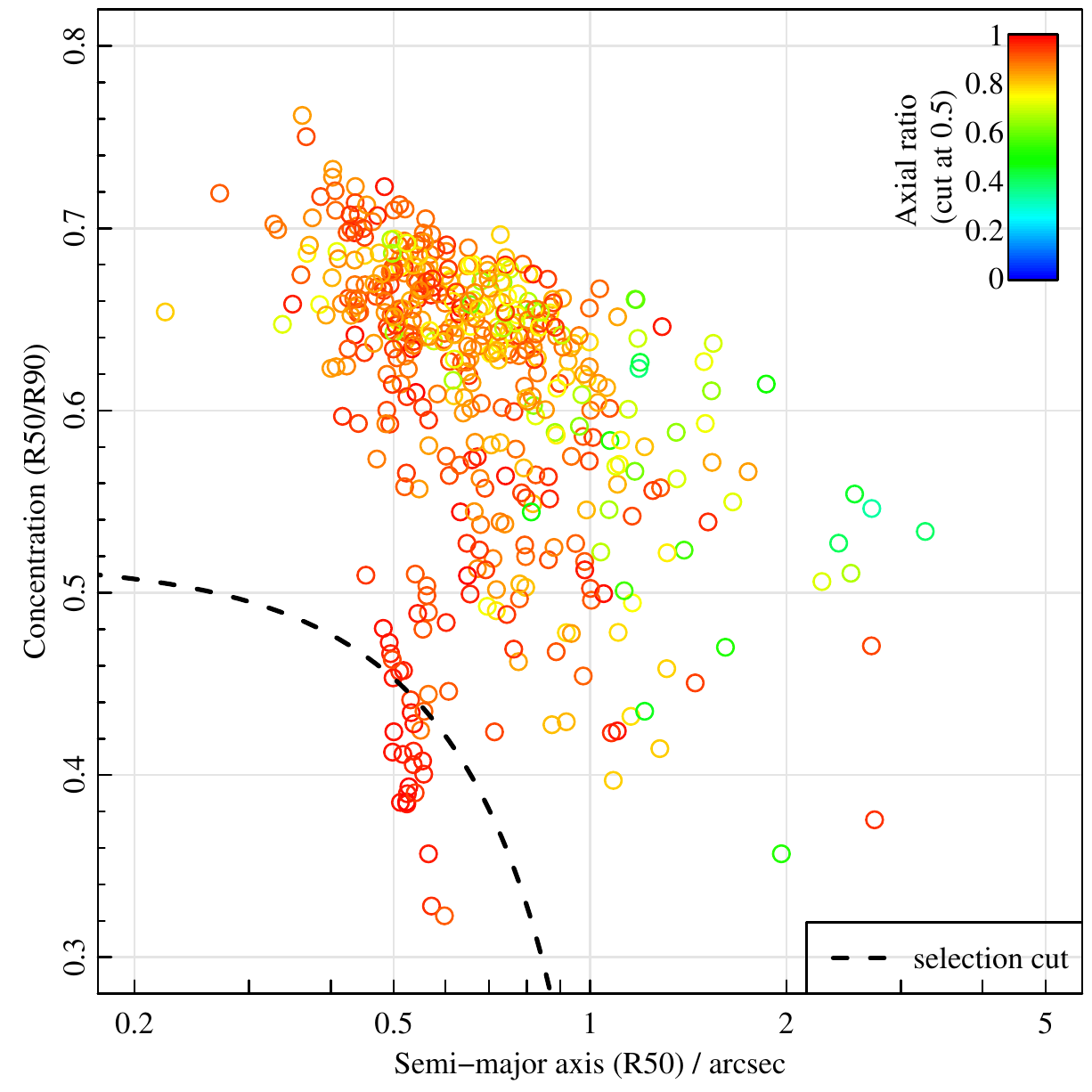}
    \caption{A diagnostic plot of the selection of star candidates for example galaxy 396740. Coloured points show the concentration \texttt{R50/R90} against \texttt{R50} (semi-major axis containing half of the total flux) for all segments in the 400\arcsec\,$\times$\,400\arcsec\ cutout around the galaxy; coloured by axial ratio. The joint selection cut in \texttt{R50} and \texttt{R50/R90} is indicated with a dashed line.}
    \label{fig:examplepsfseg}
\end{center}
\end{figure}

Of these relatively round and isolated objects, a given fraction (depending on the depth of the image and source extraction, typically 4-8\,\%) are identified as star candidates via a joint cut in \texttt{R50} (semi-major axis containing half the flux) and the concentration (\texttt{R50/R90}, where \texttt{R90} is the semi-major axis containing 90\,\% of the total flux). A diagnostic plot of this step is returned with an example shown in Figure~\ref{fig:examplepsfseg}, again for galaxy 396740. We chose this cut based on the notion that we would expect stars to be small (i.e. low \texttt{R50}) and highly concentrated (i.e. low \texttt{R50/R90}). The plot of these two quantities coloured by axial ratio (example in Figure~\ref{fig:examplepsfseg}) typically shows a rather distinct population of objects that are constant in (small) \texttt{R50} and (high) axial ratio, with low \texttt{R50/R90}. To capture most of this population of objects without including too many others, we then empirically devised the joint cut in \texttt{R50} and \texttt{R50/R90} as indicated in Figure~\ref{fig:examplepsfseg}. 

We found the best results using a fixed percentage of objects to accommodate both crowded fields and those with only few objects and/or large masked areas. The chosen percentage depends on the depth of the source extraction (i.e. the \texttt{skycut} value in \texttt{profoundProFound}) and the depth of the image. For the stacked $gri$ KiDS images and a \texttt{skycut} value of 2, we found that classifying 4\,\% of all segments as star candidates delivers reasonable results for the vast majority of cutouts. While this does not always capture all stars in the frame, it is usually enough to allow for a robust PSF estimate. Conversely, for some frames it may include a few objects that are not actually stars, but as long as they are not the majority, they will be screened out at a later stage in the PSF estimation when we reject outliers.  

Note that since the number of segments and the fraction of how many of those are stars are determined by the segmentation map alone, this value can be used for images of varying depth as long as they use the same segmentation map. In other words, even though for example the $i$-band images are shallower than the $gri$ stacks; and the segment properties (including \texttt{R50} and \texttt{R90}) are re-calculated in the $i$-band, the 4\,\% selection cut is appropriate because we originally defined the segments on the $gri$ stacked image. Hence from \texttt{v03} of the pipeline onwards, when we started using the $gri$ stacks to define common segments for all bands and fixed \texttt{skycut} to two, we always use the 4\,\% selection cut (0.02\,$\times$\,\texttt{skycut}). Earlier (single-band) versions of the pipeline use 0.03\,$\times$\,\texttt{skycut} for $g$ and $r$, 0.04\,$\times$\,\texttt{skycut} for $i$ and 0.06\,$\times$\,\texttt{skycut} for $u$. 

Around each of these star candidates, a smaller cutout is taken (side length equal to 2\,$\times$\,\texttt{R100}; 6\,$\times$\,\texttt{R100} if a background is to be fit as well) and a subsample selected:
\begin{itemize}
\item objects brighter than the 5-sigma point source detection limit and fainter than the saturation limit (both taken from the headers of the corresponding KiDS tile)
\item objects where less than 10\,\% of pixels in the cutout belong to other segments (rising to 30\,\% for the larger cutouts when fitting a background) 
\item objects where the star cutout does not overlap with the edge of the large cutout
\item objects with a positive sum of the cutout (excluding poorly background-subtracted and/or purely noise-dominated objects)
\end{itemize}
The cutout size is large enough to include the entire star segment, which are the pixels that we use for fitting the star. When additionally fitting a background, we need larger cutouts to include a sufficient number of sky pixels (see Section~\ref{sec:backgroundstudies}). This only affects a number of test runs related to Section~\ref{sec:backgroundstudies}; none of the full runs used additional background fitting. The remaining selection criteria then ensure that the stars are neither saturated nor too faint to produce a reliable fit; and that they are isolated and fully contained within the image. We also experimented with constraining the magnitude range further (e.g. fitting only stars that are at least 2\,mag brighter than the 5-sigma point source detection limit), but found that this does not significantly improve the fitting quality and instead can limit the number of stars too much for fields in which only few star candidates are available. 

The order in which we apply these selection cuts was essentially defined by trial and error: the cut in size and concentration (Figure~\ref{fig:examplepsfseg}) is best applied after a pre-selection of isolated relatively round objects, since otherwise the number of stars in crowded fields is reduced too much. This is because in these fields, the fraction of stars is usually higher, which we can at least partially account for by the pre-selection of round objects. However, the magnitude cuts should be applied after the concentration and size cuts since otherwise the assumed fraction of stars in the frame (4\,\%) needs to be carefully tuned to the chosen magnitude cuts. The reason here is that stars tend to belong to the brightest objects in the frames, so for example cutting 2\,mag off the faint end will increase the fraction of stars substantially. 

If no stars pass the above criteria (usually this only happens when the vast majority of the large cutout is masked), then a default perfectly round PSF is created with the FWHM taken from the KiDS header and a Moffat concentration index of 2.1 (the median of all measured $r$-band concentrations). If this is the case, a warning is given and an appropriate flag value added to the PSF quality flag, such that galaxies for which only this default (dummy) PSF exists are skipped during the fitting procedure. If more than 32 star candidates pass the above criteria, we limit the fitting to the 32 objects closest to the galaxy in order to save computational time. 

Ideally, we would use only those stars that are on the same detector chip as the galaxy of interest; or at least stars that have the same number of dithers as the galaxy. However, the KiDS instrumental setup and dithering pattern result in frequent changes of the number of dithers across a tile (cf. the darker stripes in the weight map of Figure~\ref{fig:exampleseg}). Trying to identify the corresponding stars is hence non-trivial and usually results in only very few star candidates being left. Moreover, different parts of the galaxy itself are often covered by different numbers of dithers, which would - to be consistent - then require the galaxy to be fit with different PSFs for different regions. These difficulties could be avoided by working at the pawprint level (like we do for VIKING, Section~\ref{sec:pipelineupdates}) instead of using the stacked science tiles. For reasons outlined in Section~\ref{sec:otherprepworkchoices}, we however decided to use the science tiles for KiDS. Since all dithers of a tile were taken in close temporal succession, they should have very similar PSFs in general. Therefore the stacking and the changing number of dithers across the field will not have a large effect on the PSF. Indeed we do find the PSF to vary only slowly across tiles, with no abrupt changes (cf. Figure~\ref{fig:tileoverview}).

\subsubsection{Star fitting}

\begin{figure}
\begin{center}
	\includegraphics[width=0.8\textwidth]{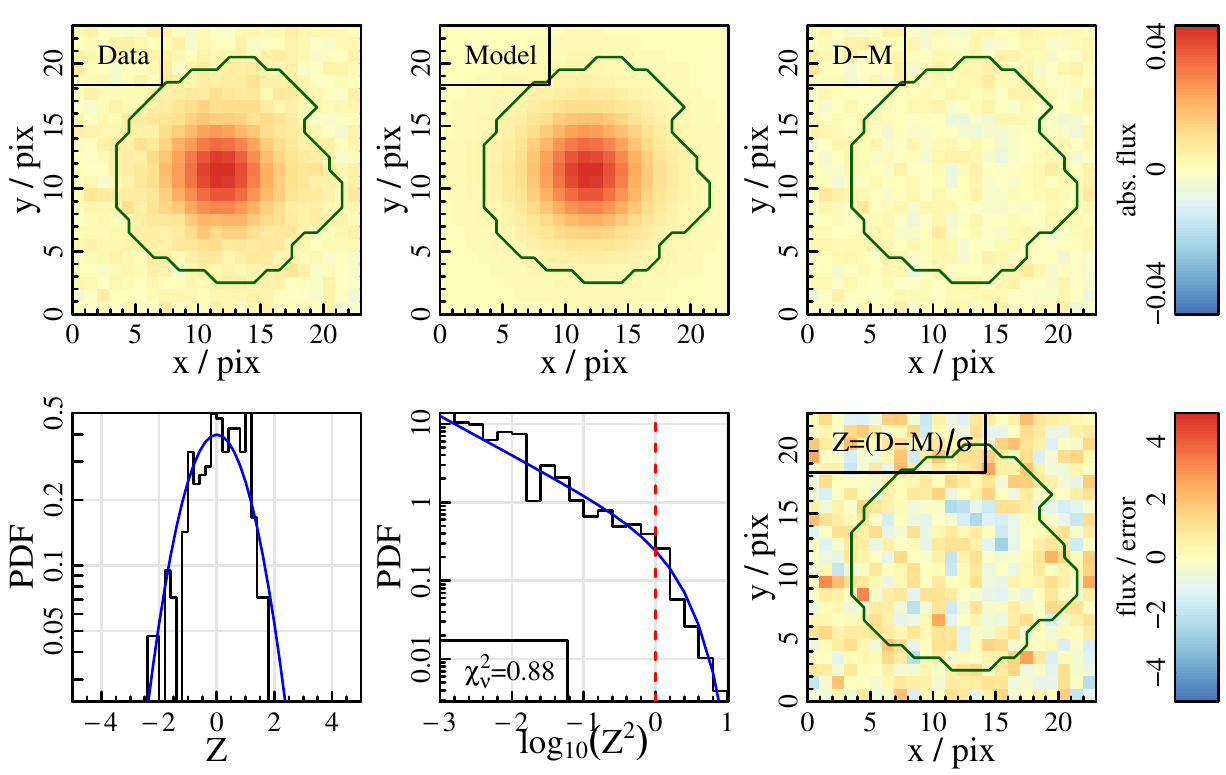}
    \caption{The Moffat function fit to an example star near galaxy 396740. Panels are the same as the top two rows in Figure~\ref{fig:examplefit}.}
    \label{fig:examplestarfit}
\end{center}
\end{figure}

In the next step, the star candidate cutouts are normalised to a magnitude of 0, masked appropriately and fitted with a Moffat function using \texttt{ProFit} (see fitting details in Section~\ref{sec:preparatorysteps}). We choose a Normal likelihood function since we need near-perfect fits for adequate PSF representation and so the residuals are expected to be distributed Normally. The normalisation of the PSF to a magnitude of 0 brings the sum of the cutout close to 1 (for the KiDS magnitude zeropoint of 0) and so individual pixels mostly have values around a few $\times$\,$10^{-4}$. We found this to be close to the optimum in terms of fit accuracy when using the default tolerance levels for convergence in \texttt{optim} (it is also where the fitting time is longest). If no background is fitted, only the segment around the star is used for fitting; with a background fit the entire cutout is used but with all objects (except the segment of interest) masked. The diagnostic plot of the fit returned by \texttt{ProFit} is saved for each star and an example shown in Figure~\ref{fig:examplestarfit}. 

As briefly mentioned above, we introduce this fitting step (instead of a purely empirical estimate based on, e.g., simply averaging appropriately scaled cutouts around suitable stars) since it reduces the noise in our PSF estimate. This is particularly important given that we only use a relatively small number of stars to obtain the model PSF. Additionally, it gives an opportunity to easily remove outliers, so the star candidate selection process need not be perfect. It does, however, require finding a modelling function that can adequately represent the PSF, since otherwise one will introduce systematic biases. We chose to fit a \citet{Moffat1969} function since that is known to parameterise telescope PSFs well and, in fact, was devised for that exact purpose.

As one can see, a Moffat function is a perfect fit for the star shown in Figure~\ref{fig:examplestarfit}. The normalised residual $Z$ (bottom right panel) shows pure noise and its distribution follows a Gaussian function (bottom left panel), while $Z^2$ follows a chi-squared function as it should (bottom middle panel). The fact that the reduced chi-squared ($\chi^2_\nu$ given in the bottom middle panel) is below the ideal value of one is because the KiDS weight maps tend to slightly overestimate the true pixel error, see Section~\ref{sec:otherprepworkchoices}. 

\begin{figure}[t!]
\begin{center}
	\includegraphics[width=0.8\textwidth]{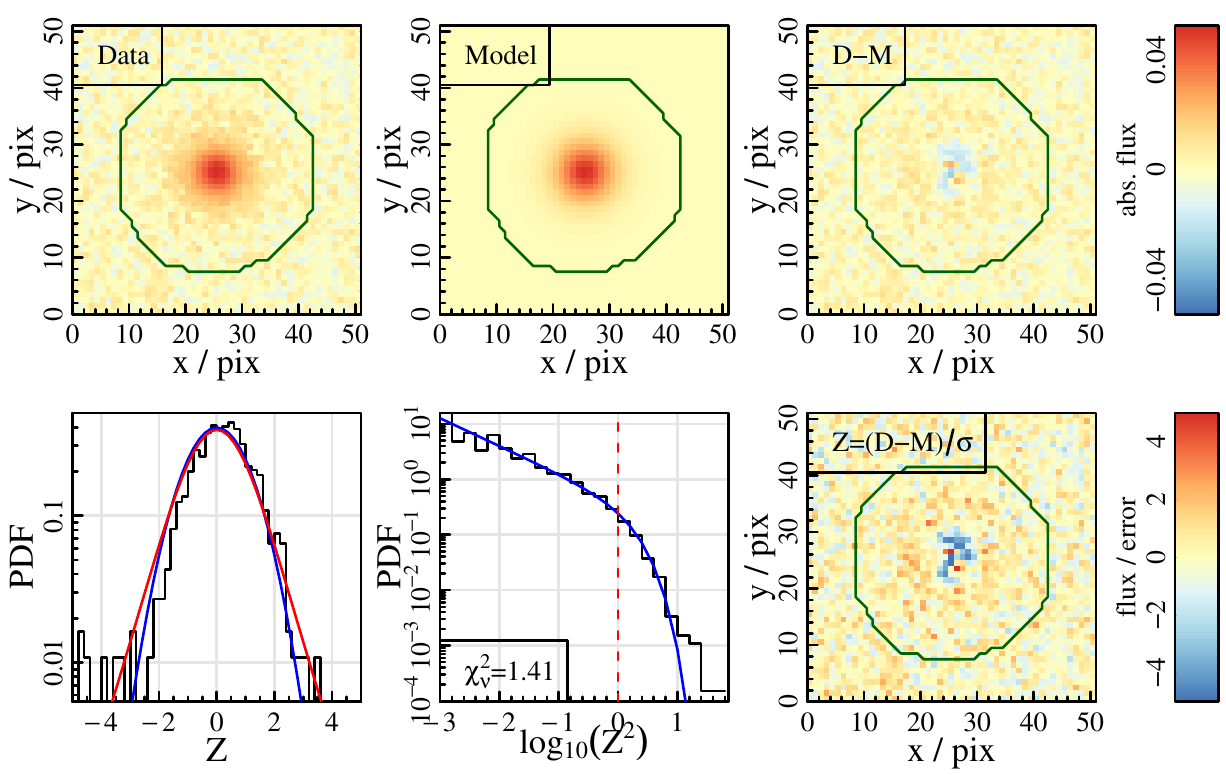}
	\includegraphics[width=0.8\textwidth]{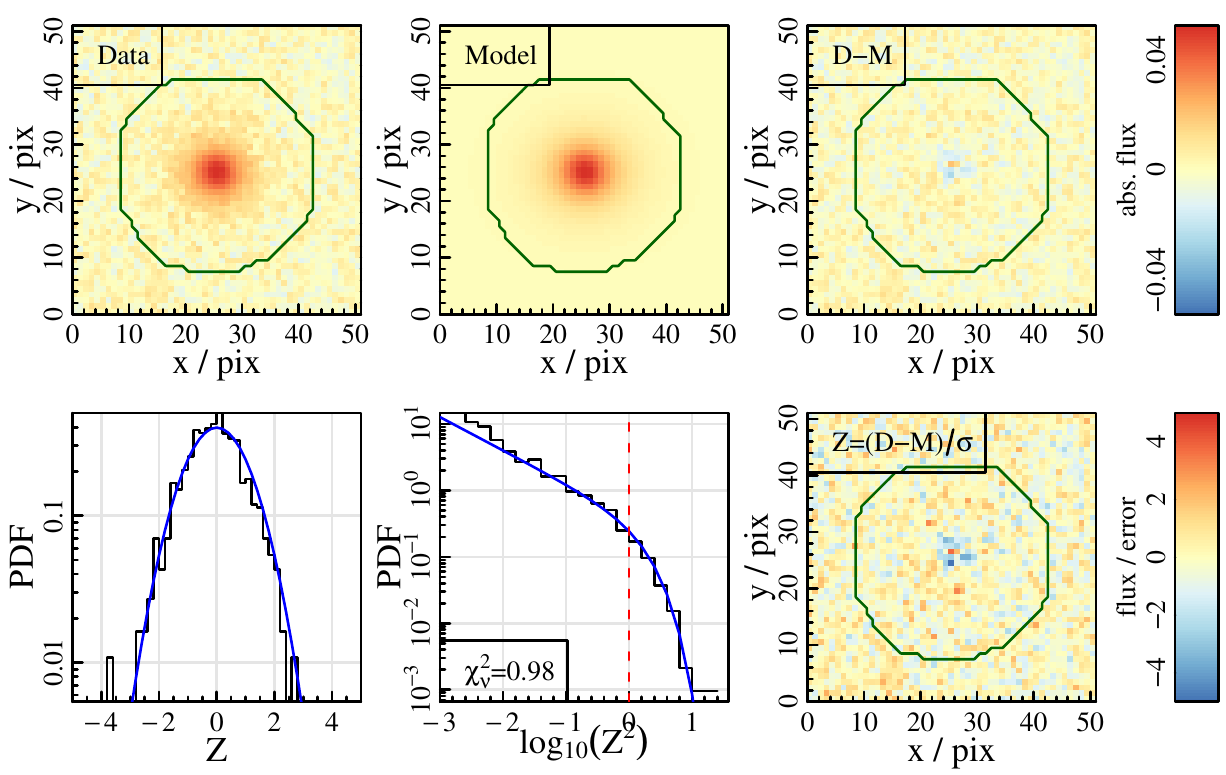}
    \caption{\textbf{Top two rows}: the Moffat function fit to a relatively bright example star near galaxy 7623. Panels are the same as the top two rows in Figure~\ref{fig:examplefit}. \textbf{Bottom two rows}: the double Moffat function fit to the same star.}
    \label{fig:examplestarfitbright}
\end{center}
\end{figure}

Most isolated stars are represented very well by Moffat functions (like the example shown), with two exceptions. First, very bright stars sometimes show residuals at their cores, which can be systematic in the sense that all bright stars in a field show similar residuals. This indicates that a Moffat function is not able to capture the true shape of the PSF at the cores of bright stars. The residual pattern usually shows a red core surrounded by a blue ring or arc, often not entirely symmetric, see the example in the top panels of  Figure~\ref{fig:examplestarfitbright}.
Nonetheless, the parameters fitted to these bright stars are indistinguishable from the parameters fitted to fainter stars of the same field. In addition, these bright stars reach surface brightnesses close to the saturation limit, where CCD detectors are known to have a non-linear response function. It is therefore unclear whether the residuals are due to actual deviations of the seeing PSF from a Moffat function or are simply due to the CCD response function. 

The former case would mean that we introduce a systematic bias into our galaxy parameters from systematically wrong PSFs, while the latter would not affect the galaxy fitting since the galaxies in our sample have much lower surface brightnesses, typically far from the saturation regime. While we cannot be sure, we assumed that the latter is the case and we need not worry about the brightest stars since they lie in a different regime of the dynamic range of the CCD than our galaxies. This belief was formed mainly by the observation that stars of intermediate brightness typically show no such systematic residuals (i.e. the residuals are not just weaker but disappear completely), while at the same time we obtain very consistent results for the fitting parameters across the stars of a field. We therefore focused on stars of intermediate brightness for which we can obtain perfect fits with a Moffat function. 

\begin{figure}[t!]
\begin{center}
	\includegraphics[width=0.8\textwidth]{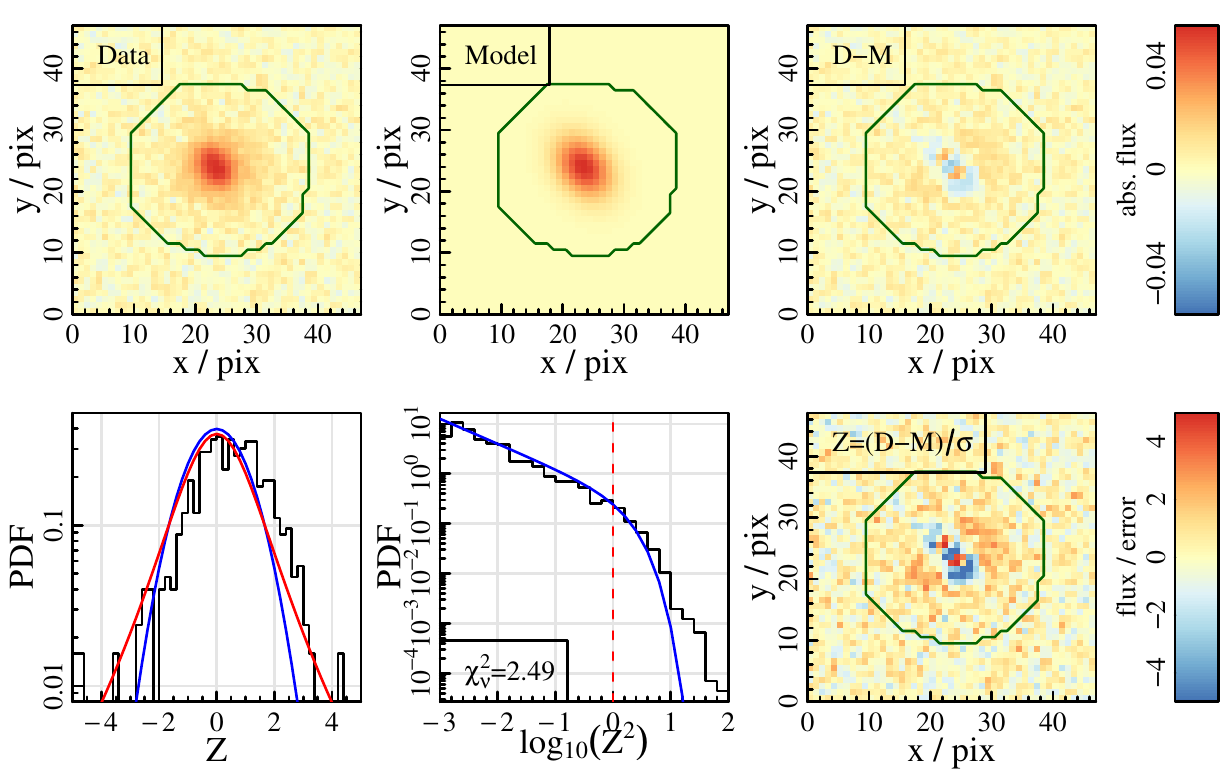}
	\includegraphics[width=0.8\textwidth]{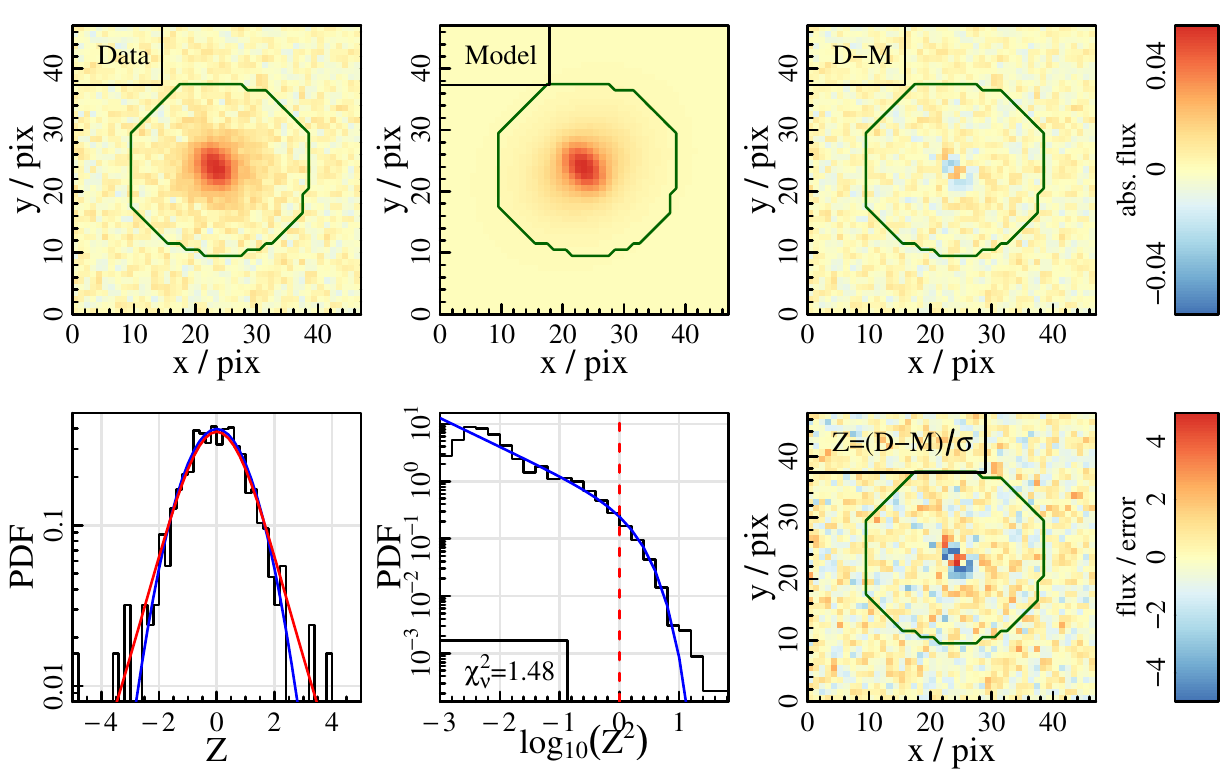}
    \caption{\textbf{Top two rows}: the Moffat function fit to an elongated, egg-shaped example star near galaxy 105600. Panels are the same as the top two rows in Figure~\ref{fig:examplefit}. \textbf{Bottom two rows}: the double Moffat function fit to the same star.}
    \label{fig:examplestarfitegg}
\end{center}
\end{figure}

The second case where we see clear systematic residuals in the star fits is for very elongated objects near the corners of KiDS tiles. An example is shown in the top two rows of Figure~\ref{fig:examplestarfitegg}. 
Here, there are clear asymmetries in the residuals for stars of all brightnesses, which show that the true PSF is not perfectly elliptical but instead egg-shaped, with the wider part pointing towards the tile corner. Using the perfectly elliptical Moffat PSF will introduce a systematic uncertainty for the parameters of the corresponding galaxy. In particular we might bias the position angle and axial ratio, which are most severly influenced by incorrect PSFs (see Section~\ref{sec:systematics}). Due to the symmetry of the problem, however, only galaxies with a major axis oriented at approximately 45\degr\ with respect to the PSF elongation will be significantly affected. In addition, not all tile corners show elongated PSFs and for those that do the region is usually limited to the very outer edges of a tile. This also means that there are often multiple matches (i.e. the galaxy belongs to the overlap sample), not all of which are necessarily compromised. Therefore, the overall number of galaxies affected is low and limited to those objects suffering from the worst data quality in many other respects, too. 

In an attempt to improve the fits to very bright and/or elongated stars, we experimented with fitting double Moffat functions instead of just a single Moffat function to each star (following a suggestion by Dan Taranu, who uses this in the \texttt{AllStarFit} package). However, the success was limited: although the fit does improve owing to the increased number of parameters, in both cases residuals remained (see the bottom panels in Figures~\ref{fig:examplestarfitbright} and ~\ref{fig:examplestarfitegg}). In particular, the systematic residual due to the egg-shape of the elongated PSFs is still present and also the bright star still shows hints of the residual pattern visible in the single Moffat fit.
Conversely, the double Moffat fits (apart from increasing the computational time for PSF estimation by a factor of $\sim$\,20) often produced unreliable results for those stars that are perfectly represented by a single Moffat function, leaving the second Moffat function unconstrained. Fitting some stars with a double Moffat and some with a single Moffat function bears its own problems, starting with having to decide which star to fit with which function and ending with having to devise a method to combine the different estimates into a single model PSF for the galaxy. Additionally, there are always parameter degeneracies when fitting two identical functions. 

For these reasons, we concluded that fitting double Moffat functions to KiDS stars is only viable when jointly fitting all stars of a field. This increases the signal-to-noise ratio with respect to individual fits, so all parameters can be reasonably constrained; and eliminates the problem of combining the resulting PSFs, since there will only be one model. While this is the approach that e.g. \citet{Taranu2017} have taken (and was the original reason for the development of the \texttt{AllStarFit} package which we have used in other contexts), we decided against it for reasons outlined in Section~\ref{sec:psfdetails}. Therefore we decided to proceed with single Moffat PSFs despite its shortcomings for a small minority of galaxies; in particular since we believe that the imperfect fits to bright stars do not affect our analysis and so only a few galaxies in tile corners with very elongated PSFs are compromised. The test runs with double Moffat PSFs did also not show any significant or systematic improvement to the galaxy fits compared to galaxy fits with single Moffat PSFs.

\subsubsection{Model PSF generation}

After fitting, suitable stars for PSF estimation are determined as follows:
\begin{itemize}
\item The fitted centre in x and y must be within $\pm$\,1 pixel of the centre of the cutout 
\item The fitted magnitude must be within $\pm$\,0.1\,mag of 0 
\item The reduced chi-squared ($\chi^2_\nu$) of the fit must be smaller than 3 (where $\chi^2_\nu$ is evaluated within the star segment only even if a larger region was fit)
\item FWHM, concentration index, angle, axial ratio and background (if fit) must not be equal to the fit limits (except for the axial ratio, which is allowed to be exactly 1 although this is the upper limit of the fit).
\item Outliers in any of FWHM, concentration index, angle, axial ratio or background are rejected via an iterative 2$\sigma$ clip (in logarithmic space where appropriate). 
\end{itemize}

The first two criteria of these are fixed hard cuts since the input image was centred on the star and normalised to a magnitude of 0 according to the segmentation statistics. Any deviation from these values indicates a difference in the position or magnitude estimated by \texttt{ProFound} and \texttt{ProFit} which likely points to bad segmentation, additional objects in the segment or a bad model fit. The cut in $\chi^2_\nu$ is also a hard cut, chosen to remove objects that visually appear as bad fits (mostly the very bright stars discussed earlier). Note that the example star shown in Figure~\ref{fig:examplestarfitbright} is well below this limit. The reason is that we show an example star from a test run here, in order to directly compare it to the double Moffat fit. In this test run, however, we also used an additional cut on the magnitude of star candidates, excluding the brightest objects (see above). Consequently, there are only stars of intermediate brightness in this test run, one of which we show in Figure~\ref{fig:examplestarfitbright}. During subsequent pipeline development, we have then changed the strategy and stopped using the magnitude cut, except for objects at the saturation limit (see discussion above). Instead, we now exclude these brightest stars with the cut in reduced chi-squared listed here. 

We also exclude stars that hit their fit limit since these parameter estimates are not reliable and all fitting intervals were chosen very generously (except for the upper bound on axial ratio, which has a physical limit at 1, meaning perfectly round). Finally, we remove outliers in the respective distribution of the main fitting parameters to exclude any remaining objects that are not true point sources. 

\begin{figure}[t!]
\begin{center}
	\includegraphics[width=0.8\textwidth]{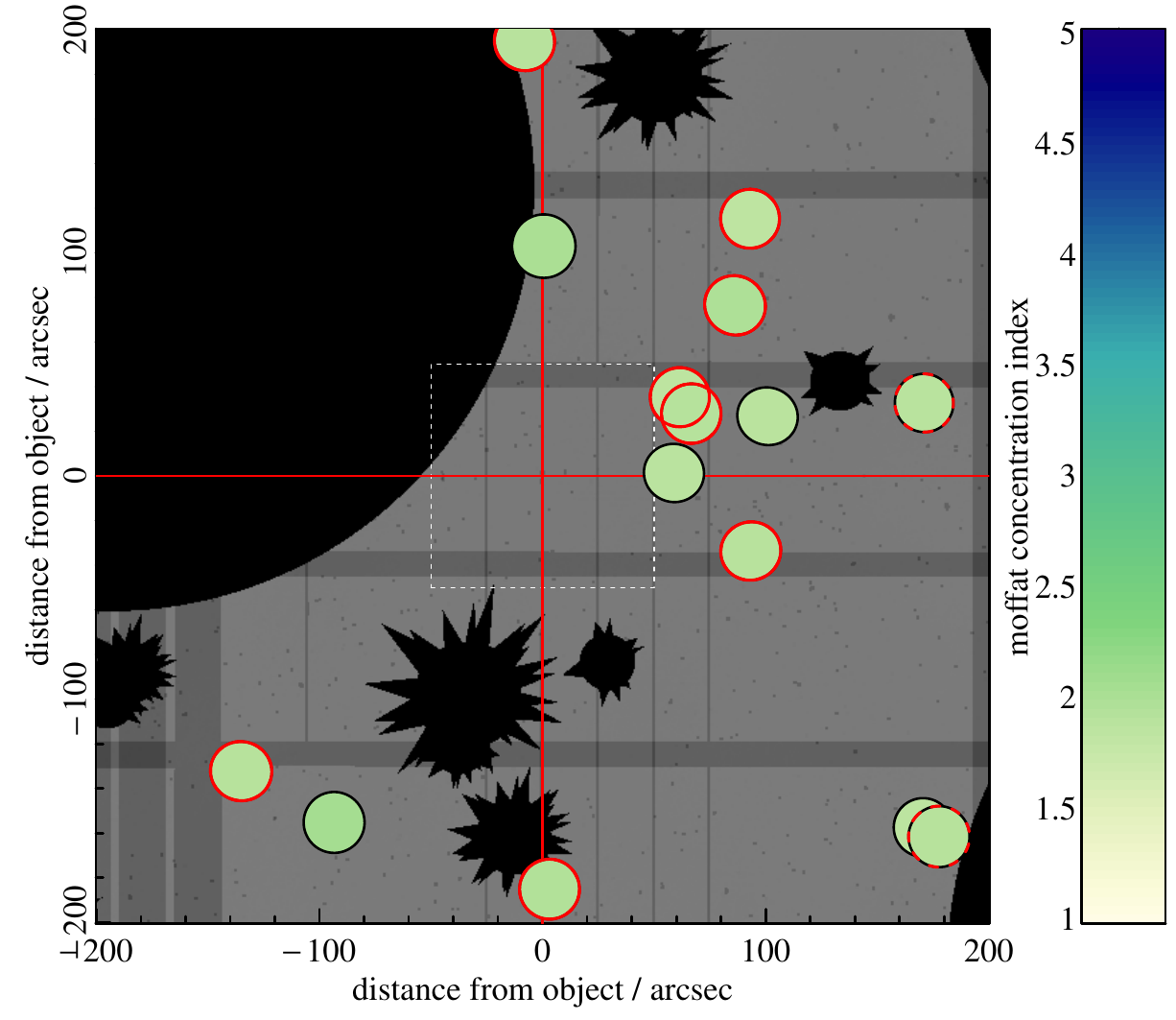}
    \caption{The result of the PSF fitting for the galaxy 396740 in the KiDS $r$-band with the dashed white square indicating the cutout shown in Figure~\ref{fig:exampleseg}. The greyscale image shows the $r$-band weight map with lighter colours meaning higher weight. Masked areas from the stacked $gri$-masks are shown in black (zero weight). The vertical and horizontal red line indicate the position of the object of interest (galaxy 396740) and split the image into its 4 quadrants. All fitted PSFs are shown as coloured ellipses with the size (FWHM multiplied by 20), axial ratio, orientation angle and concentration index (colour) taken from the fitted Moffat parameters. Stars selected for estimating the final model PSF have red borders; dashed red borders mean a fit was classified as suitable, but not selected because the maximum of 8 stars was already reached.}
    \label{fig:examplepsfoutput}
\end{center}
\end{figure}

\begin{figure}[t!]
\begin{center}
	\includegraphics[width=0.8\textwidth]{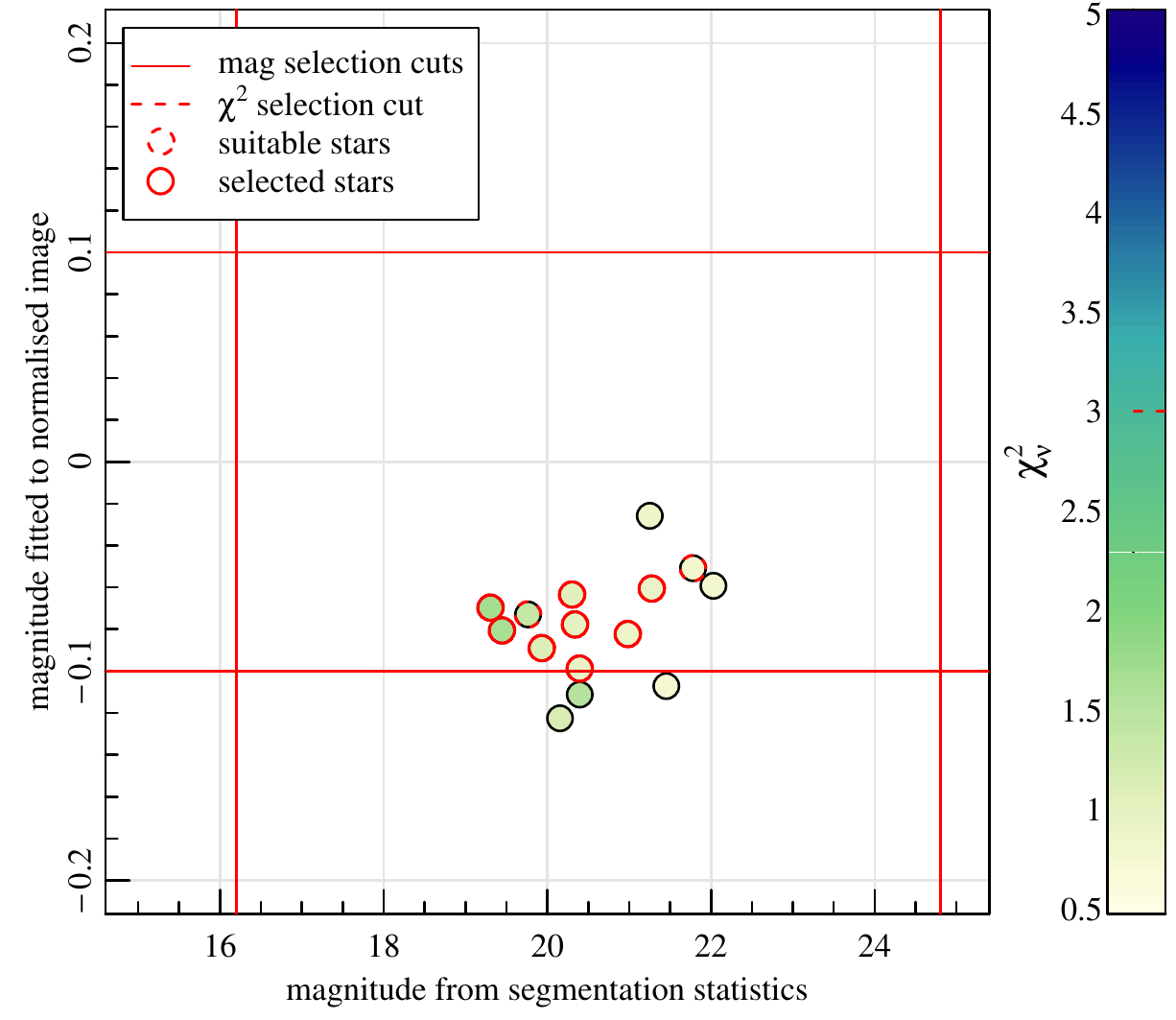}
    \caption{The second main diagnostic plot for the PSF fitting result, again for galaxy 396740 in the KiDS $r$-band. Here, we show the magnitude fitted to the normalised image of each star on the $y$-axis against the magnitude from the \texttt{ProFound} segmentation statistics on the $x$-axis, with symbols coloured by the reduced chi-squared of the fit. The magnitude and $\chi^2_\nu$ selection cuts are shown as solid and dashed red lines respectively (the latter on the colour bar); stars classified and suitable and selected are indicated with dashed and solid red borders.}
    \label{fig:examplepsfmags}
\end{center}
\end{figure}

The stars fulfilling these criteria are classified as suitable, from which the selection is made: 
\begin{itemize} 
\item The closest two from each quadrant (8 in total) are selected to make sure they are roughly evenly distributed around the galaxy. 
\item If one or more quadrant contains less than two stars, the closest stars from any other quadrant (which are not already used) are taken instead to give 8 stars in total. A warning is given and the quality flag adjusted.
\item If there are less than 8 stars in total, all of them are used with a warning (and a quality flag adjustment)
\item If there are no stars classified as suitable, the default PSF mentioned above (round, FWHM from KiDS header, default Moffat concentration index) is used, with a warning and the corresponding flag.
\end{itemize}
There are two diagnostic plots summarising these selection steps (three if the background was fit as well), examples are shown in Figures~\ref{fig:examplepsfoutput} and ~\ref{fig:examplepsfmags}. A list of stars with the relevant segmentation parameters and the fit parameters is saved for later reference. 

From Figure~\ref{fig:examplepsfoutput} it is obvious that the Moffat models fitted to all star candidates in the field are very consistent in all parameters (shape, orientation, size and concentration). This is true even for stars which were not classified as suitable for PSF estimation (symbols with black borders). We make the same observation in most cutouts, even though the PSFs themselves can vary greatly between frames (see Figure~\ref{fig:tileoverview}), with only occasional outliers appearing (which are subsequently excluded from the model PSF estimation). 

Figure~\ref{fig:examplepsfmags} shows that all star candidates of this frame are well-represented by a Moffat function. This is also generally the case except for very bright objects (close to the bright magnitude limit, of which there are none in this field), very elongated ones and those that have secondary objects included in their segmentation map. The latter of these are excluded either by the $\chi^2_\nu$ cut or by the magnitude cuts since they will generally show a disagreement between \texttt{ProFound} and \texttt{ProFit} magnitudes. Note there is a slight systematic offset for all stars in this field, too, in that all stars have fitted magnitudes smaller than zero. As a reminder, we use the \texttt{ProFound} segment magnitudes to normalise the star images to a magnitude of 0 before fitting, so any deviation from zero in the fitted magnitude indicates that \texttt{ProFit} and \texttt{ProFound} estimate different magnitudes for the same object. However, small deviations below $\sim$\,0.1\,mag are no reason for concern since there are inherent differences in the two estimates: the \texttt{ProFound} estimate is simply the sum of all flux within the segment, while the \texttt{ProFit} estimate is the total magnitude of the Moffat function integrated to infinity, which can easily be somewhat brighter. However, larger deviations are usually indications of a bad model fit, bad segmentation, intruding (wings of) nearby objects or image artifacts and hence are excluded from further consideration. 

\begin{figure}[t!]
\begin{center}
	\includegraphics[width=0.8\textwidth]{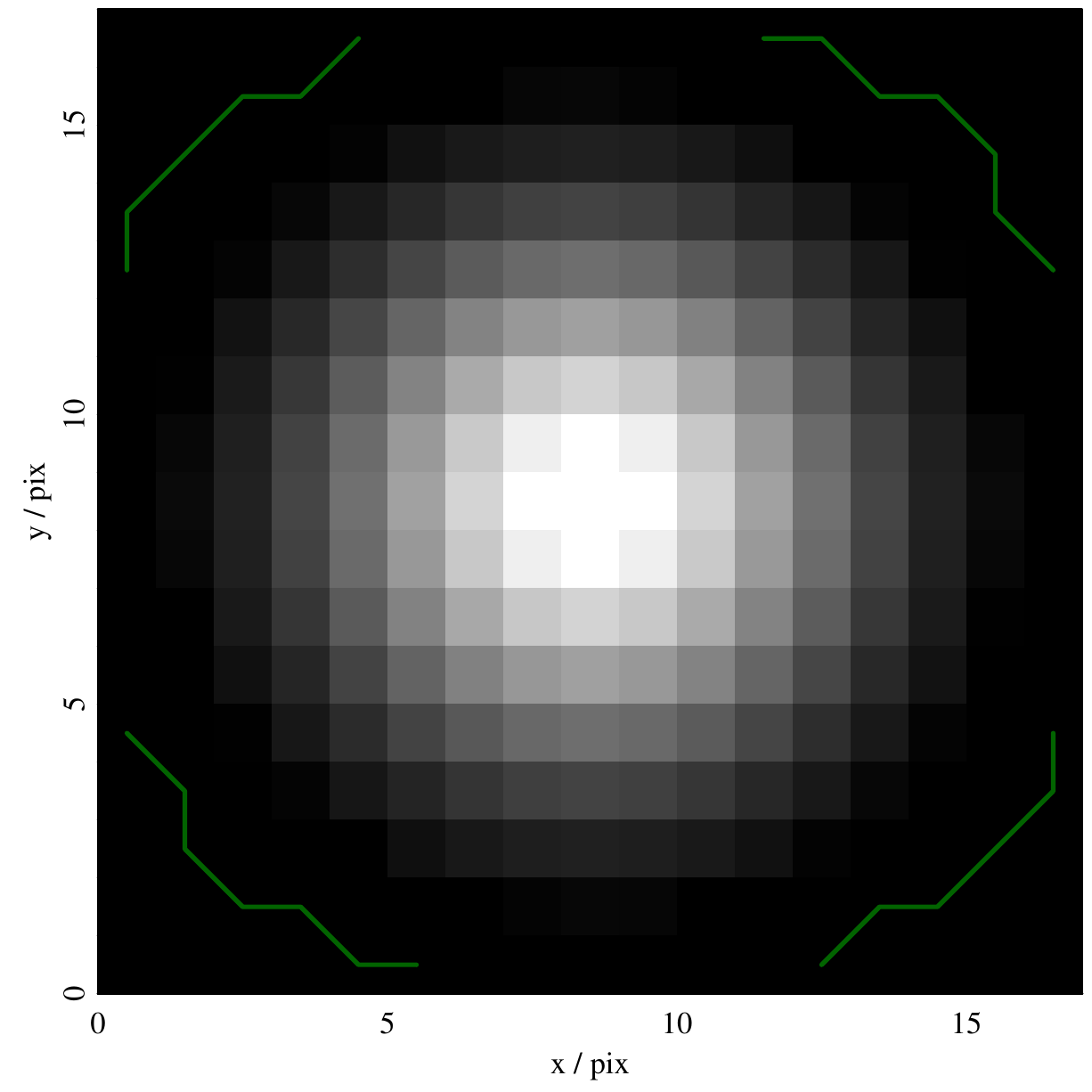}
    \caption{The model PSF obtained for galaxy 396740 in the KiDS $r$-band with asinh stretching of the flux scale. The green contours indicate the region beyond which the PSF is set to zero exactly.}
    \label{fig:examplepsf}
\end{center}
\end{figure}

Finally, the model PSF is created as an image of a two-dimensional Moffat function with parameters that are the medians of the selected stars. Exceptions are the centre of the star (in x and y), which is forced to the centre of the model image; the magnitude, which is forced to 0 and the background, which is ignored (i.e. set to 0 even if fit). The reason for forcing these values in the model PSF is to avoid systematic offsets when convolving the galaxy to be fit with its model PSF. The size of the image is adjusted to include at least 99\,\% of the flux; or to a maximum of the median segment size within which the stars were fitted (to avoid extrapolation). Pixels in the corners of the image are set to 0 to avoid having a rectangular PSF and the image is subsequently re-normalised to sum to 1. The model PSF is stored as a FITS file and an image of it is returned as well, with an example shown in Figure~\ref{fig:examplepsf}.

There are many other ways in which we could have chosen to combine the individual PSF estimates into a final model PSF. We chose the approach of simply medianing the Moffat parameters mostly since it is practical, fast and robust. Taking the median ensures that we are not susceptible to outliers in the distribution (should there be any left after all of the above treatment), while working at the parameter level ensures that the resulting PSF is a smooth Moffat function with no abrupt changes. We avoid extrapolation of the model beyond the region in which it was fitted by limiting the size of the final PSF accordingly; and in turn ensure an approximately round PSF by setting the very small values in the corners to zero exactly. We sample the PSF onto the same pixel scale as the KiDS images, so it can be used for direct convolution. 

A first potential improvement to this procedure would be to finesample the PSF (and correspondingly the galaxy image) to allow a more accurate convolution. However, even the smallest possible oversampling factor of 3 (meaning a factor 9 increase in the number of pixels to convolve) leads to a prohibitive increase in the computational time for galaxy parameter estimation since the PSF convolution needs to be repeated for each model evaluation and constitutes a large fraction of the total computational time. Also, finesampling is only effective in increasing the accuracy of the convolution if the PSF is known to the level of accuracy implied by the finesampling. This can be achieved with PSF estimates based on large numbers of stars (best combined empirically or fit jointly) and/or theoretical knowledge about instrumental effects. In our case, where we fit only a relatively low number of stars individually, we cannot ensure or verify the PSF accuracy for significant finesampling factors. 

Another potential improvement in creating the model PSF would be to consider parameter correlations. Currently, these are ignored since we take the medians of all parameters individually. One shortcoming of this is that the median of the axial ratio is taken independently of its orientation angle; which could lead to artifically elongated PSFs. For example, suppose the true PSF is perfectly round. Due to noise, the fitted axial ratios could be less than 1 (at least for some of the stars), with associated orientation angles that are randomly distributed. Taking the median of these two parameters independently of each other could therefore lead to a slightly elongated model PSF profile oriented at a random angle. In theory, this effect could lead to our PSFs being systematically too elongated, especially since it does not work the other way around (i.e. we will not have estimates which are systematically too round for truly elongated PSFs, since that would require the noise to be distributed in the same way for all stars of a field, which is highly unlikely). 

After noticing this shortcoming, we have double-checked many of our model PSFs by studying the residual left after subtracting it from all stars that it was based on; and found no evidence of the PSF being systematically elongated. We also compared the average axial ratio of all model PSFs in a tile to the average PSF ellipticity given in the KiDS header of the same tile and found our estimates to be consistent with the KiDS value for individual tiles; and slightly rounder than the KiDS value on average across all tiles. Since the above effect cannot produce PSFs that are too round, we conclude that there are no systematic biases in our PSF shape introduced by our method. The difference between the KiDS values and our averages are likely because our average is based on the model PSFs which are the medians of a highly selected set of stars; while the KiDS value is computed as the average ellipticity of all individual stars in the field which is likely to be noisier (and therefore less round). 

While we conclude that the PSF estimates are not compromised by this effect, it would still be preferable to consider parameter correlations (between angle and axial ratio, but also other parameters) when computing the final model PSF. One way to achieve this is to perform a joint fit of several stars. This has the added advantages that it increases the signal-to-noise ratio relative to fitting individual stars, allowing to improve the PSF accuracy and possibly fitting more detailed models (e.g. a double Moffat function as mentioned above). We have briefly explored this route, but unfortunately it takes much longer than fitting the stars individually since the computational time scales non-linearly with the number of parameters and there are many parameters in a joint fit since each star needs its own position ($x$ and $y$) and magnitude to be fitted, even if the PSF shape is the same for all stars. Furthermore, implementing a joint fit requires significant improvements to the star candidate selection process, which is not critical in the current approach since outliers can easily be excluded after fitting. For a joint fit, this step would have to be replaced by an alternative - or else we would have to first fit stars individually, select suitable ones and then perform a joint fit of those (instead of medianing the parameters), additionally increasing computational time. For these reasons, we did not implement such an approach in the current work. Nonetheless, it remains an interesting avenue to explore in future work since it would solve several shortcomings of the PSF estimation at once. 

As a last point of how the PSF pipeline could potentially be improved, we would like to discuss the number of stars we chose for fitting. Generally, fitting a higher number of stars will result in a more robust PSF estimate. A competing effect is that stars closer to the galaxy will reflect the PSF at the galaxy position better than those further away. A robust local PSF estimate can therefore only include stars within a field small enough such the PSF does not vary significantly across it. Larger fields would require fitting multiple PSFs and interpolating between them, introducing additional uncertainties. The size of the large cutout to identify stars in (400\arcsec\,$\times$\,400\arcsec) is chosen to balance these two effects. It is small enough such that the PSF does not usually change significantly across it (see Figure~\ref{fig:examplepsfoutput}), since the PSF in KiDS tiles does vary, but slowly (Figure~\ref{fig:tileoverview}). At the same time, the cutout is large enough to contain a reasonable number of stars for PSF fitting ($\sim$\,10-20 on average) unless large fractions of the cutout are masked. In order to make the PSF estimates comparable for crowded fields (sometimes containing more than 30 stars) and sparse ones or those with masked areas (where there are often only a few stars available), we take a maximum of 8 stars for each PSF estimate even if more would be available.

\subsubsection{PSF quality control}

\begin{figure}[t!]
	\includegraphics[width=0.5\textwidth]{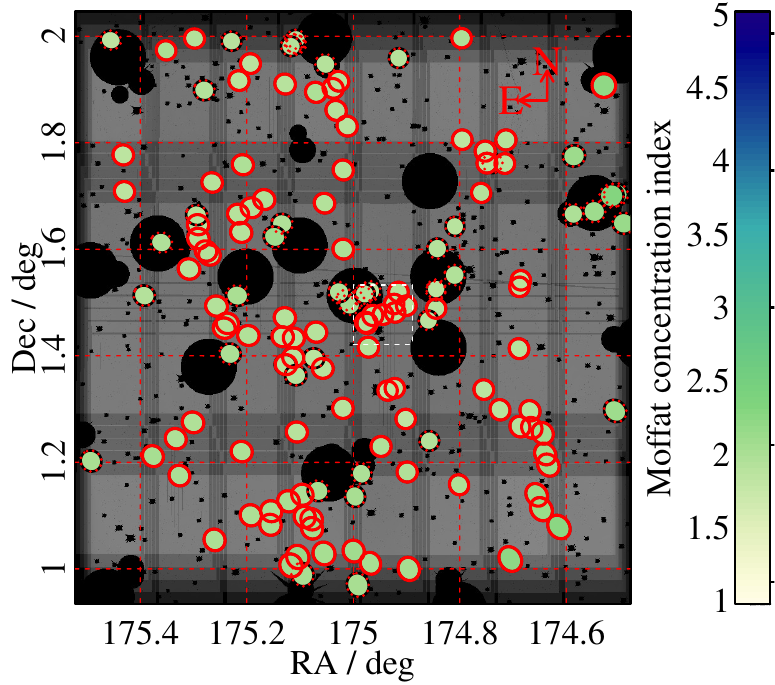}
	\includegraphics[width=0.5\textwidth]{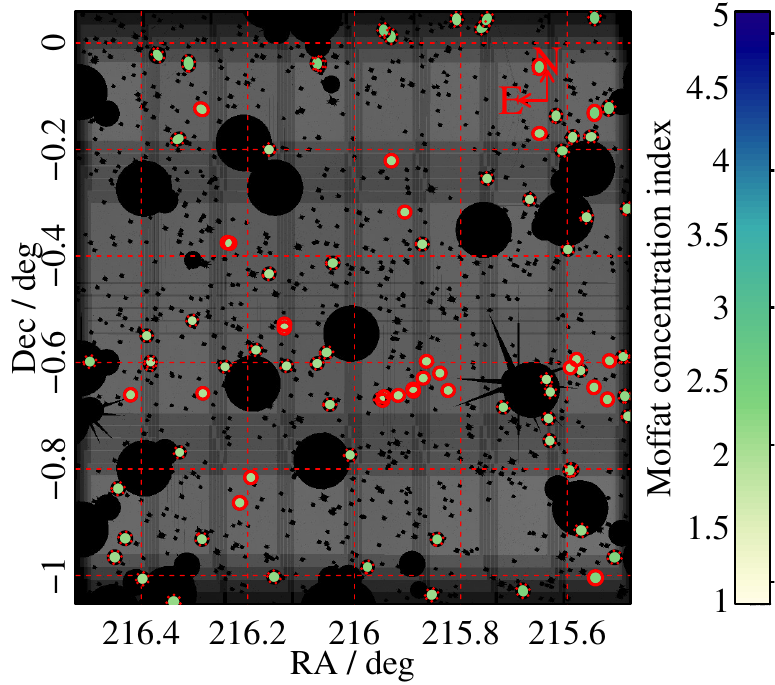}
	\includegraphics[width=0.5\textwidth]{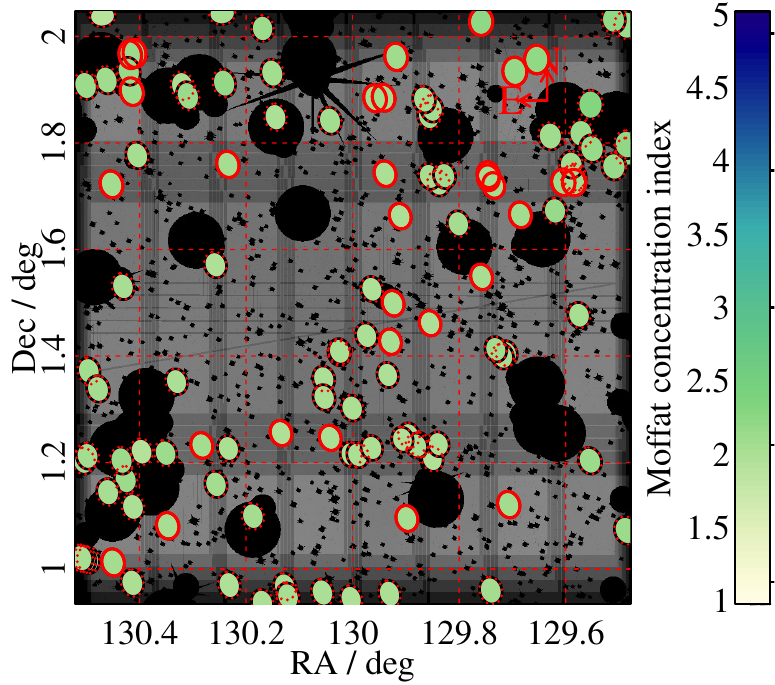}
	\includegraphics[width=0.5\textwidth]{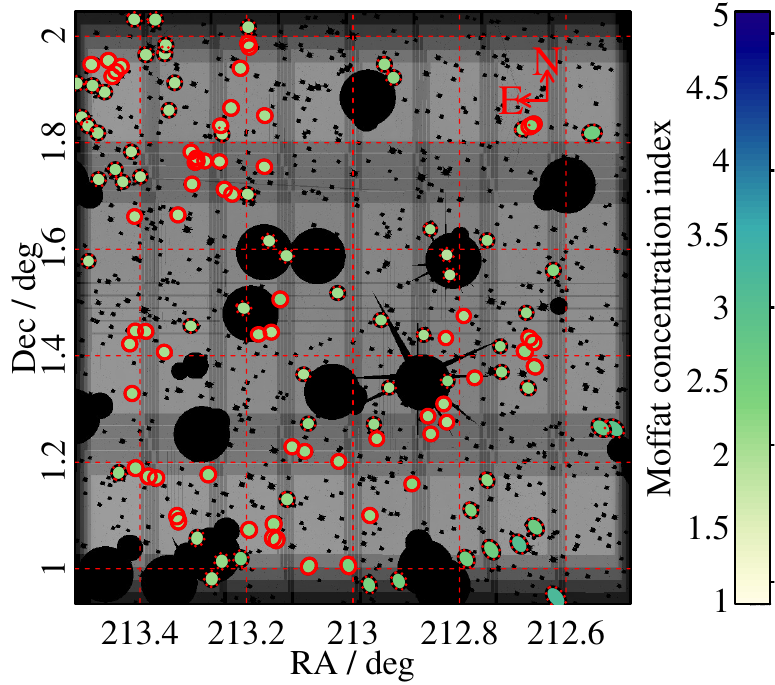}
    \caption{The model PSFs obtained for all galaxies in four different KiDS $r$-band tiles (top left: \texttt{KiDS\_DR4.0\_175.0\_1.5\_r\_sci.fits}, top right: \texttt{KiDS\_DR4.0\_216.0\_-0.5\_r\_sci.fits}, bottom left: \texttt{KiDS\_DR4.0\_130.0\_1.5\_r\_sci.fits}, bottom right: \texttt{KiDS\_DR4.0\_213.0\_1.5\_r\_sci.fits}). Similar to Figure~\ref{fig:examplepsfoutput}, the greyscale image shows the weight map with masked areas shown in black; while the coloured ellipses represent the PSFs (FWHM\,$\times$\,100, axial ratio, orientation angle and concentration index). The major difference to Figure~\ref{fig:examplepsfoutput} is that we now do not show fits to individual stars, but instead the model PSFs derived for the galaxy positions. Solid red contours now mean that the model PSF is based on 8 stars, while dotted red contours are drawn around PSF estimates based on less than 8 stars. Black contours mean this is a default (dummy) PSF and should not be considered further. The white rectancle in the top left tile indicates the region shown in Figure~\ref{fig:examplepsfoutput}.}
    \label{fig:tileoverview}
\end{figure}

As mentioned above, we produce a number of PSF quality control plots for each galaxy (Figures~\ref{fig:examplepsfseg}, \ref{fig:examplestarfit}, \ref{fig:examplepsfoutput}, \ref{fig:examplepsfmags}, \ref{fig:examplepsf}) which are stored on the GAMA file server for reference. In addition, we investigated the robustness of our estimates by comparing model PSFs for different galaxies in the same KiDS tile against each other. The main diagnostic plot of this comparison is shown in Figure~\ref{fig:tileoverview} for four different KiDS tiles. Similar to Figure~\ref{fig:examplepsfoutput}, we show the weight map as greyscale in the background with masked areas in black; and the PSFs as coloured contours with the size (FWHM\,$\times$\,100), axial ratio, orientation angle and concentration index (colour) taken from the model PSF estimated for the galaxy at this position. Note that in contrast to Figure~\ref{fig:examplepsfoutput}, the PSFs shown here are not fits to individual stars, but Moffat models created from the combination of several suitably selected stars. Therefore, the meaning of the red contours changed: solid red indicates model PSFs based on 8 stars, while dotted red means that less than 8 suitable stars were available. Black contours are drawn around default (dummy) model PSFs that had no suitable stars available and so should be ignored. Also, we now multiply the FWHM by 100 instead of 20 to enhance visibility in the much larger area ($\sim$\,1\,deg$^2$ instead of 400\arcsec\,$\times$\,400\arcsec). 

The top left panel shows the tile \texttt{KiDS\_DR4.0\_175.0\_1.5\_r\_sci.fits} in which galaxy 396740 (our usual example galaxy) resides near the centre. For orientation, we indicate the 400\arcsec\,$\times$\,400\arcsec\ cutout that is shown in Figure~\ref{fig:examplepsfoutput} as a white rectangle. The top right panel shows a tile with generally much smaller PSFs, \texttt{KiDS\_DR4.0\_216.0\_-0.5\_r\_sci.fits}. On the bottom left we see \texttt{KiDS\_DR4.0\_130.0\_1.5\_r\_sci.fits}, which shows larger and more elongated PSFs and finally, the bottom right image shows \texttt{KiDS\_DR4.0\_213.0\_1.5\_r\_sci.fits} which has more pronounced PSF variations and is also the tile in which the example elongated star of Figure~\ref{fig:examplestarfitegg} resides (near the bottom right corner). 

Comparing all tiles we can see that the PSF differences between tiles are much larger than the variations within a tile. The latter are generally relatively small and smooth with no abrupt changes. This is what one would expect for KiDS data, since it is seeing-limited and the seeing will vary greatly with time (i.e. between tiles); while all dithers composing a tile were taken in close temporal succession. The fact that we can observe this in the overview plots is therefore reassuring. Remember that each PSF in Figure~\ref{fig:tileoverview} is a model PSF that was estimated entirely independently from all other model PSFs in the tile (except for galaxies very close to each other which could have some stars overlapping). The generally similar appearance of all model PSFs in a tile (with no clear outliers) combined with the slow variation across it is a sign of our PSF estimates being robust. 

Investigating the variation across the individual tiles further, it can be seen that generally, the PSFs tend to become more elongated towards tile edges and corners, with the concentration and FWHM also increasing. This effect is more pronounced in some tiles (e.g. bottom right) than in others. Also, even within a tile, not all corners are equally affected, our impression was that the bottom right and top right corners show elongated PSFs more often than the top left and bottom left corners. The orientation angle seems to rotate such that the major axis always points towards the corners. For tiles that have intrinsically elongated PSFs, this can result in PSFs actually becoming rounder towards corners (e.g. in the top right corner of the bottom left panel). 

\begin{figure}[t!]
	\includegraphics[width=0.5\textwidth]{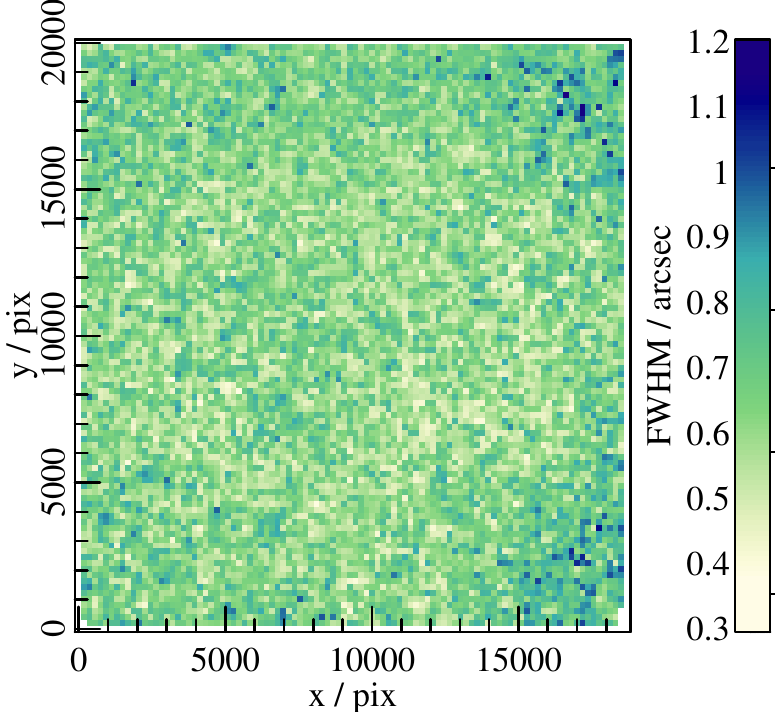}
	\includegraphics[width=0.5\textwidth]{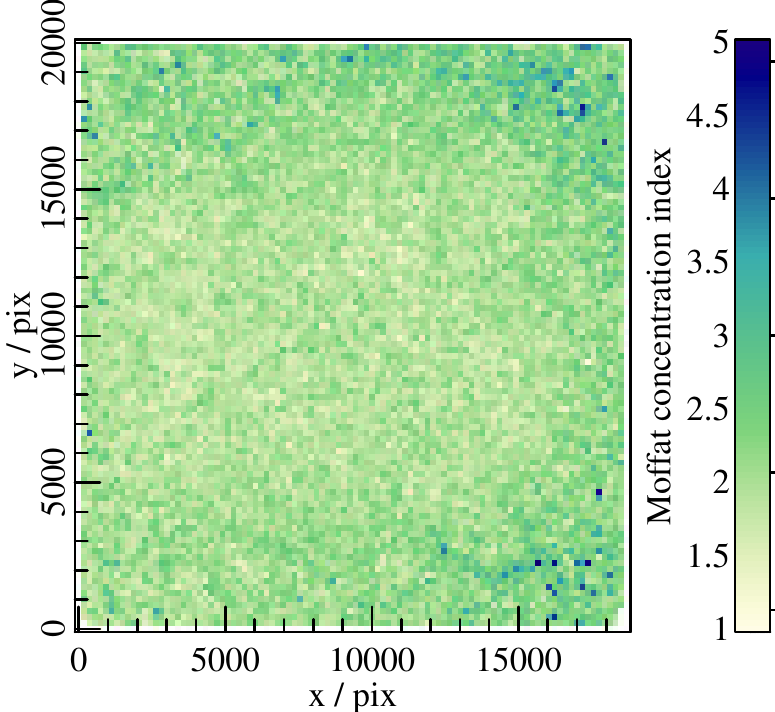}
	\includegraphics[width=0.5\textwidth]{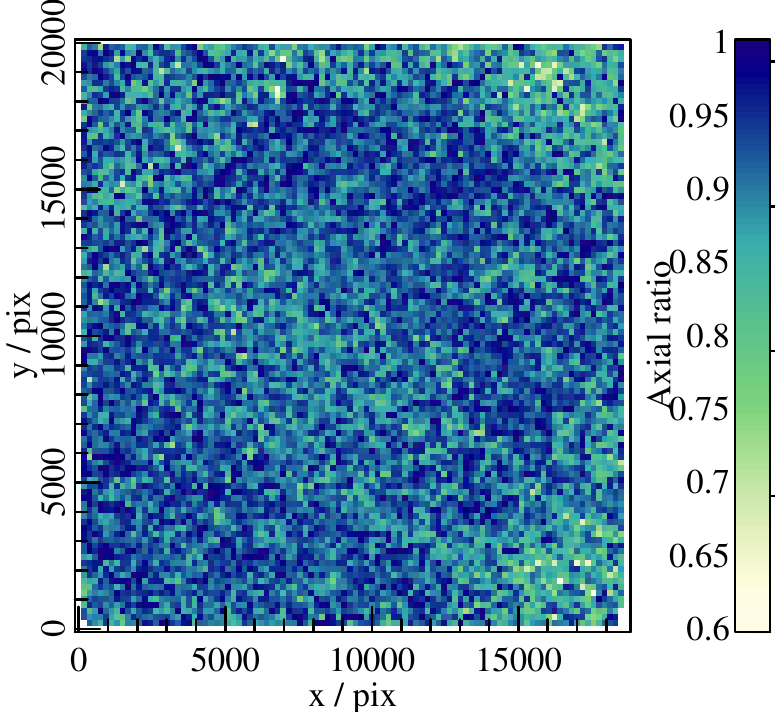}
	\includegraphics[width=0.5\textwidth]{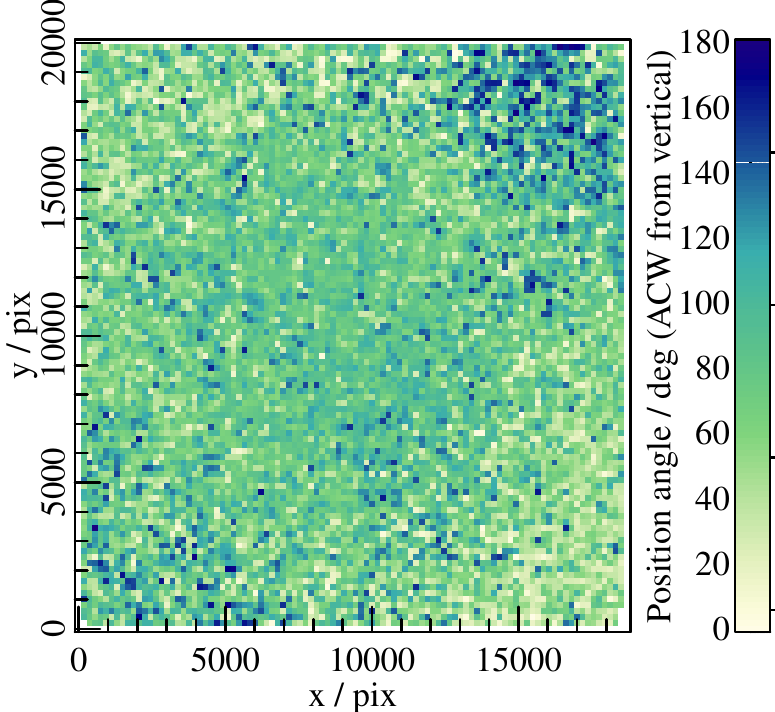}
    \caption{The variation of the FWHM (top left), Moffat concentration index (top right), axial ratio (bottom left) and position angle (bottom right) for model PSFs averaged across all KiDS $r$-band tiles. The position angle is measured anticlockwise (ACW) from the vertical axis and limited to the range between 0 and 180\degr\ due to symmetry.}
    \label{fig:tilevary}
\end{figure}

To gain a better understanding of these systematic trends, we stacked the results for all KiDS $r$-band tiles, the result of which can be seen in Figure~\ref{fig:tilevary}. As expected from the examples shown in Figure~\ref{fig:tileoverview} above, the stacked plots show that the FWHM and concentration index generally increase towards the corners of tiles, while the axial ratio decreases and the angle rotates such that the major axis always points towards the nearest corner. Our visual impression that the right side of the tiles seems to be more affected than the left side is also confirmed. Interestingly, there is also a region near the centre of the tiles where the FWHM and concentration index increase while the axial ratio decreases slightly, but with no clear corresponding trend in the position angle. The latter instead seems to be randomly distributed in the central tile region, such that averaging over all tiles results in a value of approximately 90\degr, i.e. the centre of the possible range of angles. 

We have also visually compared these tile overview plots to the PSF diagnostic plots available from the KiDS database and found the general trends to agree well.

\subsubsection{PSF effects on galaxy fitting}

\begin{figure}
	\includegraphics[width=\textwidth]{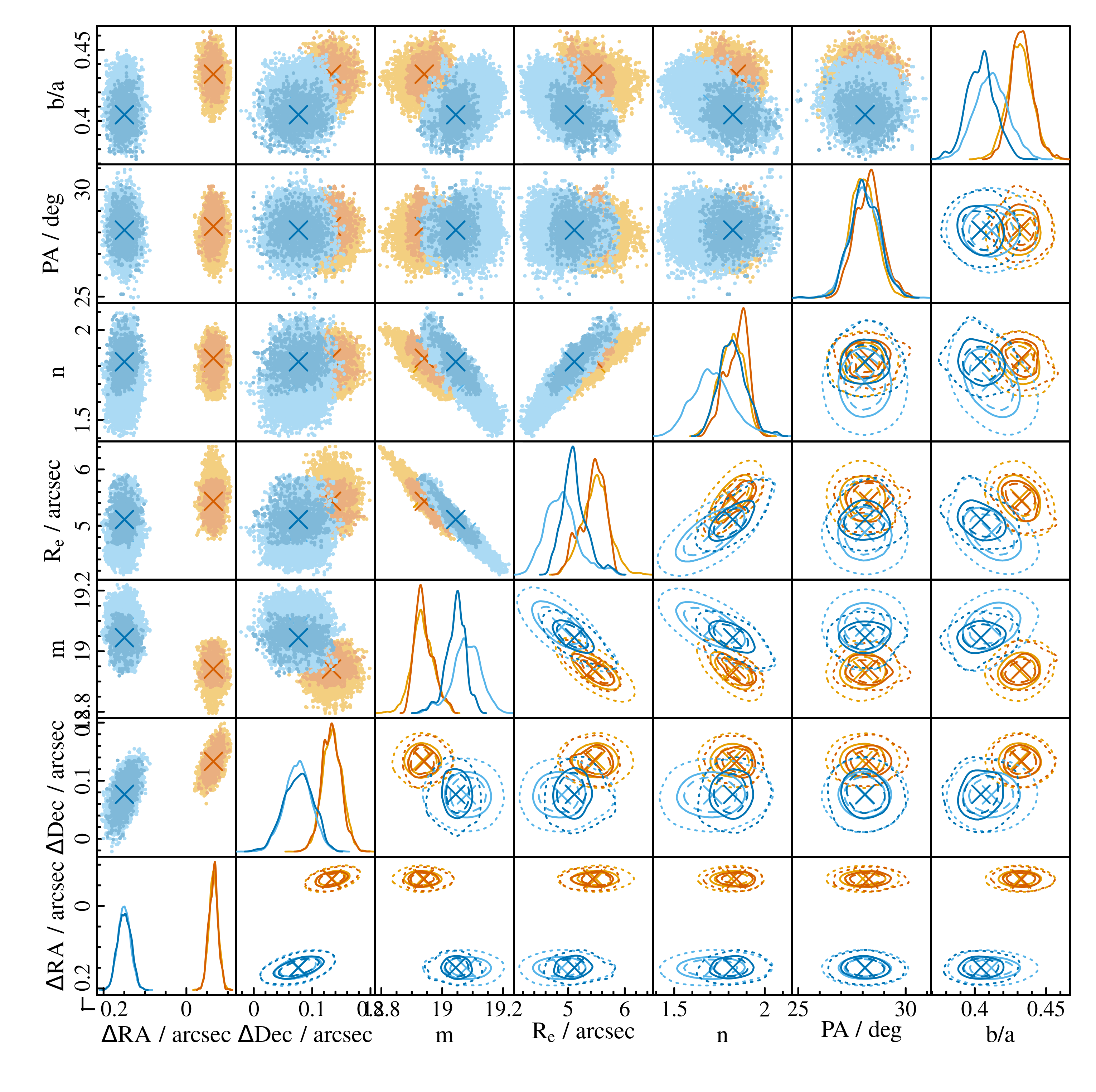}
    \caption{The effect of different PSFs on the fitted galaxy parameters for the galaxy 105723, which is part of the overlap sample. For each combination of the seven single S\'ersic parameters we show the distribution of the MCMC chain points in the upper left half of the plot; while the lower right half shows the corresponding contours including 50\,\%, 68\,\% and 95\,\% of the data for dashed, solid and dotted lines respectively. The diagonal shows the one-dimensional distribution for each parameter. Yellow and orange lines and points show the parameters fitted for the first match of this galaxy, where the former indicate the full distribution obtained by sampling from the PSFs and re-fitting the galaxy 8 times (16000 MCMC samples in total), while the latter shows the result using just the model PSF (i.e. the medianed version, 2000 MCMC samples). Light and dark blue lines and points show the same for the second match, i.e. for the same physical object but a different image since the galaxy sits in the overlap region between KiDS tiles. Crosses show the mean of each distribution. For RA and Dec, instead of the absolute values, we show the difference to the GAMA input positions in arcseconds (for readability).}
    \label{fig:islandplotpsf}
\end{figure}

Besides the internal PSF consistency checks presented above, we investigated the effects that using different PSFs has on the fitted galaxy parameters. To this end, we fitted a test sample of galaxies with various different PSFs, sampled from the stars classified as suitable. An example can be seen in Figure~\ref{fig:islandplotpsf}. 

We show the result of the fitted parameters for four different ``versions" of the same galaxy: yellow shows the result for an image of the galaxy using 8 different PSFs sampled from the suitable stars (2000 MCMC samples each, i.e. 16000 points in total). Orange is for the same image of the galaxy, just using the usual (medianed) model PSF (2000 points). Light blue and dark blue show the same for a different image of the same galaxy, since the galaxy sits in the overlap region between KiDS tiles. The top left half of the plot then shows the MCMC samples for each combination of the seven single S\'ersic parameters: position in RA and Dec, magnitude $m$, effective radius $R_e$, S\'ersic index $n$, the position angle PA and the axial ratio $b/a$. The bottom right half shows the corresponding contours, with dashed, solid and dotted lines including 50\,\%, 68\,\% and 95\,\% of the data. On the diagonal we show the one-dimensional distributions for each parameter. Crosses indicate the mean of each distribution. 

There is a clear discrepancy in the fitted RA and Dec centres for the two images, which is caused by the accuracy of the KiDS astrometric solution, see Section~\ref{sec:systematics}. Apart from that, all distributions overlap, with varying levels of consistency for the different parameters. The distributions sampling from the PSFs are broader than those using just the medianed model PSF. This shows that the PSF introduces systematic uncertainties which are not accounted for in the MCMC errors. The difference is larger for the second of the two images (blue), where the sample using the model PSF is also not centred on the same values as the sample obtained by varying the PSF. Nonetheless, the parameters fitted to the same image with different PSFs are in closer agreement than those fitted to different images. This is because for different images, there are additional sources of systematic uncertainties caused by, e.g., the background subtraction and different segment sizes. We investigate all of these systematic uncertainties in more detail in Section~\ref{sec:systematics}, where we derive the average systematic errors for each parameter using the overlap sample and point out parameter biases using simulations.

\subsection{Segment sizes}
\label{sec:segchoices}
We have briefly discussed image segmentation in Sections~\ref{sec:preparatorysteps} and~\ref{sec:backgroundstudies}. One of the conclusions drawn there was that it is best to use two different segmentation maps for the background subtraction and the galaxy fitting due to the different aims that we would like to achieve with them. We make use of the different settings in \texttt{profoundProFound} to produce these segmentation maps. In the following, we detail the definition of the segmentation maps and how it has evolved over the time of pipeline development. We focus on the fitting segmentation map rather than the background subtraction segmentation map since that has already been discussed in Section~\ref{sec:backgroundstudies}. The latter part of the section then describes our own routine for segmentation map fixing, which also makes use of the fitted galaxy parameters. Hence this section is somewhat at the intersection between the preparatory work and the galaxy fitting choices. 

\subsubsection{\texttt{skycut} and dilation}

The most important parameter that influences the segmentation in \texttt{profoundProFound} is \texttt{skycut}, which determines how deep the source extraction pushes into the noise (see Section~\ref{sec:preparatorysteps} and \citealt{Robotham2018} for details). A lower value of \texttt{skycut} means that more (fainter) sources are detected and segments for individual sources are larger. In addition, already-defined segments can be dilated (expanded) with the \texttt{profoundMakeSegimDilate} function, which is also used internally by \texttt{profoundProFound} to achieve flux convergence during the segment growing phase and to obtain a more aggressive object mask by further dilating all segments after convergence. There are additional parameters that influence the segmentation, most notably \texttt{pixcut, tolerance} and \texttt{sigma} (minimum number of pixels needed for segment detection, the tolerance to use for merging segments and the radius within which to search for neighbouring objects). We experimented with varying all of those parameters, but their effect is secondary to \texttt{skycut}. In the end we reverted to using the \texttt{profoundProFound} defaults for these parameters and will not discuss their effects further. 

For the background subtraction, described in detail in Section~\ref{sec:backgroundstudies}, we use the rather low \texttt{skycut} value of one and the aggressive object mask returned by \texttt{profoundProFound} after additionally dilating the converged segments. This choice was mainly driven by theoretical considerations and ensures that we obtain a clean sample of sky pixels without (wings of) undetected objects for background subtraction. It may lead to an increased fraction of spurious detections and a significant number of sky pixels included in the segmentation maps, but this does not affect the background estimation as long as there are enough sky pixels remaining to obtain robust statistics. 

Up to \texttt{v02} of the pipeline, we used the same segmentation map for fitting objects, just without the additional dilation for aggressive object masking. However, this often resulted in very jagged segment borders which include noisy, slightly positive pixels but exclude slightly ne\-ga\-tive pixels, thus artificially biasing the flux in galaxy outskirts. This is particularly true for large, bright objects where the number of curve of growth iterations performed to reach flux convergence is often zero or one. For the background subtraction this problem is alleviated by the additional dilation performed to obtain the aggressive object mask, but it sometimes caused issues for the galaxy fitting where no such additional dilation is performed. 

From \texttt{v03} onwards, we therefore decided to run \texttt{profoundProFound} a second time to obtain the fitting segments. Here we use a higher \texttt{skycut} value of two resulting in smaller segments with smoother borders. We then perform an additional dilation of the galaxy segment, increasing its area by typically 30\,\%, which ensures that the edges are smooth and unbiased since the dilation is independent of the pixel values. These dilated segments starting with a \texttt{skycut} value of two are then approximately the same size as the previous undilated segments with a \texttt{skycut} value of one. An example of the difference can be seen in the last two panels of Figure~\ref{fig:segfixplot}, where we discuss the segmentation map fixing. At the same time we started defining all segments on the stacked $gri$ images instead of in individual bands, to ensure that the fitted pixels are exactly the same in all bands to make the measurements most directly comparable. 

We double-check the chosen \texttt{skycut} value based on the fraction of objects returned. Assuming a Normal distribution, the limit at which approximately 50\,\% of the object pixels actually belong to objects rather than the sky background is given by the Normal quantile of one minus the fraction of pixels identified as objects. \texttt{skycut} should be larger than this limit to ensure that the majority of ``object" pixels belong to real objects rather than spurious detections or overly extended segments pushing well beyond the object wings. But \texttt{skycut} should also not be much larger than this limit if one desires to detect faint sources and low surface brightness wings of extended objects. We therefore add a warning and raise a quality flag if the chosen value of \texttt{skycut} is below the ``50-50-limit", or if it is more than 1$\sigma$ above it. 

The computed ``50-50-limit" is around unity in most cases, confirming that our assumption of Normal statistics is valid. For the early version of the pipeline, using a \texttt{skycut} value of one, the flag was therefore raised frequently due to small random fluctuations. Raising the \texttt{skycut} value to two decreased the percentage of objects with this flag raised from approximately 65\,\% to 0.3\,\%. Visual inspection of those galaxies revealed they are in regions where large fractions of the background are masked, such that the object statistics are not reliable anymore.

We conclude that our approach for defining fitting segmentation maps is suitable for our purposes. The initially higher \texttt{skycut} value of two prevents noise fluctuations from being detected, while the additional segment dilation ensures that the majority of the flux from the galaxy wings is included. Note that the segments defined in this way are still relatively tight in that they do not include many sky pixels. This is a deliberate choice we made to obtain the best possible fit in the high signal to noise regions of the galaxy, at the expense of not necessarily normalising the fitted profile to zero at large radii. We discuss the benefits and drawbacks of this approach in detail in Section~\ref{sec:postprocessing}, also in comparison to previous works.

\subsubsection{Systematics from segment sizes}

Complementary to the statistical discussions in Section~\ref{sec:postprocessing} and Section~\ref{sec:systematics}, we present a case-study on the systematic effect of different segment sizes here. 

\begin{figure}[t!]
	\includegraphics[width=\textwidth]{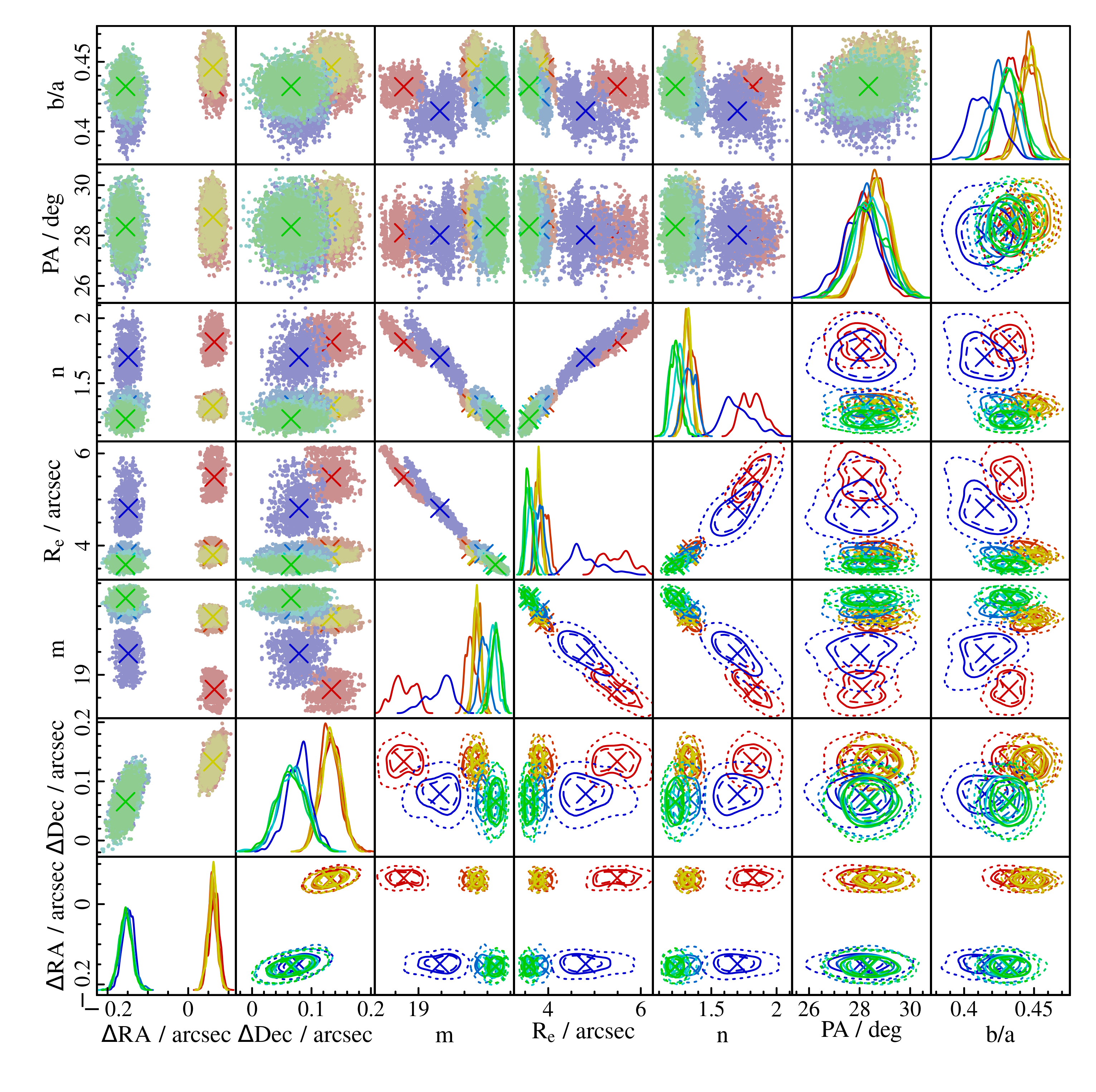}
    \caption{The effect of different segment sizes on the fitted galaxy parameters for the galaxy 105723, which is well-represented by a single S\'ersic fit. The general layout of the plot is the same as for Figure~\ref{fig:islandplotpsf}. Red to orange to yellow colours represent the parameters fitted for the first match of the galaxy for continuously increasing segment sizes. Blue through to green colours show the same for the second match of the galaxy.}
    \label{fig:islandplotseg}
\end{figure}

Figure~\ref{fig:islandplotseg} shows a corner plot similar in nature to the one presented in Figure~\ref{fig:islandplotpsf}, also for the same galaxy 105723. The distributions of the fitted parameters are shown for all combinations of the seven single S\'ersic parameters, with MCMC samples in the top left half of the plot, corresponding contours in the bottom right half and one-dimensional distributions on the diagonal. Lines and points in colours ranging from red to orange to yellow show the parameters fitted to the first image of that galaxy for continuously increasing segment sizes. Blue through to green colours show the same for the second image of the galaxy. In both cases, the results are shown for five different segment sizes: no additional dilation after the \texttt{profoundProFound} run (red and blue) and dilations which increase the segment diameter by approximately factors of 1.5, 2, 2.5 and 3 respectively. 

Note that for our final \texttt{v04} pipeline, as described above, we settled on an additional dilation based on increasing the area of the segment by approximately 30\,\%. This corresponds to a diameter increase of less than 1.5, i.e. it is not shown in the plot. The reason we do not show the final version here is that the segment size studies have been carried out on a relatively early test run, with many changes in the subsequent pipeline development both for the preparatory work (e.g. background subtraction) and the fitting routines which were still under development at that stage. We therefore only use this case study to gain an intuitive understanding of the effect of different segment sizes and return to a more quantitative analysis of the final segment choice in Figure~\ref{fig:islandplotsegsim} using our simulations. 

The most obvious feature in Figure~\ref{fig:islandplotseg} is that the RA and Dec positions fitted to the two different images of the galaxies disagree, as already seen in Figure~\ref{fig:islandplotpsf}. We believe this to be due to slightly different astrometric solutions for the two different KiDS tiles based on our investigation of the same issue in Section~\ref{sec:systematics}. For a given image, the fitted positions are in perfect agreement for all segment sizes. This is different for most other parameters, in particular for the highly correlated magnitude, effective radius and S\'ersic index: the fits using the smallest, undilated segments (shown in red and blue) are inconsistent with those using larger segments. As a general trend, also observed in other galaxies, fits to these small undilated segments tend to result in brighter and larger models with higher S\'ersic indices. 

This re-inforces our belief that for undilated segments, the galaxy flux in the outskirts is biased positive, increasing the fitted S\'ersic index. Extrapolating this to infinity to obtain the total S\'ersic magnitude then results in a model that is too bright since it includes significant amounts of flux beyond the segment borders where it is unconstrained, with a corresponding increase in effective radius. The position angle and axial ratio are uncorrelated to these three primary parameters and less affected by the segment size. For increasing segment sizes (with additional unbiased dilations) the models rapidly converge onto common values, with small differences between the two images due to other sources of systematic uncertainties. 

This rapid convergence is not achieved for all galaxies. Instead, it is only the case when the model is an adequate representation of the data, i.e. the galaxy (in this case) truly follows a single S\'ersic profile. For objects that have intrinsically different shapes than the profile(s) we try to fit to them, there is a continuous evolution of the fitted parameters with the segment size, as the focus of the fit is shifted more and more from the inner to the outer regions. The convergence regime is potentially never reached in these cases; or at a much later stage when the segment size is so large that the background normalisation dominates the fit rather than the galaxy itself. 

\begin{figure}[t!]
	\includegraphics[width=\textwidth]{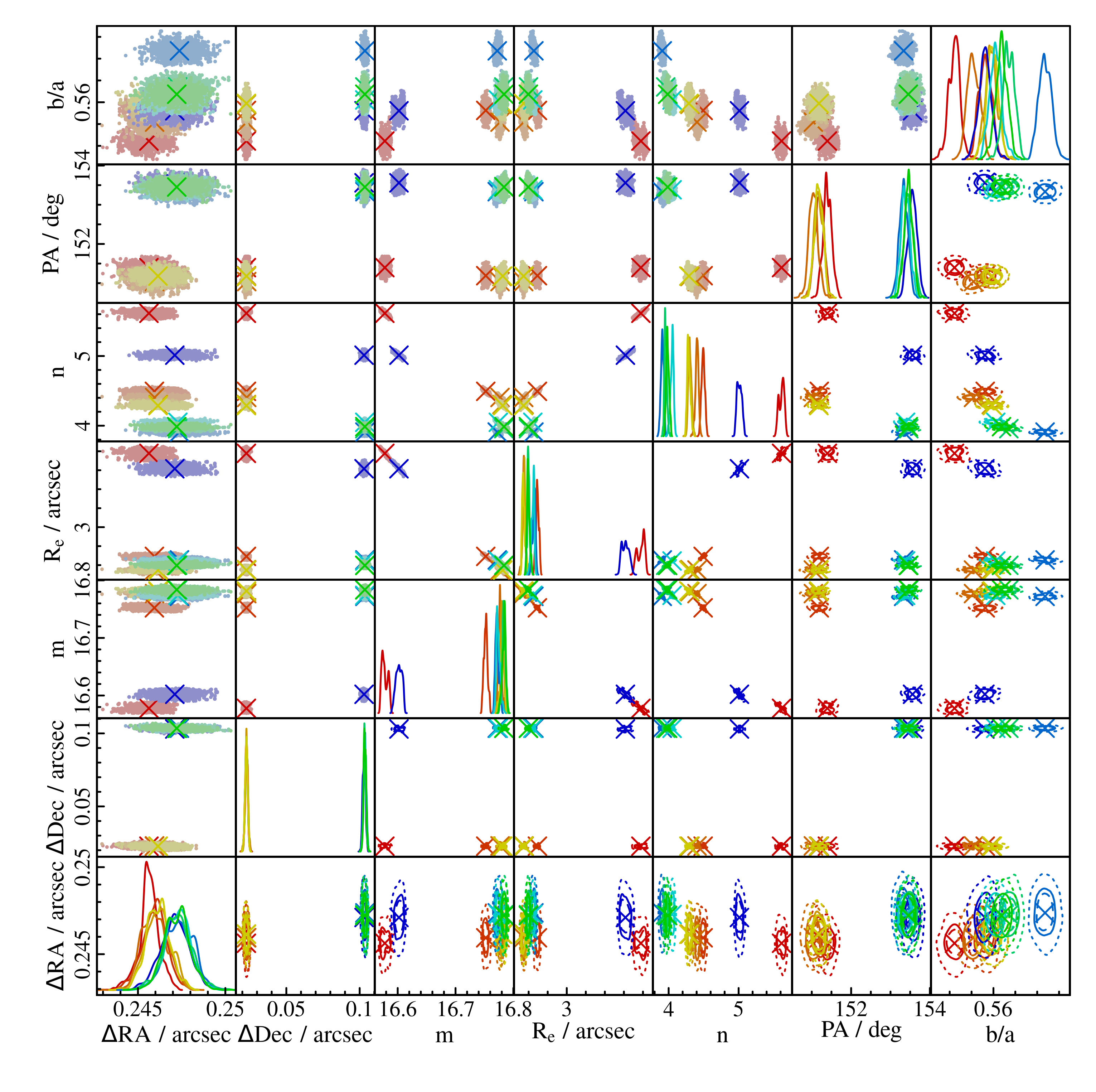}
    \caption{The same as Figure~\ref{fig:islandplotseg} for galaxy 107016, which is not well-represented by a single S\'ersic model.}
    \label{fig:islandplotseg2}
\end{figure}

\begin{figure}[t!]
	\includegraphics[width=\textwidth]{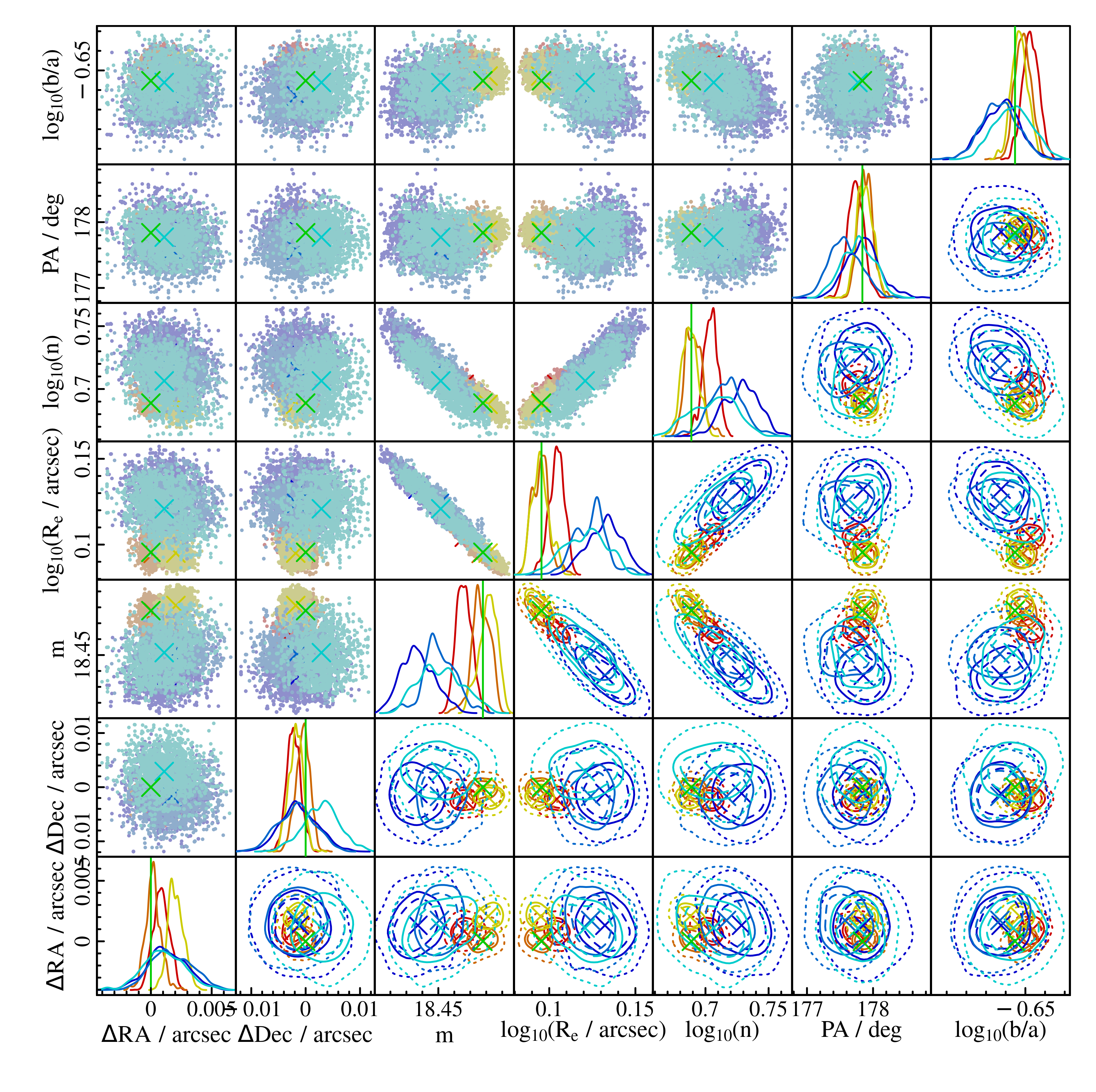}
    \caption{Similar to Figures~\ref{fig:islandplotseg} and~\ref{fig:islandplotseg2}, but now showing a random single S\'ersic simulated galaxy. Again, red through to yellow and blue through to turquoise colours show the fits obtained for the first and second match respectively for increasing segment sizes. The exact segment sizes shown differ from those in Figures~\ref{fig:islandplotseg} and~\ref{fig:islandplotseg2}, see text for details. Green crosses and lines indicate the true values used for generating the galaxy.} 
    \label{fig:islandplotsegsim}
\end{figure}

Figure~\ref{fig:islandplotseg2} shows an example of such a fit. The layout of the plot and the meaning of the colours are exactly the same as for Figure~\ref{fig:islandplotseg}, only that now we show a different galaxy, namely 107016. This galaxy is not well-represented by a single S\'ersic fit and instead needs (at least) two components.\footnote{The double component fit of the object is acceptable, although there are still slight residuals visible at the centre that may point to the existence of a bar.}. There is therefore no unambiguous, correct solution for a single S\'ersic fit to that galaxy and the fitted parameters will always depend on the exact segment size used. 

This is a point we already made in Section~\ref{sec:postprocessing}, but it is re-inforced here. Choosing the optimal segment size is therefore non-trivial and depends on the science aim. Since our focus - as explained before - is on obtaining the best possible fits of the high signal to noise regions of galaxies, we do not want to make the segments very large. Nonetheless, they should be large enough to reach the convergence regime at least for cases where there is an unambiguous solution, such as that shown in Figure~\ref{fig:islandplotseg}. 

The best way to determine where the convergence regime begins is using simulations. These have two major advantages: we know that the fake galaxy is well-represented by a single S\'ersic fit if it was created as such; and we know the true parameter values used to create the galaxy. In addition, we know the true PSF used to convolve the generated galaxy with, eliminating PSF uncertainties from the analysis. We have created such simulations for our studies of systematic uncertainties in Section~\ref{sec:simulations}. More details on their creation can be found there, for the analysis in this section it suffices to say that they are all perfect single S\'ersic objects, but realistic in all other properties. After creation, we run them through our pipeline as if they were real galaxies. 

Figure~\ref{fig:islandplotsegsim} shows an example similar to Figures~\ref{fig:islandplotseg} and~\ref{fig:islandplotseg2} for a simulated galaxy. Since the simulations were run at a later stage during pipeline development than the segment size studies shown above, the colours have changed in meaning slightly: red to yellow still shows the fits to the first match of the simulated galaxy (which we also chose to lie in the overlap region) with increasing segment size; while blue to turquoise shows the same for the second match. There are now only three segment sizes shown: no additional dilation (red and blue), dilation such that the area increases by approximately 30\,\% (i.e. the version also used for the final \texttt{v04} pipeline, orange and light blue) and dilation with a kernel five times larger than the value used for \texttt{v04}, corresponding to an area increase by about a factor of 2.5 (yellow and turquoise). Green lines and crosses show the true parameter values.

In contrast to Figures~\ref{fig:islandplotpsf} to~\ref{fig:islandplotseg2}, the RA and Dec values for both matches in Figure~\ref{fig:islandplotsegsim} are in perfect agreement since the fake galaxy was inserted after the astrometric solution had already been fixed. The agreement between the two matches and the true value for all other parameters is reasonable: the first match (red, orange, yellow) is generally closer to the truth than the second. It is also the deeper one of the two images, evidenced by its higher constraining power (narrower distributions). The second, shallower, image seems to suffer from some systematic uncertainty, possibly from the background subtraction (it cannot be due to the PSF since we use the true PSF used for generating the galaxy also for fitting). We do not investigate this in detail here but refer to Section~\ref{sec:systematics}, where we explore systematic uncertainties in more detail.

The differences between the different segment sizes are generally small. In particular, both versions of the dilated segments show a high degree of overlap despite their very different sizes (30\,\% area increase vs. a factor of 2.5). The undilated segment again tends to converge onto a solution that is brighter, larger and with higher S\'ersic index and differs more from the other two. This is despite it being relatively close in size to the first one of the dilated versions and again emphasises the importance of a dilation step that is independent of the pixel values to obtain an unbiased segment solution. 

In conclusion, an unbiased dilation step after the \texttt{profoundProFound} curve of growth convergence is necessary. A relatively small dilation, increasing the area only by approximately 30\,\% is sufficient to reach the parameter convergence regime for cases where there is an unambiguous solution, i.e. where the model represents the data well. At the same time it minimises influence from, e.g., background subtraction uncertainties or neighbouring objects that can become more dominant for larger segment sizes. It also ensures that when working with a model that is not a perfect fit to the data, the focus is on correctly representing the regions dominated by the galaxy flux rather than the low signal to noise outskirts or the sky background.

\subsubsection{Segmentation map fixing}

Early versions of \texttt{ProFound} had a tendency to split up large, well-resolved galaxies with prominent substructure into several smaller segments. To combat this unwanted shredding effect, we devised a segmentation map fixing procedure using \texttt{ProFit}. This routine was used up to \texttt{v02} of the \texttt{BDDecomp} DMU. It then became redundant since we upgraded \texttt{ProFound} to its newest version which saw major changes in how the segmentation was performed. The updated version of \texttt{ProFound} was much less likely to shred galaxies, even for large and well-resolved cases with a lot of substructure. We therefore turned off our own segmentation map fixing from \texttt{v03} onwards since it was consuming unnecessary computational time and had the unwanted side effect of sometimes including faint and small sources in the segmentation map for the galaxy. Nonetheless, we briefly describe the segmentation map fixing routine here. 

After the \texttt{profoundProFound} run during the preparatory work, the galaxy is fitted with a single component S\'ersic model using the same fast downhill gradient algorithm used for the star fitting in PSF estimation (Section~\ref{sec:preparatorysteps}). Like for the ``real" galaxy fits (Section~\ref{sec:galaxyfitting}), we fit all parameters except boxyness, use only the pixels within the galaxy segment for fitting and take initial guesses from the segmentation statistics.

For each segment in the cutout around the galaxy, the flux it contains is then compared against the flux it should contain according to the fitted model: if the model flux accounts for more than 50\,\% of the total flux in the segment and the segment contains more than 1\,\% of the total model flux, then the segment is added to the galaxy segment. Point sources within the segment of interest are also identified by comparing to a smoothed version of the image and masked out. The galaxy is then re-fitted and the process repeated until no more segments are added. A diagnostic plot is returned and the fitted parameters are written to the header of the segmentation map to serve as better initial guesses in the galaxy fitting process.

Since the success of this segmentation map fixing crucially depends on the model fit, we repeat the process several times during the actual galaxy fitting. Therefore, after the first round of single S\'ersic MCMC fits, the new (better) model image is used to check again whether any segments need to be added to the galaxy segment. If this is the case, the object is flagged, the segmentation map is updated and the fit is repeated. The updated segmentation map and initial guesses (in the header) are saved along with a diagnostic plot and the final fits. This process is repeated a maximum of five times, if the galaxy segment still changes in the last round the object is flagged. 

The segmentation maps are then checked and fixed again on the basis of the model images resulting from the double component fits (using the same procedure). Galaxies where the segmentation map changed during this step are then re-fitted with single S\'ersic models to ensure that the single S\'ersic and double component fits are directly comparable (i.e. using the same segmentation map). The segmentation maps are not fixed again during this re-fitting step nor during any subsequent (1.5-component) fits of the galaxy. 

\begin{figure}
	\includegraphics[width=\textwidth]{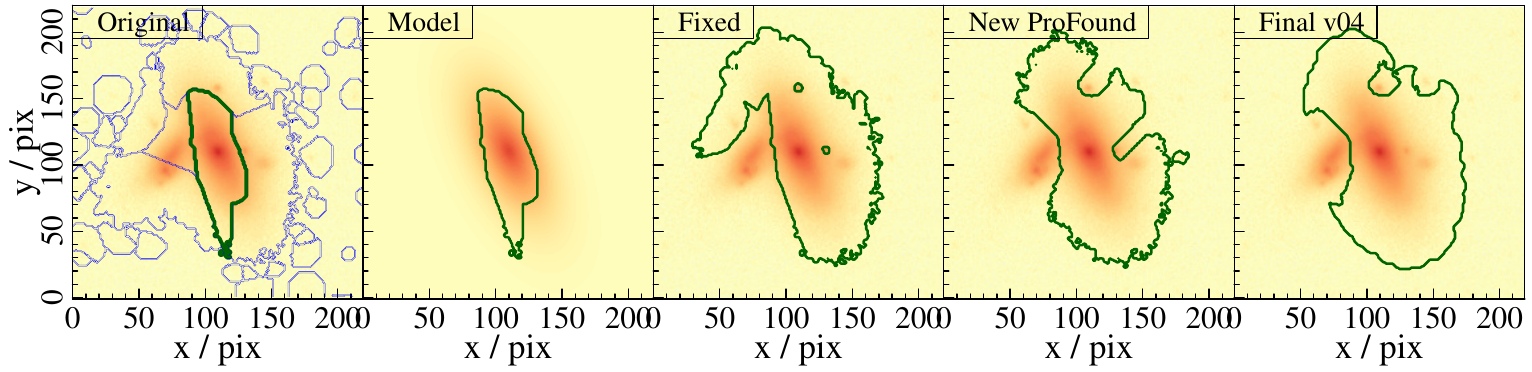}
    \caption{An example of the segmentation map fixing for galaxy 544891. \textbf{First panel:} the galaxy segment (green) and all neighbouring segments (blue) obtained with the version of \texttt{ProFound} used up to \texttt{v02} of our pipeline overlaid on the galaxy image. \textbf{Second panel:} the model image fitted to the original galaxy segment during the segmentation map fixing procedure. \textbf{Third panel:} the galaxy segment obtained after two iterations of our segmentation map fixing procedure again overlaid on the galaxy image. \textbf{Fourth panel:} the segmentation obtained with the version of \texttt{ProFound} used from \texttt{v03} of the pipeline onwards, with all other parameters unchanged with respect to the third panel. \textbf{Fifth panel:} the final \texttt{v03} and \texttt{v04} segmentation after further updates to the procedure (see text for details). Note that for clarity we do not show neighbouring segments in panels two to five. }
    \label{fig:segfixplot}
\end{figure}

Figure~\ref{fig:segfixplot} shows an example for a galaxy with a nearby companion and two small point sources that led to it being split into several segments in \texttt{v02} of the pipeline. This can be seen in the first (leftmost) panel, where the galaxy segment is shown as a green contour and all other segments in blue. The second panel then shows the model fitted to that image during the segmentation fixing process, using the original galaxy segment only. In the third panel, we see the fixed segmentation map obtained after two iterations. Three segments have been added to the galaxy segment and two point sources masked out, such that the segmentation is now acceptable. It includes the majority of the galaxy flux, but excludes the neighbouring source. These first three panels are what is also shown in the segmentation map fixing diagnostic plot mentioned above. 

The fourth panel shows the result of the segmentation directly after upgrading \texttt{ProFound} with our own segmentation fixing routine turned off. While the segmentation is not perfect, it is much better than that shown in the first panel. The potential improvement obtained by running an additional segmentation map fixing would therefore not warrant the associated effort. Finally, the last panel shows the segmentation map obtained after all other updates and represents the one that was used for \texttt{v03} and \texttt{v04}. The main change here was the switch from a \texttt{skycut} value of one to two and the addition of a further segment dilation after flux convergence is reached (see above). The segment now has a smoother, less biased border than that in panel four, while still being of comparable size. 

\subsection{Modelling decisions}
\label{sec:modellingdecisions}
Now that we have all the inputs defined, we move on to galaxy modelling. This is the second major step in the bulge-disk decomposition pipeline and described in some detail in Section~\ref{sec:galaxyfitting} and \citet{Casura2022}. An introduction to Bayesian analysis, the S\'ersic function and the most important reasoning for which models we choose is given in Sections~\ref{sec:bayesiananalysis} and~\ref{sec:sersicmodels}. Here, we expand on these sections by pointing out details that we did not previously cover. We assume the reader to be familiar with the contents of Sections~\ref{sec:bayesiananalysis}, ~\ref{sec:sersicmodels} and~\ref{sec:galaxyfitting} before reading this section.

\subsubsection{Free fitting parameters}

From a model selection perspective, it is advisable to have nested models, where the single S\'ersic model is entirely contained within the double component model (i.e. the double component model is an extension of the single S\'ersic fit). If this is not the case, then for some galaxies, the single S\'ersic fit may be more appropriate in some regions, while the double component fit is better in others, leaving the question of which region of the galaxy is ``more important" to fit. For nested models, the double component fit will always be at least as good as the single S\'ersic model and the only decision to make remains whether it is significantly better (i.e. so much better that the increased number of parameters is justified) or not. This can be determined by the Bayes factor (Section~\ref{sec:bayesiananalysis}) for models that accurately represent all features of the galaxy; or by simulations which capture all of said features; or by visual inspection as we did since we have neither perfect models nor perfect simulations for our diverse sample of galaxies. Having said that, we also considered some versions of non-nested models during pipeline development, as we outline below. 

The single S\'ersic function in two dimensions has a total of eight parameters. Six of those, namely position in $x$ and $y$, magnitude, effective radius, position angle and axial ratio definitely need to be fitted for all galaxies in the single S\'ersic fit, leaving only the S\'ersic index and the \texttt{boxyness} for consideration. Since our galaxy sample is diverse, containing classical elliptical galaxies as well as (pure) disk galaxies, double component galaxies and irregulars, we decided to leave the S\'ersic index of the single component fit as a free parameter for fitting. This allows to capture all of those different galaxy morphologies in just one fitting round and still leads to rapid convergence of the model parameters. 

We briefly experimented with fitting the \texttt{boxyness} as well, i.e. allowing deviations from an elliptical shape of the profile in two dimensions, at least for the single S\'ersic profile. However, this increased the time needed to achieve convergence by a factor of a few without major improvements in the fit quality determined from visual inspection. In addition, we would then need to allow for \texttt{boxyness} in the double component fits as well to keep the models nested, further increasing the fitting time and the problem of parameter degeneracies. Fixing the \texttt{boxyness} to zero leaves seven free parameters for the single S\'ersic fits, as mentioned previously.

\begin{figure}
\begin{center}
\includegraphics[width=0.8\textwidth]{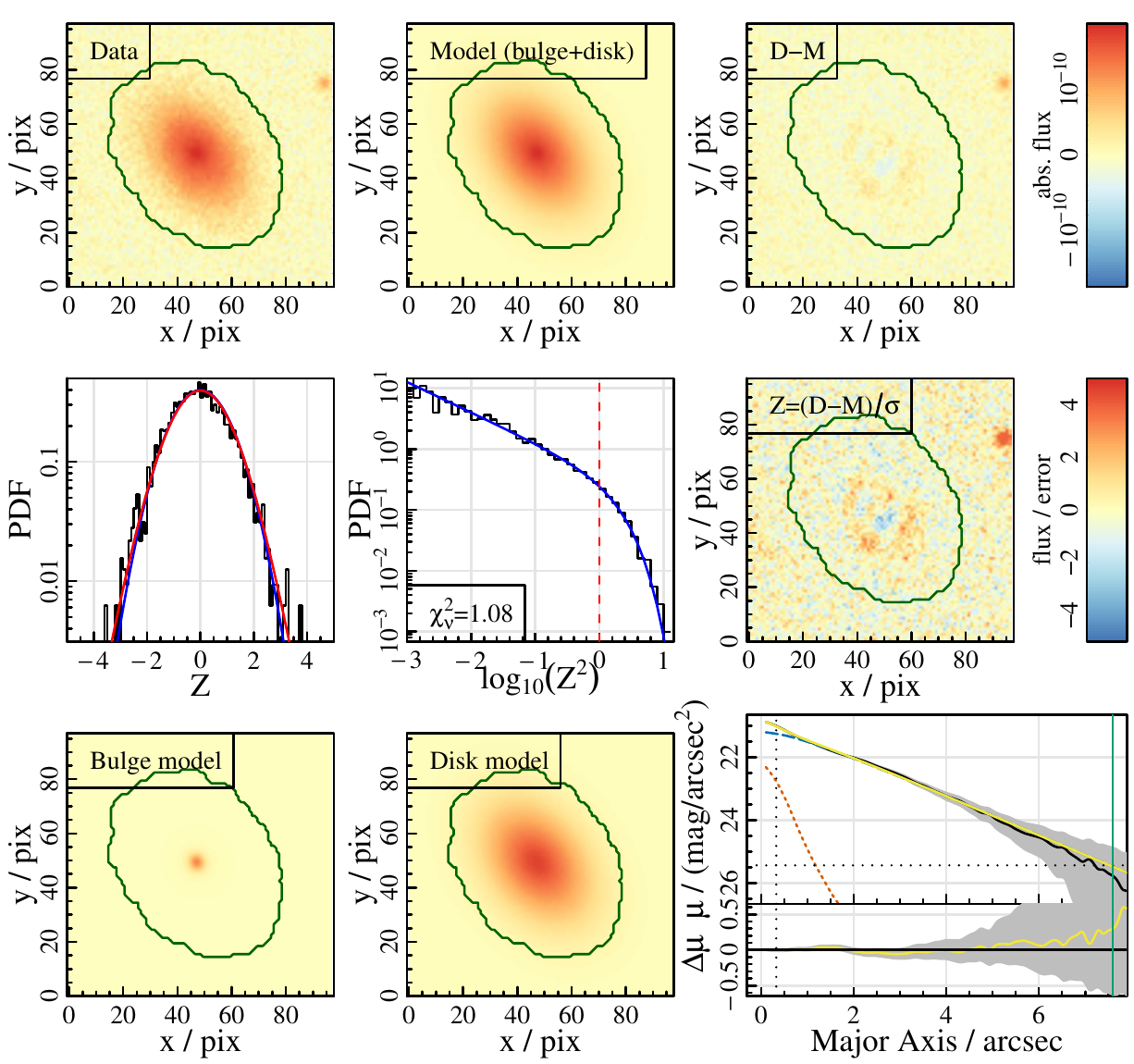}
\caption{An example of a 1.5-component fit for galaxy 124052 in the KiDS $r$-band. Panels are the same as those in Figure~\ref{fig:examplefit}.}
\label{fig:examplefit1.5}
\end{center}
\end{figure}

\begin{figure}[t!]
\begin{center}
\includegraphics[width=0.8\textwidth]{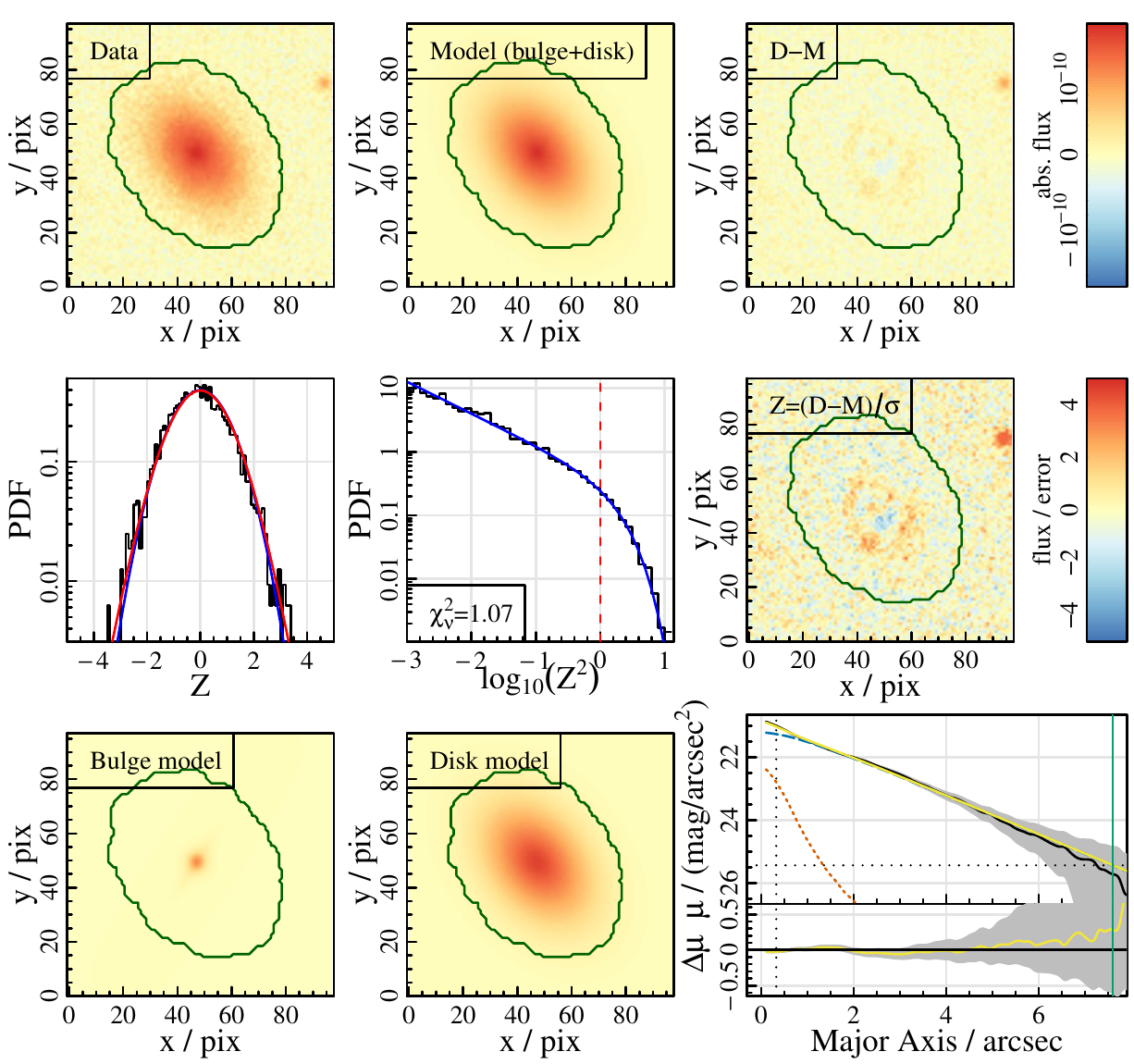}
\caption{The double component fit for galaxy 124052 (classified as 1.5-component object) in the KiDS $r$-band, for direct comparison to the 1.5-component fit in Figure~\ref{fig:examplefit1.5}. }
\label{fig:examplefit1.5b}
\end{center}
\end{figure}

The next complicated one of our three models is the 1.5-component model, which has a total of 11 parameters (eight for the S\'ersic function plus three for the point source). For the same reason as above, we did not consider \texttt{boxyness} and - as explained below for the double component model - we fixed the bulge and disk positions to lie on top of each other. Leaving the S\'ersic index free then results in an extension of the single S\'ersic model by just one parameter, namely the point source magnitude. However, the reason why we introduced the 1.5-component model was to fit unresolved bulges for double component galaxies, i.e. it was meant as a simplification of the double component fits and not an extension of the single S\'ersic fit. In fact, we originally added the 1.5-component fits after the single and double component fits for \texttt{v01} of the \texttt{BDDecomp} DMU were already processed, after noticing the $\sim$\,20\,\% of bulges with unconstrained parameters (very small or unrealistically large effective radii combined with excessively high or low S\'ersic indices and sometimes also low axial ratios, often hitting the fit limits in at least one of those parameters; see also Figures~\ref{fig:examplefit1.5} and~\ref{fig:examplefit1.5b}). 

To make the 1.5-component fits most directly comparable to the double component fits, we therefore fixed the disk S\'ersic index to 1. While it is still not strictly nested in the double component model, the 1.5-component model in this way simply replaces the S\'ersic bulge by a point source bulge (with just one parameter), thus serving its purpose of ``saving" the double component fits for which the bulge is ill-constrained. For the same reason, we also left the 1.5-component fits last in the order of processing even though in general it is advisable to start with simpler models and then increase complexity. Processing the 1.5-component fits last has the additional advantage that we need not worry about swapping of the bulge and disk since that has already been taken care of for the doubles (see below) and we can directly use the double component initial guesses. In summary, our 1.5-component model consists of a point source plus an exponential disk, tied to the same position. Note that this means the 1.5-component fit also has just 7 parameters and often results in a worse fit than the single S\'ersic model (e.g. for elliptical galaxies that cannot be represented well with an exponential disk).

An example of such a 1.5-component fit is shown in Figure~\ref{fig:examplefit1.5}, with panels the same as in Figure~\ref{fig:examplefit}. For direct comparison, we show the double component fit in Figure~\ref{fig:examplefit1.5b}. The two fits are visually indistinguishable and their disk parameters are very similar. However, the double component fit has a bulge S\'ersic index of 20 and an effective radius of 0.1\arcsec\ (0.5\,pix), combined with an axial ratio of 0.05, thereby hitting all three of those fit limits. These values (as well as the position angle) are meaningless, but they do influence the fitted magnitude, which is 21.5\,mag for the double component model. The point source model does not fit any of the meaningless parameters and instead produces a more robust magnitude estimate of 22.4\,mag. 

Both the single S\'ersic and 1.5-component fits tend to converge rapidly and without problems as there are only few parameter correlations. As explained above, the decisions for which parameters to fit were mostly done on a theoretical basis and straightforward. Most of the remaining discussion focuses on the double component models, where there are many more parameter correlations and degeneracies and therefore also more options to consider in how to limit those. 

The double component model has a total of 15 parameters (8 for the S\'ersic function and 7 for the exponential). Based on our experience with the single S\'ersic fits, we did not consider \texttt{boxyness} for either of the two components. Another natural choice to make is to tie the bulge and disk positions together due to the symmetry of ``typical" galaxies. Nonetheless, we also considered leaving the bulge and disk positions independent from each other to better capture disturbed morphologies. Unfortunately, this frequently led to one of the components wandering off to fit overlapping point sources or wings of other objects, specifically when the main part of the galaxy can be adequately represented by a single component. Conversely, it only significantly increased the fit quality for objects with an irregular morphology that are difficult to represent with our symmetrical models even when allowing offset bulges. Hence, we decided to tie the bulge and disk positions together to lie exactly on top of each other. This leaves a total of 11 free parameters for the double component model, which we also used in the end despite numerous considerations to limit the degrees of freedom further (see below).

\subsubsection{Constraints on bulges}

Some of the considerations to reduce the degrees of freedom of the double component fit involved tying the position angle of the bulge to that of the disk or forcing the bulge to be round (i.e. setting the axial ratio to 1, or >\,0.8 or similar); and/or setting the S\'ersic index of the bulge to 4 or limiting it to be larger than, e.g., 2. All of these measures can be used to ensure that the bulge component does indeed fit a classical bulge and not other features such as pseudo-bulges or bars. However, as explained in Section~\ref{sec:sersicmodels}, it is not our aim to fit classical bulges only and instead we use the S\'ersic component of the model to explicitly fit all morphological features near the centre of the galaxy. Constraining the bulge to be round or have a high S\'ersic index would therefore be counterproductive.

\begin{figure}
\begin{center}
\includegraphics[width=0.8\textwidth]{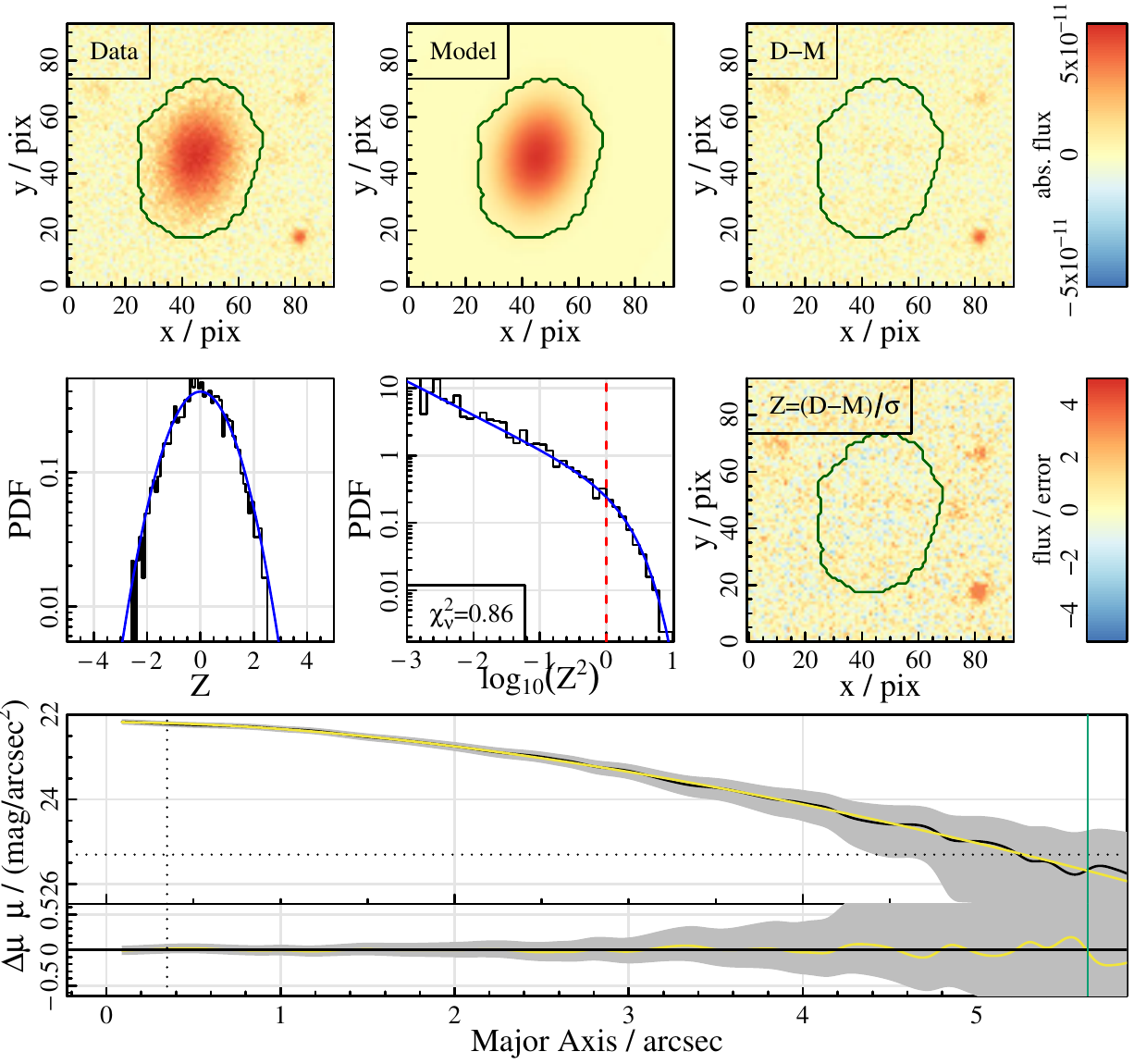}
\caption{An example of a single S\'ersic fit with a low S\'ersic index (0.6) for galaxy 220084 in the KiDS $r$-band. Panels in the top two rows are the same as those in Figure~\ref{fig:examplefit}, while the bottom row shows the one-dimensional fit only, corresponding to the rightmost panel of the bottom row in Figure~\ref{fig:examplefit}.}
\label{fig:examplefitlown}
\end{center}
\end{figure}

\begin{figure}[t!]
\begin{center}
\includegraphics[width=0.8\textwidth]{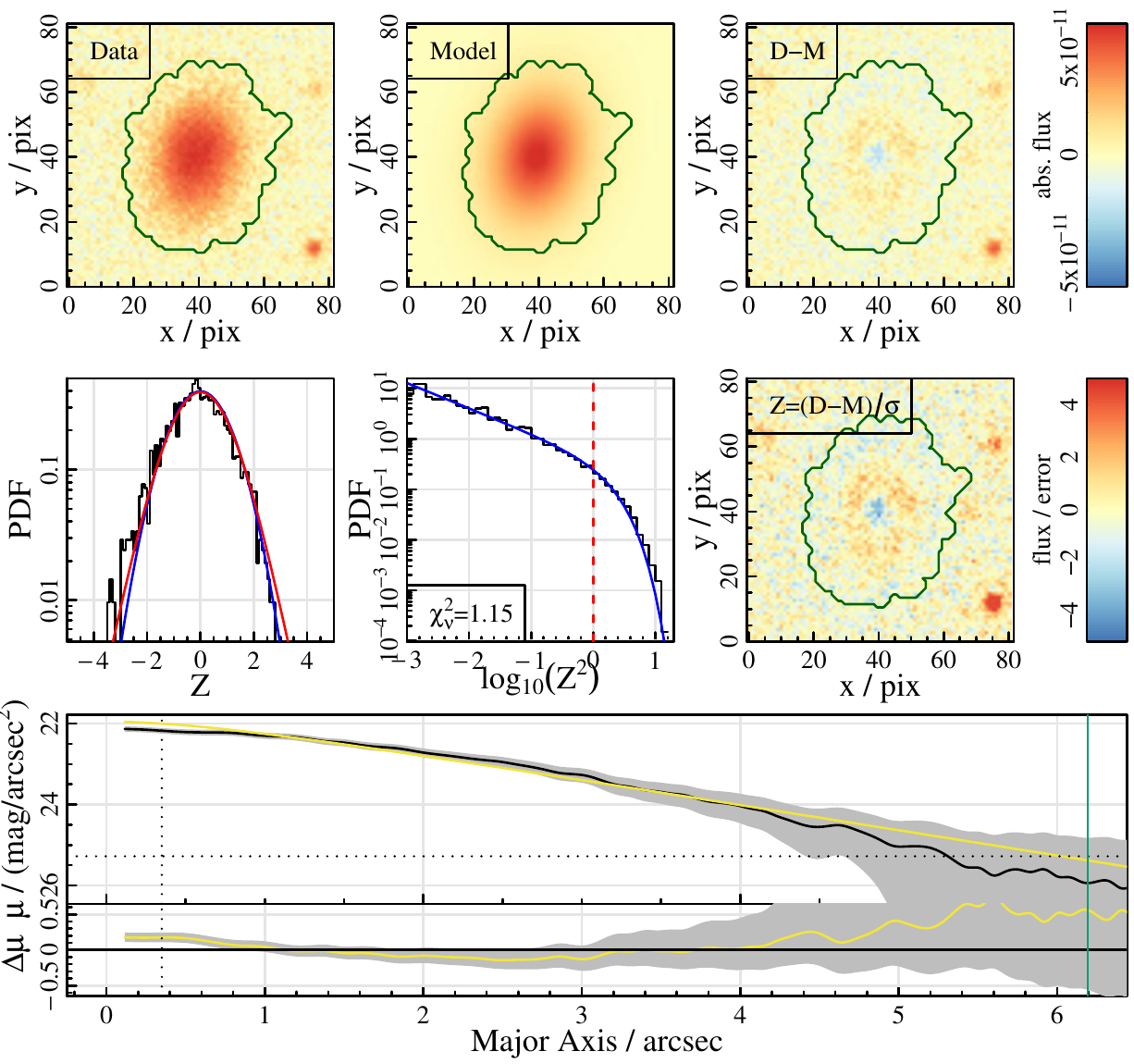}
\caption{The \texttt{v01} single S\'ersic fit to galaxy 220084, where the S\'ersic index converged onto its (then) lower limit of unity; for direct comparison to Figure~\ref{fig:examplefitlown}.}
\label{fig:examplefitlown2}
\end{center}
\end{figure}

Nonetheless, for versions of the pipeline up to \texttt{BDDecomp v02}, we used a lower limit on the bulge S\'ersic index of 1. The idea was that the bulge profile is then always steeper than the disk (or of the same steepness), so it is more likely that it will dominate the inner parts of the galaxy and not the outskirts (i.e. to avoid swapped fits). However, there were several problems with that approach. First of all, a significant fraction of bulges converged onto the lower limit ($\sim$\,18\,\%, with a further 33\,\% being consistent with the lower limit within errors, i.e. 51\,\% in total), thereby producing unreliable fits with unrealistic error estimates. Second, a number of pseudo-bulges, bars and broad cores (often galaxies with slightly disturbed morphologies, for which, however, acceptable fits are still achievable) could not be appropriately represented as they showed flattened profiles towards the centre. Third, we imposed the same lower limit on the single S\'ersic fits to keep the models nested, which led to $\sim$\,11\,\% of these converging onto the lower limit as well (with another 13\,\% consistent with the lower limit within errors, so 24\,\% in total). An example of a single S\'ersic object with a flattened core (S\'ersic index of 0.6) is shown in Figure~\ref{fig:examplefitlown}. For comparison, Figure~\ref{fig:examplefitlown2} shows the fit to the same galaxy in \texttt{v01} of the catalogue, where the S\'ersic index converged onto its lower limit of 1 and is clearly not able to capture the profile of the galaxy accurately. 

Finally, limiting the bulge S\'ersic index to be larger than one only helped in part against swapped fits since the flux ratio of the components is not solely determined by their respective S\'ersic indices but also by their relative effective radii, magnitudes and axial ratios; and high S\'ersic index bulges in particular can also dominate both the inner and the outer parts of the galaxy at the same time. From \texttt{v03} of the pipeline onwards, we therefore adjusted the S\'ersic index lower limit to 0.1 for all models. At the same time, we increased the upper limit on all S\'ersic indices from 12 to 20 to avoid fits hitting the limits (even though the difference in the actual profile is small at such high S\'ersic indices). This reduced the number of fits hitting either of the S\'ersic index limits to 0.3\,\% for the single S\'ersic models (including those consistent with the limit within errors), of which approximately two thirds were classified as outliers and the last third as 1.5-component fits. For the double S\'ersic models, 15\,\% still hit their bulge S\'ersic index limits, of which 60\,\% were classified as single S\'ersic fits, 14\,\% as 1.5-component fits and the remaining 26\,\% as outliers.

In another attempt to stop the bulge and disk components from swapping, we also ex\-pe\-ri\-men\-ted with limiting the effective radius of the bulge to be smaller than that of the disk. Unfortunately, this did also not have the desired effect, as the precise meaning of the effective radius strongly depends on the S\'ersic index and the two parameters (as well as the magnitude) are highly correlated. For example, a bulge with a high S\'ersic index can easily dominate the flux in the central parts of the galaxy even if its effective radius is comparable to or larger than that of the disk (with a S\'ersic index of 1). 

\begin{figure}[t!]
\begin{center}
\includegraphics[width=0.8\textwidth]{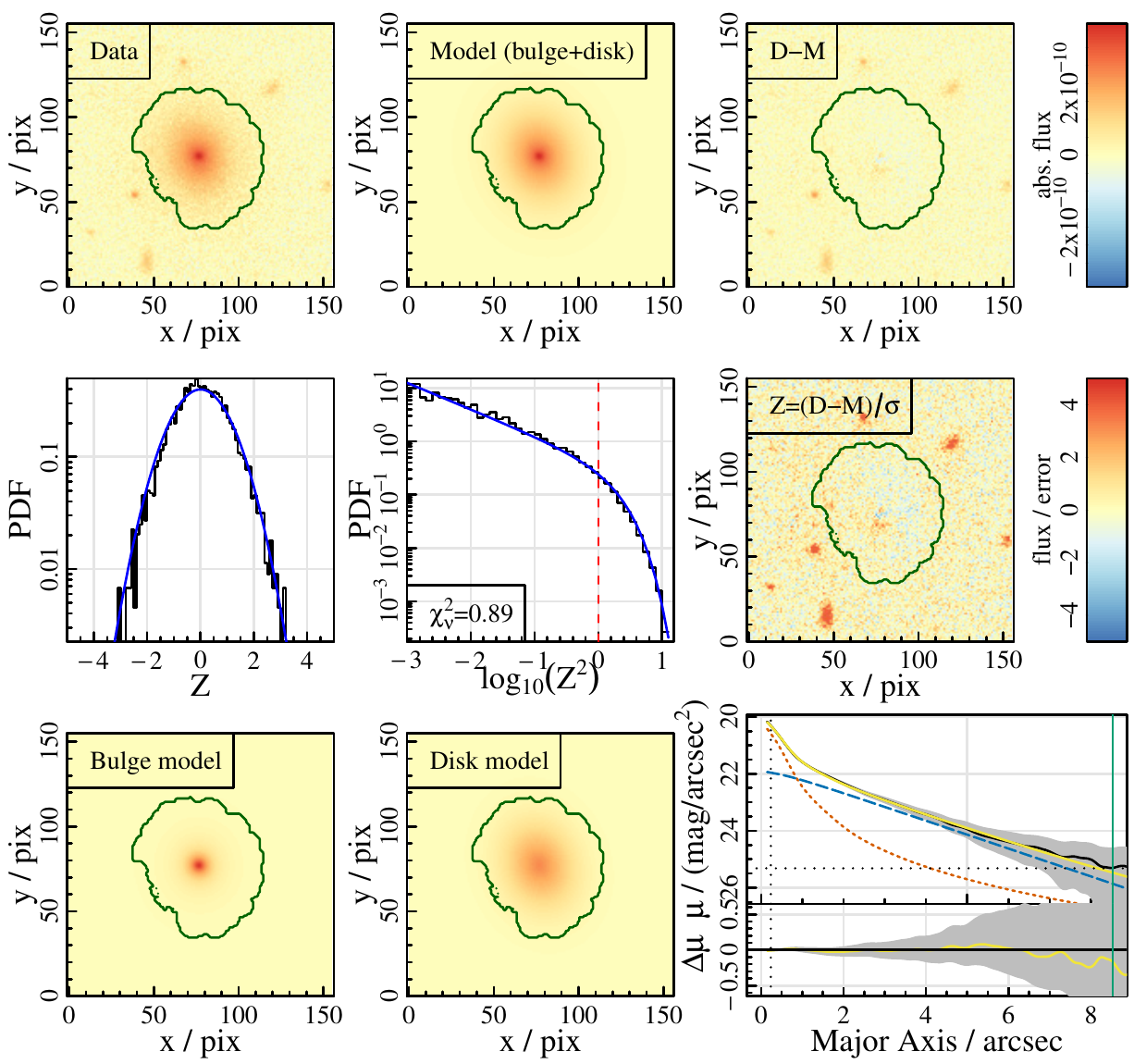}
\caption{The double component fit to galaxy 136866, where the bulge effective radius is much larger than that of the disk. Panels are the same as in Figure~\ref{fig:examplefit}}
\label{fig:examplefitbulgere}
\end{center}
\end{figure}

An example is shown in Figure~\ref{fig:examplefitbulgere}. The bulge and disk clearly dominate the inner and outer parts of the galaxy respectively. Nonetheless, the bulge effective radius is a factor of more than six larger than that of the disk due to its high S\'ersic index of 15.8. This is frequently the case especially since we fit only the pixels in relatively small segments, so the extended wings of a high $n$ S\'ersic function fall beyond the fitting region and are effectively ignored.\footnote{On a side note, when limiting the effective radii to the segment values, the disk $R_e$ is a factor of about 2 larger than that of the bulge. However, since we truncate to segment radii in a post-processing step, we cannot use this during the fit itself.} 

Given the limited success of our attempts to avoid swapped components by constraining the bulge parameters, we have instead devised a separate routine to swap the components of galaxies where necessary, see Sections~\ref{sec:galaxyfitting} and~\ref{sec:swappingandoutliers}.

\subsubsection{Priors and initial guesses}

One other route towards influencing the fit without imposing hard cuts is by using differing initial guesses, as we do in our swapping routine (Section~\ref{sec:galaxyfitting}). However, one of the main strengths of an MCMC analysis is that it is less dependent on initial guesses than e.g. downhill gradient fitting. When starting very close to a local maximum (as we do for the swapped fits), the fit is more likely to converge onto that. Similarly, when starting very far from any maximum (i.e. in the flat part of the likelihood space), convergence takes much longer and fails more frequently. For any ``reasonable" initial guesses, however, the fit result will not depend strongly on them. We obtain such reasonable initial guesses from the segmentation output of \texttt{profoundProFound} and further improve them via a fast downhill gradient algorithm before starting the MCMC (see Section~\ref{sec:fittingspecifics}). The details of, e.g., how to convert the \texttt{profoundProFound} concentration to a S\'ersic index initial guess, how to split the total flux between the bulge and disk initially, what ratio of effective radii to use and the starting value of the bulge S\'ersic index were the subject of several investigations but were found to have little effect on the fit results. 

Arguably the most ``Bayesian" way to constrain parameters is via priors. These can be hard cuts - like the fixed intervals that we use for fitting parameters - or soft, like for example Gaussian functions of a certain width centred on the most likely value. However, when imposing such informative priors, they should be based on previous knowledge (e.g. from previous data) rather than general trends or notions since they will directly affect the posterior and can even dominate over the likelihood if the prior is strong and/or the likelihood is weak due to a limited quality or quantity of data. To get the most unbiased estimate of the parameters based on the data alone it is therefore wise to choose uninformative, broad priors. 

We do this by imposing fixed limits, but using flat priors within these limits (in logarithmic space where appropriate). The limits are listed in Table~\ref{tab:singlefitlimits} and are deliberately very broad. Most of them are based on physical considerations (e.g. the axial ratio cannot be larger than 1) or limitations of the data (e.g. if the effective radius is less than 0.5\,pix, more than half of the total light of the profile is contained within a single pixel, at which point the accuracy of the PSF is very likely not sufficient anymore). The only exception to this is the S\'ersic index, for which we discuss the limits in some detail above. 

Note that as we mention in Section~\ref{sec:galaxyfitting}, we do not normalise our priors, leading to unnormalised posteriors. In theory, normalised posteriors are important to obtain meaningful Bayes' factors for model selection. In practice, we found that in our case even with normalised posteriors the model selection needs manual calibration due to the model inadequacy. Leaving the priors unnormalised has the advantage that the prior ranges do not influence the posterior and so we do not need to re-calibrate the model selection every time for different test runs with different prior intervals. It also allows to easily put infinity as the upper limit on component magnitudes and instead constrain the total magnitude of the double component fit, which otherwise would require a somewhat non-trivial joint prior to be normalisable.

\subsection{Fitting specifics}
\label{sec:fittingspecifics}
After the models are defined, there are a number of decisions to be made with regards to the actual fitting. These are summarised in Section~\ref{sec:galaxyfitting}. In the following, we point out additional details regarding the logarithmic fitting of scale parameters, the choice of likelihood function, the fitting algorithms and routine and the assessment of convergence. 

\subsubsection{Logarithmic fitting}

From a theoretical point of view, scale parameters (in our case those are the effective radius, S\'ersic index and axial ratio) are best fitted in logarithmic space, while location parameters (position, magnitude and angle) should be fitted in linear space. This is because a ``step size" for location parameters remains constant across the entire parameter range, so, e.g., the difference between a position angle of 5\degr\ and 10\degr\ is the same as that between 170\degr\ and 175\degr. For scale parameters, the step size changes across the parameter range (in linear space), for example the difference in profile shape between a S\'ersic index of 0.5 or 1 (Gaussian or exponential) is much larger than that between a S\'ersic index of 18.5 or 19 (nearly indistiguishable). Converting scale parameters into logarithmic space, i.e. fitting $\log_{10}(X)$ instead of $X$ for scale parameters $X$, equalises the step size across the (now logarithmic) parameter range again.\footnote{For the same reason, relative errors are more meaningful than absolute errors for scale parameters; and vice versa for location parameters.} Note that intensity or brightness are also scale parameters, but since magnitude is already a logarithmic measure of flux, it becomes a location parameter.\footnote{Hence why colours are calculated as magnitude differences rather than ratios.}

Given this theoretical background and the functionality of \texttt{ProFit} to easily specify which parameters should be fitted in logarithmic space (with conversions to and from logarithmic space performed internally), it seemed natural to fit scale parameters in logarithmic space. Nevertheless, we tested the effect of fitting in linear space for one or several of the scale parameters during our test runs. Fitting all parameters in linear space approximately doubled the computational time needed for single S\'ersic fits compared to logarithmic fitting of all scale parameters, while no systematic differences in the estimated parameters could be observed. 

The difference in computational time highlights the importance of choosing similar step sizes in all parameters for rapid convergence. With the logarithmic fitting of scale parameters and our chosen fitting intervals, we arrive at comparable step sizes for all parameters with one exception: the position angle. To ensure convergence of the position angle we therefore (linearly) re-scaled this parameter into units of 30\degr. All fitting parameters then have similar ranges in absolute terms (in their respective, potentially logarithmic, fitting units), as also evident from Figure~\ref{fig:differrnorm} (the absolute values of all parameters vary in a range of about $\pm$\,0.05 around the true value; and the absolute errors are on comparable scales for all parameters).

\subsubsection{Likelihood function}

In Bayesian modelling, the probability of the observed data given an (assumed) model is called the likelihood (Section~\ref{sec:bayesiananalysis}). Calculating it requires knowing or assuming a likelihood function that takes the data and model as input and returns the corresponding probability. This can in theory be any function, but is most often a ``standard" function appropriate to the statistical problem at hand, such as a Gaussian, chi-squared, Poisson or Student's t-distribution, all of which are supported by \texttt{ProFit}. 

KiDS data is expected to have random Normal errors with pixel-specific standard deviations expressed in the sigma map (Section~\ref{sec:preparatorysteps}), which in turn is a combination of the KiDS weight maps and the object shot noise. For a perfect model, the log-likelihood can therefore simply be calculated as the sum of the Gaussian probabilities to obtain each measured flux value given the model flux value of the corresponding pixel and its error from the sigma map (\texttt{ProFit} does this internally). The correct likelihood function to use is therefore the Normal one, provided the model is a perfect fit to the data. 

To accommodate models that are not perfect fits to the data at hand (i.e. most of our models for the majority of our galaxies), \texttt{ProFit} supports the use of a Student's t-likelihood instead. This evaluates the probability of the data given the model and errors under the assumption of a Student's t-distribution rather than a Gaussian one, optimising the degrees of freedom of the Student's t-distribution at the same time. Due to the broader wings of the Student's t-distribution, this effectively downweights the regions of the galaxy that cannot be captured by the model during the fit. This makes it more robust for the purpose of galaxy fitting. 

\begin{figure}[t!]
\begin{center}
\includegraphics[width=0.8\textwidth]{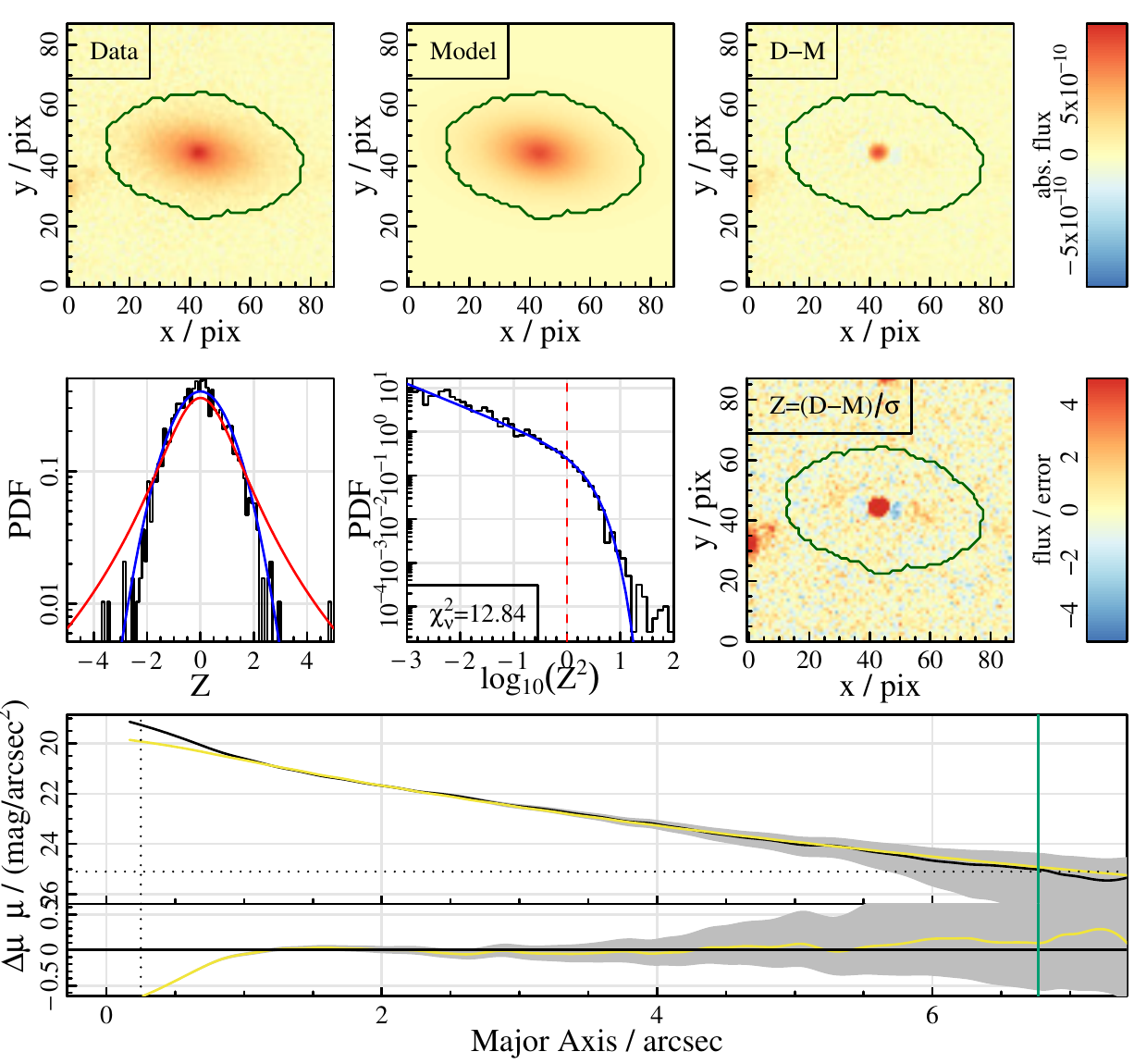}
\caption{The Student's t-likelihood function single S\'ersic fit to galaxy 177705 (a double component object) in the KiDS $r$-band. Panels in the top two rows are the same as those in Figure~\ref{fig:examplefit}, while the bottom row shows the one-dimensional fit only, corresponding to the rightmost panel of the bottom row in Figure~\ref{fig:examplefit}.}
\label{fig:examplefitnormvst1}
\end{center}
\end{figure}

\begin{figure}[t!]
\begin{center}
\includegraphics[width=0.8\textwidth]{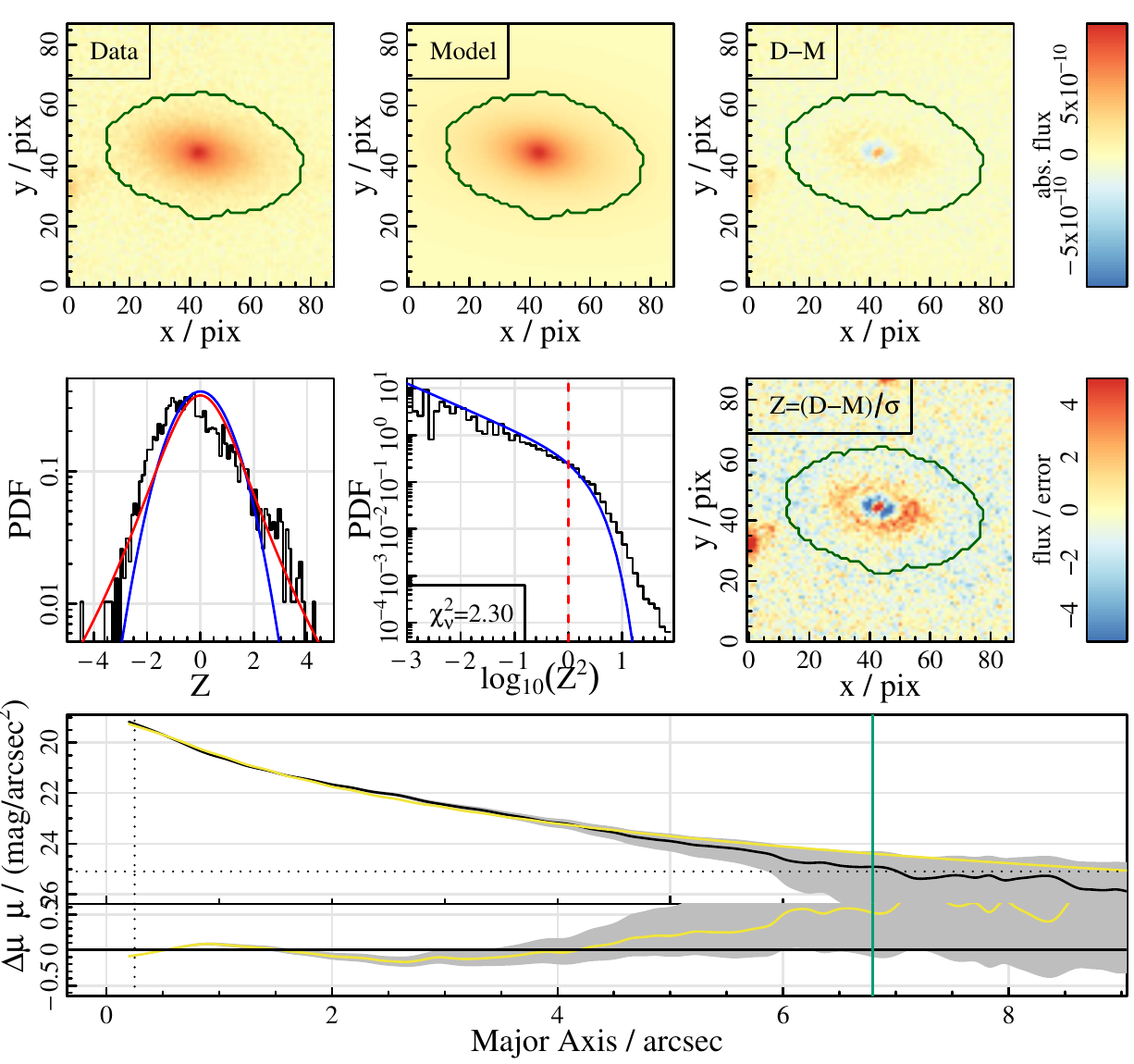}
\caption{The Normal likelihood function single S\'ersic fit to galaxy 177705, for direct comparison to Figure~\ref{fig:examplefitnormvst1}.}
\label{fig:examplefitnormvst2}
\end{center}
\end{figure}

As an example, we show the single S\'ersic fit to galaxy 177705 using a Student's t-likelihood in Figure~\ref{fig:examplefitnormvst1} and the fit to the same galaxy with a Normal likelihood in Figure~\ref{fig:examplefitnormvst2}. This galaxy needs two components to be accurately represented. Fitting it with just a single S\'ersic function gives different results for the two likelihood functions: the Student's t-likelihood is more likely to ``tolerate" a few pixels that are in high tension with the model (i.e. pixels belonging to the bulge) if this allows a better fit on average to the remainder of the galaxy. A normal likelihood instead ``tries harder" to avoid strong outliers and compromises more between fitting both components, even if that results in a generally slightly worse fit over larger areas such as the disk. While in the example here it is debatable which option is to be preferred, a Student's t-distribution is definitely prefereable if the pixels in high tension are - e.g. - an overlapping foreground point source or a clumpy component of a star-forming spiral arm. 

Since our models are generally a simplification of the true complexity of galaxies, we would expect a Student's t-distribution to be more suitable to our analysis. This was therefore the default likelihood we used initially. Unfortunately, though, it turned out to have several problems as already outlined in Section~\ref{sec:galaxyfitting}. For the final pipeline, we therefore use a Normal likelihood function for all galaxies. Since this decision was made entirely during the test runs, the \texttt{BDDecomp} DMU has only Normal likelihood function fits from \texttt{v01} onwards already. However, we still like to point out the development over the test runs, since this was a major point of concern and one aspect in which the final decision does not meet the theoretical expectations. 

\begin{figure}[t!]
\begin{center}
\includegraphics[width=0.8\textwidth]{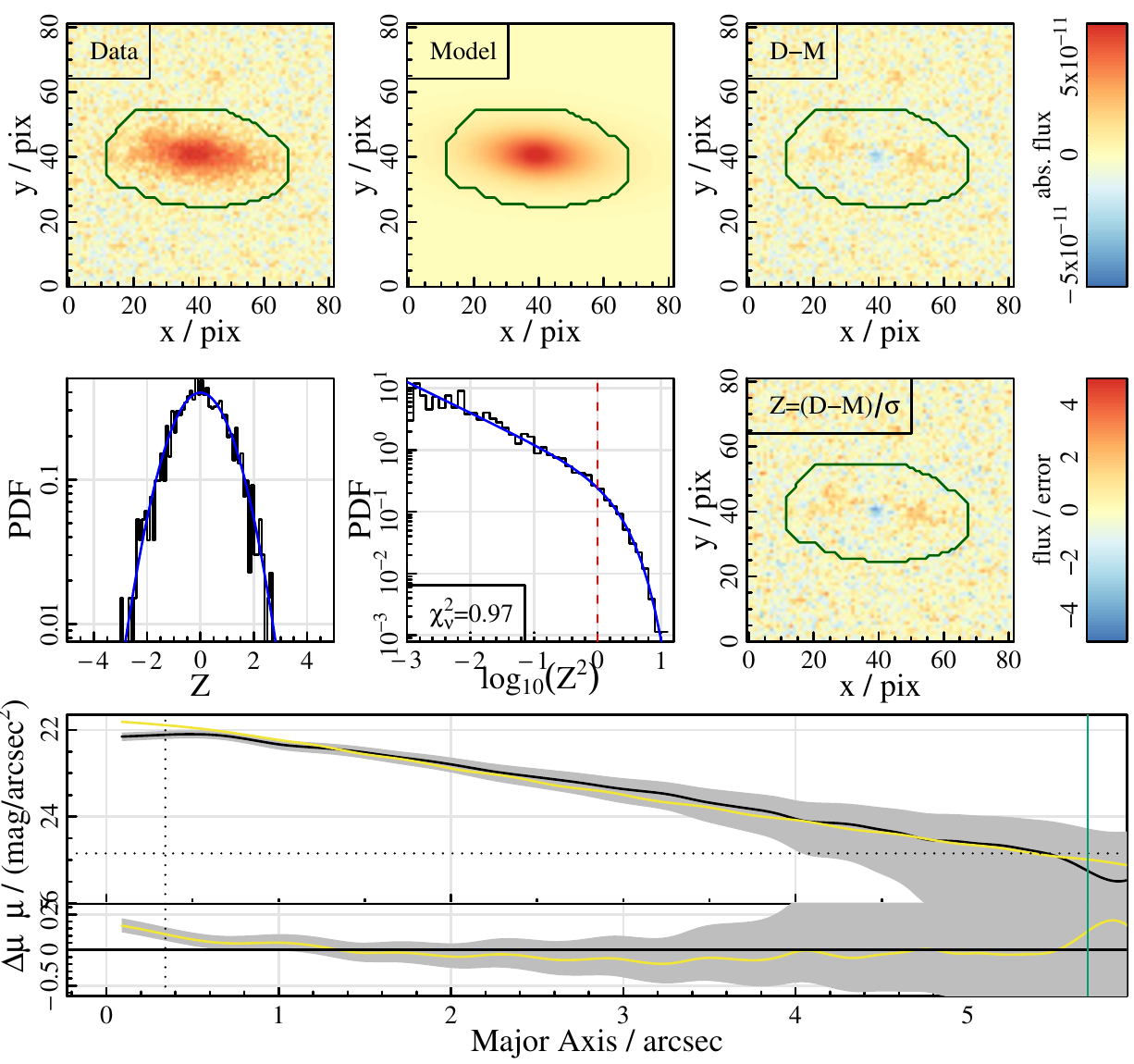}
\caption{The Student's t-likelihood function single S\'ersic fit to galaxy 534802, which is well-represented by a single S\'ersic profile, in the KiDS $r$-band. Panels in the top two rows are the same as those in Figure~\ref{fig:examplefit}, while the bottom row shows the one-dimensional fit only, corresponding to the rightmost panel of the bottom row in Figure~\ref{fig:examplefit}.}
\label{fig:examplefitnormvst3}
\end{center}
\end{figure}

\begin{figure}[t!]
\begin{center}
\includegraphics[width=0.8\textwidth]{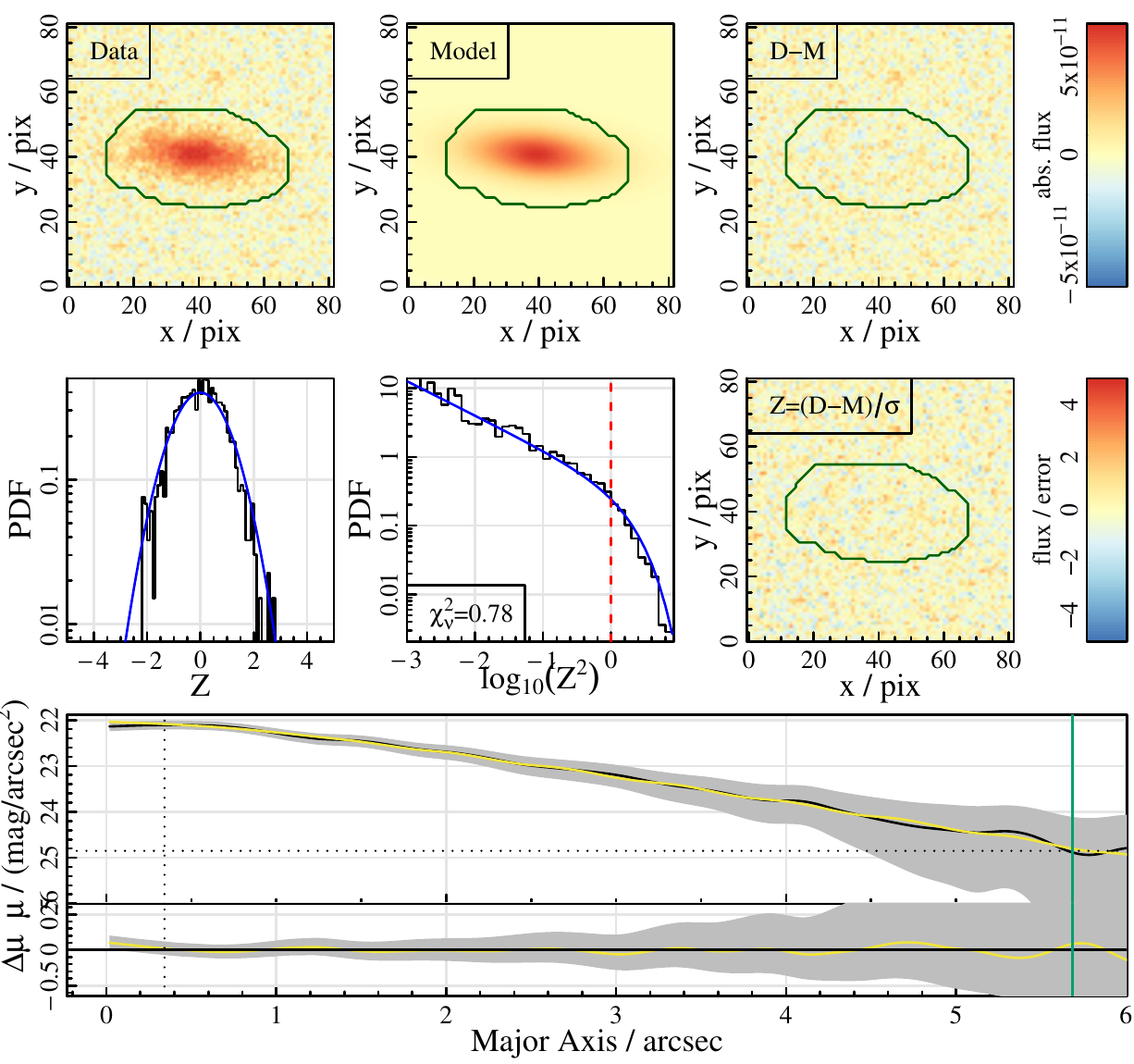}
\caption{The Normal likelihood function single S\'ersic fit to galaxy 534802, for direct comparison to Figure~\ref{fig:examplefitnormvst3}.}
\label{fig:examplefitnormvst4}
\end{center}
\end{figure}

\begin{figure}[hb!]
\begin{center}
\includegraphics[width=0.8\textwidth]{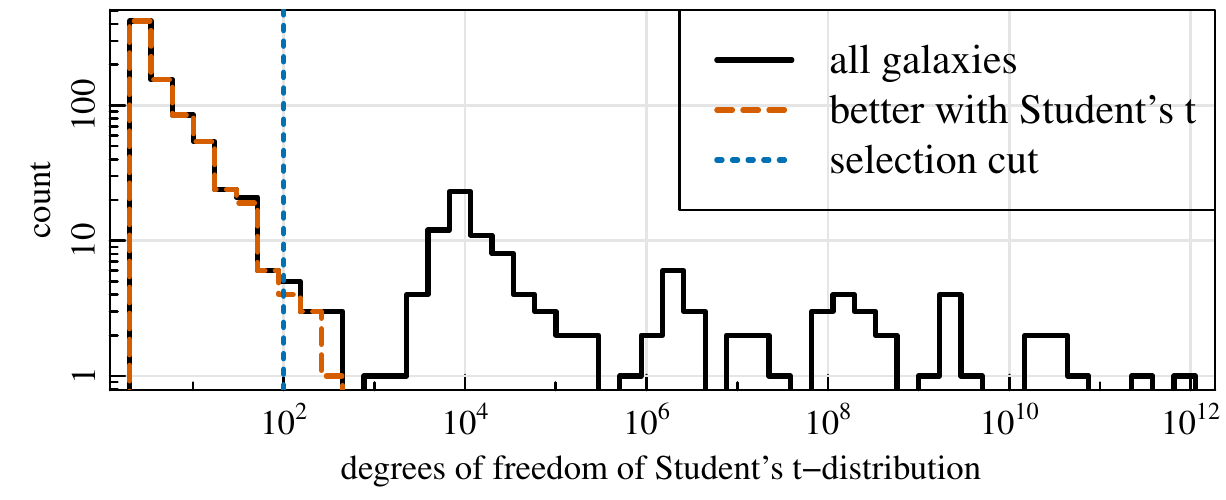}
\caption{The distribution of the Student's t-distribution degrees of freedom for a set of test galaxies (solid black line) and for the subset of those that achieves a better fit with a Student's t-likelihood compared to a Normal likelihood (dashed orange line). The dotted blue line shows the selection cut for deciding which likelihood function to use.}
\label{fig:normvsthist}
\end{center}
\end{figure}

The first problem of using Student's t-distribution likelihood functions became apparent for galaxies that are perfectly represented by their respective model. In this case, as explained above, the errors truly are distributed Normally. The Student's t-distribution, however, ``expects" a certain fraction of pixels that are in tension with the model and - if they are not present - will artificially produce them by making the fit unnecessarily worse. An example for this is shown in Figures~\ref{fig:examplefitnormvst3} and~\ref{fig:examplefitnormvst4} for galaxy 534802. This object can be perfectly represented by a single S\'ersic component, as evident from the fit with a Normal likelihood function. The fit with the Student's t-likelihood, though, shows considerable residuals. 

We concluded that fitting all galaxies with a Student's t-likelihood is unsuitable and instead a certain fraction, namely those approximately 20\,\% of galaxies that are perfectly captured by one of our models, need a Normal likelihood function. The correct way to decide which likelihood function to use would be to fit each galaxy twice (once with each likelihood function) and then choose the one which achieves the better fit, i.e. the higher likelihood. However, we noticed that the degrees of freedom estimated for the Student's t-likelihood during the fit can provide a shortcut to save computational time. 

As the degrees of freedom of the Student's t-function approach infinity, its distribution approaches that of a Gaussian. It is therefore not surprising that the degrees of freedom converge onto a large value for those galaxies that would be better fitted by a Normal likelihood. First fitting all galaxies with a Student's t-likelihood therefore allows to select only a subset of galaxies for re-fitting with a Normal likelihood based on a cut in the degrees of freedom of the Students's t-function. 

We demonstrate this in Figure~\ref{fig:normvsthist} for a test sample of KiDS $r$-band galaxies that we fitted with both likelihood functions. The solid black histogram shows the degrees of freedom of the Student's t-distribution for all galaxies in the test sample. There is a clear peak encompassing the majority of galaxies at low values of degrees of freedom and a long tail to very large values (note the logarithmic scaling of both axes). With a dashed orange line, we overplot the subset of galaxies that are better fitted with a Student's t-likelihood, i.e. those that achieve a higher likelihood with a Student's t-likelihood than with a Normal one. This encompasses essentially all of the galaxies with low degrees of freedom and none of those with high values (for which a Normal likelihood function achieves a better fit). 

The dotted blue line indicates our selection cut for deciding which galaxies to re-fit with a Normal likelihood function, namely those for which the degrees of freedom of the Student's t-likelihood function fit converged to values larger than 100. Note that this cut is de\-li\-be\-ra\-te\-ly more to the left, since galaxies to the left of this line will only be fitted with a Student's t-\\\\distribution (therefore missing the best fit for any galaxies in that sample for which a Normal likelihood would be better suited), while those to the right of the line will be fitted with both likelihoods such that the optimal likelihood can still be chosen based upon which achieves the better fit (therefore only wasting a small amount of computational time on the galaxies in that sample for which a Student's t-likelihood is better suited).  

These distributions look similar for double component fits and in other bands, although the relative numbers of galaxies which are better fitted with a normal or a t-distribution likelihood function changes due to the different depths and seeing. In summary, at early stages during pipeline development, we first fitted all galaxies using a Student's t-likelihood function. Then, galaxies for which the degrees of freedom of the Student's t-distribution converged onto values larger than 100 were fitted again with a Normal likelihood function (for all models independently). For all galaxies fitted twice with the same model, we selected the better fit according to which achieved the higher likelihood. 

The exact order of processing (including the segmentation map fixing steps used in earlier versions of the pipeline, cf. Section~\ref{sec:segchoices}) was as follows:\footnote{We subsequently extended this procedure to include 1.5-component fits; and then simplified it again by first dropping the Student's t-likelihood fits and later also the segmentation map fixing; see Section~\ref{sec:galaxyfitting} for the final \texttt{v04} procedure.}
\begin{enumerate}[label=(\roman*)]
\item Run the galaxies through the preparatory work pipeline, including downhill-gradient single S\'ersic fits for the segmentation map fixing. 
\item Do single component Student's t-likelihood function fits on all galaxies using the output of the preparatory work as initial guesses (more precisely the fits from the segmentation fixing) and the corresponding segmentation maps. Check whether segmentation maps need fixing after the fit and iterate if needed.
\item Do double component Student's t-likelihood function fits on all galaxies now using the outputs of the previous step as initial guesses and also using the updated segmentation map. Again check whether the segmentation maps need updating and iterate if needed.
\item Re-run the single component Student's t-likelihood function fits for the galaxies for which the segmentation map changed again during the double component fits. This uses the output of the previous single component fit as initial guess, but takes the updated segmentation map (after the double component fit) to make all segmentation maps used consistent for model selection. Do not change the segmentation maps anymore in this run.
\item Run single component Normal likelihood function fits for the galaxies for which the degrees of freedom of the single component Student's t-likelihood fit were greater than 100 or the fit failed (which happened in a number of cases when the degrees of freedom reached infinity). Use the previous single component fits as initial guesses and do not change the segmentation maps. 
\item Run double component Normal likelihood function fits for the galaxies for which the degrees of freedom of the double component Student's t-likelihood fit were greater than 100 or the double component fit failed. Use the previous double component fits as initial guesses and do not change the segmentation maps.
\item Repeat these steps for all desired bands.
\end{enumerate}

This procedure solved the first problem that we encountered using Student's t-likelihood functions (the unnecessary residuals for perfect fits). There was, however, one other major problem which eventually prompted us to use Normal likelihood functions for all fits: the Student's t-likelihood function fits frequently missed the bulges in double component fits. 

\begin{figure}[t!]
\begin{center}
\includegraphics[width=0.8\textwidth]{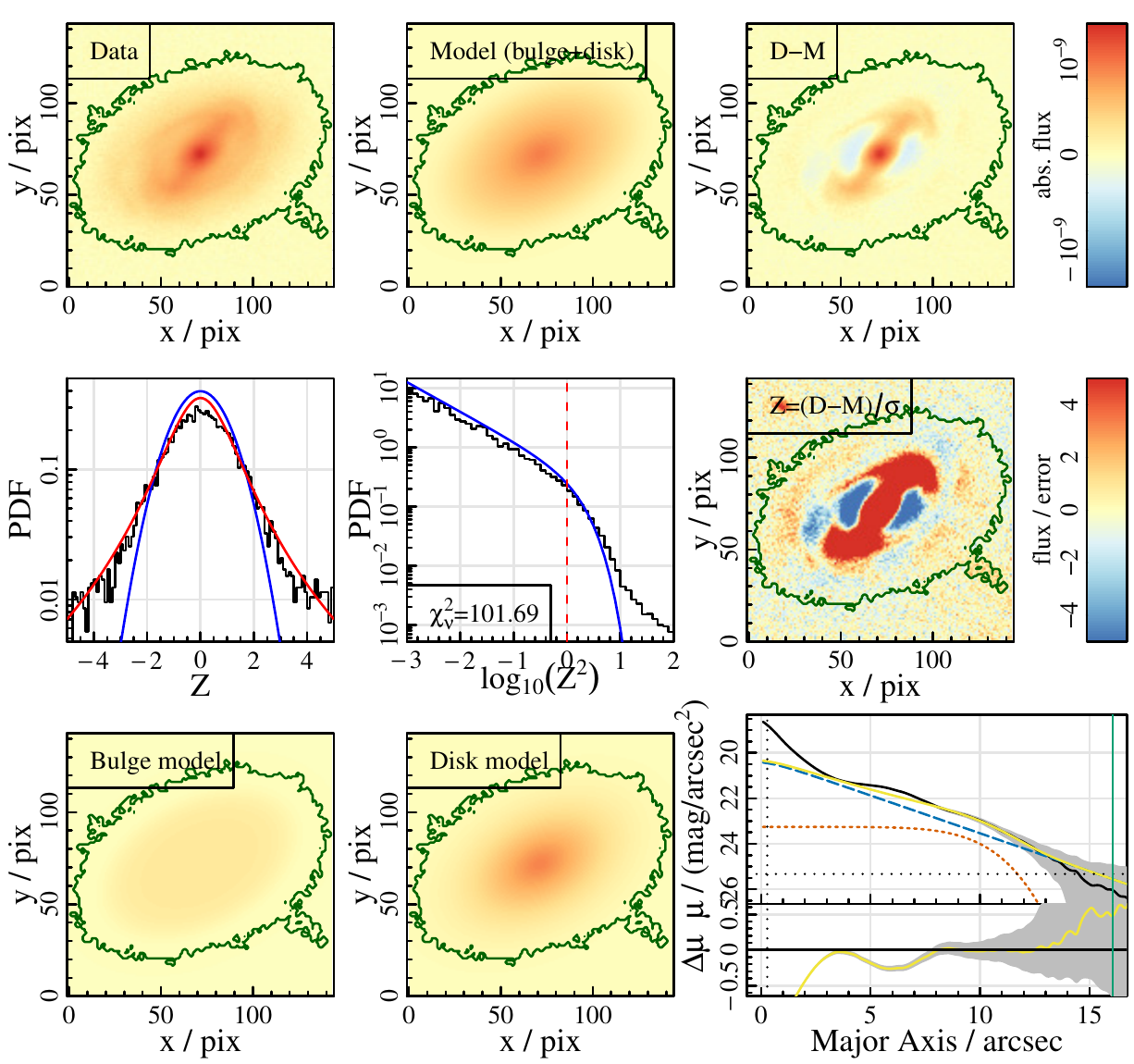}
\caption{The Student's t-likelihood function double component fit to galaxy 595088 in the KiDS $r$-band. Panels are the same as those in Figure~\ref{fig:examplefit}.}
\label{fig:examplefitnormvst5}
\end{center}
\end{figure}

\begin{figure}[t!]
\begin{center}
\includegraphics[width=0.8\textwidth]{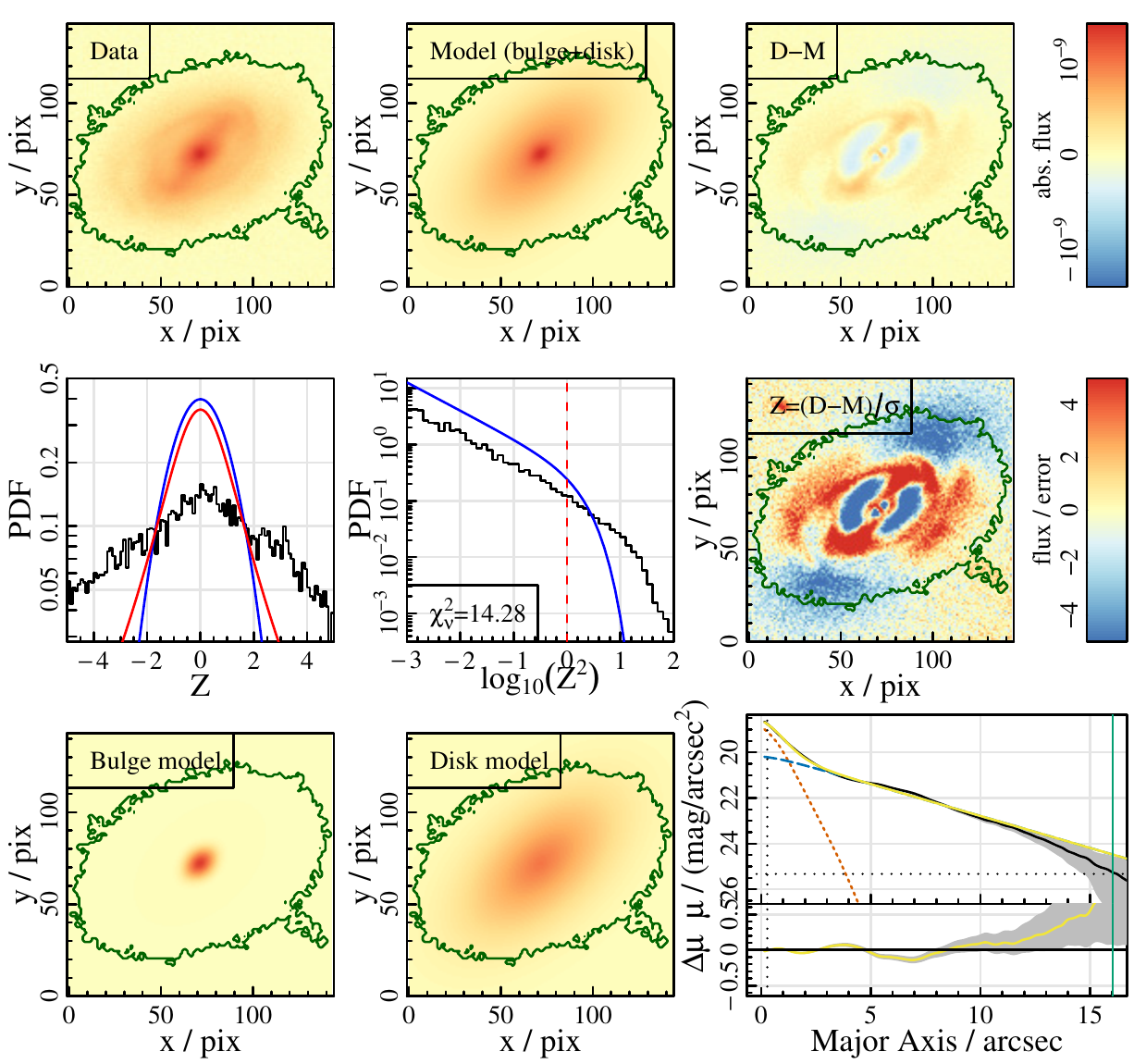}
\caption{The Normal likelihood function double component fit to galaxy 595088, for direct comparison to Figure~\ref{fig:examplefitnormvst5}.}
\label{fig:examplefitnormvst6}
\end{center}
\end{figure}

An example of this is shown in Figures~\ref{fig:examplefitnormvst5} and ~\ref{fig:examplefitnormvst6} for galaxy 595088. This is a bright and well-resolved object with clear evidence of a bulge, bar, disk, ring and spiral arms. Fitting only two components will not allow to capture all of these details, and the results obtained with the two different likelihood functions differ greatly. Similar to the single S\'ersic example in Figures~\ref{fig:examplefitnormvst1} and \ref{fig:examplefitnormvst2}, the Student's t-likelihood fit ignores the bulge and instead uses the freedom of the S\'ersic component to fit disk features that cannot be captured by the exponential model. This is because the Student's t-function ``prefers" a few strong outliers over many weak ones, which can be an advantage as explained above. In this case, though, this behaviour is highly undesirable since we explicitly intend the S\'ersic component of the double component fit to represent the central regions of the galaxy and not to capture deviations of the disk from an exponential profile. The Normal likelihood function fit instead fits both components, as desired. 

Many of our experiments with initial guesses, priors and constraints on bulge parameters (Section~\ref{sec:modellingdecisions}) were attempts to recover those ``missing" bulges. As explained above, they were not successful and in the end, the solution was to use a Normal likelihood function for all fits. This is therefore what we do in the current pipeline (and earlier versions including \texttt{v01} of the \texttt{BDDecomp} DMU). It has the added advantage that each model only needs to be fitted with one likelihood function, thereby saving computational time. In addition, it facilitates the manual calibration of the model selection, which is non-trivial if different galaxies and potentially also different models for the same galaxy use different likelihood functions.

\subsubsection{Optimisation algorithm}

One of the main assets of \texttt{ProFit} (cf. Section~\ref{sec:profit}) is that its pixel integration is faster and more accurate than that of other commonly used algorithms, which allows us to use a more robust but also more computationally expensive MCMC algorithm instead of simple downhill gradient optimisers even for our relatively large sample of galaxies. As briefly mentioned before, \texttt{ProFit} allows great flexibility and can be combined with a variety of optimisation algorithms, including more than 60 variants of MCMC from the \texttt{LaplacesDemon} package as well as various downhill gradient optimisers and genetic algorithms. While we definitely want to make use of the advantages of the MCMC options, the questions remained whether it would be best to combine this with some of the other options, which MCMC variant(s) to use and how to ensure and assess convergence. 

Instead of devising our own routine, we opted to use the \texttt{convergeFit} function from the \texttt{AllStarFit} package \citep{AllStarFit}, which has already been optimised to achieve convergence within reasonable computational times by \citet{Taranu2017}. As briefly mentioned in Section~\ref{sec:galaxyfitting}, this function uses a combination of different downhill gradient optimisers followed by several MCMC fits until its internal convergence criterion is met. 

The downhill gradient optimisers are taken from the \texttt{nloptr} package \citep{nloptr} and are used first to improve the initial guesses with little computational effort. The MCMC chain is not very sensitive to the initial guesses, but converges much faster if starting close to the peak of the likelihood. In most cases, the downhill gradient algorithm will improve the initial guesses by moving closer to the peak of the likelihood, saving computational time. There can be cases where the downhill gradient moves in the ``wrong" direction and/or becomes stuck in a local maximum, in which case it may then take the MCMC chain slightly longer to leave the local maximum and find the globally best fit. However, since this only happens in a minority of cases and the time loss is small compared to the time savings in successful cases, we still save large amounts of computational time overall. Apart from that, even an MCMC algorithm would tend to first explore the local maximum if starting close to it, so skipping the downhill gradient step may not actually shorten fitting times for galaxies with initial guesses close to local maxima. 

After the downhill gradient algorithms have converged, \texttt{convergeFit} starts an MCMC chain with 500 iterations using the Hit-and-Run-Metropolis (HARM)\footnote{Both HARM and CHARM are variants of the Hit-And-Run algorithm introduced by \citet{Turchin1971}.} algorithm in the \texttt{LaplacesDemon} package \citep{LaplacesDemon}. This process is repeated until convergence is reached, where convergence is defined as the fractional change in the log-likelihood between two runs being less than $e$. We compare this criterion to other measures of convergence and stationarity below. 

As a last step, the Componentwise Hit-and-Run-Metropolis (CHARM), also from \texttt{LaplacesDemon}, is run with 2000 iterations and again repeated until convergence (usually only once since the chains have converged already using HARM). This algorithm is more robust but also a lot slower than HARM, hence it is only called in the end to collect likelihood samples around the peak which has already been found in previous steps. For further analysis of the galaxy it is then assumed that all samples returned by CHARM are stationary, see below for a test of this assumption.

On average, for the $r$-band fits of \texttt{v04} of the \texttt{BDDecomp} DMU, the entire fitting procedure takes 7.6\,minutes per galaxy for the single S\'ersic fits, 28.9\,minutes for the initial double component fits, 27.5\,minutes for the swapped double component fits (only performed on approximately one third of all galaxies) and 6.0\,minutes for the 1.5-component fits. In total, the computational time to fit all models and including the preparatory work and post-processing is just under one hour per galaxy, although with strong variations for individual objects depending on their size and complexity (i.e. how rapidly convergence can be reached) and whether or not they enter the second round of double component fits for swapping.

We also tested just using 10\,000 CHARM iterations, without any of the downhill gradient or HARM steps and simply assuming convergence is reached after such a large number of iterations. This is the approach taken in some other works based on \texttt{ProFit} fitting of galaxies, e.g. \citet{Cook2019} and \citet{Hashemizadeh2022}. Compared to the \texttt{convergeFit} function, this procedure drastically increased the computational time for the vast majority of our galaxies (by a factor of about four on average) with no systematic differences in the fit results and no improvements to the convergence or stationarity criteria.

\subsubsection{MCMC convergence}

Judging the convergence (or equivalently, stationarity) of a chain is one of the main challenges in MCMC analysis with no universal solution.\footnote{We use the terms ``convergence" and ``stationarity" interchangeably here, although in detail they are not exactly the same.} Hence, all algorithms require user input regarding the stopping criterion; in the HARM and CHARM algorithms this is simply the number of iterations to perform. However, depending on the galaxy morphology, the number of iterations required to achieve chain convergence can be vastly different, so to achieve stationary results for all galaxies of a large sample while still retaining some computational efficiency, a fixed number of iterations is not suitable and a more flexible stopping criterion is needed. In the \texttt{convergeFit} function, this is implemented via repeated chains of 500 (or 2000 for CHARM) iterations each, which are judged for convergence comparing to the previous batch of 500 (2000) iterations: a chain is considered stationary if the fractional change in the log-likelihood obtained from the current batch is less than $e$ compared to the previous batch. The threshold ($e$) is somewhat arbitrary and has been chosen based on visual inspection (Dan Taranu, private communication). According to this criterion, all of our fits have converged. 

There are many other ways to assess stationarity, such as the BMK diagnostic \citep{Boone2014}, the Geweke Diagnostic \citep{Geweke1992}, the Heidelberger Diagnostic \citep{Heidelberger1981, Heidelberger1983}, or the Kolmogorov-Smirnov Convergence Diagnostic \citep{Brooks2003} to name just a few of the tools available in the \texttt{LaplacesDemon} package. All of them have in common that they are ultimately based on an arbitrary threshold of some form; and that there can be difficulties with multivariate chains, especially if correlations between parameters exist (which they do in our case, in particular the magnitude, S\'ersic index and effective radius are generally (anti-)correlated). Nonetheless, we investigated the first two of those in more detail. 

The BMK diagnostic is based on calculating the Hellinger distances \citep{Hellinger1909} between consecutive batches of the chain and is automatically computed when running \texttt{LaplacesDemon}. According to this criterion, 29\,\% of our $r$-band single component fits are not converged; and the majority of the double component fits. The Geweke diagnostic looks for trends or changes in moments in the given samples to assess stationarity and can easily be computed from the MCMC chain. According to this, 87\,\% of the $r$-band single component fits (and almost all double component fits) are non-stationary. Cross-comparing the two criteria shows that 30\,\% of the Geweke-non-stationary single component fits are not converged according to the BMK diagnostic; and vice versa 89\,\% of the BMK-non-converged single component fits are not stationary according to the Geweke diagnostic. These fractions are very similar to the fractions in the full sample (30\,\% vs. 29\,\% and 89\,\% vs. 87\,\%), so the two criteria seem only marginally (if at all) correlated. Also, the percentages are very similar when increasing the number of iterations; when re-fitting the galaxies using the previous fits as initial guesses and when using simulated galaxies for which we know that parameters are well-recovered (cf. Section~\ref{sec:simulations}). 

\begin{figure}[t!]
	\includegraphics[width=\textwidth]{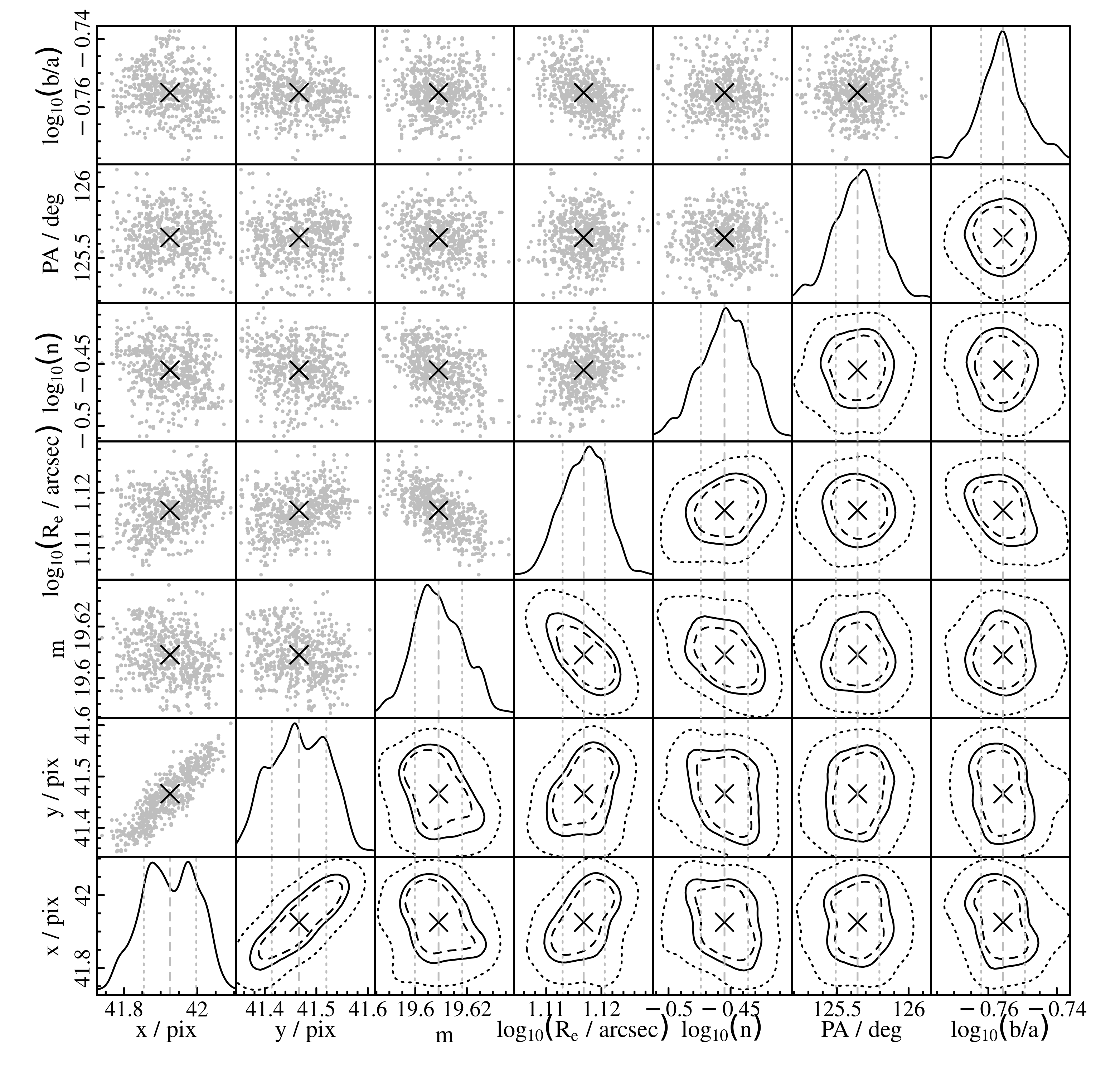}
    \caption{The MCMC chain for the $r$-band single component fit of galaxy 136524 that is neither converged according to the BMK diagnostic nor stationary according to the Geweke diagnostic. Panels are the same as in Figure~\ref{fig:islandplotpsf}.}	
    \label{fig:islandplotconv1}
\end{figure}

\begin{figure}[t!]
	\includegraphics[width=\textwidth]{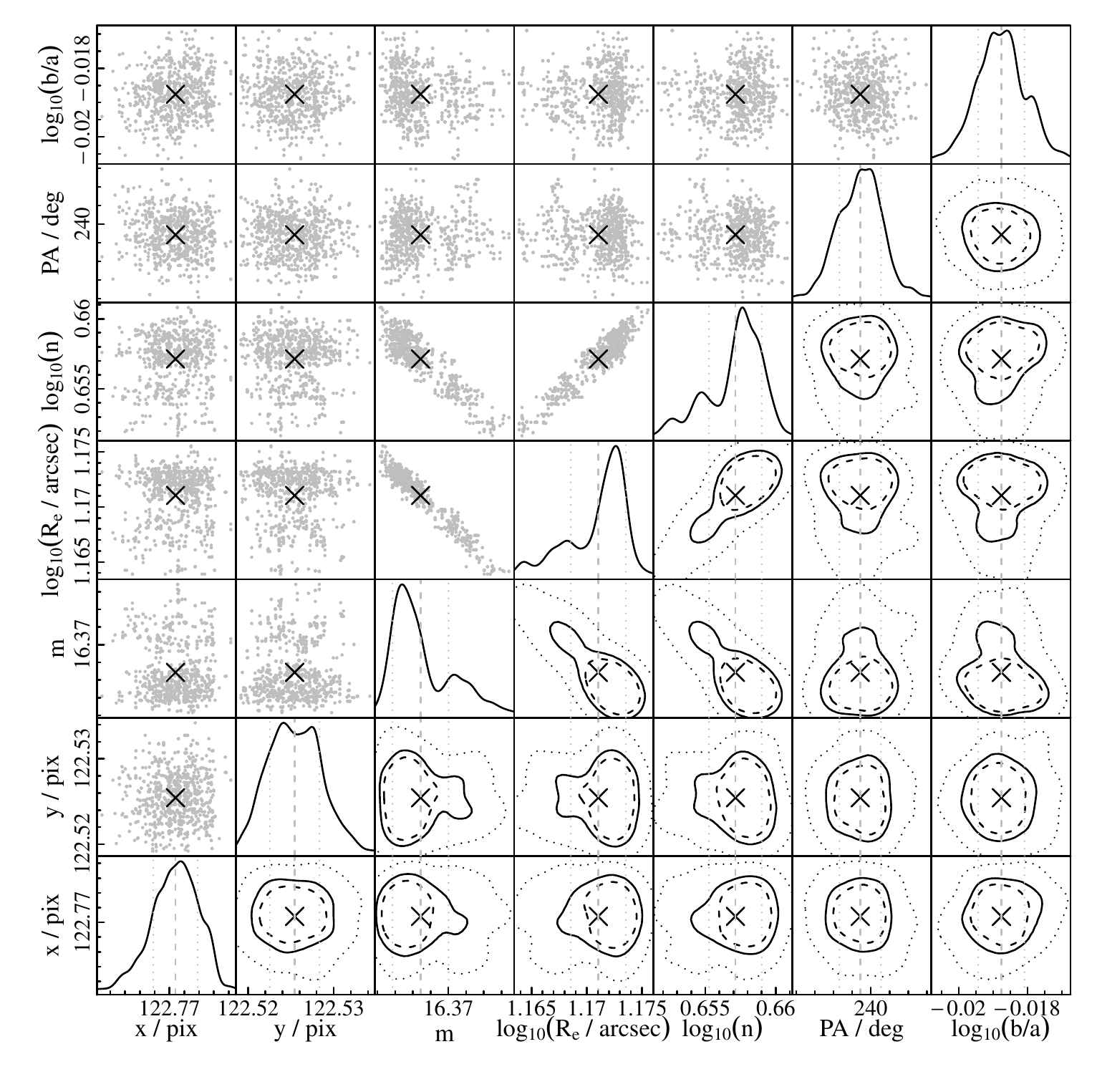}
    \caption{The MCMC chain for the $r$-band single component fit of galaxy 136851 that is converged/stationary according to both the BMK diagnostic and the Geweke diagnostic, for direct comparison to Figure~\ref{fig:islandplotconv1}.}	
    \label{fig:islandplotconv2}
\end{figure}

In addition, we visually inspected some of the chains that were flagged as converged/stationary by both criteria as well as some that are non-converged/non-stationary according to both criteria and could not see an obvious difference between the two samples. As a demonstration, we show the chains for the single component fits for two example galaxies in Figures~\ref{fig:islandplotconv1} and~\ref{fig:islandplotconv2}, where the chain that is not converged according to either criterion visually appears to be more stable than the one that is converged according to both criteria. \\

For all of these reasons, we decided to keep the default convergence criterion used in the \texttt{convergeFit} function rather than implementing a more complicated diagnostic with no obvious advantages. We assume the last 2000 CHARM samples are sufficiently stationary for our needs based on the visual inspection of a number of example chains and the simulations (Section~\ref{sec:simulations}).

\subsection{Manual calibrations}
\label{sec:swappingandoutliers}
While the \texttt{v04} pipeline is fully automated, it depends on several tuning parameters that required manual calibrations based upon visual inspection during pipeline development. We point these out here as they may not be directly transferable to other bands or types of data with different depths and resolutions and therefore require special attention. We consequently also consider them separately in Section~\ref{sec:pipelineupdates}, where we introduce the VIKING data and hence need to re-calibrate some of the visual inspections (Section~\ref{sec:manualcalibrationchanges}). 

The first notable manual calibration is the selection of star candidates for the PSF estimation, which we describe in Section~\ref{sec:psfdetails}. It depends on the depth of the data used for the segmentation map and the exact segmentation procedure. We hence re-calibrated this for each of the individual bands as well as for the segmentation maps obtained from the stacked $gri$ images and double-checked the calibrations (adjusting when needed) every time when there were major changes in the segmentation procedure. See details in Section~\ref{sec:psfdetails}. 

The second manual calibration of the preparatory work pipeline is the $\chi^2_\nu$-cut to exclude stars which visually appear as a bad fit from the model PSF creation, see Section~\ref{sec:psfdetails}. Which cut is appropriate depends mainly on the segment size, since the reduced chi-squared within the segment will generally decrease for larger segments due to the larger number of background pixels with little deviations from the model. This effect is amplified by the fact that the KiDS weight maps are generally conservative, such that the true pixel errors tend to be smaller than the standard deviations quoted in the sigma map (cf. Section~\ref{sec:otherprepworkchoices}). We therefore re-calibrated this cut whenever there were major changes in the segmentation procedure, such as between \texttt{v02} and \texttt{v03} of the pipeline. 

Since we have described both of these manual calibrations in the preparatory work pipeline in detail before, we do not elaborate on them further here. However, there are three more procedures based upon visual inspection during the fitting and post-processing, namely the criteria for swapping components, those for flagging bad fits and the calibration of the DIC cuts for the model selection. These are listed in Sections~\ref{sec:galaxyfitting} and~\ref{sec:postprocessing}, with more details and diagnostic plots given below.

In addition to these manual calibrations, our estimation of systematic errors (Section~\ref{sec:systematics}) is based on simulations that were tuned to reproduce KiDS $r$-band single S\'ersic fits for our sample of galaxies with our pipeline. They may not be transferable to other bands, other models, a different pipeline or different samples of galaxies. 

\subsubsection{Swapping of components}

The two components of the double component model are nearly identical, with the only difference being the S\'ersic index, which is a free fitting parameter for the bulge and fixed to 1 (exponential) for the disk. This similarity of the profiles leads to frequent swapping of the two components, such that around 20-30\,\% of the double component fits have the ``disk" dominating the flux in the inner regions of the galaxy and the ``bulge" the outskirts. This is a common problem in galaxy fitting and usually requires some form of post-processing, see e.g. the logical filter in \citet{Allen2006}. We attempted to reduce the fraction of swapped fits by constraining the bulge parameters in various ways, see Section~\ref{sec:modellingdecisions}. These methods, which would solve the problem at the fitting stage, had very limited success. We therefore developed a separate routine for reducing the number of swapped components that is somewhat at the interface between fitting and post-processing: the double component models are fitted, then the swapping routine is applied, including an additional fitting step for a subsample of galaxies (about one third of the full sample). This routine is outlined in Section~\ref{sec:galaxyfitting} and results in the number of swapped fits being reduced to $\sim$\,1-2\,\%. We supplement Section~\ref{sec:galaxyfitting} here with more information on how the routine was developed and some diagnostic plots. 

Note that the 1.5-component fits do not suffer from swapped components for two reasons. First, they are last in the order of processing, such that the double component models have already been fitted and swapped where necessary. The 1.5-component fits benefit from these results as initial guesses and are therefore very unlikely to swap the components again. Second, and maybe even more importantly, the 1.5-component fits have much fewer degrees of freedom than the double component fits. The point source bulge is defined by just a single parameter - the magnitude - and offers very little flexibility to fit deviations of the disk from an exponential profile (which is the most common reason why the S\'ersic component in the double component model fits the disk rather than the bulge). We therefore only apply the swapping routine to the double component fits. 

As a reminder, the basic idea that solved the problem of swapping components is very simple: our two components are very similar and nearly interchangeable. In general, therefore, there will be two high maxima in likelihood space that are far apart. Moving from one to the other is statistically unlikely given limited run times as it requires changing 9 of the 11 double component parameters at once. The code - even the MCMC - is hence more likely to converge onto the maximum that is closer to the initial guesses. By swapping the initial guesses and re-fitting a galaxy, we can assist the code in finding the ``correct" maximum (not necessarily the statistically better one but the one we find physically more appropriate, i.e. with the bulge at the centre). 

Manual calibrations enter at two stages in this process, namely for selecting galaxies to enter the re-fit (to avoid having to fit all galaxies twice) and for selecting the better of the two fits after the re-fit. Both selection criteria were determined by visual inspections of $r$-band galaxies during pipeline development; and double-checked for their suitability to $g$- and $i$-band fits. 

\begin{figure}[t!]
\begin{center}
	\includegraphics[width=0.8\textwidth]{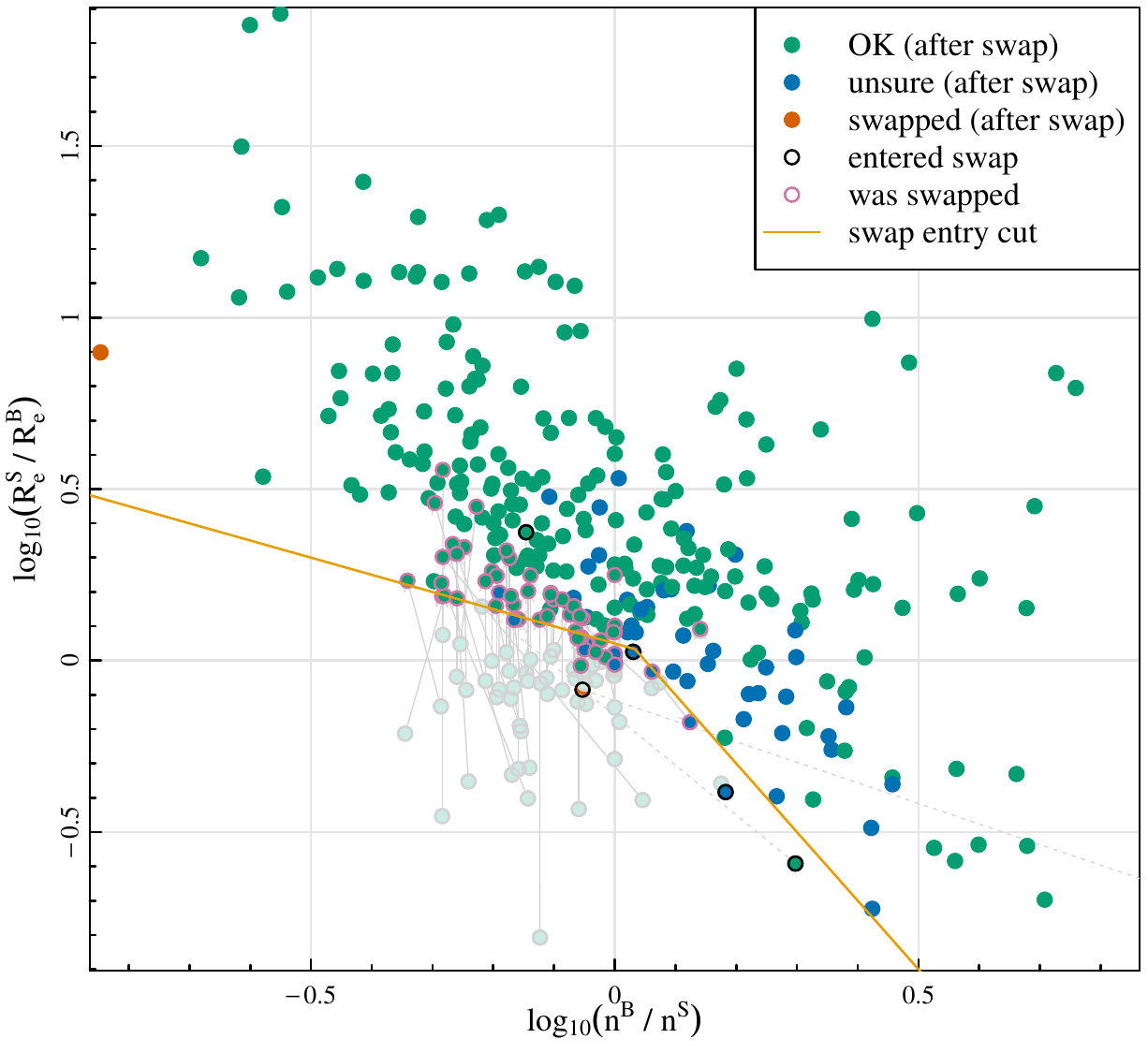}
    \caption{A diagnostic plot of the swapping procedure for a test sample of visually inspected $r$-band galaxies classified as double component systems. The $x$-axis shows the logarithm of the ratio of the bulge S\'ersic index to that of the single S\'ersic fit; while the $y$-axis shows the logarithm of the ratio of the single S\'ersic effective radius to that of the bulge in the double component fit. The yellow line show our selection cut for determining which galaxies to re-fit (those below the line). The colours of the points indicate the result of the visual inspection (performed after the swapping procedure) as detailed in the legend. Points encircled in pink represent galaxies that were swapped (i.e. they entered the swapping procedure and the new fit was considered better). Lighter points show the corresponding original fits (that were replaced during the swapping procedure), connected by grey arrows. Points encircled in black show galaxies that entered the swapping procedure but were not swapped (i.e. the original fit was considered better). They are connected to the points where they would have ended up if they were swapped by dashed grey arrows.}	
    \label{fig:swapdiagnostic}
\end{center}
\end{figure}

Figure~\ref{fig:swapdiagnostic} shows a diagnostic plot of the swapping results. We show the logarithm of the ratio of the single S\'ersic effective radius to that of the bulge in the double component fit on the $y$-axis against the logarithm of the ratio of the bulge S\'ersic index to that of the single S\'ersic fit on the $x$-axis. This is the plane in which we select galaxies to enter the swapping procedure (i.e. to be re-fitted). For un-swapped fits, we would expect both of these quantities to be larger than zero. To allow for some scatter, we do not cut exactly at zero, but instead re-fit all galaxies that have 2$x$\,+\,$y$\,<\,0.1 and 2$y$\,+\,$x$\,<\,0.1. This cut is indicated with the yellow line, any galaxies for which the original fit falls below the cut enter the re-fitting stage. 

We have experimented with many other parameter combinations also including the bulge-to-total flux ratio, the relative bulge and disk effective radii, the bulge and disk axial ratios and the chi-squared value of the central pixel, but have found the current procedure to produce the best results. Note that the exact cuts were refined several times during pipeline development, we only show the final \texttt{v04} versions here. Cuts for earlier version of the pipeline were made in the same plane and generally similar. We only adjusted them slightly based on new visual inspections (also making use of visual inspections done in the context of the other manual calibrations) to minimise the number of galaxies entering the swapping stage while at the same time minimising the number of galaxies with swapped components in the final catalogue. 

The coloured points show where the fits fall in this plane after the swapping procedure, with the colours indicating the result of the visual inspection (which was performed on the final fits, i.e. after the swapping procedure): green indicates unswapped fits, orange shows swapped fits and blue indicates those that were unclear, e.g. if both profiles are very similar, or the galaxy may better be represented by a single S\'ersic fit (i.e. a model selection fail) or because the galaxy is very irregular. 

The points encircled in pink are the ones that entered the swapping procedure and were swapped, i.e. these coloured points represent the new (better) fits after the swapping. The corresponding original fits for the same galaxies are shown as lighter points (of the same colour) connected by light grey arrows. All of the original fits lie below the selection cuts, while most of the re-fits lie above it; i.e. generally galaxies are moved towards the ``good" areas of the plot during the swapping procedure, even though this criterion is not applied in the selection of the better fit (see below). After the swapping, all fits are also considered un-swapped or at least unsure. Note that we did not do a visual calibration on the fits before swapping (so the colours of the lighter points just represent the final result for the galaxy after the swapping), but given that they changed their parameters significantly during the procedure it is very likely that they indeed had swapped components before.

The points encircled in black are those that entered the swapping procedure but were not swapped, i.e. the original fit was considered better. Light grey dotted arrows connect them to their corresponding re-fits in lighter colours (that were then discarded). There are only five such objects in the entire sample of 300 visually inspected galaxies, i.e. the number of re-fits done in vain is very small. Note one of these is actually above the selection cuts and only entered the re-fits because it is the second match of a galaxy in the overlap sample for which the first match required swapping (for ease of processing, we enter the entire galaxy for re-fitting if any of its matches were identified as being potentially swapped). Three more of these objects are close to the selection cut and either did not change their parameters much during the re-fit (for two of them) or were fine before the re-fit (the third). 

Finally, the last object that entered the swapping procedure and did not get swapped is one of the two objects (of a total of 300) that remain swapped after our procedure, i.e. a ``failed swap". For this object (partly covered by another object so the orange colour is difficult to see), the re-fit seems to have failed as indicated by the dotted grey arrow pointing far beyond the plotting region. The second of the failed swaps is the leftmost point in the plot and did not enter the swapping procedure. Reducing the number of swapped components from around one third to only two out of 300 can be seen as a success though. Further visual inspections indicate that the number of failed swaps is generally around $1-2\,\%$ in all bands. 

\begin{figure}[t!]
\begin{center}
\includegraphics[width=0.8\textwidth]{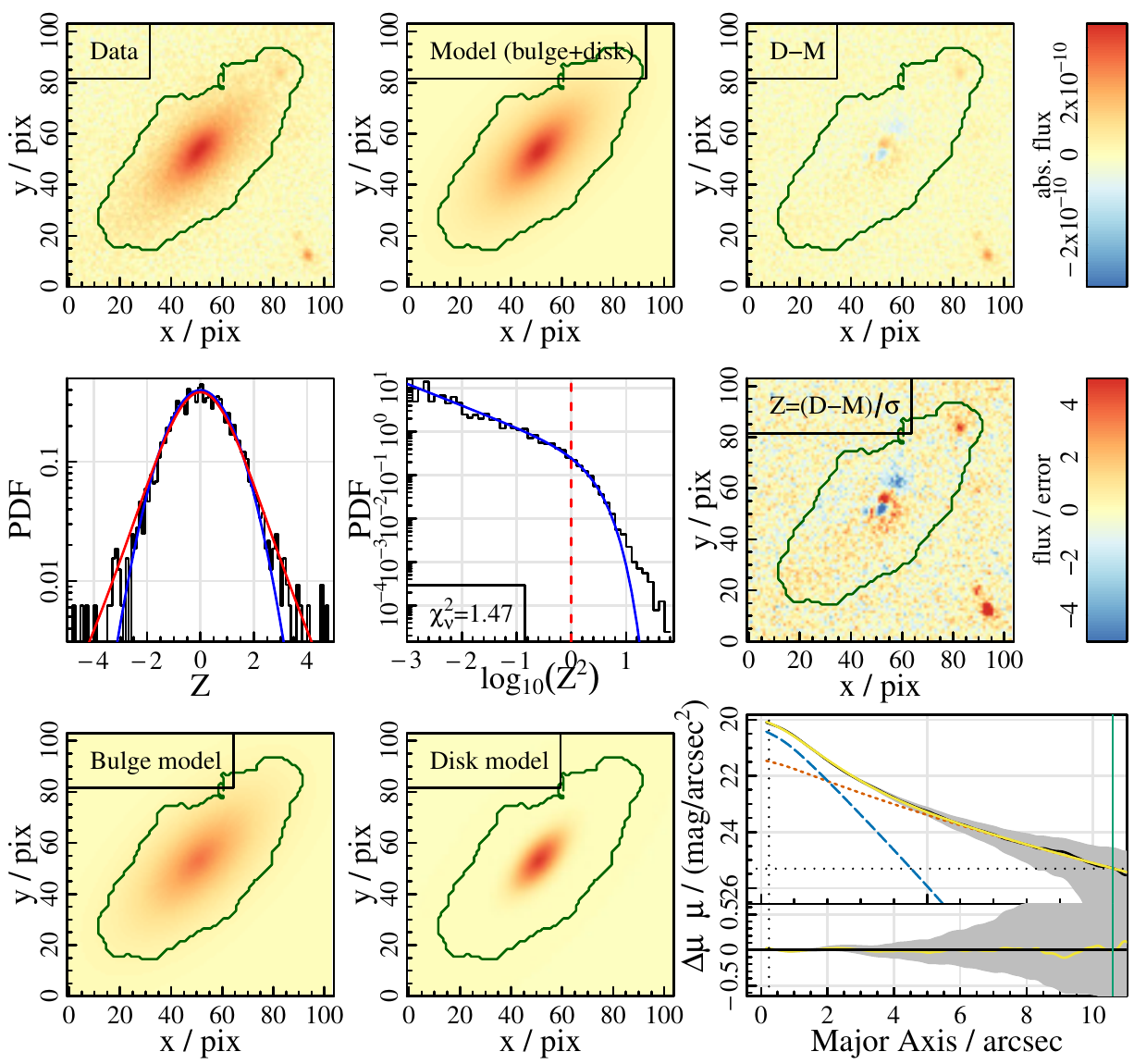}
\caption{The double component fit to galaxy 593119 in the KiDS $r$-band before it entered the swapping procedure. Panels are the same as those in Figure~\ref{fig:examplefit}.}
\label{fig:examplefitswap1}
\end{center}
\end{figure}

\begin{figure}[t!]
\begin{center}
\includegraphics[width=0.8\textwidth]{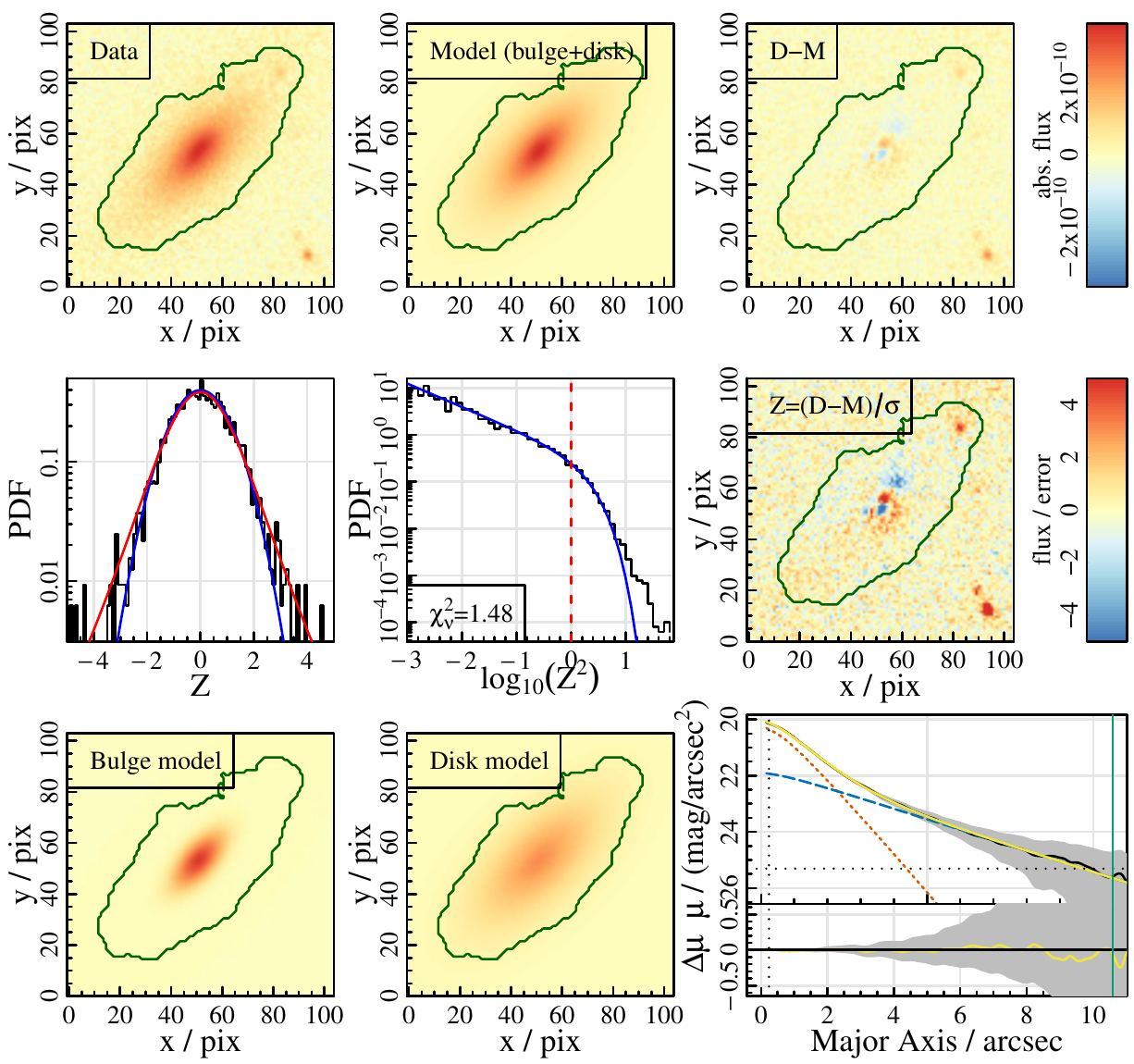}
\caption{The double component fit to galaxy 593119 after the swapping procedure, for direct comparison to Figure~\ref{fig:examplefitswap1}.}
\label{fig:examplefitswap2}
\end{center}
\end{figure}

The last point that remains to be explained is how we select the better one of the re-fits. This criterion is stated in Section~\ref{sec:galaxyfitting} and is a combination of physical considerations and visual inspections. To recap, if the ``bulge" component is close to exponential, larger than the disk and contains the majority of the flux for only one of the two fits, then we choose the other one. For all galaxies for which this criterion is met by both or neither of the fits (which is the majority), we select the one with the higher absolute value of bulge flux in the central pixel. We devised and calibrated these criteria during visual inspection of samples of galaxies that were fitted twice (once with initial guesses swapped), comparing the two fits against each other and flagging the physically most appropriate one. 

We also experimented with many different criteria for this selection, e.g. the higher absolute value of flux in central regions of differing sizes (instead of only one pixel), higher values of bulge flux relative to the disk flux (rather than absolute), chi-squared values in the central pixel(s), combinations of other parameters such as S\'ersic index, bulge-to-total ratio and effective radii or selections that are statistically better motivated, for example choosing the fit with the higher likelihood. In the end, the rather simple selection of the fit with the higher absolute value of bulge flux in the central pixel proved to be most suitable to our needs.

Figures~\ref{fig:examplefitswap1} and~\ref{fig:examplefitswap2} show an example galaxy pre- and post-swapping. The resulting fit after the swapping procedure is not perfect (galaxies that can be perfectly represented by our models rarely suffer from swapped components) and in fact has a slightly higher reduced chi-squared value than the original fit, i.e. it is statistically worse. Still, the fit post-swapping is physically more appropriate than the original one as the bulge now dominates the flux in the central region and the disk that in the outskirts. 

\subsubsection{Outlier flagging}

Similar to the swapping procedure, the flagging of bad fits (outliers) proceeded in a very iterative way, experiencing multiple refinements during pipeline development. The basic procedure was to visually inspect a random sample of galaxies and flag them as outlier or not; then visualise the distributions of outliers and good fits as a function of many (combinations of) metrics that we would expect to capture some aspects of ``bad fits" either from a theoretical point of view or from the visual impression gained while inspecting our sample of galaxies. We then proceeded to visually inspect more (non-random) samples of galaxies in the regions where we identified possible metrics, to better define the exact cuts. 

This process or parts thereof were repeated several times when we changed the preparatory work procedure (in particular the segmentation) or revised some modelling decisions such as the fitting limits or parameter constraints. The resulting metrics that we used for outlier flagging in \texttt{v04} of the \texttt{BDDecomp} DMU are listed in Section~\ref{sec:postprocessing}, including an indication of how many fits they affect. They were all calibrated on the $r$-band fits, but double-checked and found to be suitable for the $g$ and $i$ band fits as well. 

Initially, the outlier flagging was heavily linked to the swapping and model selection calibration, since all three processes are interconnected. For these reasons, many of the calibrations are in fact based on the same visual inspections, where we flagged numerous categories such as ``the double component fit is swapped but could probably be a good fit if it were not", ``borderline between single/double component fit and outlier", ``this galaxy would need at least three components" or ``failed segmentation map (could maybe be a good fit otherwise)" along with more straight-forward categories for galaxies which can be appropriately represented by one of our models\footnote{The 1.5-component fits were not considered during most of this iterative process since they were added at a later stage during pipeline development. The outlier flagging was then only adapted slightly to accommodate the new model fits.}.

Once we had finalised the swapping procedure, the number of swapped components was very low such that this effect was de-coupled from the outlier flagging and model selection. The latter two remain connected though, since the decision of whether or not a given galaxy can be appropriately represented by one of the models depends on the available model fits. The visual inspection for the model selection therefore always included an ``outlier" category, which we in turn could also use to define the outlier criteria (with further refinements based on non-random samples as stated above; which did not enter the model selection calibration). Further details on the final order of processing of the model selection and outlier rejection routines are given in Section~\ref{sec:postprocessing}. 

Note that in contrast to the swapping procedure, the outlier flagging and model selection are entirely post-processing routines, i.e. they do not require any re-fits and also did not require specific test runs to be calibrated during pipeline development. Nonetheless, some of the visual inspections performed in the context of model selection and outlier flagging also influenced our modelling decisions such as the definition of the fitting limits for the S\'ersic index (cf. Section~\ref{sec:modellingdecisions}) and not least the addition of the 1.5-component models. 

\begin{figure}[t!]
\begin{center}
	\includegraphics[width=0.8\textwidth]{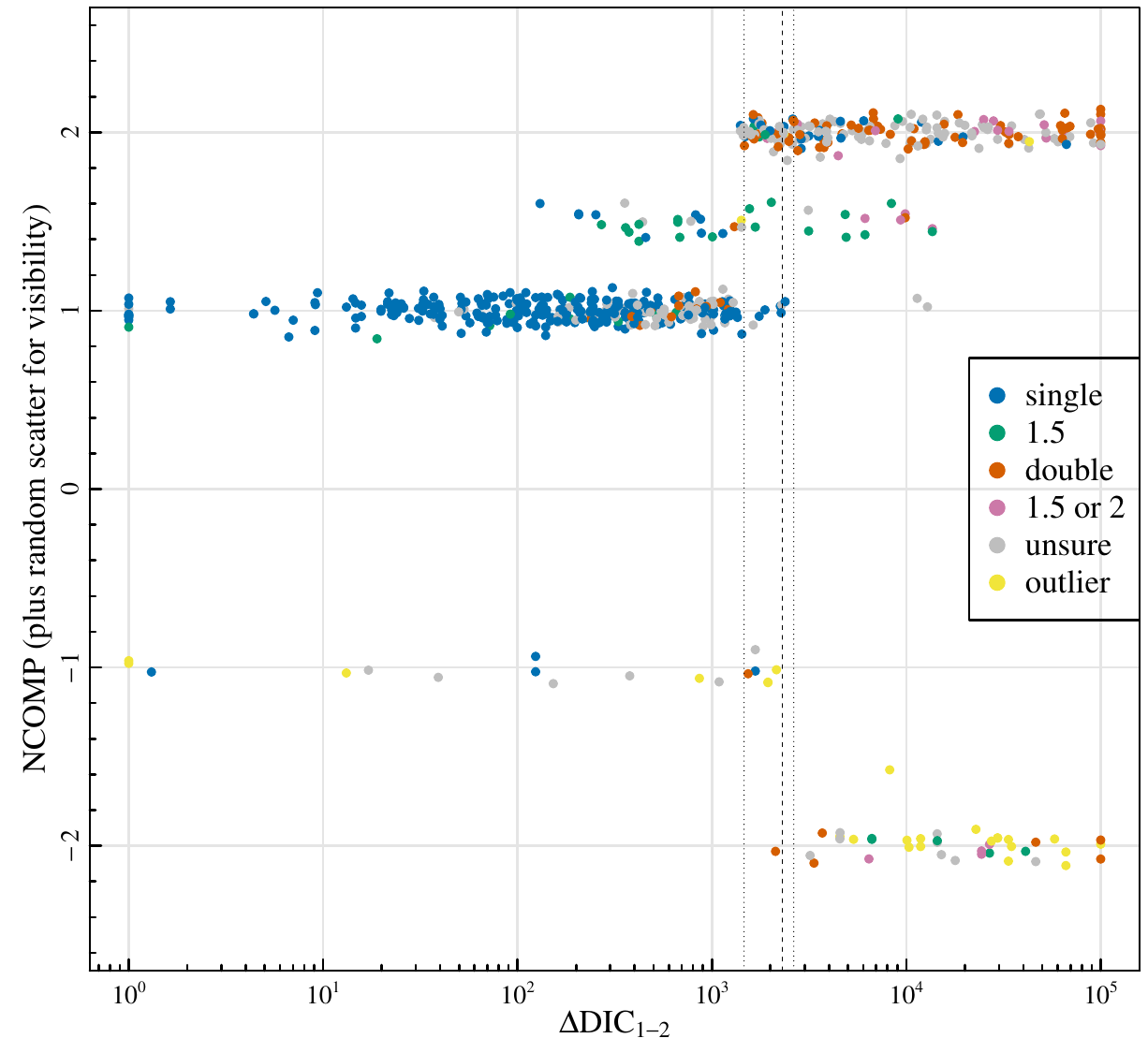}
    \caption{A diagnostic plot of the outlier rejection and model selection routines. We show the number of components assigned by the automated routine (\texttt{NCOMP}) on the $y$-axis against the DIC difference between the single and double component fits on the $x$-axis. Each point represents one of the $\sim$\,700 galaxies used during model selection calibration, coloured by the result of the visual inspection as indicated in the legend. The vertical black dashed and dotted lines indicate the calibrated DIC cut with its bootstrap uncertainties.} 
    \label{fig:modelseldiagnostic}
\end{center}
\end{figure}

Figure~\ref{fig:modelseldiagnostic} shows a diagnostic plot of the combined outlier flagging and model selection. It shows the number of components assigmed by the automated routine against the DIC difference between the single and double component models, clipped to the plot limits (for simplicity, we do not show the other two DIC differences). As a reminder, \texttt{NCOMP} values of 1, 1.5 and 2 mean that the object was classified as single, 1.5- and double component fit respectively, while objects with negative \texttt{NCOMP} are outliers, where the absolute value indicates the category that the object would have been assigned to if it were not an outlier. Each coloured point is one of the $\sim$\,700 galaxies used for \texttt{v04} model selection calibration and sorted into one of the six categories indicated by the legend during visual inspection (cf. Section~\ref{sec:postprocessing}). If all routines were perfect, we would expect all yellow points to be in the lower part of the plot; while all blue, green and red points would be on the \texttt{NCOMP}\,=\,1, 1.5 and 2 lines respectively (except for the random scatter that we added for visibility). Pink points can either be on the 1.5 or 2 lines and grey points (unsure; excluded from all analyses) are allowed to be anywhere, though preferentially in the positive region. 

\begin{figure}[t!]
\begin{center}
	\includegraphics[width=0.8\textwidth]{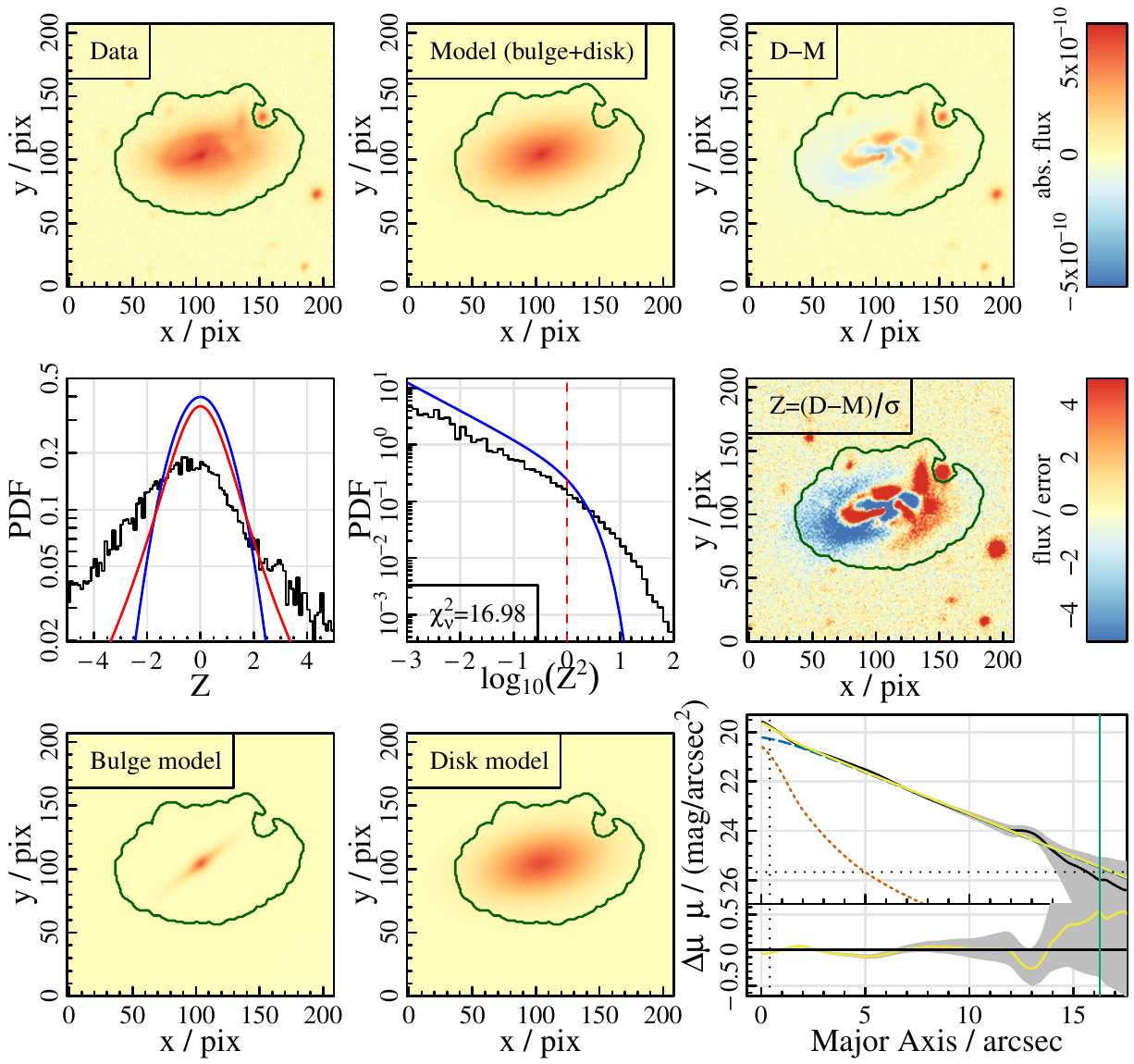}
    \caption{The double component fit to galaxy 278760, classified as outlier (\texttt{NCOMP}\,=\,$-2$) in the KiDS $r$-band. Panels are the same as those in Figure~\ref{fig:examplefit}.} 
    \label{fig:examplefit-2}
\end{center}
\end{figure}

We further investigate the model selection accuracy below. Focusing on just the outlier flagging, there are a total of three objects that were visually classified as outliers but are considered good fits in the automated selection. Vice versa, there are 20 objects classified as acceptable fits which are flagged as outliers by the automated routine. This amounts to a total of approximately 3\,\% of all objects being classified wrongly by the outlier flagging routine. A further 14 objects (2\,\%) are visually classified as ``unsure" and ended up being flagged as outlier. The ``unsure" category includes galaxies for which it was unclear which model is the best (since several models delivered equally good fits) as well as those for which it was unclear whether or not they are an outlier (borderline cases and/or several models delivered equally bad fits). The majority of the grey points should hence lie in the positive regions of the plot, but we cannot judge how many exactly are erroneously flagged as outliers. 

Depending on whether the ``unsure" category is considered or not, the outlier flagging routine has an overall accuracy of 95-97\,\% compared against visual inspection for the KiDS $r$-band. The corresponding fractions in the $g$ and $i$ bands are 94-97\,\% and 95-97\,\% respectively. An example of a galaxy classified as outlier is shown in Figure~\ref{fig:examplefit-2}.

\subsubsection{Model selection calibration}

The third and arguably most important calibration based on visual inspection is that of the DIC difference cuts for model selection. As mentioned above, it is intrinsically linked to the flagging of outliers and to a lesser extent also to the swapping routine. The theoretical background for Bayesian model selection is given in Section~\ref{sec:bayesiananalysis}. Our model selection routine (including its visual calibration) is then explained on that basis in Section~\ref{sec:postprocessing}. 

We emphasize again that purely statistical measures cannot achieve the aim of our model selection, namely to select the physically most appropriate fit, especially in the context of our models being statistically inadequate for the majority of galaxies in our sample. A purely visually-based classification is possible (see, e.g., \citealt{Hashemizadeh2022} or \citealt{Driver2022}), but time-consuming for large samples of galaxies such as ours. This is particularly true because the visual classification would need to be repeated for each band individually as long as bands are also fitted individually. Classifying galaxies from e.g. colour images to determine the number of physically distinct components they contain, sometimes even before (or without subsequent) model fitting, is common in the galaxy fitting community (see e.g. \citet{Hashemizadeh2022}) and certainly desirable from a scientific perspective. From a statistical perspective though, if the single-band image used for the fit is too shallow to constrain two components, then a double component fit will not produce reliable parameters even for galaxies which physically do consist of two components. For this reason, we calibrate the model selection on single-band diagnostic plots comparing the fits achieved by our three models and selecting the best (or - if there are several equally good fits - simplest) one.

We then use the result of the visual inspection of a subsample of galaxies to calibrate a cut based on statistical measures (the DIC) that we subsequently apply to the full sample. This limits the number of galaxies that need visual inspection while overcoming the unsuitability of a statistical measure alone. Nonetheless, the model selection is limited by the statistical measure to some extent, as the statistical measure will never be able to distinguish between different ``kinds" of bad model fits. For example, the automated procedure based on a DIC cut will always select the double component fit if it is significantly better than the single component fit (where ``significantly better" is the part that we calibrate during visual inspection), irrespective of whether the bulge and disk components are swapped or for example the bulge dominates both the centre and the outskirts of the galaxy. 

To further limit these shortcomings of a DIC difference cut, we experimented with combining it with more physical measures, e.g. a $\Delta$DIC cut as a function of bulge S\'ersic index or bulge to total flux ratio (and many other versions). None of these more complicated joint cuts improved the model selection significantly, so we chose the simplest version of just a (one-dimensional) cut in $\Delta$DIC. The accuracy that we achieve compared against visual inspection (see confusion matrices in Section~\ref{sec:postprocessing} and below) is sufficient for our needs despite the intrinsic limitations of the method. The only viable technique to overcome such limitations in the future is machine learning, as investigated by e.g. \citet{Dimauro2018}. However, the majority of recent efforts to apply machine learning to the problem of galaxy classifications is focused on reproducing visual morphological classifications based on images alone (e.g. \citealt{Nolte2019, Turner2021, Tarsitano2022}), which are to be distinguished from model selection. 

Figure~\ref{fig:modelseldiagnostic}, which was already introduced above, shows the result of the model selection compared against visual calibration. Most single component galaxies (blue points) are correctly classified as such, especially for low $\Delta$DIC; while those with high $\Delta$DIC are correctly identified as being the double component fits (orange dots). The $\Delta$DIC cut with its uncertainty region (from bootstrapping) is indicated with vertical dashed and dotted black lines; and can also be inferred from the distribution of points in the \texttt{NCOMP}\,=\,1 and \texttt{NCOMP}\,=\,2 classes. Most confusion between the single and double component models is near those cuts, where coincidentally (but not entirely surprisingly) we can also find the highest number of galaxies classified as ``unsure". Interestingly, the 1.5-component models have intermediate values of $\Delta$DIC$_{1-2}$ (which is the quantity we show on the $x$-axis; we do not show $\Delta$DIC$_{1-1.5}$ or $\Delta$DIC$_{1.5-2}$ which are the relevant quantities for 1.5-component model selection). 

The last point to note about Figure~\ref{fig:modelseldiagnostic} is that most galaxies that are wrongly classified as outliers (i.e. blue, green, orange and pink points in the lower half of the plot) are at least classified in their correct model category, i.e. the majority of blue points in the lower half of the plot lies on the \texttt{NCOMP}\,=\,$-1$ line, while the majority of orange points is at \texttt{NCOMP}\,=\,$-2$. For this reason, and in an attempt to characterise the accuracy of the model selection alone rather than in combination with the reliability of the outlier rejection routine, we use absolute values of \texttt{NCOMP} for all model selection statistics and ignore the ``outlier" category (essentially adding the yellow points to the grey population). 

The full confusion matrix for the $r$-band model selection is given in Table~\ref{tab:modelselconfusionr} in Section~\ref{sec:postprocessing}. For completeness, we add the corresponding confusion matrices for the $g$ and $i$ bands and the joint model selection in Table~\ref{tab:modelselconfusiongijoint}, although the statistics are generally similar between bands. For reference, we also list the calibrated DIC difference cuts between all three models for all three bands and the joint model selection in Table~\ref{tab:v04diccuts}. As one would expect (due to Ockham's factor), the cuts are generally higher for models with larger differences in the numbers of parameters and for bands with better data quality. 

\begin{table}
\centering
\caption{The confusion matrices for our model selection based on a $\Delta$DIC cut compared against visual inspection for the $g$ and $i$ bands and the joint model selection. All values are in percent of the total number of visually inspected galaxies in the respective band(s). Bold font highlights galaxies classified correctly, while grey shows those that were ignored during the calibration.}
\label{tab:modelselconfusiongijoint}
\begin{subtable}{0.5\textwidth}
\centering
\caption{The $g$-band model selection confusion matrix.}
\label{tab:modelselconfusiong}
	\begin{tabu}{lcrrrc} 
		\hline
		 & \multicolumn{5}{r}{number of components} \\
		visual class. & & 1 & 1.5 & 2 &\\
		\hline
		``single" && \textbf{48.9} & 0.4 & 1.0 &\\
		``1.5" && 2.1 & \textbf{3.7} & 0.6 &\\
		``double" && 1.8 & 0.1 & \textbf{6.0} &\\
		``1.5 or double" && 0.4 & \textbf{1.5} & \textbf{1.9} &\\
		\rowfont{\color{gray}}
		``unsure" && 19.5 & 0.9 & 7.3 &\\
		\rowfont{\color{gray}}
		``unfittable" && 1.3 & 0.3 & 2.1 & \\
        \hline
	\end{tabu}
\end{subtable}%
\begin{subtable}{0.5\textwidth}
\centering
\caption{The $i$-band model selection confusion matrix.}
\label{tab:modelselconfusioni}
	\begin{tabu}{lcrrrc} 
		\hline
		 & \multicolumn{5}{r}{number of components} \\
		visual class. & & 1 & 1.5 & 2 &\\
		\hline
		``single" && \textbf{51.8} & 0.4 & 1.5 &\\
		``1.5" && 1.9 & \textbf{3.0} & 1.5 &\\
		``double" && 0.7 & 0 & \textbf{8.9} &\\
		``1.5 or double" && 0 & \textbf{1.0} & \textbf{1.9} &\\
		\rowfont{\color{gray}}
		``unsure" && 10.0 & 0.7 & 12.4 &\\
		\rowfont{\color{gray}}
		``unfittable" && 1.2 & 0.3 & 2.2 & \\
        \hline
	\end{tabu}
\end{subtable}%

\bigskip
\begin{subtable}{\textwidth}
\centering
\caption{The joint model selection confusion matrix.}
\label{tab:modelselconfusionjoint}
	\begin{tabu}{lcrrrc} 
		\hline
		 & \multicolumn{5}{r}{number of components} \\
		visual class. & & 1 & 1.5 & 2 &\\
		\hline
		``single" && \textbf{47.4} & 0.5 & 1.5 &\\
		``1.5" && 2.2 & \textbf{2.4} & 1.4 &\\
		``double" && 2.2 & 0.4 & \textbf{7.3} &\\
		``1.5 or double" && 0.3 & \textbf{1.1} & \textbf{2.1} &\\
		\rowfont{\color{gray}}
		``unsure" && 15.0 & 0.7 & 11.1 &\\
		\rowfont{\color{gray}}
		``unfittable" && 1.0 & 0.4 & 2.4 & \\
        \hline
	\end{tabu}
\end{subtable}%

\end{table}

\begin{table}
	\centering
	\caption{The calibrated DIC difference cuts used for \texttt{v04} of the \texttt{BDDecomp} DMU. For each band, we show the lower bound (LB), the actual cut and the upper bound (UB) for each of the three DIC differences between our models. For the joint model selection, the cuts refer to the differences between the summed DICs of all bands (cf. Section~\ref{sec:postprocessing}).}
	\label{tab:v04diccuts}
	\begin{tabu}{lrrrrrrrrrrrr} 
		\hline
		 && \multicolumn{3}{c}{$\Delta$DIC$_{1-1.5}$} && 
		 \multicolumn{3}{c}{$\Delta$DIC$_{1-2}$} &&
		 \multicolumn{3}{c}{$\Delta$DIC$_{1.5-2}$}\\
		band && LB & cut & UB && LB & cut & UB && LB & cut & UB\\
		\hline
		$g$ && 43 & 68 & 93 && 1760 & 2273 & 4856 && 551 & 817 & 940\\
		$r$ && 57 & 172 & 248 && 1463 & 2298 & 2636 && 690 & 853 & 947\\
		$i$ && 33 & 57 & 127 && 396 & 473 & 520 && 308 & 336 & 438\\
		joint && 183 & 255 & 341 && 5707 & 5751 & 5797 && 1650 & 1824 & 3104\\
        \hline
	\end{tabu}
\end{table}

\section{Pipeline updates}
\label{sec:pipelineupdates}
This section describes the updates to the pipeline performed when adding the processing of VIKING data, i.e. in between \texttt{v04} and \texttt{v05} of the \texttt{BDDecomp} DMU, the latter of which is to be released on the GAMA database alongside the publication of this thesis. At the same time we included the KiDS $u$-band, such that now our catalogue covers all 9 optical and near-infrared bands from KiDS and VIKING ($u$, $g$, $r$, $i$, $Z$, $Y$, $J$, $H$, $K_s$). This work has not been presented to date. 

\subsection{VIKING data products}
\label{sec:vikingdataproducts}
As for KiDS (see Section~\ref{sec:otherprepworkchoices}), we needed to decide which of the VIKING data products to use. VIKING provides data products at two different levels: pawprints and tiles, the latter of which are made of six pawprints stacked together to close the gaps between detector chips. In addition, for the $J$-band, there are ``deep stacks" composed of two tiles each, since the $J$-band tiles have been observed twice in the VIKING observing strategy (see Section~\ref{sec:viking}). All of these data products are astrometrically and photometrically calibrated. In contrast to KiDS, however, they are not re-gridded, so they have various pixel sizes (close to $0\farcs$34), are only approximately aligned in RA and Dec and are calibrated to differing Vega zeropoints given in the image headers. The frames are also not corrected for atmospheric extinction and the flux units still contain the exposure time. The sky background variations are subtracted, but only on large scales \citep{Edge2020}. 

The most important difference to KiDS images for our analysis, however, is that the pawprints making up a tile are not taken in direct succession but can be taken hours or even months apart. This leads to seeing variations and hence considerable PSF differences between the pawprints making up a tile \citep{Edge2020}. For accurate photometry, it is therefore necessary to apply an aperture correction that is a function of position in the tile, termed ``grouting". This is done by the VIKING team to obtain their photometric catalogue, but with various different pipeline versions as their data reduction procedure evolved during the time of the survey. The tiles themselves cannot easily be corrected for this and are hence left with potentially strongly and abruptly varying PSFs since different pairs of pawprints contribute to different areas of the tile given the VIKING dithering pattern, see \citet{Edge2020}.

This drove our decision to work at the pawprint level instead of using stacked image tiles, since a reliable PSF estimation is vital for an accurate galaxy fit. Very similar considerations led \citet{Wright2019} and \citet{Driver2016} to the same conclusion. Instead of downloading the pawprints and re-calibrating them ourselves, we decided to directly use the preprocessed detector chips of \citet{Wright2019} (who were following \citealt{Driver2016} in their analysis). They are not publicly available, but the KiDS team kindly provided them to us for this work. Given that the KiDS team use these re-processed chips for their own photometric analysis combining KiDS and VIKING data in \citet{Wright2019}, they seemed highly appropriate to use for our work as well. 

Wright (private communication) provided individual detector chips (i.e. each pawprint split into its 16 chips) with a size of $\sim$\,2200\,$\times$\,2200\,pix$^2$ or approximately 750\arcsec\,$\times$\,750\arcsec\. Apart from the data reduction already performed by VIKING, they have been re-calibrated with a multiplicative correction factor to account for atmospheric extinction, remove the exposure time from the image units and convert the images onto a common AB magnitude zeropoint of 30 (see \citealt{Wright2019} for details). In addition, the chips are rotated slightly such that their $x$ and $y$ axes align exactly in RA and Dec and at the same time the background is subtracted and a weight map created. This is performed using the SWarp software \citep{SWarp} after truncating the chip edges slightly to exclude unreliable pixels \citep[][and Wright, private communication]{Wright2019}. It results in a background-subtracted chip image with a corresponding weight map that is zero around the edges of the chips and has a uniform value elsewhere, giving the average inverse variance over the chip. Chips with unreliable photometry are then identified via a cut in the recalibration factor and discarded.

Note that \citet{Wright2019} do not use the VIKING confidence maps in their analysis. They exist for pawprints and tiles and give the pixel-by-pixel variation in exposure time, with a value of 100 referring to the median exposure time in the given band. They also include masking for detector artifacts such as bad pixels and the two ``bad patches" that do not flat-field well \citep{Edge2020}. Apart from these bad patches, the confidence maps are relatively uniform for individual chips, with variations for the tiles mostly caused by the dithering pattern. Since \citet{Wright2019} exclusively work at the detector level and use a weighted combination of the individual detector fluxes with outlier rejection, there was no need to consider the confidence maps in their analysis. 

Following the same arguments, we decided to simply use the uniform-valued weight maps of \citet{Wright2019} for our analysis as well. While we do not perform a weighted combination of fluxes with outlier rejection, each of our galaxy is covered by typically 3-4 VIKING exposures and up to 40 in the $J$-band overlap regions. Should any of these (partly) fall into a ``bad patch", it will likely not dominate the galaxy fit. The same applies to satellite tracks and similar artefacts that only affect individual chips. Nonetheless, a potential improvement of our analysis of the VIKING data would be to re-scale the \citet{Wright2019} weight maps with the corresponding confidence maps, after rotating the latter onto the same pixel grid. 

The last difference to the KiDS data that we would like to mention is that VIKING do not provide masks for bright stars and their possible reflection halos. These could be created following, e.g., the procedure in \citet{Barnett2021}, which is another potential improvement to our analysis of VIKING data. For the current work, we decided against such a procedure and instead rely on the rather conservative KiDS masks that were considered during image segmentation (see Section~\ref{sec:v05segmentation}). 

In summary, we opt to use the individual VIKING chips recalibrated by \citet{Wright2019} with associated uniform-valued weight maps and no bright star masks. This results in many data matches for each galaxy (a median of 3 in $Z, Y, H, K_s$ and 6 in $J$; although it can be more than 20 in overlap regions) with significantly different PSFs, making a joint fit necessary to avoid stacking. This became possible with the multi-frame fitting functionality of \texttt{ProFit v2.0.0}, released in February 2021.

\subsection{Segmentation changes}
\label{sec:v05segmentation}
Most of the preparatory work remained unchanged between \texttt{v04} and \texttt{v05} of the pipeline, except for the technical details to enable the analysis of the VIKING data (e.g. differing naming conventions). We still take cutouts around each object for each data match, estimate the sigma maps and do the background subtraction and PSF estimation separately for each image. The only major difference is in the treatment of the segmentation maps and masks. 

For the KiDS $g$, $r$ and $i$ bands, we used a stacked image to define a joint segmentation map in all three bands, including a joint mask (see Section~\ref{sec:preparatorysteps}). This was straight-forward since all tiles are registered to the same pixel grid across all KiDS bands, so there is always an exact correspondence of a tile in the $r$-band to one in the $g$- and one in the $i$-band; and they can simply be added (apart from very slight variations in the tile sizes that are easily taken account of). The VIKING chips generally also cover similar areas of sky in the different bands (except for the ``missing chips", see Section~\ref{sec:jointfitting}), but they are not registered to the same grid and have differing pixel sizes. 

\begin{figure}
	\includegraphics[width=\textwidth]{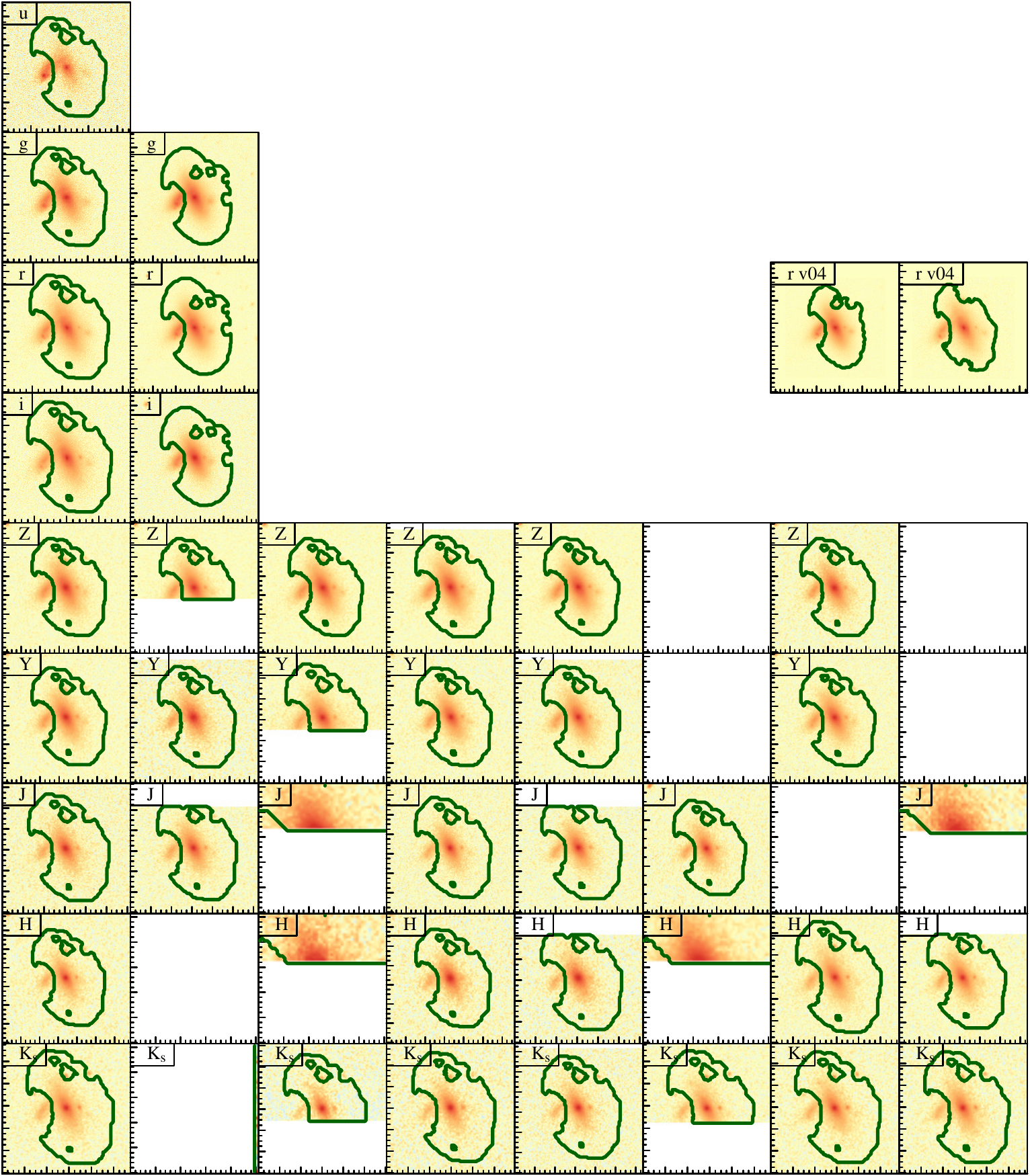}
    \caption{An example of the images and \texttt{v05} segmentation maps (green contours) used for fitting galaxy 544891 in all nine bands (the same object as that shown in Figure~\ref{fig:segfixplot}). This galaxy returned one data match in the $u$-band, two in $g, r, i$ (of which none were skipped), 14 in $Z$ (5 of which were skipped), 13 in $Y$ (5 skipped), 21 in $J$ (9 skipped), 8 in $H$ (5 skipped) and also 8 in $K_s$ (1 skipped). For the VIKING bands, we only show the first 8 matches each. White regions indicate missing data beyond the image edge or in the region of the chip that has zero weight assigned (cf. Section~\ref{sec:vikingdataproducts}). For the $r$-band, we also show the \texttt{v04} segmentation maps for direct comparison (the second of which corresponds to the rightmost panel in Figure~\ref{fig:segfixplot}). These are shown in the same cutout size as their corresponding \texttt{v05} versions for easier visual comparison. All other cutouts are shown in their original size (i.e. with differing numbers of pixels) and with individual flux scaling for best visibility (due to the different pixel scales). We omit axis labels and colour bars due to the qualitative nature of this plot.} 
    \label{fig:v05matchseg}
\end{figure}

A joint analysis of all bands is still possible, but requires more care to correctly account for these effects. In addition, due to the large number of matches and the much smaller sizes of VIKING chips compared to KiDS tiles (750\arcsec\ vs. 1\degr\ per side), galaxies are more frequently covered only partly by individual exposures. Considering only the area covered (and not masked) by all exposures in all bands would unnecessarily decrease the amount of available data for fitting and result in no data being left in many cases (see also Figure~\ref{fig:v05matchseg}). Nonetheless, we want to have the same fitting regions in all bands, to avoid systematic effects due to different segment sizes (see Sections~\ref{sec:postprocessing} and~\ref{sec:segchoices}). 

For these reasons, we decided to define the KiDS $g$, $r$ and $i$ bands as our ``core" bands (cf. Section~\ref{sec:kids}) and perform the segmentation on the $gri$ stacks as for \texttt{v04}. For the $u, Z, Y, J, H$ and $K_s$ bands, we then simply transfer these $gri$ segmentation maps (one for sky subtraction and one for galaxy fitting) onto the corresponding world coordinates and pixel grid of the target image using \texttt{profoundSegimWarp}. In case there are several segmentation maps available for the same galaxy (i.e. for the overlap sample in $gri$), we use the first match for transferring to other bands.\footnote{We actually intended to use the largest segmentation map instead of the first match, but this did not work due to a small bug in our code. The bug is fixed for any future runs, but we did not consider this to be a problem that would justify re-processing the entire \texttt{v05} of the \texttt{BDDecomp} DMU.} Should the segmentation map extend beyond the edge of the target image, it is truncated accordingly. 

For the $u$-band, we then additionally apply the $u$-band mask, although the vast majority of areas masked in the $u$-band are already included in the stacked $gri$ masks that were considered during the segmentation. As mentioned before, we do not use any additional masks for VIKING data (apart from those included in the weight maps, see Section~\ref{sec:vikingdataproducts}). Stars that are saturated in KiDS data, as well as reflection haloes, are conservatively masked during our analysis (Section~\ref{sec:otherprepworkchoices}) and will automatically be excluded from the segmentation. We consider it very unlikely that a star would be saturated and/or produce significant reflection haloes in one of the VIKING bands while being unmasked in all of the KiDS $g$, $r$ and $i$ bands. This is particularly true since we use individual detector chip images for VIKING and stacked (hence much deeper) tiles for KiDS. 

This procedure ensures that the fitting regions across all bands are as similar as possible (and exactly identical in our three core bands), while maximising the available data for fitting. The result for an example galaxy can be seen in Figure~\ref{fig:v05matchseg}, where we give a qualitative impression of the matches available for each band and the corresponding segmentation maps. All images are shown on their native pixel scales and for this reason, the cutout sizes and flux scalings are also different. Since the figure is only meant to give a visual impression, we omit axis labels and colour bars. 

The galaxy shown is part of the overlap sample in the KiDS $g, r, i$ bands and also has many matches in the VIKING bands (we show the first eight for each band). The two segmentation maps for the two matches in $gri$ are similar, but not identical. For all other bands, the first of those was transferred to all matches. Matches for which the galaxy centre falls into a white region (missing data) are skipped, i.e. not used during the joint fit. Some further matches are skipped due to PSF failures, see Section~\ref{sec:jointfitting}. For the $r$-band, we additionally show the \texttt{v04} segmentation maps on the same axis scale as their corresponding \texttt{v05} versions for direct comparison (see below for reasons why the segment size increased in \texttt{v05}). 

Potential improvements to this segmentation procedure would be to jointly treat multiple matches in $gri$ (i.e. for the overlap sample, where we still produce separate segmentation maps for each match) and/or use the correct corresponding match in the $u$-band, which is also pixel-matched to the other KiDS bands (instead of simply transferring the larger segmentation map to all matches). Ideally, we would in the future consider a stack\footnote{Stacking is not a problem in this context, since PSF variations do not affect the segmentation heavily.} of all bands used for analysis to define the segmentation maps, with regions missing from individual frames (because they are masked or beyond the edge of the corresponding image) downweighted appropriately without excluding them entirely. This is not possible with the current version of \texttt{ProFound} and would be non-trivial to implement.

One last change in the \texttt{v05} segmentation maps came from two default changes in \texttt{profoundProFound}: the \texttt{skycut} default changed from 1 to 1.5 and the \texttt{SBdilate} default changed from \texttt{NULL} to 2. 

For the meaning and effect of \texttt{skycut}, see Section~\ref{sec:segchoices}. We use two different versions of segmentation maps for the background subtraction and galaxy fitting with explicitly different aims (see also Sections~\ref{sec:preparatorysteps}, \ref{sec:backgroundstudies} and \ref{sec:segchoices}) that are produced with \texttt{skycut} values of 1 and 2 respectively. This was the optimal procedure we found after numerous experiments with different \texttt{skycut} values, including 1.5. We therefore considered it best to discard the new default and instead set \texttt{skycut} to 1 explicitly in all places where it was previously left on its default. 

The \texttt{SBdilate} option specifies how many magnitudes to push beyond the surface brightness limit in segment dilation if the number of pixels to be added is larger than \texttt{SBN100} (with a default of 100 that we left unchanged). After defining initial segments, \texttt{profoundProFound} expands those until its convergence criterion is met. By default, the convergence criterion is that the total flux within the segment increases by less than 5\,\% during the dilation \citep{Robotham2018}. For bright and extended galaxies, this means the number of dilations is often zero or one (cf. Section~\ref{sec:segchoices}) since the flux already contained in the segment is large, so the fractional change during dilation is small. For these objects, the \texttt{SBdilate} option can be used to dilate the segments further and capture the low surface brightness flux in the extended wings. In detail, if the dilated annulus contains at least \texttt{SBN100} pixels and has a mean surface brightness brighter than the sky surface brightness limit plus \texttt{SBdilate}, then the segment dilation continues (i.e. the annulus will be added to the current segment even if it contributes less than 5\,\% additional flux). For the new default values this means that we assume that 100\,pix are sufficient to safely push the detection 2\,mag beyond the sky surface brightness limit of individual pixels.  

We had not previously considered the \texttt{SBdilate} option since it was not available in the early versions of \texttt{ProFound} that we used for most of the preparatory work pipeline development; and we did not re-visit this after \texttt{SBdilate} was introduced. After the most recent \texttt{ProFound} update, we considered simply setting it back to its previous default of \texttt{NULL}, meaning that the \texttt{SBdilate} routine is not triggered. However, we decided to adopt the new defaults for several reasons. First, the logic behind \texttt{SBdilate} seemed reasonable and we could indeed observe that it typically does not change the star segments (for PSF estimation) but does dilate most galaxy segments, especially those of bright and extended objects. For these objects we had also noticed (during the detailed analysis of the \texttt{v04} results) that the azimuthally averaged one-dimensional flux profiles often continue to be positive and well-defined beyond the \texttt{v04} segment region, see e.g. Figure~\ref{fig:examplefit}. This is an indication that while individual pixels are below the surface brightness limit, their sum is still above it (and would be sufficient to constrain the model profile to larger radii), i.e. we have not reached the actual sky background yet. While our aim is to have relatively tight segments that do not include many background pixels, we did not intend to discard the extended wings of galaxies. 

In addition, since we now use the segments defined on the $gri$ images also for the other bands, we need to allow for potential variations of object sizes as a function of wavelength. There can also be variations in depth between bands and even within the same band (in particular in the overlap regions) due to the fact that we now jointly fit all matches but still define segments on individual matches of the $gri$ stacks. For all of these reasons, it seemed safer to use somewhat larger, more dilated segments for \texttt{v05} compared to \texttt{v04} and we decided to adopt the new default of \texttt{SBdilate}.

An example of the resulting \texttt{v05} segmentation map can be seen in Figure~\ref{fig:v05matchseg} for galaxy 544891. This is the same object that we also showed in Figure~\ref{fig:segfixplot} to allow for a visual impression of the evolution of segmentation maps during pipeline development. For a direct comparison to the \texttt{v04} segmentation maps, we also show these in Figure~\ref{fig:v05matchseg} in the rightmost columns as indicated. They are shown in the same cutout size as their corresponding \texttt{v05} version. Clearly, the segment size increased due to the \texttt{SBdilate} procedure.

\subsection{Joint fitting}
\label{sec:jointfitting}
\begin{figure}[t!]
\begin{center}
	\includegraphics[width=0.8\textwidth]{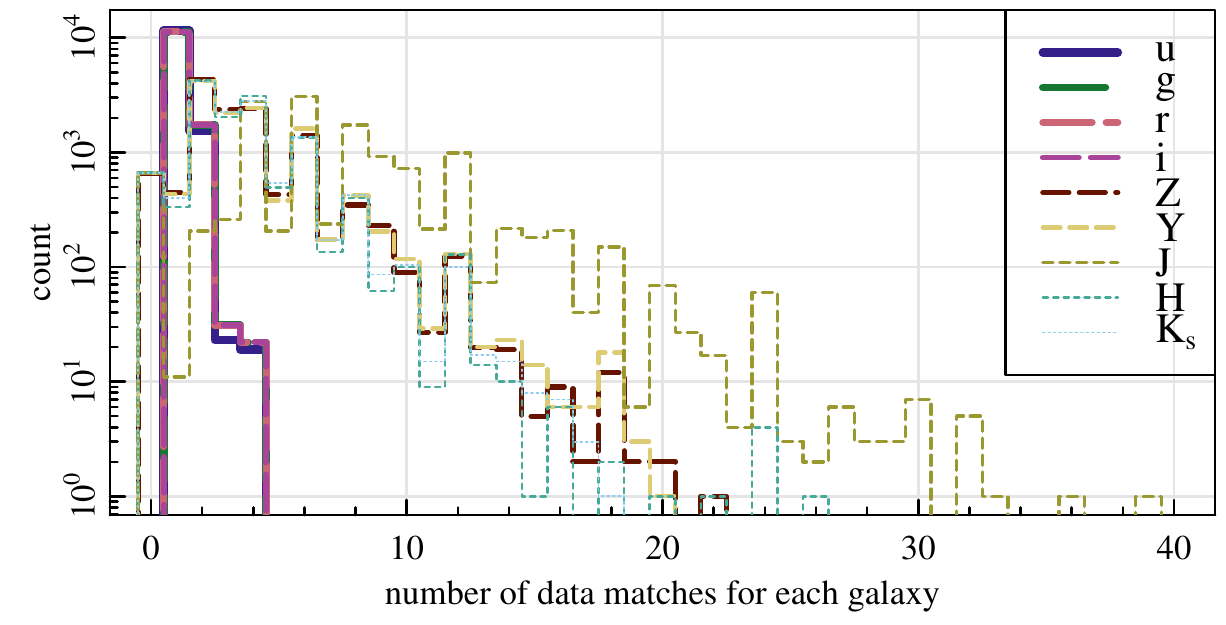}
	\includegraphics[width=0.8\textwidth]{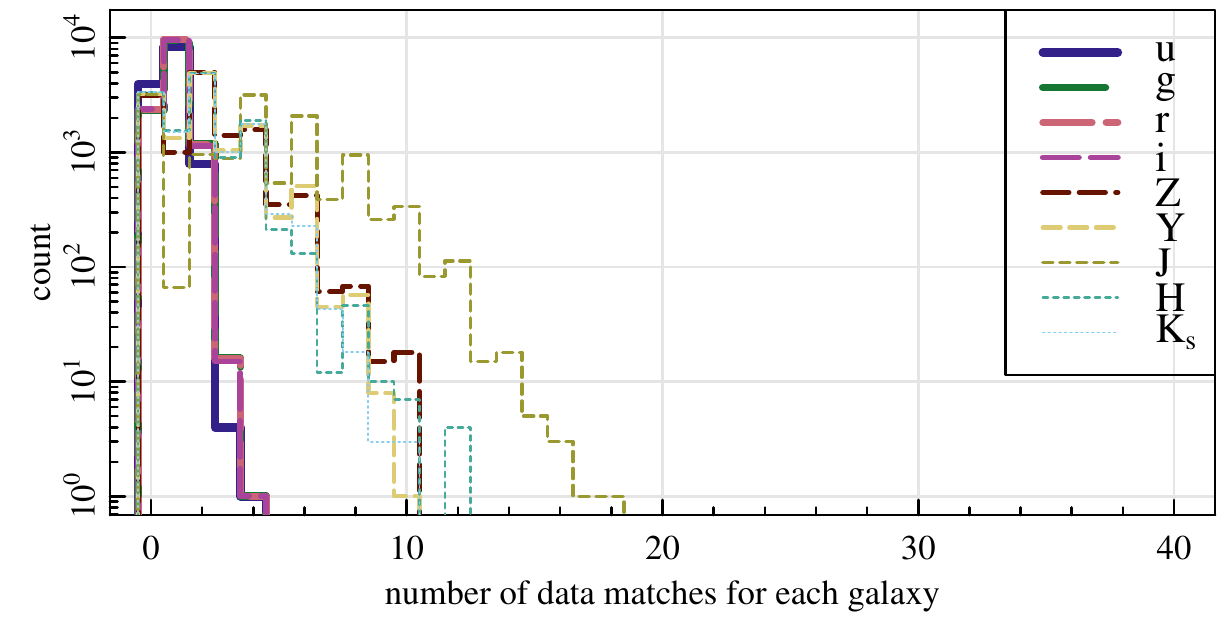}
    \caption{\textbf{Top:} the number of data matches to each of our 13096 galaxies for all nine KiDS and VIKING bands as indicated in the legend. Note the logarithmic scaling on the $y$-axis. \textbf{Bottom:} the same but showing only non-skipped matches (i.e. those that were actually used for fitting).} 
    \label{fig:nmatchviking}
\end{center}
\end{figure}

Since we work at the individual exposure level for VIKING data, galaxies frequently have multiple data matches as can be seen in the top panel of Figure~\ref{fig:nmatchviking}. Due to our joint processing of the $g$, $r$ and $i$ bands (cf. Section~\ref{sec:v05segmentation}), these have the exact same number of matches (with a minimum and median of 1 and a maximum of 4). The $u$-band follows the $g$, $r$ and $i$ bands closely (also with a minimum and median of 1 and a maximum of 4 matches), while the VIKING bands have many more matches on average but also some missing matches. The $Z$, $Y$, $H$ and $K_s$ bands show similar distributions ranging from 0 to around 20 with medians of 3. The $J$ band has typically twice as many matches (a median of 6 and a maximum of 39 matches) due to the VIKING observing strategy, see Section~\ref{sec:viking}.

The number of missing (zero) matches is very similar in all VIKING bands. They are due to small gaps in the VIKING data caused by a combination of incomplete sky coverage (see \citealt{Edge2020} and also figure 1 in \citealt{Wright2019}) and the quality control cuts applied by the VIKING team as well as by \citet{Wright2019}, see Section~\ref{sec:vikingdataproducts}. This affects approximately 5\,\% of the galaxies in our sample which are largely the same in all VIKING bands (cf. Table~\ref{tab:resultsv05}) and preferentially distributed around the edges of the three equatorial GAMA regions.  

Except for these ``missing objects", all matches pass through the preparatory work pipeline. However, some may be skipped at the fitting stage if the galaxy centre falls within a masked region or the PSF estimation failed. The number of images actually used for fitting each galaxy is shown in the bottom panel of Figure~\ref{fig:nmatchviking}. For the $g, r, i$ bands, around 20\,\% of individual matches are skipped due to the KiDS masking and a further $\sim$\,1\,\% due to PSF fails (cf. Table~\ref{tab:results}). Approximately 18\,\% of galaxies have no matches available for fitting and are skipped entirely (Table~\ref{tab:resultsv05}). 

For the $u$-band, the number of galaxies with no matches available is higher due to the additional masking, the decreased data quality and the slightly smaller footprint of the $u$-band tiles compared to the $g$, $r$ and $i$ bands (since the $u$-band tiles consist of only four dithers, while for the other bands it is five, cf. Section~\ref{sec:kids}). The smaller footprint by itself does not exclude any objects (i.e. there is no missing data in the $u$-band like in the VIKING bands), but it decreases the overlap regions between tiles. Hence galaxies in these regions are more often covered by only one exposure and/or less dithers (i.e. shallower data). These effects combined - plus the additional $u$-band masking - lead to more galaxies being skipped due to masking ($\sim$\,2\,\%) and PSF failures ($\sim$\,11\,\%), see also Table~\ref{tab:resultsv05} in Chapter~\ref{chap:results}. The latter effect in particular is very significant in the $u$-band due to the shallower data (not only in the overlap regions) in combination with our usage of the $gri$ segmentation maps: many candidate stars are too faint in $u$ to reliably estimate a PSF. Those that would be bright enough are masked, since they are saturated in the other bands. The median and maximum numbers of matches remain one and four respectively for all KiDS bands. 

Due to our use of the $gri$ segmentation maps, all galaxies that are skipped due to masking in the core bands (i.e. around 18\,\%) also have zero images available for fitting in the VIKING bands. Additionally, there are the $\sim$\,5\,\% of galaxies for which no matches were returned in the first place, such that the number of objects skipped entirely during the fit in the VIKING bands is slightly higher than that in the KiDS $g$, $r$ and $i$ bands, although still lower than in the $u$-band. The number of images available for fitting is, however, reduced significantly in all bands, especially for those objects with many data matches. This is mainly due to the truncation of chip edges by \citet{Wright2019}, whereby unreliable pixels around the edges of chips are assigned zero weight (Section~\ref{sec:vikingdataproducts}). A galaxy falling within this region of the chip is hence counted as a match, but may by covered only partly or not at all by the actual data. We count these objects as being ``masked" in all subsequent analysis. 

Since this affects the chip edges only (there are no additional masks for the VIKING data), it exclusively affects the VIKING overlap regions. In addition, some matches may be skipped due to PSF failures. This also preferentially affects the VIKING overlap regions where there are less stars available for fitting (both due to the missing data around the edge of each chip as well as the large cutouts used for PSF estimation extending beyond the actual edge of the chip). As we typically have many data matches available in the overlap regions, it is relatively rare that a galaxy ends up with zero matches due to either of these reasons (both combined contribute around 1-2\,\% in addition to the KiDS masking, cf. Table~\ref{tab:resultsv05}). The number of matches available for fitting therefore remains higher than for KiDS data, with values typically around two to four in the $Z, Y, H, K_s$ bands and a few more in the $J$-band (with medians of two and four respectively). The maximum number of matches is reduced to 18 in the $J$-band and around 10 in all other bands. An example of a galaxy with many matches is shown in Figure~\ref{fig:v05matchseg}. 

This high number of data matches made treating each match individually (as we did for \texttt{v04}) impractical from a computational point of view and - more importantly - unwise from a scientific point of view as we would lose too much depth of the data. Since we want to avoid stacking due to the different PSFs in VIKING frames, we perform a multi-frame fit whereby we jointly fit all images of the same galaxy in each band. Note the bands are still treated individually, though, i.e. we perform multi-frame rather than multi-band fits. For consistency, we also re-process the KiDS $g$, $r$ and $i$ bands.

This means that we pass all (individually sky-subtracted) images for the same physical object in the same band - with their corresponding masks, segmentation maps, sigma maps and PSFs - to \texttt{ProFit} at once. The model is then fitted to all images jointly, with appropriate weighting and taking account of the different PSFs. Since we treat each band individually, we do not allow any model parameters to vary between images, i.e. there is only one model applied to all images. The only additional user input needed is an additive offset between the individual images in $x$ and $y$ (in pixels) and in rotation angle (in degrees); as well as a multiplicative offset in pixel scale. 

The rotation angle offset is zero in our case since all KiDS tiles as well as all VIKING chips re-processed by \citet{Wright2019} are exactly aligned in RA and Dec. For KiDS data, the pixel scale offset is also zero. The VIKING chips have slightly varying pixels scales ranging between 0\farcs337 and 0\farcs341. We do not consider this since the early versions of the \texttt{ProFit} multi-frame fitting mode did not offer support for varying pixel scales (it does now, so we can improve this in the future). This means that the estimates of the effective radius (the only parameter that depends on the pixel scale) are inconsistent at the 1\,\% level. This is well within its uncertainty for the vast majority of galaxies, with the median relative error on single S\'ersic effective radius ranging from 4\,\% in the $Z$-band to 10\,\% in $K_s$. 

The only non-zero offsets in our analysis are the differences between the $x$ and $y$ positions of the galaxy centre in the different frames. These can be calculated easily from the world coordinate system headers of the corresponding images, assuming the astrometric solutions to be exact. The KiDS DR4.0 astrometric solutions show a scatter of approximately 0\farcs04 each in RA and Dec \citep{Kuijken2019}. As we discuss in Section~\ref{sec:systematics}, this - although it is well within the pixel scale of 0\farcs2 - is a factor of approximately four larger than the median MCMC error on position when fitting individual images (see also Figure~\ref{fig:differrnorm}; and examples of the same effect can be seen in Figures~\ref{fig:islandplotpsf} and~\ref{fig:islandplotseg}). For the joint fit, it is therefore possible to see visual offsets of the galaxy centre in the different matches for extreme cases. 

\begin{figure}
\begin{center}
\includegraphics[width=0.8\textwidth]{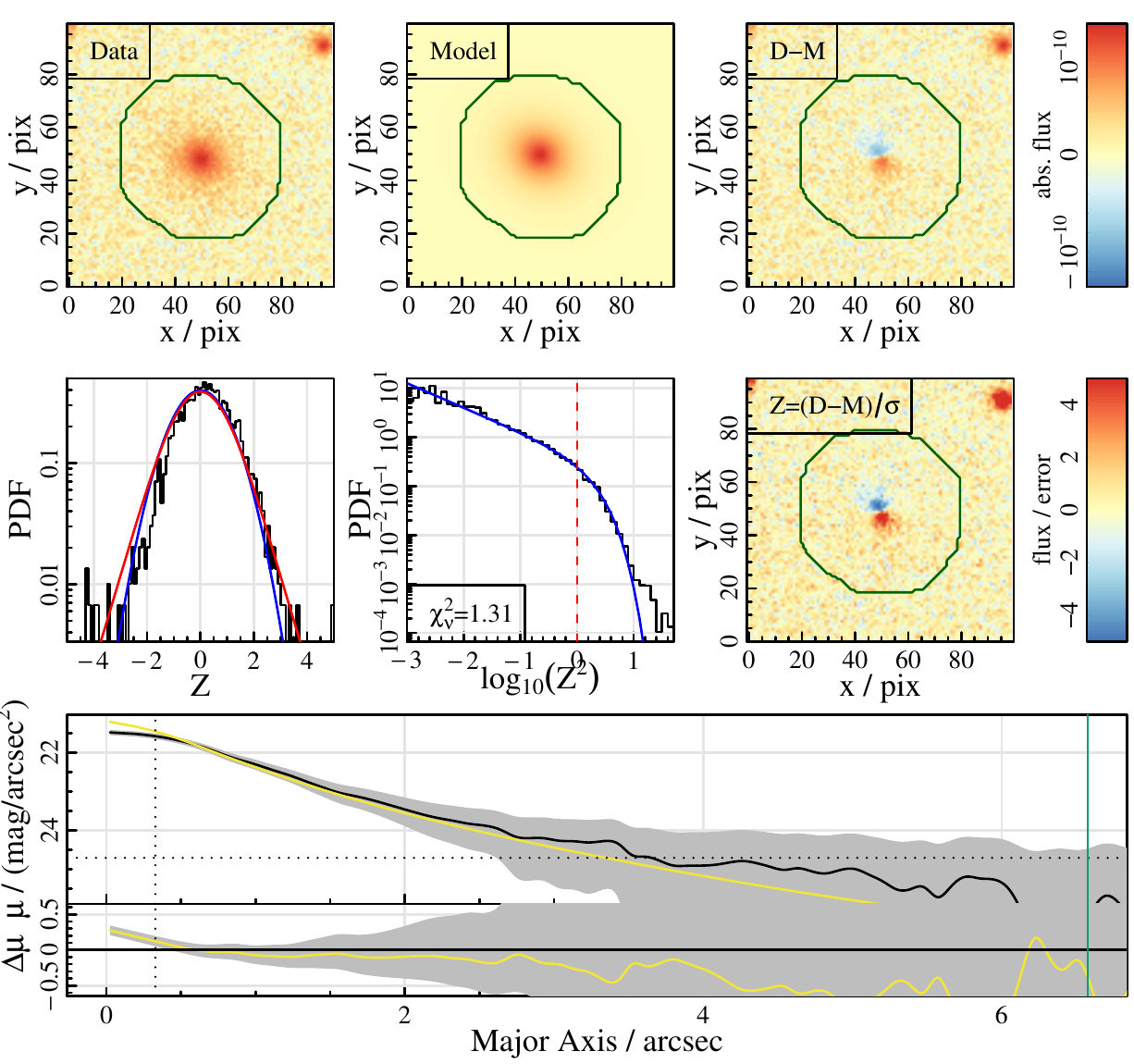}
\caption{The (joint \texttt{v05}) single S\'ersic fit to the first image of galaxy 345472 in the KiDS $r$-band. Panels in the top two rows are the same as those in Figure~\ref{fig:examplefit}, while the bottom row shows the one-dimensional fit only, corresponding to the rightmost panel of the bottom row in Figure~\ref{fig:examplefit}.}
\label{fig:examplefitoff1}
\end{center}
\end{figure}

\begin{figure}[t!]
\begin{center}
\includegraphics[width=0.8\textwidth]{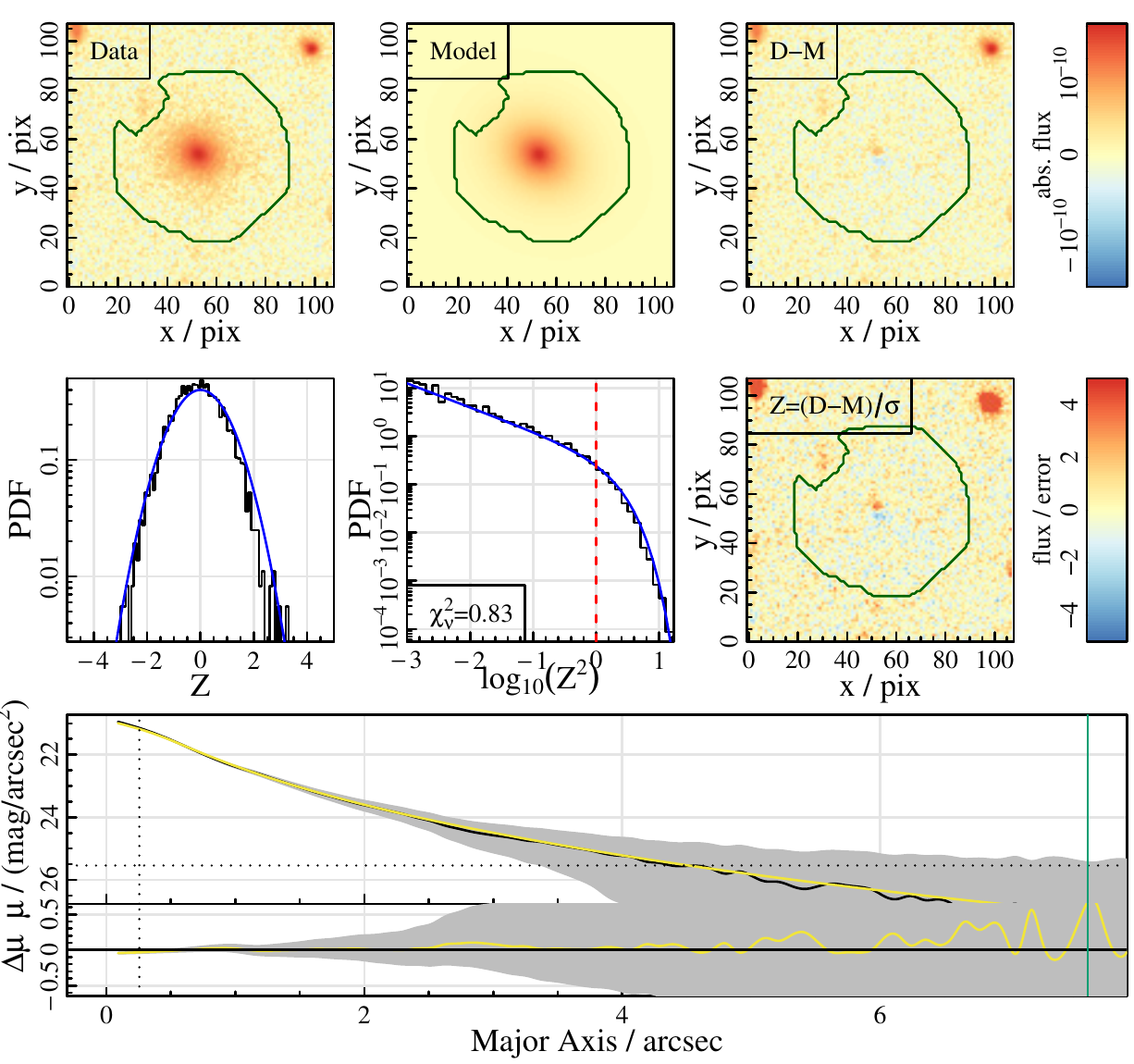}
\caption{The (joint \texttt{v05}) single S\'ersic fit to the second image of galaxy 345472 in the KiDS $r$-band, for direct comparison to Figure~\ref{fig:examplefitoff1}.}
\label{fig:examplefitoff2}
\end{center}
\end{figure}

An example for such an extreme case in the $r$-band is shown in Figures~\ref{fig:examplefitoff1} and~\ref{fig:examplefitoff2}. There is a discrepancy in the astrometric solution between the two images: for the individual fits in \texttt{v04}, the difference in the fitted RA and Dec positions for the two images are 0\farcs1 and 0\farcs4 respectively. For the joint fit in \texttt{v05}, this leads to the shallower of the two fits in particular being offset with respect to the actual centre of the galaxy, showing up as red and blue regions in Figure~\ref{fig:examplefitoff1}. The fit to the other image (Figure~\ref{fig:examplefitoff2}) shows a corresponding offset in the other direction, although not as clearly visible since it is the deeper image and dominates the fit. This adds onto the systematic uncertainties. The vast majority of offsets are much smaller than the extreme case shown here and cannot be visually seen in the model fits. It also only affects the overlap sample in KiDS where there is more than one match; and at the same time the astrometric solutions as well as the overall data quality are worst. 

For VIKING data, the scatter of the astrometric solution is 0\farcs09 \citep{Edge2020}. This is a factor of about two larger than for KiDS. However, the pixel size is also a factor of approximately two larger and the median MCMC uncertainty on position ranges from 0\farcs03 in $Z$ to 0\farcs05 in $K_s$ in both $x$ and $y$. The systematic uncertainty from the astrometric solution is therefore only a factor of two to three larger than the random uncertainty, compared to a factor of four for KiDS. It is therefore likely that the astrometric uncertainty affects VIKING bands less than KiDS bands despite the increased number of matches for each galaxy. Our visual impression was also that offsets visible by eye (such as those shown in Figures~\ref{fig:examplefitoff1} and~\ref{fig:examplefitoff2}) are less frequent in VIKING bands than in the KiDS $g$, $r$ and $i$ bands. 

To account for the uncertainty in the astrometric solution, we would need to allow for varying $x$ and $y$ centres to be fitted to each image of the same galaxy. This was not supported by early versions of the \texttt{ProFit} multi-frame fitting mode and can have several caveats as it would need to be ensured that the fitted offset is indeed due to the astrometric solution and not due to differing noise, image artefacts or similar issues. We therefore did not implement this for \texttt{v05} of the pipeline. However, it could be added in future work, minimising the caveats for example by imposing a (Gaussian) prior with a width taken from the known scatter of the astrometric solution.

There are no changes with respect to \texttt{v04} on the modelling side, except for some technical details, where, e.g., we needed to modify the \texttt{convergeFit} function to make it compatible with the \texttt{ProFit} multi-frame fitting mode. Other than that, we use the same three models, the same initial parameter guesses and priors, the Normal likelihood function and the same fitting algorithm and convergence criteria. The only difference is that we now obtain just a single fit for each physical object, referring to the first (non-skipped) data match where necessary (i.e. for $x$ and $y$ positions in our case; and potentially also for the position angle and effective radius for non-zero offsets in rotation angle and pixel scale). 

The advantage of this approach is that it makes full use of the data available for each physical object; and there is only one result for each object so there is no need to combine or choose between different measurements of the same galaxy. The disadvantage is that we lose the overlap sample for internal consistency checks and estimation of systematic uncertainties, as we have done in Section~\ref{sec:systematics} for the \texttt{v04} results. In addition, the method relies heavily on the accuracy of the astrometric solutions (see above). 

The natural next step would be to move to multi-band fits, i.e. fitting all matches from all bands for the same physical object jointly. This is the approach taken by the \texttt{MEGAMORPH} project team \citep{Haeussler2013, Vika2013, Haeussler2022} and has many advantages since it ensures smooth wavelength trends while preserving physical variation and additionally allows more robust fits to fainter magnitudes. We did not consider this approach initially due to the lack of support for multi-frame fitting in early versions of \texttt{ProFit}. With \texttt{ProFit v2.0.0} this is now possible, and certainly an interesting option for future work. We did not implement multi-band fitting for \texttt{v05} yet as it requires a major effort to be implemented computationally and above all requires the definition of how parameters can vary across wavelength. 

The newly developed package \texttt{ProFuse} \citep{Robotham2022}, combining \texttt{ProFit} with the spectral energy distribution fitting package \texttt{ProSpect} \citep{Robotham2020} is a first step towards fitting physically meaningful variations as a function of wavelength. However, like other works using simultaneous multi-band fits \citep[e.g.][]{Haeussler2022}, it (for now) fixes the structural parameters and only leaves the magnitudes free to vary. This is not an option for our work since variations in size as a function of wavelength are crucial for our science aim (Section~\ref{sec:scienceaims}). Hence the upgrade from \texttt{v04} to \texttt{v05} of our catalogue focused on expanding our previous work to all nine bands and to multi-frame fitting. It is a first step towards multi-band fitting and at the same time complementary to current works in that area; and can even serve to inform choices of ``allowed" variations in other parameters.

\subsection{Manual re-calibrations}
\label{sec:manualcalibrationchanges}
We point out the tuning parameters of the \texttt{v04} pipeline that are based on manual calibration steps in Section~\ref{sec:swappingandoutliers}. These may not all be directly transferable to VIKING data and hence we re-calibrated some of them for \texttt{v05}. There are five such calibrations in total. 

The first one is the cut chosen for the candidate star selection during PSF estimation (the percentage of all segments that are likely to be stars, see Section~\ref{sec:psfdetails}). It depends on the depth of the data used for the segmentation map and the total number of segments returned by the segmentation procedure. Since we use the $gri$ segmentation maps for the VIKING bands (and the \texttt{SBdilate} option only changes the size and not the number of segments), we do not need to re-calibrate this cut. We visually inspected a random sample of the diagnostic plots of this step (example shown in Figure~\ref{fig:examplepsfseg}) to ensure that the 4\,\% selection cut is indeed appropriate for the KiDS $u$ and all five VIKING bands provided that the $gri$ segmentations are used. 

The second visually tuned parameter is the cut in reduced chi-squared to exclude bad star fits from the model PSF creation (Section~\ref{sec:psfdetails}), which is influenced mainly by the segment size. While the galaxy segments for \texttt{v05} generally increased in size due to the \texttt{SBdilate} option (cf. Section~\ref{sec:v05segmentation}), the star segments did not change since stars are compact and not extended objects. We therefore also do not need to adjust this cut for \texttt{v05}.

The third manually tuned process is the swapping of bulge and disk components where ne\-ces\-sary. This procedure was developed with many visual inspections and test runs of $r$-band fits. However, the resulting procedure, described in detail in Sections~\ref{sec:galaxyfitting} and~\ref{sec:swappingandoutliers}, does not require any tuning parameters in absolute terms and instead relies only on relative quantities such as the ratio of the double component bulge effective radius to that of the single S\'ersic fit. It is therefore expected to work well for other bands, too, as long as the galaxies in that band show comparable properties as those in the $r$-band. We confirmed this in detail using the $g$ and $i$ band fits in \texttt{v04} by repeating all visual inspection steps. For the bands added in \texttt{v05}, we checked the most important statistics: the fraction of objects entering the swapping procedure and the fraction of those actually being swapped are very similar to the $r$ band fractions. The fraction of objects that remain swapped after the procedure is also very low in all bands based on the visual inspection of galaxies in the context of the model selection re-calibration (see below). We conclude that the swapping procedure is appropriate for all nine bands and does not require re-calibration. 

The fourth set of criteria that were developed with a large number of visual inspections are those for flagging bad fits (listed in Section~\ref{sec:postprocessing} with further details in Section~\ref{sec:swappingandoutliers}). Similar to the swapping procedure, we calibrated this on the $r$-band but found it to be appropriate for the $g$ and $i$ bands as well. We therefore assumed that it would work reasonably well for the $u$ and VIKING bands, too, and did not perform any re-calibration for \texttt{v05}. However, in contrast to the swapping procedure, there are a number of hard cuts in the outlier flagging that depend on the pixel scale and segment size or can be influenced by the depth of the data. We investigate this in detail in Section~\ref{sec:outlierstats}. In summary, we find that the outlier flagging is adequate for \texttt{v05} and only the flag value ``very irregular segment" could potentially have benefitted from a re-calibration. This can easily be added since the outlier flagging is a pure post-processing procedure and we provide the paramters for all model fits irrespective of their outlier status. We also provide full details of the criteria used for flagging outliers in the corresponding flag of the \texttt{BDDecomp} DMU. Users are therefore free to ignore certain flag values or discard our flagging entirely and perform their own identification of outliers.

The fifth and last manual calibration is that of the DIC difference cut for model selection, which is described in Sections~\ref{sec:postprocessing} and~\ref{sec:swappingandoutliers}. These are cuts in absolute values that are strong functions of data quality (depth; and to a lesser extent also seeing) and need to be re-calibrated for each band individually. We therefore visually inspected 200 randomly selected galaxies for each band (1800 objects in total) and re-calibrated all $\Delta$DIC cuts. The procedure is exactly the same as that used for \texttt{v04}, so we refer the reader to Sections~\ref{sec:postprocessing} and~\ref{sec:swappingandoutliers} for a description. Due to the differences in processing for \texttt{v04} and \texttt{v05} and for maximal consistency of the manual calibration between all nine bands, we also re-calibrated the three core bands. The differences between the $g$, $r$ and $i$ calibrations in \texttt{v04} and \texttt{v05} are investigated further in Section~\ref{sec:modelseldiffs}. Here, we focus on \texttt{v05} only.

\begin{figure}[t!]
\begin{center}
\includegraphics[width=0.8\textwidth]{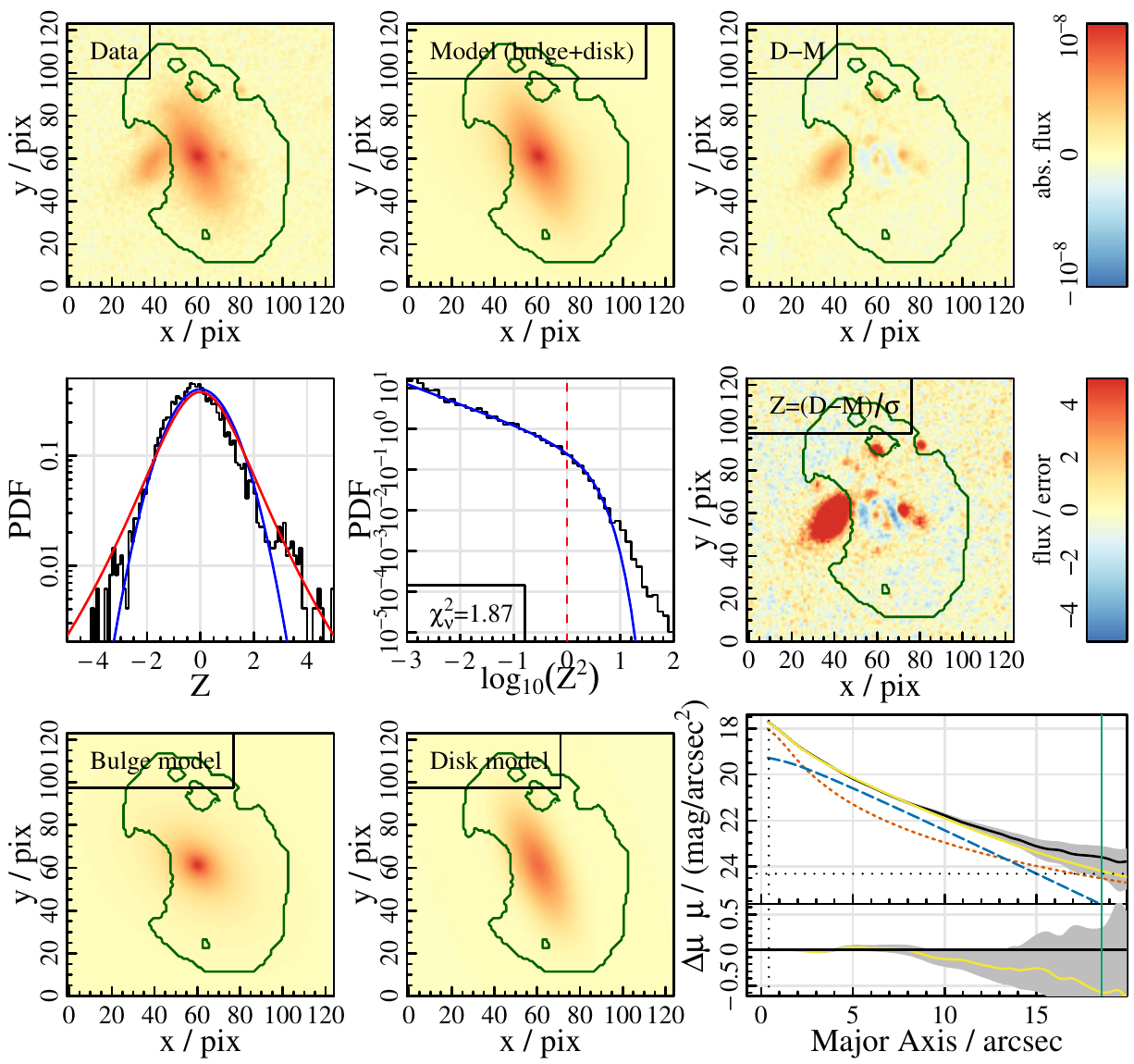}
\caption{The (joint \texttt{v05}) double component fit to the galaxy 544891 in the VIKING $J$-band. The diagnostic plot is composed of 12 individual frames for both the data and the model that are stacked in a consistent way, taking PSF differences into account. This is the same galaxy as that shown in Figure~\ref{fig:v05matchseg}, so the stacked version here can be compared to the individual frames shown in that figure. Panels are the same as those in Figure~\ref{fig:examplefit}.}
\label{fig:examplefitstack}
\end{center}
\end{figure}

Since we perform joint simultaneous fits on all images of the same object in a given band, the most consistent way to calibrate the model selection is on combined images composed of a stack of all individual frames. Otherwise it may be difficult to evaluate the model fit for objects that have many data matches (apart from having to look at 10 diagnostic plots for a single object), since the individual frames may be too shallow to visually identify two components, but the combination of all allows to constrain them during the fit. Note that since the stacking is performed after model fitting, we can avoid the problems with the different PSFs that led us to use joint fits in the first place. To this end, we first create model images for each individual frame using its native pixel scale and PSF. We then use the \texttt{magwarp} function to re-map these PSF-convolved model images, corresponding data images and sigma maps onto the world coordinate system (pixel grid) of the first data match. Subsequently, we stack all individual data frames and model frames, both weighted appropriately by the sigma maps. Like this, the model images and data images have undergone the exact same procedure and can be directly compared. An example of such a stacked diagnostic plot composed of 12 $J$-band images is shown in Figure~\ref{fig:examplefitstack} for galaxy 544891 (which we also show in Figure~\ref{fig:v05matchseg}).

We use the visual inspections to calibrate individual model selections for each band and two versions of joint model selections: one for the core bands only ($gri$, using the summed DICs from these three bands) and one for all bands ($ugriZYJHK_s$, summing all nine DICs). The resulting confusion matrices for all bands are listed in Table~\ref{tab:modelselconfusionv05} and the calibrated DIC cuts are given in Table~\ref{tab:v05diccuts}. The overall percentages of fits classified wrongly compared to visual inspection amount to $\sim$\,3, 15, 17, 9, 9, 3, 3, 5, 5, 17 and 18\,\% for the $u, g, r, i, Z, Y, J, H, K_s$ individual, $gri$ joint and $ugriZYJHK_s$ joint model selections respectively. The high fractions in the $g$, $r$ and joint selections are driven by large numbers of galaxies visually classified as double component fit that ended up in the \texttt{NCOMP}\,=\,1 category. We comment on this further in Section~\ref{sec:modelseldiffs}. For all other bands, the confusion rate is acceptable.

\begin{table}
\centering
\caption{The confusion matrices for the \texttt{v05} model selection based on a DIC difference cut compared against visual inspection for all nine KiDS and VIKING bands and the joint model selections in the core bands only as well as in all bands. All values are in percent of the total number of visually inspected galaxies in the respective band(s). Bold font highlights galaxies classified correctly, while grey shows those that were ignored during the calibration.}
\label{tab:modelselconfusionv05}
\begin{subtable}{1\textwidth}
\centering
	\begin{tabu}{lcrrrcrrrcrrrc} 
	& & \multicolumn{4}{c}{$u$-band} 
	& \multicolumn{4}{c}{$g$-band} 
	& \multicolumn{4}{c}{$r$-band}\\
	\hline
	& & \multicolumn{4}{c}{number of comps.} 
	& \multicolumn{4}{c}{number of comps.} 
	& \multicolumn{4}{c}{number of comps.}\\
		 
		visual class. & 
		& 1 & 1.5 & 2 &
		& 1 & 1.5 & 2 &
		& 1 & 1.5 & 2 &\\
		\hline
		
		``single" &
		& \textbf{91.5} & 0 & 0 &
		& \textbf{58.5} & 0 & 2.5 &
		& \textbf{47.5} & 0.5 & 2.5 &\\
		
		``1.5" &
		& 0 & \textbf{1.0} & 0 &
		& 1.5 & \textbf{0} & 1.5 &
		& 0.5 & \textbf{0.5} & 2.5 &\\
		
		``double" &
		& 2.5 & 0 & \textbf{0.5} &
		& 7.5 & 0.5 & \textbf{12.5} &
		& 10.5 & 0 & \textbf{23.0} &\\
		
		``1.5 or double" &
		& 0 & \textbf{0.5} & \textbf{0} &
		& 1.5 & \textbf{0.5} & \textbf{2.0} &
		& 0.5 & \textbf{0.5} & \textbf{2.0} &\\
		
		\rowfont{\color{gray}}
		``unsure" &
		& 2.5 & 0 & 0 &
		& 6.5 & 0 & 3.0 &
		& 4.0 & 0 & 3.5 &\\
		
		\rowfont{\color{gray}}
		``unfittable" &
		& 1.5 & 0 & 0 &
		& 1.5 & 0 & 0.5 &
		& 1.0 & 0 & 1.0 &\\
        \hline
	\end{tabu}
\end{subtable}%

\bigskip
\begin{subtable}{1\textwidth}
\centering
	\begin{tabu}{lcrrrcrrrcrrrc} 
	& & \multicolumn{4}{c}{$i$-band} 
	& \multicolumn{4}{c}{$Z$-band} 
	& \multicolumn{4}{c}{$Y$-band}\\
	\hline
	& & \multicolumn{4}{c}{number of comps.} 
	& \multicolumn{4}{c}{number of comps.} 
	& \multicolumn{4}{c}{number of comps.}\\
		 
		visual class. & 
		& 1 & 1.5 & 2 &
		& 1 & 1.5 & 2 &
		& 1 & 1.5 & 2 &\\
		\hline
		
		``single" &
		& \textbf{69.0} & 0.5 & 2.5 &
		& \textbf{63.5} & 2.0 & 1.5 &
		& \textbf{73.0} & 0 & 0 &\\
		
		``1.5" &
		& 1.0 & \textbf{1.0} & 1.0 &
		& 1.5 & \textbf{4.0} & 1.0 &
		& 0.5 & \textbf{4.0} & 0 &\\
		
		``double" &
		& 4.0 & 0 & \textbf{11.5} &
		& 2.5 & 0 & \textbf{12.0} &
		& 1.5 & 1.0 & \textbf{12.5} &\\
		
		``1.5 or double" &
		& 0 & \textbf{0} & \textbf{1.0} &
		& 0 & \textbf{1.5} & \textbf{1.0} &
		& 0 & \textbf{0} & \textbf{1.5} &\\
		
		\rowfont{\color{gray}}
		``unsure" &
		& 3.0 & 0 & 3.0 &
		& 4.0 & 1.0 & 2.0 &
		& 1.0 & 0.5 & 2.5 &\\
		
		\rowfont{\color{gray}}
		``unfittable" &
		& 2.0 & 0 & 0.5 &
		& 1.5 & 0 & 1.0 &
		& 1.0 & 0 & 1.0 &\\
        \hline
	\end{tabu}
\end{subtable}%

\bigskip
\begin{subtable}{1\textwidth}
\centering
	\begin{tabu}{lcrrrcrrrcrrrc} 
	& & \multicolumn{4}{c}{$J$-band} 
	& \multicolumn{4}{c}{$H$-band} 
	& \multicolumn{4}{c}{$K_s$-band}\\
	\hline
	& & \multicolumn{4}{c}{number of comps.} 
	& \multicolumn{4}{c}{number of comps.} 
	& \multicolumn{4}{c}{number of comps.}\\
		 
		visual class. & 
		& 1 & 1.5 & 2 &
		& 1 & 1.5 & 2 &
		& 1 & 1.5 & 2 &\\
		\hline
		
		``single" &
		& \textbf{76.5} & 0 & 0.5 &
		& \textbf{79.0} & 0.5 & 1.0 &
		& \textbf{75.0} & 0 & 1.0 &\\
		
		``1.5" &
		& 0.5 & \textbf{1.0} & 1.0 &
		& 0.5 & \textbf{5.5} & 1.0 &
		& 2.5 & \textbf{2.0} & 0.5 &\\
		
		``double" &
		& 0.5 & 0.5 & \textbf{14.0} &
		& 0.5 & 0.5 & \textbf{7.0} &
		& 0 & 0.5 & \textbf{13.5} &\\
		
		``1.5 or double" &
		& 0 & \textbf{0} & \textbf{1.0} &
		& 0.5 & \textbf{0} & \textbf{0.5} &
		& 0 & \textbf{0.5} & \textbf{0} &\\
		
		\rowfont{\color{gray}}
		``unsure" &
		& 1.5 & 0.5 & 1.0 &
		& 1.0 & 0 & 2.0 &
		& 2.0 & 0 & 2.0 &\\
		
		\rowfont{\color{gray}}
		``unfittable" &
		& 0.5 & 0.5 & 0.5 &
		& 0.5 & 0 & 0 &
		& 0.5 & 0 & 0 &\\
        \hline
	\end{tabu}
\end{subtable}%

\bigskip
\begin{subtable}{1\textwidth}
\centering
	\begin{tabu}{lcrrrcrrrc} 
	& & \multicolumn{4}{c}{joint $gri$} 
	& \multicolumn{4}{c}{joint $ugriZYJHK_s$}\\
	\hline
	& & \multicolumn{4}{c}{number of comps.} 
	& \multicolumn{4}{c}{number of comps.}\\
		 
		visual class. & 
		& 1 & 1.5 & 2 &
		& 1 & 1.5 & 2 &\\
		\hline
		
		``single" &
		& \textbf{59.4} & 0.3 & 1.5 &
		& \textbf{71.3} & 0.7 & 0.5 &\\
		
		``1.5" &
		& 0.8 & \textbf{1.2} & 1.2 &
		& 3.6 & \textbf{0} & 0.1 &\\
		
		``double" &
		& 10.9 & 1.7 & \textbf{10.7} &
		& 11.6 & 0.1 & \textbf{4.1} &\\
		
		``1.5 or double" &
		& 0.7 & \textbf{0.5} & \textbf{1.5} &
		& 1.4 & \textbf{0} & \textbf{0.1} &\\
		
		\rowfont{\color{gray}}
		``unsure" &
		& 5.5 & 0 & 2.2 &
		& 4.4 & 0.1 & 0.4 &\\
		
		\rowfont{\color{gray}}
		``unfittable" &
		& 1.5 & 0 & 0.5 &
		& 1.4 & 0 & 0.1 &\\
        \hline
	\end{tabu}
\end{subtable}%
\end{table}

\begin{table}
	\centering
	\caption{The calibrated DIC difference cuts used for \texttt{v05} of the \texttt{BDDecomp} DMU. For each band, we show the lower bound (LB), the actual cut and the upper bound (UB) for each of the three DIC differences between our models. For the joint model selection, the cuts refer to the differences between the summed DICs of all bands.}
	\label{tab:v05diccuts}
	\begin{tabu}{lrrrrrrrrrrrr} 
		\hline
		 && \multicolumn{3}{c}{$\Delta$DIC$_{1-1.5}$} && 
		 \multicolumn{3}{c}{$\Delta$DIC$_{1-2}$} &&
		 \multicolumn{3}{c}{$\Delta$DIC$_{1.5-2}$}\\
		band && LB & cut & UB && LB & cut & UB && LB & cut & UB\\
		\hline
		$u$ && 66 & 77 & 370 && 265 & $10^{18}$ & $10^{40}$ && 54 & 73 & 101\\
		$g$ && 32 & 63 & 106 && 2096 & 2612 & 3867 && $10^{-11}$ & 66 & 260\\
		$r$ && 29 & 39 & 64 && 645 & 1881 & 3146 && $10^{-13}$ & 116 & 228\\
		$i$ && 137 & 180 & $10^{17}$ && 1051 & 1247 & 1616 && 481 & 512 & 576\\
		$Z$ && 28 & 31 & 281 && 464 & 650 & 819 && 171 & 306 & 944\\
		$Y$ && 21 & 52 & 65 && 256 & 316 & 382 && 53 & 119 & 175\\
		$J$ && 106 & 116 & 223 && 524 & 743 & 944 && 61 & 77 & 442\\
		$H$ && 47 & 52 & 60 && 389 & 523 & 642 && 165 & 200 & 734\\
		$K_s$ && 44 & $10^{12}$ & $10^{35}$ && 190 & 264 & 447 && 329 & 395 & 520\\		
		$gri$ && 175 & 330 & 27391 && 15145 & 17827 & 18885 && $10^{-54}$ & $10^{-24}$ & $10^{-4}$\\
		$ugriZYJHK_s$ && $10^4$ & $10^8$ & $10^{19}$ && 59994 & 79722 & 82856 && $10^{-53}$ & $10^{-23}$ & $10^{-4}$\\
        \hline
	\end{tabu}
\end{table}

For the shallower bands (especially KiDS $u$), the data quality is not sufficient to constrain more than one component in most cases, so the fraction of single component fits increases drastically. In part, this is the reason for the low confusion rates in these bands: essentially all fits are (correctly) classified as single S\'ersic objects. This is enhanced even more by the fact that we minimise the absolute number of fits classified wrongly during the model selection calibration, resulting in a bias against the rarer categories (cf. Section~\ref{sec:postprocessing}). Additionally, the corresponding DIC difference cuts are often poorly constrained due to the low number of objects available for calibration in those rare categories. This is also apparent from Table~\ref{tab:v05diccuts}, where some of the cuts show large uncertainty regions or take on unreasonably large or small values (essentially tending to zero or infinity).\footnote{The lower cuts tend to zero rather than negative infinity because we calibrate the $\Delta$DIC cuts in logarithmic space.}
The latter problem could be alleviated by visually inspecting a larger subsample of galaxies in each band. 

The joint model selection encompassing all nine bands benefits from the high number of 1800 visually inspected galaxies for calibration. However, this does not solve the underlying problem of the bands being too shallow to constrain two components. Also, joint model selection is always a compromise between all bands by necessity. Since most bands have relatively high numbers of single S\'ersic fits, the joint model selection does, too. This means that for the deeper bands with the best seeing, many fits are classified as single S\'ersic objects in the joint model selection even though they do have two well-defined components. Vice versa, an object classified as double component fit in the joint model selection may well show unconstrained components in the shallowest bands. For this reason, the joint model selection should be used with care. The compromise becomes worse for greater differences in data quality between bands.

For this reason, we also added the second version of the joint model selection, concentrating on the core bands only (like in \texttt{v04}). These are most directly comparable not only in their treatment throughout our pipeline, but also in their data quality, so there are less caveats in their joint model selection (although it is still a compromise, see also Section~\ref{sec:postprocessing}). To avoid such inconsistencies, some works (e.g. \citealt{Lackner2012, Kim2016}) take the structural parameters from one band (e.g. the $r$-band) and simply impose them onto the other fitting bands without further adjustments; only fitting the component magnitudes. Others (e.g. \citealt{Simard2011, Kennedy2016}) use simultaneous multi-band fits but again only allow the magnitudes to vary between bands. While this may result in robust colours, it does not capture physical variations of structural parameters with wavelength, which are a key quantity we want to obtain. The only way to avoid these issues and successfully use shallower bands for multi-component fits (other than deeper observations) are simultaneous multi-band fits with varying parameters as a function of wavelength. Constraints on the ``allowed" structural variation between bands can be motivated by theory, simulations or previous work fitting bands individually, such as this one.

\clearpage
\newpage
\chapter{Results}
\label{chap:results}

In this chapter, we present the results of our bulge-disk decomposition pipeline. This is mainly a catalogue of parameters for the (components of) galaxies in our sample along with a wealth of ancillary data and quality control metrics. We investigate some of these metrics here. A more detailed quality control of the catalogue is then presented in Chapter~\ref{chap:QC}, where we compare our work to previous works in the field and perform a detailed study of systematic uncertainties. Since the quality control focuses on \texttt{v04} of the \texttt{BDDecomp} DMU, we concentrate on this version here, too (following \citealt{Casura2022}). However, we expand on \texttt{v04} with the new results from \texttt{v05} that have not been presented previously, highlighting and discussing differences where relevant. This anchores the \texttt{v05} results to those of \texttt{v04}, thereby benefitting from the detailed quality control of that version presented in Chapter~\ref{chap:QC}. Results from DMU versions prior to \texttt{v04} are not presented since they were somewhat preliminary with limited quality control and were superseded entirely by \texttt{v04}.

\section{\texttt{BDDecomp} DMU}
\label{sec:bddecompdmu}

Our main result is the \texttt{BDDecomp} DMU on the GAMA database. It contains two catalogues per band: \texttt{BDInputs} with the most important outputs of the preparatory work pipeline (segmentation, PSF estimation, initial guesses) and \texttt{BDModelsAll} with the output from the galaxy fitting and post-processing (model selection, flagging of bad fits and truncating to segment radii). From \texttt{v03} onwards, where we include several bands in the analysis, the catalogue names also specify the band to which they refer (e.g. \texttt{BDModelsAllR}). Up to \texttt{v04}, all images of the same object in the same band (data matches) are listed separately since they were also treated individually. For \texttt{v05}, they were fitted jointly so each galaxy only has one entry in the \texttt{BDModelsAll} table. The matches remain separate in the \texttt{BDInputs} tables since the preparatory work was still carried out on each image individually. 

In addition to these band-specific tables, \texttt{BDModels} gives the most important columns of the \texttt{BDModelsAll} table(s), combining results from all bands from \texttt{v03} onwards. It also has a few additional joint columns (mainly joint model selection). Finally, the table \texttt{BDModelsAlt} (added from \texttt{v03} onwards) presents the same information as \texttt{BDModels} just with the different bands arranged in rows instead of columns. This results in a total of three tables up to \texttt{v02} ($r$-band only), eight tables for \texttt{v03} and \texttt{v04} ($g$, $r$, $i$) and 20 tables for \texttt{v05} ($u, g, r, i, Z, Y, J, H, K_s$). 

Each table is accompanied by comprehensive documentation including descriptions of all columns, details on the processing steps and practical tips for using the catalogue. The DMU also provides all input data used for the fitting (i.e. image cutouts, masks, error maps, segmentation maps, sky estimates, PSFs) as well as various diagnostic plots of the fit results on the GAMA file server, where detailed descriptions of these files can be found.

Sections~\ref{sec:v01} to~\ref{sec:v06} summarise the key changes between the different DMU versions (also listed on the GAMA database) and give the basis for each DMU version, as labelled in the directory tree of our local machines. Furthermore, they provide the release date and the versions of \texttt{ProFit} and \texttt{ProFound} used. Individual aspects of this pipeline evolution are pointed out throughout Chapter~\ref{chap:pipeline} and we refer the reader to this chapter for details. 

In the remainder of this chapter we then present an overview over the contents of the main catalogue, \texttt{BDModels}, in \texttt{v04} and \texttt{v05}.

\subsection{\texttt{BDDecomp v01}}
\label{sec:v01}

\texttt{v01} is the first release of our bulge-disk decomposition catalogue, published on 2019-02-08. It is limited to the $r$-band only and is based on the preparatory work run labelled \texttt{"run3"} and the galaxy fitting \texttt{"run5"}. All previous runs as well as numerous test runs were not published on the GAMA database. \texttt{ProFit} and \texttt{ProFound} versions used were from 2018-04-24.

\subsection{\texttt{BDDecomp v02}}
\label{sec:v02}

\texttt{v02} is only a minor update of \texttt{v01} (published 2019-02-27) with two small mistakes fixed in the post-processing of the \texttt{BDModels} and \texttt{BDModelsAll} tables:
\begin{itemize}
\item The \texttt{*\_OUTLIER\_FLAG}s were fixed for 297 single component fits, 
   2211 double component fits and 3437 1.5-component fits.
\item The \texttt{P\_BDQUAL\_FLAG}s were fixed for all 1.5-component fits.
\end{itemize}
The preparatory work and galaxy fitting results did not change; and both are based on the same runs as \texttt{v01}. 

\subsection{\texttt{BDDecomp v03}}
\label{sec:v03}

\texttt{v03} is a major update which was released on 2019-12-03 and is based on preparatory work \texttt{"run4"} and fitting \texttt{"run6"}. We included the $g$ and $i$ bands in addition to the $r$-band, increasing the number of catalogues from three to eight (two per band plus two joint ones). This also resulted in a slight re-labelling of the catalogues themselves and their column names to distinguish all values between the three bands. At the same time, we upgraded to KiDS DR4.0 (which had become available in the meantime), including the photometric homogenisation given in the \texttt{DMAG} header keyword. We also updated \texttt{ProFit} and \texttt{ProFound} to the versions from 2019-08-19, which saw major changes since the \texttt{v01}/\texttt{v02} DMU release. All of the above, as well as our own investigations and feedback from both collaborators and users of the first catalogue versions resulted in numerous changes to the pipeline.

In the preparatory work pipeline (see the first half of Section~\ref{sec:pipelinedevelopment} for details):
\begin{itemize}
\item We turned off our own routine for segmentation map fixing since the new version of \texttt{ProFound} was much less prone to ``shredding" galaxies and our routine tended to include faint secondary objects in the segments.
\item We increased the \texttt{skycut} value in \texttt{profoundProFound} from 1 to 2 resulting in smaller segments with smoother borders.
\item We added one additional dilation of the galaxy segment after the \texttt{profoundProFound} run, which ensures that the edges are smooth and unbiased (without this, the segment border can be very jagged including noisy, slightly positive pixels but excluding slightly negative pixels; especially for large bright objects where the number of curve of growth iterations is often zero or one). The dilated segments are then approximately the same size as they were in the previous run due to the higher \texttt{skycut} value.
\item We defined the segmentation maps on stacked images of the $g, r, i$ bands (with corresponding stacked masks) and used these segmentation maps in all bands. 
\item We defined and used two different segmentation maps for the sky estimation and object fitting; the one used for sky estimation is more dilated to exclude faint sources and extended wings from the sky (aggressive object masking and lower \texttt{skycut}). 
\item The star fraction cut to identify candidate stars and the chi-squared cut to exclude bad fits from PSF modelling were adjusted slightly to account for the new segments.
\item We stopped using the \texttt{profoundSkySplitFFT} routine and instead take the sky estimate from \texttt{profoundProFound} directly, as we found that to be more robust. 
\end{itemize}

In the galaxy fitting and post-processing (see the second half of Section~\ref{sec:pipelinedevelopment} for details):
\begin{itemize}
\item We increased the limits for the S\'ersic index fitting to 0.1 to 20 (previously 1 to 12) allowing for flatter bulges and causing fewer fits hitting the limits.
\item We improved the swapping criteria slightly (both for the selection of fits to enter the swapping procedure and for selecting the better of two swapped fits) based on new visual inspections.
\item We updated the outlier rejection criteria to reflect the new S\'ersic index limit; and we included a new criterion based on the difference between the magnitude within the segment and within the segment radius (also based on new visual inspections). 
\item We re-calibrated the model selection for each band (new visual inspections) and additionally provided a joint model selection that is based on the summed DICs for all bands. 
\item We added several alternative measurements of the magnitudes, effective radii and bulge-to-total (B/T) ratios for each fit, resulting in the catalogues having more columns: 
\begin{itemize}
\item The S\'ersic magnitude and effective radius.
\item The magnitude contained within the segment and the corresponding effective radius (containing half of the segment flux).
\item The magnitude and effective radius within the ``segment radius", which is defined as the maximum distance between the centre of the fit and the edge of the segment. This is the upper limit to which our model fits are valid, everything beyond that is extrapolated and unconstrained. We recommend using these values for magnitude, effective radius and bulge-to-total flux ratios because many of our fits have high S\'ersic indices with unphysically large wings beyond the segment borders. This is because we opted for relatively tight segments focusing on fitting the central regions well rather than forcing the models to zero at large radii (Section~\ref{sec:postprocessing}). For reproducibility and ease of comparison to other catalogues, the segment radius (\texttt{RAD\_SEG}) is provided in a separate column as well. 
\item The magnitude and effective radius within 10 S\'ersic effective radii. Note that these values may still include unphysical results (see above) and were provided for completeness only. 
\end{itemize}
\end{itemize}

\subsection{\texttt{BDDecomp v04}}
\label{sec:v04}

\texttt{v04} is again a relatively minor update with no changes to the statistical properties of the galaxy parameters. It is based on preparatory work \texttt{"run5"} and fitting \texttt{"run7"}, used \texttt{ProFit} and \texttt{ProFound} versions from 2019-08-19 and was released on 2021-06-25. Its major effort was in better characterising the systematic uncertainties. This is the version that \citet{Casura2022} is based on and is the currently newest version on the GAMA database. 

In detail:
\begin{itemize}
\item Parameter errors labelled \texttt{*\_ERR} now include our best estimate of systematic uncertainties. The purely random MCMC errors have been re-labelled to \texttt{*\_ERR\_MCMC}. The \texttt{*\_ERR} values are obtained from the \texttt{*\_ERR\_MCMC} values by multiplying with the ``error underestimate" factors for each parameter listed in table 4 of \citet{Casura2022} and reproduced here in Table~\ref{tab:errorunderestimate}. We do not apply the bias corrections listed in the same table since they are only applicable to the averages of large statistical samples, not individual galaxies. Since the error understimates are slightly different for \texttt{*\_SEGRAD} values, these values now also have associated uncertainties (\texttt{*\_SEGRAD\_ERR}). Note the error underestimates are derived from single component $r$-band fits. We apply them to the $g$ and $i$ bands and bulge and disk parameters for 1.5 and double component fits as well, since that should result in more realistic errors than the random ones alone. However, it is likely that the true errors are even larger for component parameters (in particular for the non-dominant component) and/or the $g$ and $i$ bands (also for single S\'ersic fits, since the $r$-band is the deepest and best-resolved).
\item B/T values now also have errors (*\_BT\_ERR) given for convenience. They are derived from the bulge and disk magnitude errors assuming independent variables (which is not actually true) and hence should be taken as lower limits.
\item We fixed a small bug in the preparatory work pipeline where in \texttt{v03} we accidentally performed the KiDS zeropoint homogenisation correction (with the \texttt{DMAG} header keyword) twice, resulting in initial guesses for magnitudes being biased by about 0.03 mag on average. Since the MCMC fits are not strongly dependent on the initial guesses, this did not affect the fits (the zeropoint was correct during the actual fitting). Nonetheless, we re-ran the entire preparatory work and fitting pipelines. 
\item Position angles are now all in the interval 0 to 180\degr\ (which they were already supposed to be in \texttt{v03}, but were not in all cases). 
\item Galaxies are now also flagged as outliers if the single S\'ersic fit failed (even if the 1.5- or double component fits were successful) because then the segment radius (and thus all \texttt{*\_SEGRAD} columns) are missing; and selecting good single S\'ersic fits with \texttt{NCOMP}\,>\,0 fails. This affects only one fit in the $g$-band and none in the other bands. 
\item For consistency we renamed the columns \texttt{*\_FLUX\_ERR\_SEG} to \texttt{*\_FLUX\_SEG\_ERR} and \texttt{*\_RAD\_SEG} to \texttt{*\_SEGRAD}.
\item We added a quick-start guide to the DMU description on the GAMA database to cover the most important aspects in catalogue usage. 
\end{itemize}

\subsection{\texttt{BDDecomp v05}}
\label{sec:v05}

\texttt{v05} is another major update, to be released on the GAMA database alongside this thesis. It uses preparatory work \texttt{"run6"} and fitting \texttt{"run8"} with \texttt{ProFit} and \texttt{ProFound} versions last updated on 2022-02-11. In addition, we moved from Ubuntu 16.04 to Ubuntu 20.04 on all of our local machines, which necessitated updating \texttt{R} along with all of its packages and replacing the \texttt{astro} package by the (relatively new) \texttt{Rfits} package for writing FITS files. On the scientific side, the major upgrades are that we now include the KiDS $u$ and VIKING $Z, Y, J, H$ and $K_s$ bands and do multi-frame fitting. The corresponding changes are described in detail in Section~\ref{sec:pipelineupdates}. 

In summary:
\begin{itemize}
\item We now process all nine KiDS and VIKING bands, using the science tiles for KiDS but individual detector chips preprocessed by \citet{Wright2019} for VIKING. This increases the number of catalogues from eight to 20 (two per band plus two joint ones). 
\item We now do multi-frame fitting, i.e. when there is more than one data match to the same galaxy in the same band, they are fitted jointly rather than separately (assuming the astrometric solutions to be the ground truth). For the preparatory work, the matches are still treated individually and also listed individually in the corresponding \texttt{BDInputs} tables; but for the \texttt{BDModels} tables, there is now only one entry per galaxy. This means that \texttt{CATAID} is a unique identifier again (along with the \texttt{OBJECTID}s which we keep for compatibility with earlier versions). 
\item To reflect these changes, the \texttt{BDQUAL\_FLAG}s and \texttt{CUTOUT\_FLAG}s have changed in meaning slightly, the \texttt{BEST\_IMG} column (identifying the best image to use in the case of multiple matches) has been replaced with the \texttt{N\_MATCH} and \texttt{N\_FIT} columns (giving the number of data matches and the number of images actually used for fitting). The pixel scale (which varies for VIKING) has been added in the \texttt{PIXSC} column to give meaning to all columns with units of pixels. \texttt{JOINT\_*} columns have been renamed into \texttt{GRI\_*} (or whichever bands exactly they refer to, to differentiate between the different versions of joint model selection). 
\item We define the KiDS $g, r$ and $i$ bands as our core bands and transfer the segmentation maps defined on a stack of those core bands to all other bands. 
\item The galaxy segment sizes are generally larger than in \texttt{v04} due to a default change in \texttt{profoundProFound}. We use the new default of \texttt{SBdilate}\,=\,2, but discarded the \texttt{skycut} default change. 
\item We re-calibrated the model selection with new visual inspections in all bands. 
\end{itemize}

\subsection{\texttt{BDDecomp v06}}
\label{sec:v06}

We envision future versions of the DMU to provide simultaneous multi-band fits for all KiDS and VIKING observations of our sample.

\section{Catalogue statistics}
\label{sec:statistics}

We begin the presentation of the contents of the \texttt{BDModels} catalogue with an overview of the numbers of galaxies classified in each model selection category, including outliers and skipped fits. For this, Section~\ref{sec:statsoverview} presents two large tables (Table~\ref{tab:results} for \texttt{v04} and~\ref{tab:resultsv05} for \texttt{v05}) detailing the evolution of the numbers of galaxies at each step in our pipeline, with the most important information condensed in Figure~\ref{fig:ncompstats}. The \texttt{v04} versions of this are heavily based on \citet{Casura2022}. In Sections~\ref{sec:outlierstats} and~\ref{sec:modelseldiffs} we then investigate the outlier statistics and model selection statistics in more detail, with a particular focus on comparing \texttt{v04} and \texttt{v05}. 

\subsection{Statistics overview}
\label{sec:statsoverview}

Table~\ref{tab:results} gives an overview of the fit and post-processing results for \texttt{v04}. Table~\ref{tab:resultsv05} presents the analogous table for \texttt{v05}. Starting with our full sample (13096 galaxies from the combination of our main and SAMI samples, see Section~\ref{sec:sampleselection}), we show how the number of galaxies evolves through all steps of the pipeline in both versions. The results are split per-band and per-model where necessary. At some steps, we also include percentages of galaxies lost or remaining (grey font). In short, for \texttt{v04} we lose nearly 20\,\% of our sample to masking and a further almost 10\,\% to the flagging of bad fits; where the former is a random subset while the latter preferentially affects certain types of galaxies (e.g. mergers and irregulars). These fractions are similar for \texttt{v05} in the core bands, but increase for the KiDS $u$ and the VIKING bands - especially the longer wavelength ones - due to missing data (VIKING only), additional masking, a higher fraction of PSF failures ($u$-band only) and a larger fraction of outliers mainly caused by the shallower data. The former three effects are explained in Section~\ref{sec:jointfitting}, while the latter one is investigated in more detail in Section~\ref{sec:outlierstats}. For \texttt{v05}, the numbers of fit failures also increases significantly in all bands (including the core bands), but this does not affect the final statistics since preferentially those fits failed that would not have been selected in the model selection anyway (Section~\ref{sec:modelseldiffs}). 
 
\afterpage{%
\clearpage%
\begin{landscape}
\begin{table*}
	\centering
	\caption{Fit results for \texttt{v04} of the \texttt{BDDecomp} DMU: numbers (black) and percentages (grey) of galaxies remaining or lost at each step in our pipeline, split per-band and per-model where necessary.$^1$ Individual steps are numbered through, see text for details.}
	\label{tab:results}
	\setlength{\tabcolsep}{4pt} 
	\begin{tabu}{l rrr c rrr c rrr c rrr} 
	    \hline
	    band
	    & \multicolumn{3}{c}{$g$}
	    && \multicolumn{3}{c}{$r$}
	    && \multicolumn{3}{c}{$i$}
	    && \multicolumn{3}{c}{joint $gri$}
	    \\
	    model (components)
	    & \multicolumn{1}{c}{1} & \multicolumn{1}{c}{1.5} & \multicolumn{1}{c}{2} 
	    && \multicolumn{1}{c}{1} & \multicolumn{1}{c}{1.5} & \multicolumn{1}{c}{2} 
	    && \multicolumn{1}{c}{1} & \multicolumn{1}{c}{1.5} & \multicolumn{1}{c}{2} 
	    && \multicolumn{1}{c}{1} & \multicolumn{1}{c}{1.5} & \multicolumn{1}{c}{2} 
	    \\
	    \hline
	    number of: 
	    \\
	    1) unique objects (galaxies)
	    & \multicolumn{15}{c}{13096}
	    \\
	    2) images (independent fits)
	    & \multicolumn{15}{c}{14966}
	    \\
	    3) images not masked
	    & \multicolumn{15}{c}{11989}
	    \\
	    \rowfont{\color{gray}}
	    lost due to masking (\%)
	    & \multicolumn{15}{c}{20}
	    \\
	    \\	    
	    4) successful PSFs
	    & \multicolumn{3}{c}{11838}
	    && \multicolumn{3}{c}{11872}
	    && \multicolumn{3}{c}{11946}
	    && \multicolumn{3}{c}{11683}
	    \\
	    \rowfont{\color{gray}}
	    lost due to PSF fails (\%)
	    & \multicolumn{3}{c}{1}
	    && \multicolumn{3}{c}{0.8}
	    && \multicolumn{3}{c}{0.3}
	    && \multicolumn{3}{c}{2}
	    \\
	    \\
		5) successful fits 
		& 11837 & 11837 & 11831 
		&& 11872 & 11870 & 11861 
		&& 11946 & 11943 & 11945
		&& 11682 & 11678 & 11665
		\\		
		\rowfont{\color{gray}}
		lost due to fit fails (\%)
		& <\,0.01 & <\,0.01 & 0.05 
		&& 0 & 0.01 & 0.07 
		&& 0 & 0.02 & <\,0.01
		&& <\,0.01 & 0.03 & 0.12
		\\		
		6) fits not flagged 
		& 10951 & 7122 & 8022 
		&& 11025 & 8164 & 8759 
		&& 11086 & 7620 & 7775
		&& 10680 & 6446 & 5870
		\\		
		\rowfont{\color{gray}} 
	    not flagged/successful (\%) 
		& 93 & 60 & 68 
		&& 93 & 69 & 74 
		&& 93 & 64 & 65 
		&& 91 & 55 & 50
		\\		
		7) selected fits 
		& 8294 & 740 & 1743 
		&& 7061 & 585 & 2935 
		&& 7411 & 662 & 2663 
		&& 7308 & 621 & 2009
		\\		
		\rowfont{\color{gray}} 
		selected/successful (\%)
		& 70 & 6 & 15 
		&& 59 & 5 & 25 
		&& 62 & 6 & 22 
		&& 63 & 5 & 17
		\\		
		\\
		total number (per band) of:\\
		8a) good\,|\,flagged\,|\,skipped fits
		& \multicolumn{3}{c}{10777\,|\,1061\,|\,3128} 
		&& \multicolumn{3}{c}{10581\,|\,1291\,|\,3094} 
		&& \multicolumn{3}{c}{10736\,|\,1210\,|\,3020} 
		&& \multicolumn{3}{c}{9938\,|\,1745\,|\,3283} 
		\\
		\rowfont{\color{gray}} 
		good\,|\,f.\,|\,s./all images (\%)
		& \multicolumn{3}{c}{72\,|\,7\,|\,21} 
		&& \multicolumn{3}{c}{71\,|\,9\,|\,21} 
		&& \multicolumn{3}{c}{72\,|\,8\,|\,20} 
		&& \multicolumn{3}{c}{66\,|\,12\,|\,22}
		\\
		\\
		8b) good\,|\,flagged\,|\,skipped gal.
		& \multicolumn{3}{c}{9722\,|\,935\,|\,2439}
		&& \multicolumn{3}{c}{9545\,|\,1145\,|\,2406} 
		&& \multicolumn{3}{c}{9687\,|\,1059\,|\,2350} 
		&& \multicolumn{3}{c}{8998\,|\,1559\,|\,2539} 
		\\
		\rowfont{\color{gray}} 
		good\,|\,f.\,|\,s./unique objects (\%) 
		& \multicolumn{3}{c}{74\,|\,7\,|\,19} 
		&& \multicolumn{3}{c}{73\,|\,9\,|\,18} 
		&& \multicolumn{3}{c}{74\,|\,8\,|\,18} 
		&& \multicolumn{3}{c}{69\,|\,12\,|\,19}
		\\
        \hline
        \multicolumn{16}{l}{$^1$Based on information given in the \texttt{*\_BDQUAL\_FLAG}, \texttt{*\_OUTLIER\_FLAG} and \texttt{*\_NCOMP} columns of the \texttt{BDModels} catalogue.}
        \\
	\end{tabu}
 \end{table*}
\end{landscape}
\clearpage
}

\afterpage{%
\clearpage%
\begin{landscape}
\begin{table*}
	\centering
	\caption{Fit results for \texttt{v05} of the \texttt{BDDecomp} DMU: numbers (black) and percentages (grey) of galaxies remaining or lost at each step in our pipeline, split per-band and per-model where necessary.$^1$ Individual steps are numbered through, see text for details.}
	\label{tab:resultsv05}
	\setlength{\tabcolsep}{4pt} 
	\begin{tabu}{l rrr c rrr c rrr c rrr} 
	    \hline
	    band
	    & \multicolumn{3}{c}{$u$}
	    && \multicolumn{3}{c}{$g$}
	    && \multicolumn{3}{c}{$r$}
	    && \multicolumn{3}{c}{$i$}
	    \\
	    model (components)
	    & \multicolumn{1}{c}{1} & \multicolumn{1}{c}{1.5} & \multicolumn{1}{c}{2} 
	    && \multicolumn{1}{c}{1} & \multicolumn{1}{c}{1.5} & \multicolumn{1}{c}{2} 
	    && \multicolumn{1}{c}{1} & \multicolumn{1}{c}{1.5} & \multicolumn{1}{c}{2} 
	    && \multicolumn{1}{c}{1} & \multicolumn{1}{c}{1.5} & \multicolumn{1}{c}{2} 
	    \\
	    \hline
	    number of: 
	    \\
	    1) unique objects (galaxies)
	    & \multicolumn{3}{c}{13096}
	    && \multicolumn{3}{c}{13096}
	    && \multicolumn{3}{c}{13096}
	    && \multicolumn{3}{c}{13096}
	    \\
	    2a) images (data matches)
	    & \multicolumn{3}{c}{14738}
	    && \multicolumn{3}{c}{14966}
	    && \multicolumn{3}{c}{14966}
	    && \multicolumn{3}{c}{14966}
	    \\
	    2b) objects with >\,0 images
	    & \multicolumn{3}{c}{13096}
	    && \multicolumn{3}{c}{13096}
	    && \multicolumn{3}{c}{13096}
	    && \multicolumn{3}{c}{13096}
	    \\
	    \rowfont{\color{gray}}
	    lost due to missing data (\%)
	    & \multicolumn{3}{c}{0}
	    && \multicolumn{3}{c}{0}
	    && \multicolumn{3}{c}{0}
	    && \multicolumn{3}{c}{0}
	    \\
	    \\
	    3) objects not masked
	    & \multicolumn{3}{c}{10540}
	    && \multicolumn{3}{c}{10768}
	    && \multicolumn{3}{c}{10768}
	    && \multicolumn{3}{c}{10768}
	    \\
	    \rowfont{\color{gray}}
	    lost due to masking (\%)
	    & \multicolumn{3}{c}{20}
	    && \multicolumn{3}{c}{18}
	    && \multicolumn{3}{c}{18}
	    && \multicolumn{3}{c}{18}
	    \\	    
	    4) successful PSFs
	    & \multicolumn{3}{c}{9141}
	    && \multicolumn{3}{c}{10768}
	    && \multicolumn{3}{c}{10752}
	    && \multicolumn{3}{c}{10731}
	    \\
	    \rowfont{\color{gray}}
	    lost due to PSF fails (\%)
	    & \multicolumn{3}{c}{11}
	    && \multicolumn{3}{c}{0}
	    && \multicolumn{3}{c}{0.1}
	    && \multicolumn{3}{c}{0.3}
	    \\
	    \\
		5) successful fits 
		& 9087 & 8902 & 8504  
		&& 10715 & 10640 & 9805 
		&& 10706 & 10609 & 9845
		&& 10674 & 10566 & 9875
		\\		
		\rowfont{\color{gray}}
		lost due to fit fails (\%)
		& 0.4 & 2 & 5
		&& 0.4 & 1 & 7
		&& 0.4 & 1 & 7
		&& 0.4 & 1 & 7
		\\		
		6) fits not flagged 
		& 6165 & 2408 & 880 
		&& 9762 & 6096 & 5913 
		&& 9803 & 7079 & 7000
		&& 9568 & 6527 & 4865
		\\		
		\rowfont{\color{gray}} 
	    not flagged/successful (\%) 
		& 68 & 27 & 10 
		&& 91 & 57 & 60 
		&& 92 & 67 & 71 
		&& 90 & 62 & 49
		\\		
		7) selected fits 
		& 5888 & 254 & 57 
		&& 7213 & 607 & 1691 
		&& 5522 & 682 & 3201 
		&& 7207 & 537 & 1686
		\\		
		\rowfont{\color{gray}} 
		selected/successful (\%)
		& 65 & 3 & 0.7 
		&& 67 & 6 & 17 
		&& 52 & 6 & 33 
		&& 68 & 5 & 17
		\\		
		\\
		total number (per band) of:\\
		8) good\,|\,flagged\,|\,skipped fits
		& \multicolumn{3}{c}{6199\,|\,2936\,|\,3961} 
		&& \multicolumn{3}{c}{9511\,|\,1249\,|\,2336} 
		&& \multicolumn{3}{c}{9405\,|\,1345\,|\,2346} 
		&& \multicolumn{3}{c}{9430\,|\,1298\,|\,2368} 
		\\
		\rowfont{\color{gray}} 
		good\,|\,f.\,|\,s./all objects (\%)
		& \multicolumn{3}{c}{47\,|\,22\,|\,30} 
		&& \multicolumn{3}{c}{73\,|\,10\,|\,18} 
		&& \multicolumn{3}{c}{72\,|\,10\,|\,18} 
		&& \multicolumn{3}{c}{72\,|\,10\,|\,18}
		\\
        \hline
	\end{tabu}
 \end{table*}

\begin{table*}\ContinuedFloat
	\centering
	\setlength{\tabcolsep}{4pt} 
	\begin{tabu}{l rrr c rrr c rrr c rrr} 
	    \hline
	    band
	    & \multicolumn{3}{c}{$Z$}
	    && \multicolumn{3}{c}{$Y$}
	    && \multicolumn{3}{c}{$J$}
	    && \multicolumn{3}{c}{$H$}
	    \\
	    model (components)
	    & \multicolumn{1}{c}{1} & \multicolumn{1}{c}{1.5} & \multicolumn{1}{c}{2} 
	    && \multicolumn{1}{c}{1} & \multicolumn{1}{c}{1.5} & \multicolumn{1}{c}{2} 
	    && \multicolumn{1}{c}{1} & \multicolumn{1}{c}{1.5} & \multicolumn{1}{c}{2} 
	    && \multicolumn{1}{c}{1} & \multicolumn{1}{c}{1.5} & \multicolumn{1}{c}{2} 
	    \\
	    \hline
	    number of: 
	    \\
	    1) unique objects (galaxies)
	    & \multicolumn{3}{c}{13096}
	    && \multicolumn{3}{c}{13096}
	    && \multicolumn{3}{c}{13096}
	    && \multicolumn{3}{c}{13096}
	    \\
	    2a) images (data matches)
	    & \multicolumn{3}{c}{46225}
	    && \multicolumn{3}{c}{47678}
	    && \multicolumn{3}{c}{95396}
	    && \multicolumn{3}{c}{45918}
	    \\
	    2b) objects with >\,0 images
	    & \multicolumn{3}{c}{12437}
	    && \multicolumn{3}{c}{12437}
	    && \multicolumn{3}{c}{12436}
	    && \multicolumn{3}{c}{12429}
	    \\
	    \rowfont{\color{gray}}
	    lost due to missing data (\%)
	    & \multicolumn{3}{c}{5}
	    && \multicolumn{3}{c}{5}
	    && \multicolumn{3}{c}{5}
	    && \multicolumn{3}{c}{5}
	    \\
	    \\
	    3) objects not masked
	    & \multicolumn{3}{c}{9989}
	    && \multicolumn{3}{c}{9969}
	    && \multicolumn{3}{c}{9961}
	    && \multicolumn{3}{c}{9945}
	    \\
	    \rowfont{\color{gray}}
	    lost due to masking (\%)
	    & \multicolumn{3}{c}{19}
	    && \multicolumn{3}{c}{19}
	    && \multicolumn{3}{c}{19}
	    && \multicolumn{3}{c}{19}
	    \\	    
	    4) successful PSFs
	    & \multicolumn{3}{c}{9942}
	    && \multicolumn{3}{c}{9894}
	    && \multicolumn{3}{c}{9898}
	    && \multicolumn{3}{c}{9825}
	    \\
	    \rowfont{\color{gray}}
	    lost due to PSF fails (\%)
	    & \multicolumn{3}{c}{0.4}
	    && \multicolumn{3}{c}{0.6}
	    && \multicolumn{3}{c}{0.5}
	    && \multicolumn{3}{c}{0.9}
	    \\
	    \\
		5) successful fits 
		& 9899 & 9825 & 8943 
		&& 9861 & 9721 & 9148 
		&& 9848 & 9768 & 8929
		&& 9687 & 9534 & 8770
		\\		
		\rowfont{\color{gray}}
		lost due to fit fails (\%)
		& 0.3 & 0.9 & 8 
		&& 0.3 & 1 & 6
		&& 0.4 & 1 & 7
		&& 1 & 2 & 8
		\\		
		6) fits not flagged 
		& 8796 & 5305 & 3868 
		&& 8261 & 4414 & 2538 
		&& 8120 & 5081 & 2938
		&& 7368 & 4529 & 2299
		\\		
		\rowfont{\color{gray}} 
	    not flagged/successful (\%) 
		& 89 & 54 & 43 
		&& 84 & 45 & 28 
		&& 82 & 52 & 33 
		&& 76 & 48 & 26
		\\		
		7) selected fits 
		& 6363 & 949 & 1395 
		&& 6247 & 610 & 1239 
		&& 6336 & 475 & 1257 
		&& 5597 & 712 & 1024
		\\		
		\rowfont{\color{gray}} 
		selected/successful (\%)
		& 64 & 10 & 16 
		&& 63 & 6 & 14 
		&& 64 & 5 & 14 
		&& 58 & 7 & 12
		\\		
		\\
		total number (per band) of:\\
		8) good\,|\,flagged\,|\,skipped fits
		& \multicolumn{3}{c}{8707\,|\,1221\,|\,3168} 
		&& \multicolumn{3}{c}{8096\,|\,1786\,|\,3214} 
		&& \multicolumn{3}{c}{8068\,|\,1820\,|\,3208} 
		&& \multicolumn{3}{c}{7333\,|\,2405\,|\,3358} 
		\\
		\rowfont{\color{gray}} 
		good\,|\,f.\,|\,s./all objects (\%)
		& \multicolumn{3}{c}{66\,|\,9\,|\,24} 
		&& \multicolumn{3}{c}{62\,|\,14\,|\,25} 
		&& \multicolumn{3}{c}{62\,|\,14\,|\,24} 
		&& \multicolumn{3}{c}{56\,|\,18\,|\,26}
		\\
        \hline
	\end{tabu}
 \end{table*}
 
 \begin{table*}\ContinuedFloat
	\centering
	\setlength{\tabcolsep}{4pt} 
	\begin{tabu}{l rrr c rrr c rrr} 
	    \hline
	    band
	    & \multicolumn{3}{c}{$K_s$}
	    && \multicolumn{3}{c}{$gri$}
	    && \multicolumn{3}{c}{$ugriZYJHK_s$}
	    \\
	    model (components)
	    & \multicolumn{1}{c}{1} & \multicolumn{1}{c}{1.5} & \multicolumn{1}{c}{2} 
	    && \multicolumn{1}{c}{1} & \multicolumn{1}{c}{1.5} & \multicolumn{1}{c}{2} 
	    && \multicolumn{1}{c}{1} & \multicolumn{1}{c}{1.5} & \multicolumn{1}{c}{2} 
	    \\
	    \hline
	    number of: 
	    \\
	    1) unique objects (galaxies)
	    & \multicolumn{3}{c}{13096}
	    && \multicolumn{3}{c}{13096}
	    && \multicolumn{3}{c}{13096}
	    \\
	    2a) images (data matches)
	    & \multicolumn{3}{c}{45993}
	    && \multicolumn{3}{c}{14966\,$\times$\,3}
	    && \multicolumn{3}{c}{340846}
	    \\
	    2b) objects with >\,0 images
	    & \multicolumn{3}{c}{12438}
	    && \multicolumn{3}{c}{13096}
	    && \multicolumn{3}{c}{12426}
	    \\
	    \rowfont{\color{gray}}
	    lost due to missing data (\%)
	    & \multicolumn{3}{c}{5}
	    && \multicolumn{3}{c}{0}
	    && \multicolumn{3}{c}{5}
	    \\
	    \\
	    3) objects not masked
	    & \multicolumn{3}{c}{9955}
	    && \multicolumn{3}{c}{10768}
	    && \multicolumn{3}{c}{9738}
	    \\
	    \rowfont{\color{gray}}
	    lost due to masking (\%)
	    & \multicolumn{3}{c}{19}
	    && \multicolumn{3}{c}{18}
	    && \multicolumn{3}{c}{21}
	    \\
	    4) successful PSFs
	    & \multicolumn{3}{c}{9800}
	    && \multicolumn{3}{c}{10715}
	    && \multicolumn{3}{c}{8082}
	    \\
	    \rowfont{\color{gray}}
	    lost due to PSF fails (\%)
	    & \multicolumn{3}{c}{1}
	    && \multicolumn{3}{c}{0.4}
	    && \multicolumn{3}{c}{13}
	    \\
	    \\
		5) successful fits 
		& 9660 & 9516 & 8694 
		&& 10597 & 10315 & 8635 
		&& 7964 & 7149 & 4457
		\\		
		\rowfont{\color{gray}}
		lost due to fit fails (\%)
		& 1 & 2 & 8
		&& 0.9 & 3 & 16
		&& 0.9 & 7 & 28
		\\		
		6) fits not flagged 
		& 6687 & 4136 & 1910 
		&& 9371 & 5378 & 3464 
		&& 4312 & 1443 & 357
		\\		
		\rowfont{\color{gray}} 
	    not flagged/successful (\%) 
		& 69 & 43 & 22 
		&& 88 & 52 & 40 
		&& 54 & 20 & 8 
		\\		
		7) selected fits 
		& 5366 & 343 & 929 
		&& 7202 & 604 & 1097 
		&& 4051 & 37 & 141 
		\\		
		\rowfont{\color{gray}} 
		selected/successful (\%)
		& 56 & 4 & 11
		&& 68 & 6 & 13 
		&& 51 & 0.5 & 3 
		\\		
		\\
		total number (per band) of:\\
		8) good\,|\,flagged\,|\,skipped fits
		& \multicolumn{3}{c}{6638\,|\,3093\,|\,3365} 
		&& \multicolumn{3}{c}{8903\,|\,1788\,|\,2405} 
		&& \multicolumn{3}{c}{4229\,|\,3912\,|\,4955} 
		\\
		\rowfont{\color{gray}} 
		good\,|\,f.\,|\,s./all objects (\%)
		& \multicolumn{3}{c}{51\,|\,24\,|\,26} 
		&& \multicolumn{3}{c}{68\,|\,14\,|\,18} 
		&& \multicolumn{3}{c}{32\,|\,30\,|\,38} 
		\\
        \hline
        \multicolumn{12}{l}{$^1$Based on the \texttt{*\_N\_MATCH}, \texttt{*\_BDQUAL\_FLAG}, \texttt{*\_OUTLIER\_FLAG} and \texttt{*\_NCOMP} columns.}
        \\
	\end{tabu}
 \end{table*}
 
\end{landscape}
\clearpage
}

Note that for both versions, we used stacked $gri$ images for segmentation and masking (plus additional masking in the non-core bands), but then treated the galaxies independently in all bands except for the model selection, where we performed both a per-band and a joint version. Therefore, the column ``joint $gri$" in Table~\ref{tab:results} always gives the number of galaxies that were ``good" in all three bands (hence why numbers are generally lower), except for the model selection, where it shows the results of the joint model selection (cf. Section~\ref{sec:postprocessing}). The same is true for the corresponding ``$gri$" and ``$ugriZYJHK_s$" columns in Table~\ref{tab:resultsv05}, except that for step 2a) we now - somewhat arbitrarily - give the sum of all data matches in all respective bands since there is no one-to-one correspondence between data matches in the KiDS and VIKING bands, so it is impossible to define the number of joint data matches in all bands. For the remaining rows, however, we then list the numbers of objects that were ``good" in all bands again, resulting in very low numbers for the 9-band joint selection. Also note that for \texttt{v05}, all numbers except those in step 2a) refer to the numbers of galaxies (unique objects) since all images (data matches) of a galaxy were fitted jointly (per band). For \texttt{v04}, individual images were fitted separately, so the numbers in steps 2) to 8a) refer to individual data matches, not unique objects. 

Explanations of each step as numbered through in Tables~\ref{tab:results} and~\ref{tab:resultsv05}:
\begin{enumerate}
\item The full sample results from the combination of our main and SAMI samples (Section~\ref{sec:sampleselection}). 
\item For KiDS bands, some galaxies have been imaged more than once due to overlap regions between the tiles. For VIKING bands, most galaxies have been covered by more than one exposure since we work at the individual chip level. For \texttt{v04}, we treat these duplicate observations of the same physical object independently throughout our pipeline, so numbers in all subsequent rows of Table~\ref{tab:results} refer to individual images, not unique objects. For \texttt{v05}, all images of the same object in the same band are fitted jointly, so all numbers in Table~\ref{tab:resultsv05} refer to unique objects. For the VIKING bands, about 5\,\% of objects are not covered by any chip due to small gaps in the data coverage (Section~\ref{sec:jointfitting}). 
\item For the core bands, we use the associated KiDS masks, combining the three bands. \emph{Images} for which the central galaxy pixel is masked ($\sim$\,20\,\%) are skipped during the fitting (Section~\ref{sec:preparatorysteps}). This results in $\sim$\,18\,\% of \emph{unique objects} being skipped in the core bands. Since we use the $gri$ segmentation maps for all other bands, too, these objects are automatically skipped in all bands. 1-2\,\% of objects are additionally skipped in the KiDS $u$ and VIKING bands due to the masking in the respective bands (Section~\ref{sec:jointfitting}).
\item For each image in each band, a PSF is then estimated by fitting nearby stars. If the PSF estimation fails, the image is skipped during the fit (Section~\ref{sec:preparatorysteps}). Note that technically, we estimate PSFs also for galaxies that are masked in step 3, but we do not list those here. For \texttt{v05}, an object is only skipped if all of its data matches are either masked or have a failed PSF. We count those for which all matches are masked in step 3 (irrespective of whether they also have failed PSFs or not); and those for which all matches failed the PSF as well as those for for which there was a mixture between matches that were masked (but have a PSF) and failed the PSF (but were not masked) here in step 4. 
\item For each non-masked image with a successful PSF estimate, we attempt 3 fits: a single S\'ersic (1), a pointsource + exponential (1.5) and a S\'ersic + exponential (2). In \texttt{v05}, we attempt a joint fit of all images for each object that has at least one non-masked image with a successful PSF estimate. Sometimes, the fit attempts fail with an error (Section~\ref{sec:galaxyfitting}).
\item Each fit (for each model independently) is passed through our outlier flagging process, identifying bad fits (Section~\ref{sec:postprocessing}). We further assess the differences between \texttt{v04} and \texttt{v05} in Section~\ref{sec:outlierstats}. 
\item Of the non-flagged (i.e. good) fits, we then select the most appropriate one during model selection (Section~\ref{sec:postprocessing}). \texttt{v04} and \texttt{v05} model selection results are compared in detail in Section~\ref{sec:modelseldiffs}. 
\item Summing up the selected fits for each model (step 7) gives the total number of good fits. The difference between the good and successful fits (step 5) stems from the outlier flagging. Skipped fits are due to missing data, masking, PSF or fit fails (steps 2b, 3, 4, 5). For \texttt{v04}, the sum of good, flagged and skipped fits in step 8a gives the total number of independent fits (step 2). Removing duplicate fits for the same physical objects gives the number of good, flagged and skipped galaxies in step 8b, which sum to the number of unique objects (step 1). For this, we always use the best available result for each galaxy, i.e. it is counted as ``good" if at least one of the multiple fits was ``good". For \texttt{v05}, there is only one fit per galaxy, so the good, flagged and skipped fits from step 8 sum directly to the number of unique objects (step 1).
\end{enumerate}

\begin{figure}
\centering
	\includegraphics[width=0.65\textwidth]{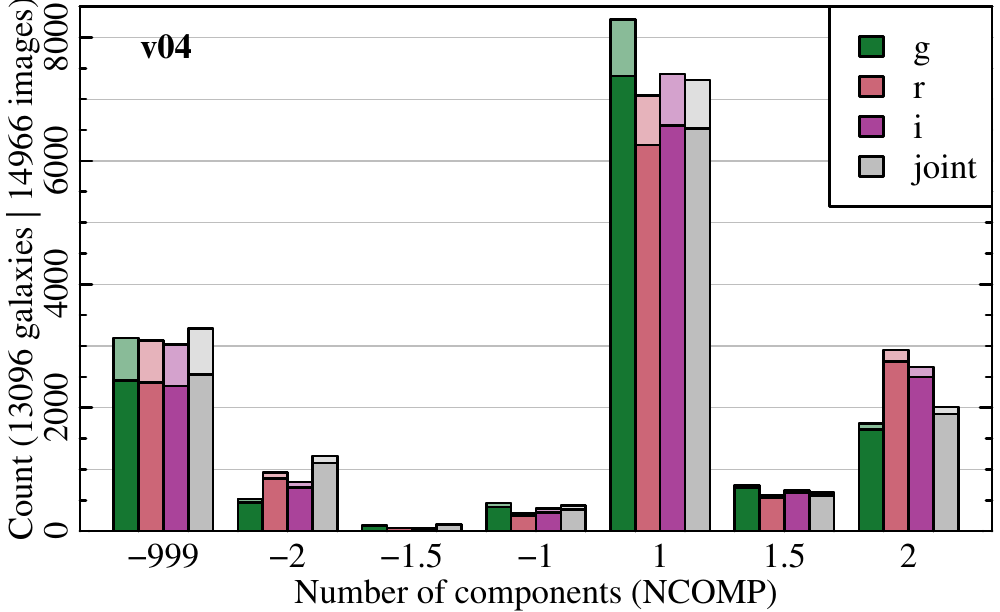}
	
	\bigskip
	\includegraphics[width=\textwidth]{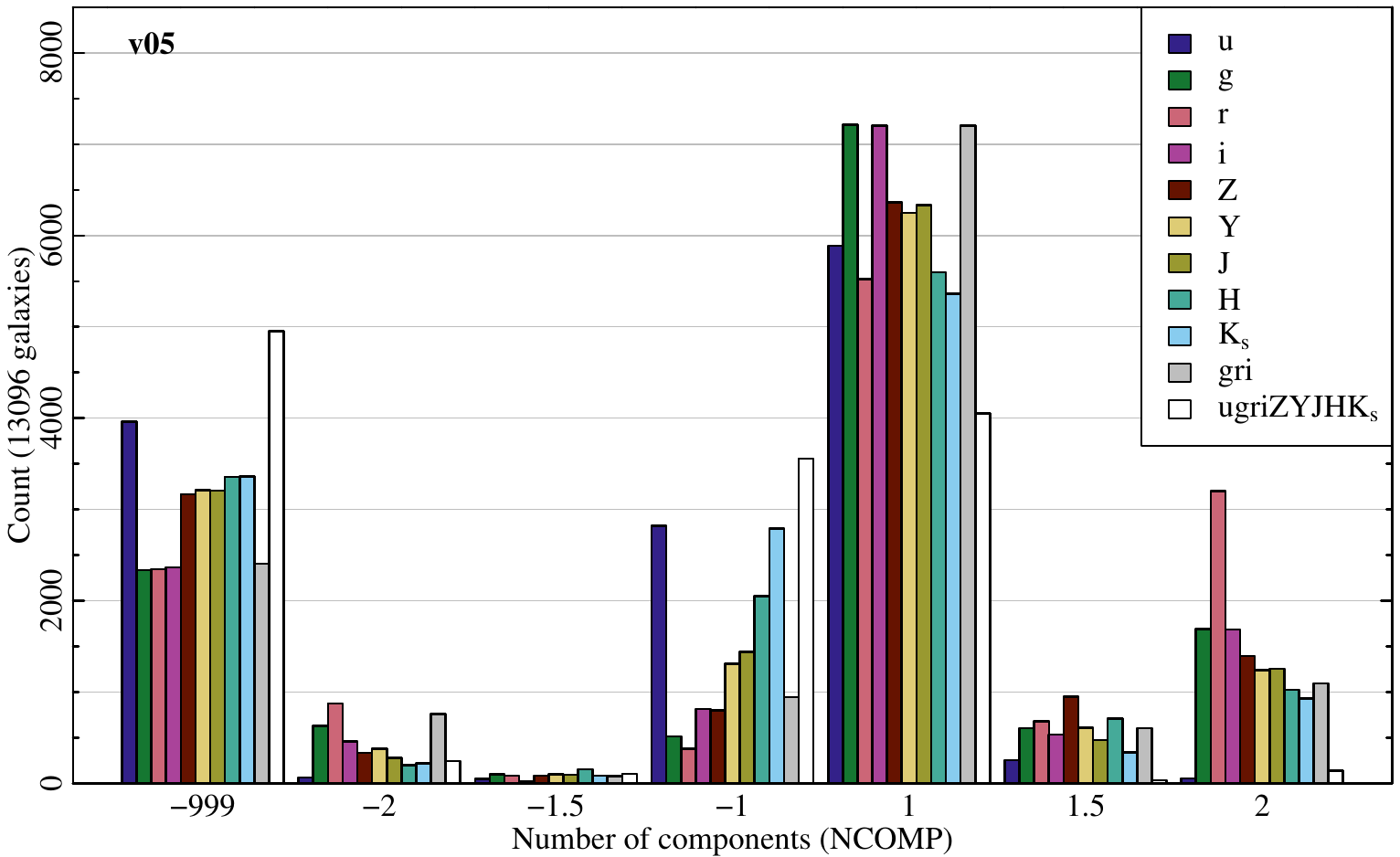}
    \caption{The number of components assigned in our model selection procedure for individual bands and the joint analysis for \texttt{v04} (\textbf{top}) and \texttt{v05} (\textbf{bottom}) of the pipeline. 1, 1.5 and 2 mean single S\'ersic, point source bulge + exponential disk and S\'ersic bulge + exponential disk models respectively. Negative values indicate that the chosen (best) fit was flagged as unreliable (mostly irregular or partly masked galaxies). -999 is assigned to skipped fits, either because the galaxy centre is masked (most cases), the galaxy falls into a gap in the data (VIKING bands only) or because the PSF estimation failed. For \texttt{v04}, the lighter (higher) bars show the number of images, whereas saturated bars indicate the number of unique objects in both figures; see text and Tables~\ref{tab:results} and~\ref{tab:resultsv05} for details. Horizontal grey lines in the background are placed at equal intervals in both figures to ease the direct comparison.}		
    \label{fig:ncompstats}
\end{figure}

Figure~\ref{fig:ncompstats} visualises the most important information given in Tables~\ref{tab:results} and~\ref{tab:resultsv05}, namely the final number of objects classified in each category. For \texttt{v04} (top panel), lighter bars in the background refer to individual fits (total 14966) with the number of unique galaxies (total 13096) overplotted. When several fits to the same galaxy were classified in different categories, we allocate it to the highest of those,\footnote{This means that a galaxy is classified as ``outlier" if \emph{all} fits to it are outliers and it is ``skipped" only if \emph{all} fits are skipped. Galaxies with good fits are allocated to the most complex model of the available fits (assuming that one of the images was deeper and allowed to constrain more components than the other(s)), while within the outlier categories we allocate it to the simplest model. Note that in Table~\ref{tab:results} we only show the total number of flagged fits and do not split them into the different outlier categories.} which is consistent with Table~\ref{tab:results}. For \texttt{v05} (bottom panel), we only show the number of unique objects, since all images were fitted jointly.

\texttt{NCOMP}\,=\,$-999$ means the object was skipped (not fitted) because it is masked or the PSF estimation failed (usually because of large masked areas in the immediate vicinity of the object). For VIKING bands, objects can also be skipped due to small gaps in the sky coverage. \texttt{NCOMP}\,=\,1, 1.5 or 2 indicates that this is a good fit classified as single, 1.5- or double component fit. \texttt{NCOMP}\,=\,$-1$, $-1.5$ or $-2$ indicates that this is a bad fit (outlier) which would have been classified as single, 1.5- or double component fit if it were not an outlier (most often these are mergers/irregular galaxies for which our models are not appropriate; or galaxies that are partly masked). We keep these three classes separate since automated outlier identification can never be perfect; and what should be considered a bad fit will depend on the use case. The flagging of fits is hence only intended as a guide and all available information in the catalogue is retained for all fitted objects. We analyse Figure~\ref{fig:ncompstats} in more detail in Sections~\ref{sec:outlierstats} and~\ref{sec:modelseldiffs}, where we compare the outlier flagging and model selection statistics for \texttt{v04} and \texttt{v05} of the \texttt{BDDecomp} DMU. 

Examples of fits classified in each category can be found throughout Chapter~\ref{chap:pipeline}, see e.g. Figure~\ref{fig:examplefit} for a double component fit, Figure~\ref{fig:examplefit1.5} for a 1.5-component object, Figure~\ref{fig:examplefitnormvst4} for a single S\'ersic galaxy and Figure~\ref{fig:examplefit-2} for an outlier.

\subsection{Outlier statistics}
\label{sec:outlierstats}

In this section, we analyse the numbers of outliers in Figure~\ref{fig:ncompstats} and the corresponding Tables~\ref{tab:results} and~\ref{tab:resultsv05} in more detail. Since the outlier flagging was calibrated on \texttt{v04} (cf. Section~\ref{sec:manualcalibrationchanges}), we put particular emphasis on comparing the \texttt{v05} results to those of \texttt{v04} for the core bands ($g$, $r$, $i$) and compare all other bands against those three within \texttt{v05}. The dif\-fe\-ren\-ces in the numbers of skipped fits (\texttt{NCOMP}\,=\,$-999$) are explained in Section~\ref{sec:jointfitting}. In short, the $g, r$ and $i$ bands show the same number of skipped fits in \texttt{v04} and \texttt{v05} since that is mainly due to masking, which did not change. The $u$-band has a higher number of skipped fits due to its smaller footprint, shallower data and the additional masking. VIKING bands have higher numbers of skipped fits mainly due to small gaps in the data coverage. All objects that are skipped in at least one of the bands are counted as skipped for the joint categories, which is the reason for the higher number of skipped objects for the 9-band joint analysis. The differences in the model selection are investigated in Section~\ref{sec:modelseldiffs}. In this section, we focus on differences in the outlier categories, i.e. \texttt{NCOMP}\,=\,$-1$, $-1.5$ and $-2$. 

\begin{table}[t!]
\caption{The percentage of fits flagged as outlier according to each criterion (see text for details) in all bands in \texttt{v04} and \texttt{v05} of the \texttt{BDDecomp} DMU. All values are in per cent of the total number of non-skipped fits, except for the ``extreme B/T ratio" criterion, which is in per cent of the total number of non-skipped 1.5- and double component fits (since it does not apply to single S\'ersic fits). Grey font indicates cautionary flags that are not considered during the final outlier flagging. Bold font in the \texttt{v05} table highlights major changes in the fractions relative to \texttt{v04} (for $g$, $r$ and $i$ bands) and relative to the (\texttt{v05}) $r$-band for all other bands.}
\label{tab:outlierstats}

\begin{subtable}{1\textwidth}
\centering
\caption{Outlier statistics in \texttt{v04}.}
\label{tab:v04outlierstats}
	\begin{tabu}{lrrr} 
		\hline
		flag & $g$ & $r$ & $i$ \\
		\hline
		irregular segment & 5.48 & 5.37 & 5.27\\ 
		extreme B/T ratio & 0.00 & 0.05 & 0.04\\ 
		numerical problems & 0.08 & 0.20 & 0.17\\ 
		rel. param hit limit & 3.67 & 5.82 & 4.94\\ 
		\rowfont{\color{gray}}
		any param hit limit & 4.60 & 6.46 & 8.09\\ 
		\rowfont{\color{gray}}
		small or large error & 2.14 & 2.06 & 4.13\\ 
		poor $\chi^2$ statistics & 0.04 & 0.14 & 0.05\\ 
		position offset >\,2\arcsec\ & 0.42 & 0.35 & 0.39\\ 
		\rowfont{\color{gray}} 
		position offset >\,1\arcsec\ & 1.50 & 1.29 & 1.41\\ 
		flux in seg <\,20\,\% & 1.42 & 1.42 & 1.29\\ 
		\rowfont{\color{gray}}
		flux in seg <\,50\,\% & 8.04 & 9.32 & 9.33\\ 
        \hline
        \textbf{total outliers} & \textbf{8.95} & \textbf{10.87} & \textbf{10.13}\\ 
        \hline
	\end{tabu}
\end{subtable}

\bigskip
\begin{subtable}{1\textwidth}
\centering
\caption{Outlier statistics in \texttt{v05}.}
\label{tab:v05outlierstats}
	\begin{tabu}{lrrrrrrrrr} 
		\hline
		flag & $u$ & $g$ & $r$ & $i$ & $Z$ & $Y$ & $J$ & $H$ & $K_s$ \\
		\hline
		irregular segment & 7.21 & \textbf{6.85} & \textbf{6.83} & \textbf{6.68} & 7.06 & 6.81 & 7.88 & 7.65 & 7.71\\ 
		extreme B/T ratio & 0.36 & 0.07 & 0.04 & 0.10 & 0.13 & 0.18 & 0.19 & 0.50 & 0.98\\ 
		numerical problems & 1.38 & 0.31 & 0.62 & 0.34 & 0.36 & 0.83 & 0.64 & 1.56 & 2.90\\ 
		rel. param hit limit & \textbf{25.40} & \textbf{4.91} & 5.60 & \textbf{5.67} & 5.25 & \textbf{11.18} & \textbf{10.99} & \textbf{17.11} & \textbf{23.60}\\ 
		\rowfont{\color{gray}}
		any param hit limit & 28.66 & 5.61 & 6.23 & 6.71 & 6.42 & 13.47 & 13.01 & 20.48 & 28.78\\ 
		\rowfont{\color{gray}}
		small or large error & 28.86 & 5.73 & 6.26 & 6.73 & 6.51 & 13.37 & 12.91 & 20.45 & 28.94\\ 
		poor $\chi^2$ statistics & 0.10 & 0.09 & 0.17 & 0.11 & 0.08 & 0.13 & 0.10 & 0.15 & 0.15\\ 
		position offset >\,2\arcsec\ & \textbf{1.02} & 0.39 & 0.32 & 0.34 & 0.48 & 0.54 & 0.59 & \textbf{1.12} & \textbf{1.93}\\ 
		\rowfont{\color{gray}} 
		position offset >\,1\arcsec\ & 3.65 & 1.49 & 1.25 & 1.40 & 1.84 & 2.08 & 2.17 & 3.20 & 4.59\\ 
		flux in seg <\,20\,\% & 1.54 & 1.09 & 1.27 & 1.14 & 0.86 & 0.84 & 1.13 & 1.51 & 1.71\\ 
		\rowfont{\color{gray}}
		flux in seg <\,50\,\% & 6.23 & 5.25 & 6.45 & 5.77 & 3.88 & 3.59 & 4.88 & 5.51 & 6.39\\ 
        \hline
        \textbf{total outliers} & \textbf{32.14} & \textbf{11.61} & \textbf{12.50} & \textbf{12.10} & \textbf{12.30} & \textbf{18.06} & \textbf{18.41} & \textbf{24.70} & \textbf{31.79}\\ 
        \hline
	\end{tabu}
\end{subtable}

\end{table}

The core bands in general have similar numbers of flagged fits in \texttt{v05} with respect to \texttt{v04}. The VIKING $Z$-band is also comparable to the core bands. For the longer wavelength bands and the KiDS $u$-band, the number of outliers steadily increases. To investigate the origin of this further, Table~\ref{tab:outlierstats} gives the percentages of fits flagged according to each criterion for all bands in \texttt{v04} and \texttt{v05}. To decouple the analysis from differences in the model selection, we show the total number of outliers, i.e. categories \texttt{NCOMP}\,=\,$-2$, $-1.5$ and $-1$ combined. A detailed description of all criteria is given in Section~\ref{sec:postprocessing}, where the corresponding $r$-band \texttt{v04} percentages are also indicated. All values are given in percent of the total number of non-skipped fits; counting only those in the respective model selection category.\footnote{E.g. the 7.21\,\% of $u$-band fits with an irregular segment are composed of the fraction of objects that had this flag raised for their single S\'ersic fit \emph{and} were classified as single S\'ersic fits or single S\'ersic outlier (absolute value of \texttt{NCOMP} equal to one), plus the corresponding fractions of 1.5- and double component fits. For fractions of flagged fits for each model independently, see step 6 of Tables~\ref{tab:results} and~\ref{tab:resultsv05}. Note also that the total number of flagged fits in step 8 of those tables is given as a fraction of the total number of objects, whereas Table~\ref{tab:outlierstats} gives all values as percentages of non-skipped objects.}
Note that the total number of outliers is smaller than the sum of the individual flag values since bad fits frequently fall into multiple outlier categories. Grey font indicates cautionary flags that are not taken as criteria for flagging outliers. We only list these for completeness and do not comment on them further. In Table~\ref{tab:v05outlierstats}, bold font highlights values that show large differences with respect to the \texttt{v04} percentages for the $g$, $r$ and $i$ bands; and with respect to the \texttt{v05} $r$-band for all other bands. 

The fractions of outliers in the $g$, $r$ and $i$ bands in \texttt{v05} are approximately 12\,\%, 13\,\% and 12\,\% respectively. These are 2\,\% to 3\,\% more outliers than in \texttt{v04}. The main driver of this is the ``very irregular fitting segment (irregular segment)" criterion (+1.5\,\%). The $g$ and $i$ bands also seem to have a higher number of (reliable) parameters hitting their fit limits (+1\,\%), which is not observed in the $r$-band. In addition, there are increases in the numerical problems and to a lesser extent in the extreme B/T ratio objects, but the overall numbers of fits classified as outliers according to these criteria are so low that they do not affect the final outlier fraction much. 

There are no differences in the data used between \texttt{v04} and \texttt{v05} and the numbers of skipped fits are nearly identical for the core bands. The only possible reasons for the increased fraction of outliers are differences in the processing, namely the increased segment sizes and the joint fitting of all images (Section~\ref{sec:pipelineupdates}). Considering only galaxies with a single data match (in the core bands) does not change any flag values significantly. This suggests that the joint fitting is not the cause of the increased outlier fractions and instead it is the slightly different (larger) segments. Since the segments used are the same in all bands, the fraction of fits flagged according to the ``irregular segment" criterion are very similar across all bands (with slight differences between bands arising from the definition of this criterion, which also depends on the fit itself, cf. Section~\ref{sec:postprocessing}). 

\begin{figure}
\begin{center}
\includegraphics[width=0.8\textwidth]{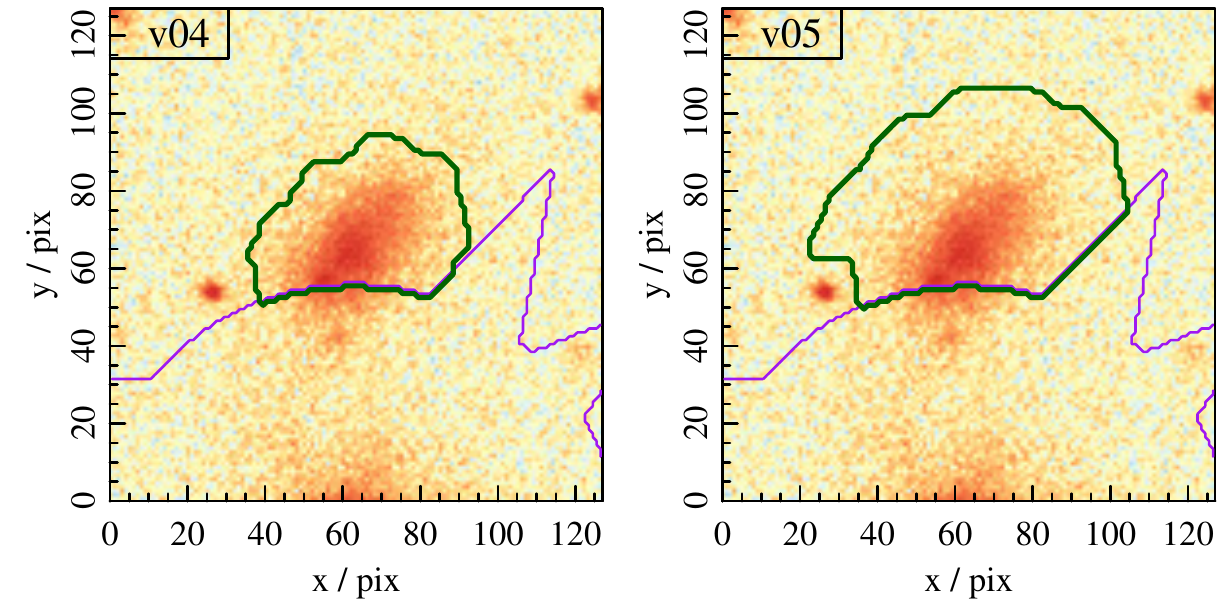}
\caption{The $r$-band segmentation map (dark green contour) for galaxy 14912, which was flagged as having an irregular fitting segment in \texttt{v05} (\textbf{right}), but not in \texttt{v04} (\textbf{left}). Purple contours indicate masked regions. }
\label{fig:examplesegflag}
\end{center}
\end{figure}

Visual inspection of a number of fits, which were flagged for having an irregular segment in \texttt{v05} but not in \texttt{v04}, showed that they were all borderline cases, mostly with large fractions of the segment cut off by a masked region or neighbouring object. An example is shown in Figure~\ref{fig:examplesegflag} for the single S\'ersic object 14912: both segments (dark green contours) are cut off by the bright star mask (purple contour). The segment for \texttt{v05} (right panel) is not ``worse" than that for \texttt{v04} (left panel), suggesting that the ``irregular segment" flag could benefit from a re-calibration in \texttt{v05}. However, since it only affects a relatively small number of objects, all of which are borderline cases, we decided against a re-calibration at this stage. Since we provide all modelling results and flag values for all objects in the catalogue, this can still be revised in the future.

\begin{figure}[t!]
\begin{center}
\includegraphics[width=0.8\textwidth]{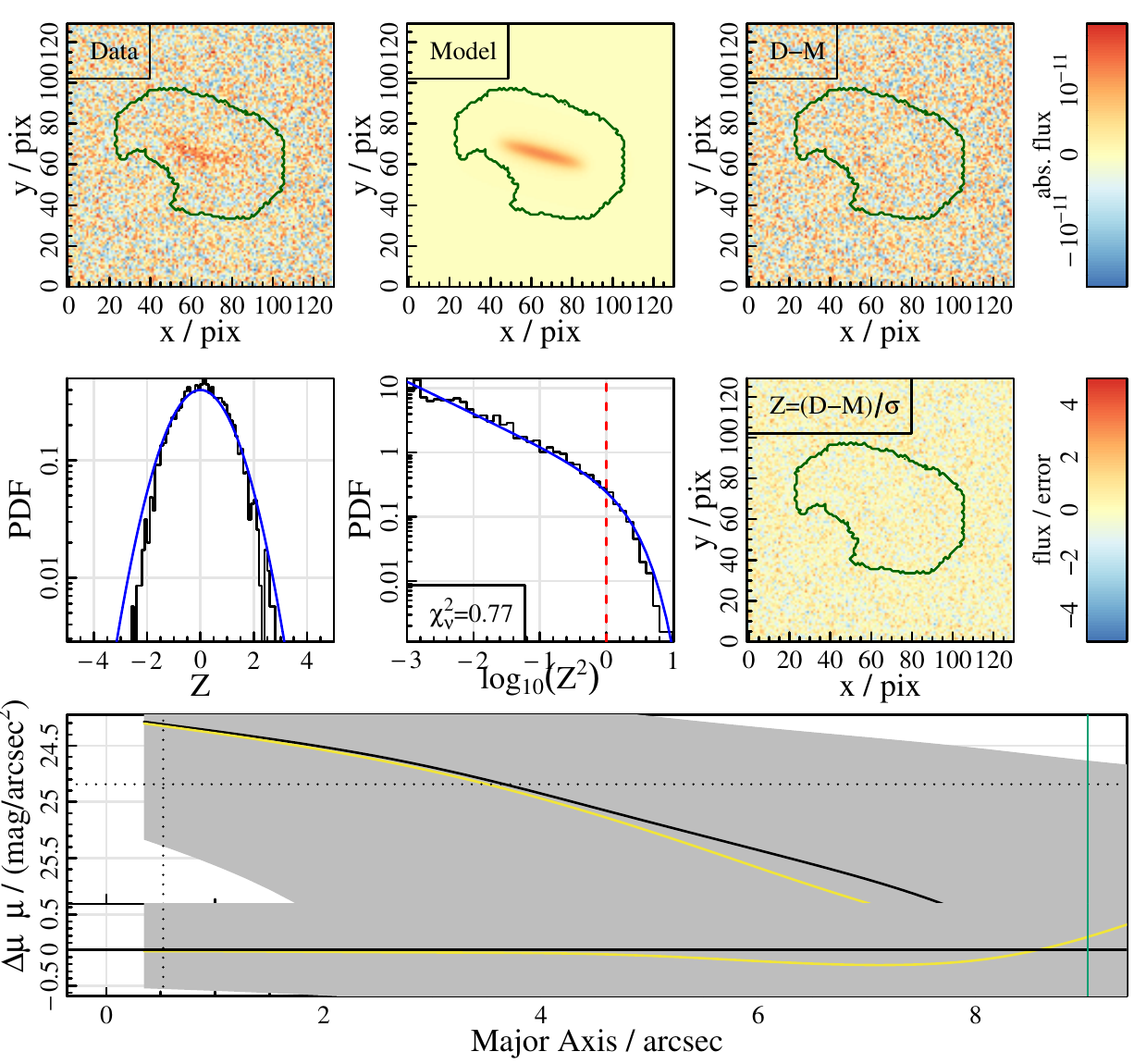}
\caption{The $u$-band single S\'ersic fit to galaxy 16537, which is classified as an outlier since it hit its fitting limits in several parameters. Panels in the top two rows are the same as those in Figure~\ref{fig:examplefit}, while the bottom row shows the one-dimensional fit only, corresponding to the rightmost panel of the bottom row in Figure~\ref{fig:examplefit}.}
\label{fig:examplefitshallow1}
\end{center}
\end{figure}

\begin{figure}[t!]
\begin{center}
\includegraphics[width=0.8\textwidth]{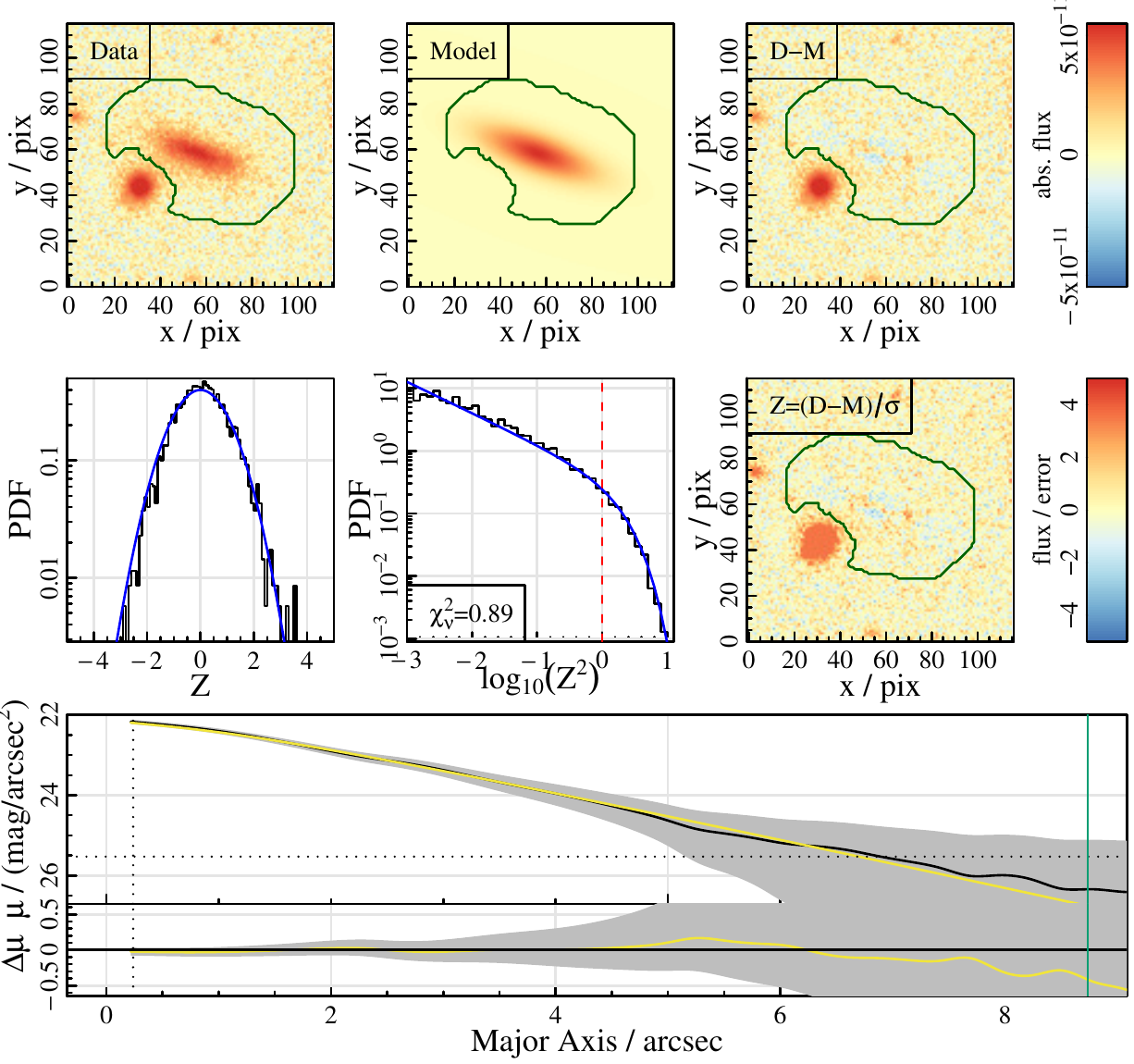}
\caption{The $r$-band single S\'ersic fit to galaxy 16537, classified as a single S\'ersic object, for direct comparison to Figure~\ref{fig:examplefitshallow1}.}
\label{fig:examplefitshallow2}
\end{center}
\end{figure}

The fraction of outliers in the VIKING $Z$-band (12\,\%) is comparable to that in the core bands, while the KiDS $u$ and the VIKING $Y, J, H, K_s$ bands have significantly higher fractions, ranging between 18\,\% and 32\,\%. Part of the reason for these higher fractions are the lower numbers of non-skipped fits (i.e. lower normalisation), but as evident from Figure~\ref{fig:ncompstats}, the absolute numbers of outliers increase, too. In all cases, the dominant reason for this are (reliable) parameters hitting their fit limits, with maybe a slight secondary contribution from the position offset flag (Table~\ref{tab:v05outlierstats}). All other flag values show only small increases that can be attributed to the lower normalisation. For the VIKING bands, the fractions of fits flagged generally increases from $Z$ through to $K_s$ for all flag values (in part again due to the higher number of skipped fits from $Z$ through to $K_s$).  

The parameter that hits its fit limit most frequently for single S\'ersic fits is the axial ratio, followed by the S\'ersic index (with a large overlap between the two) and to a lesser extent the effective radius. Almost all of the parameters hitting their fit limits also have a suspiciously large error (meaning they are an outlier in their respective error distribution), which is a cautionary flag. Both of these are indications that the parameters are ill-constrained and can be explained by the decreasing data quality (in particular depth) as a function of wavelength for VIKING (Section~\ref{sec:viking}) and the KiDS $u$-band (Section~\ref{sec:kids}; which was also the reason why the $u$-band was not considered during \texttt{v04} and is not used as a ``core" band in \texttt{v05}). The shallowest bands ($u$ and $K_s$) are hence not only too shallow to constrain two components in most cases (cf. Section~\ref{sec:manualcalibrationchanges}), but also too shallow to constrain a single S\'ersic fit for approximately one quarter of our sample and therefore become flagged as outliers. As we discuss in Section~\ref{sec:manualcalibrationchanges}, this is a fundamental problem that can only be solved with deeper data or simultaneous multi-band fits. 

For the same reason, the fractions of non-flagged fits provided in Table~\ref{tab:resultsv05} decrease drastically as a function of data quality and model complexity. For example, in the $u$-band, 90\,\% of the attempted double component fits are flagged as outliers. The vast majority of these, however, were not classified as double component objects since the data can equally well be represented by a single S\'ersic fit. What drives the higher fraction of outliers in these shallow bands are therefore mostly the objects for which even the single S\'ersic fits cannot be constrained anymore. Figure~\ref{fig:examplefitshallow1} shows an example of such a fit where the $u$-band data is of insufficient quality to constrain a single S\'ersic model. The fit was flagged as an outlier since it hit its parameter limits for the axial ratio and S\'ersic index. For comparison, we show the $r$-band version of the same galaxy in Figure~\ref{fig:examplefitshallow2}. This fit was classified as a (good) single S\'ersic fit.

\subsection{Model selection differences}
\label{sec:modelseldiffs}

We now turn to the model selection differences between \texttt{v04} and \texttt{v05}, as shown in Figure~\ref{fig:ncompstats} and Tables~\ref{tab:results} and~\ref{tab:resultsv05}. The model selection differences between individual bands in \texttt{v05} have already been discussed in Section~\ref{sec:manualcalibrationchanges} where we also give full confusion matrices against visual inspection for the model selection in all bands and the joint model selection in \texttt{v05}. Corresponding \texttt{v04} confusion matrices are provided in Sections~\ref{sec:postprocessing} and~\ref{sec:swappingandoutliers}. Here, we supplement these by a new type of confusion matrix, namely that between the \texttt{v04} and \texttt{v05} model selections for the core bands. For the most direct comparison, we consider only those galaxies that had a single data match and were not skipped in any bands in either catalogue (leaving 8900 objects), and to decouple the analysis from outlier flagging differences (Section~\ref{sec:outlierstats}), we take absolute values of \texttt{NCOMP}. 

\begin{table}
\centering
\caption{The confusion matrices between the \texttt{v04} and \texttt{v05} model selections in percent of the total number of objects that have a single data match and were not skipped in any of the core bands in either catalogue. Bold font highlights those galaxies classified in the same category for both versions. Total fractions of objects classified into each of the three cateogories (i.e. the sum of each row/column) are also given; as well as the total percentage of objects for which the model selection is in agreement between the two versions (sum of the diagonal).}
\label{tab:modelselv04vsv05}
\begin{subtable}{1\textwidth}
\centering
	\begin{tabu}{lcrrrcrcccrrrcr} 
	& & \multicolumn{5}{c}{$g$-band} & &
	& & \multicolumn{5}{c}{$r$-band}\\
	\hline
	& & \multicolumn{3}{c}{\texttt{v05 NCOMP}} &&&&
	& & \multicolumn{3}{c}{\texttt{v05 NCOMP}} &&\\
		 
		\texttt{v04 NCOMP} & 
		& 1 & 1.5 & 2 && total &&& 
		& 1 & 1.5 & 2 && total\\
		\hline
		
		1 &
		& \textbf{69.2} & 0.9 & 2.3 && 72.4 &&&
		& \textbf{53.1} & 2.1 & 5.0 && 60.2\\
		
		1.5 &
		& 0.8 & \textbf{4.2} & 2.4 && 7.4 &&&
		& 0.2 & \textbf{3.0} & 2.2 && 5.5\\
		
		2 &
		& 1.9 & 1.5 & \textbf{16.8} && 20.2 &&& 
		& 1.6 & 2.0 & \textbf{30.8} && 34.3\\\\
		
		total &
		& 71.8 & 6.7 & 21.5 && \textbf{90.2} &&& 
		& 54.8 & 7.1 & 38.1 && \textbf{86.9}\\
		
        \hline
	\end{tabu}
\end{subtable}%

\bigskip
\begin{subtable}{1\textwidth}
\centering
	\begin{tabu}{lcrrrcrcccrrrcr} 
	& & \multicolumn{5}{c}{$i$-band} & &
	& & \multicolumn{5}{c}{$gri$ joint}\\
	\hline
	& & \multicolumn{3}{c}{\texttt{v05 NCOMP}} &&&&
	& & \multicolumn{3}{c}{\texttt{v05 NCOMP}} &&\\
		 
		\texttt{v04 NCOMP} &  
		& 1 & 1.5 & 2 && total &&&
		& 1 & 1.5 & 2 && total\\
		\hline
		
		1 &
		& \textbf{63.3} & 0.1 & 0.3 && 63.8 &&& 
		& \textbf{63.4} & 0.4 & 0.8 && 64.6\\
		
		1.5 &
		& 2.5 & \textbf{3.5} & 0.2 && 6.2 &&& 
		& 1.6 & \textbf{2.8} & 2.0 && 6.3 \\
		
		2 &
		& 8.9 & 1.8 & \textbf{19.4} && 30.1 &&& 
		& 11.1 & 3.3 & \textbf{14.7} && 29.1 \\\\
		
		total &
		& 74.7 & 5.3 & 19.9 && \textbf{86.2} &&&
		& 76.0 & 6.5 & 17.5 && \textbf{80.9} \\
		
        \hline
	\end{tabu}
\end{subtable}%
\end{table}

\begin{table}
\centering
\caption{The same as Table~\ref{tab:modelselv04vsv05} but using the \texttt{v04}-calibrated $\Delta$DIC cuts for the \texttt{v05} model selection.}
\label{tab:modelselv04vsv05b}
\begin{subtable}{1\textwidth}
\centering
	\begin{tabu}{lcrrrcrcccrrrcr} 
	& & \multicolumn{5}{c}{$g$-band} & &
	& & \multicolumn{5}{c}{$r$-band}\\
	\hline
	& & \multicolumn{5}{c}{\texttt{v05 NCOMP (v04 cal.)}} &&
	& & \multicolumn{5}{c}{\texttt{v05 NCOMP (v04 cal.)}} \\
		 
		\texttt{v04 NCOMP} & 
		& 1 & 1.5 & 2 && total &&& 
		& 1 & 1.5 & 2 && total\\
		\hline
		
		1 &
		& \textbf{69.3} & 1.5 & 1.7 && 72.4 &&&
		& \textbf{57.7} & 0.7 & 1.8 && 60.2\\
		
		1.5 &
		& 0.8 & \textbf{6.1} & 0.5 && 7.4 &&&
		& 0.3 & \textbf{4.6} & 0.5 && 5.5\\
		
		2 &
		& 1.6 & 1.7 & \textbf{16.9} && 20.2 &&& 
		& 1.9 & 2.1 & \textbf{30.3} && 34.3\\\\
		
		total &
		& 71.6 & 9.3 & 19.1 && \textbf{92.3} &&& 
		& 60.0 & 7.4 & 32.6 && \textbf{92.6}\\
		
        \hline
	\end{tabu}
\end{subtable}%

\bigskip
\begin{subtable}{1\textwidth}
\centering
	\begin{tabu}{lcrrrcrcccrrrcr} 
	& & \multicolumn{5}{c}{$i$-band} & &
	& & \multicolumn{5}{c}{$gri$ joint}\\
	\hline
	& & \multicolumn{5}{c}{\texttt{v05 NCOMP (v04 cal.)}} &&
	& & \multicolumn{5}{c}{\texttt{v05 NCOMP (v04 cal.)}} \\
		 
		\texttt{v04 NCOMP} &  
		& 1 & 1.5 & 2 && total &&&
		& 1 & 1.5 & 2 && total\\
		\hline
		
		1 &
		& \textbf{61.4} & 0.9 & 1.5 && 63.8 &&& 
		& \textbf{62.1} & 1.1 & 1.4 && 64.6\\
		
		1.5 &
		& 0.5 & \textbf{5.3} & 0.4 && 6.2 &&& 
		& 0.6 & \textbf{5.4} & 0.3 && 6.3 \\
		
		2 &
		& 1.7 & 1.7 & \textbf{26.6} && 30.1 &&& 
		& 2.9 & 3.9 & \textbf{22.2} && 29.1 \\\\
		
		total &
		& 63.6 & 7.9 & 28.5 && \textbf{93.2} &&&
		& 65.6 & 10.4 & 24.0 && \textbf{89.7} \\
		
        \hline
	\end{tabu}
\end{subtable}%
\end{table}

Table~\ref{tab:modelselv04vsv05} shows the resulting confusion matrices for all bands and the joint $gri$ model selection, with bold colours highlighting the galaxies that were classified in the same category in both catalogue versions. For better overview, we also show the sums of each row and column (total numbers of objects classified in each model selection category for each version independently) and the sum of the diagonal, i.e. the total number of objects for which the model selection agrees between the two versions. All values are in percent of the total number of objects.

The overall agreement between the model selections in \texttt{v04} and \texttt{v05} ranges between $\sim$\,81\,\% for the joint model selection to 90\,\% for the $g$-band. The largest rate of confusion is observed between single and double component objects. For the $r$-band, there are more double and 1.5-component fits in \texttt{v05} than in \texttt{v04} (also evident from Figure~\ref{fig:ncompstats}). This is reversed in the $i$-band and the joint model selection, where the number of single component fits is higher in \texttt{v05}. For the $g$-band, the relative numbers of 1.5-/double component fits and single S\'ersic objects is approximately the same in both versions. 

Comparing the $\Delta$DIC$_{1-2}$ cuts shown in Tables~\ref{tab:v04diccuts} and~\ref{tab:v05diccuts}, we can see that for the $g$-band, the $\Delta$DIC cuts are similar in \texttt{v04} and \texttt{v05}.\footnote{Note that the absolute DIC values in \texttt{v05} are generally a factor of about two larger than in \texttt{v04} due to the larger segments. However, the DIC \emph{differences} are comparable since all models are equally affected by the larger segments. We therefore would expect the $\Delta$DIC cuts to be similar in both catalogue versions.} 
For the $r$-band, the \texttt{v05} cut is slightly lower than in \texttt{v04}, although still within the uncertainty region. The $i$-band and the joint $gri$ model selection both have significantly higher $\Delta$DIC$_{1-2}$ cuts in \texttt{v05} than in \texttt{v04}, with no overlap between the ``unsure" regions. This explains the low confusion rate between \texttt{v04} and \texttt{v05} in the $g$-band, the slightly higher number of double components fits in \texttt{v05} in $r$, and the significantly lower number of double component fits in the $i$-band and joint model selection. 

Since the procedure for the model selection calibration is identical in \texttt{v04} and \texttt{v05}, these differences must stem from differences in the fits themselves or in the manual (re-)calibrations. To judge the effect of differences in the fits (either from the random nature of the MCMC sampling or from systematic differences e.g. due to the larger segments), we additionally performed the model selection on \texttt{v05} using the $\Delta$DIC cuts from \texttt{v04}. The resulting confusion matrices between the model selection categories for \texttt{v04} and \texttt{v05} (with \texttt{v04} calibration) are given in Table~\ref{tab:modelselv04vsv05b}. The total fractions of fits classified in the same model selection category (sum of the diagonals) rise to 92.3\,\%, 92.6\,\%, 93.2\,\% and 89.7\,\% respectively for the $g, r, i$ individual and $gri$ joint model selection, compared to 90.2\,\%, 86.9\,\%, 86.2\,\% and 80.9\,\% in Table~\ref{tab:modelselv04vsv05}. The relative fractions of single S\'ersic and 1.5-/double component fits in the two versions are now within 0.2\,\% of each other for all bands ($\sim$\,1\,\% for the joint selection). The fractions of 1.5-component fits are about 2\,\% higher in \texttt{v05} than in \texttt{v04} for all bands (4\,\% for the joint $gri$ selection), with the double component fractions correspondingly lower, despite using the same $\Delta$DIC cuts. 

This indicates that confusion between models on the level of approximately 7-8\,\% (10\,\% for the joint selection) stems from differences between the fits themselves (to the same galaxies), either of random nature or due to the different fitting procedures in \texttt{v04} and \texttt{v05}, most likely the larger segments (see Section~\ref{sec:pipelineupdates}). This introduces no systematic differences in the classification between the single S\'ersic and 1.5-/double component models, but generally increases the numbers of 1.5-component fits relative to the double component fits. The reasons for the latter remain to be investigated.

The differences between Tables~\ref{tab:modelselv04vsv05} and~\ref{tab:modelselv04vsv05b} are caused by differences in the $\Delta$DIC cuts for \texttt{v04} and \texttt{v05}, which in turn must be caused by differences in the visual calibrations. Reasons could be statistical fluctuations due to the randomly selected calibration samples of galaxies (that differ in \texttt{v04} and \texttt{v05}) as well as human error, both amplified by the relatively small sample size of 200 objects per band in \texttt{v05}. Since the visual classifications were performed several years apart, there could also be systematic differences due to a deepened understanding of the fit results, modelling limitations, systematic uncertainties and model selection caveats based on the detailed investigations of the \texttt{v04} results in the meantime. In particular, the \texttt{v05} calibration put special emphasis on classifying as few fits as possible in the ``unsure" or ``outlier" categories since that further reduces the already relatively low number of objects available for model selection calibration. We also explicitly de-coupled model selection from outlier rejection, the identification of swapped fits and segmentation failures. If, for example, the double component fit (within the segment) is significantly better than the other two models, in the \texttt{v05} calibration we labelled this object as a double component object regardless of whether it is swapped, would be better classified as an outlier or the segmentation map failed (e.g. containing secondary objects). This means that the two categories ``unsure" and ``unfittable/outlier" became equivalent (as they always were in the $\Delta$DIC cut calibration, where both are ignored). Conversely, if the object clearly has a bulge but the 1.5- and double component models failed to fit it and are not better than the single S\'ersic fit, then the object was labelled as ``single". While the general notions were the same during the \texttt{v04} visual inspections - driven by the limitations of a model selection based on a statistical measure - the criteria were not as sharply defined. 

\begin{figure}[t!]
\includegraphics[width=0.5\textwidth]{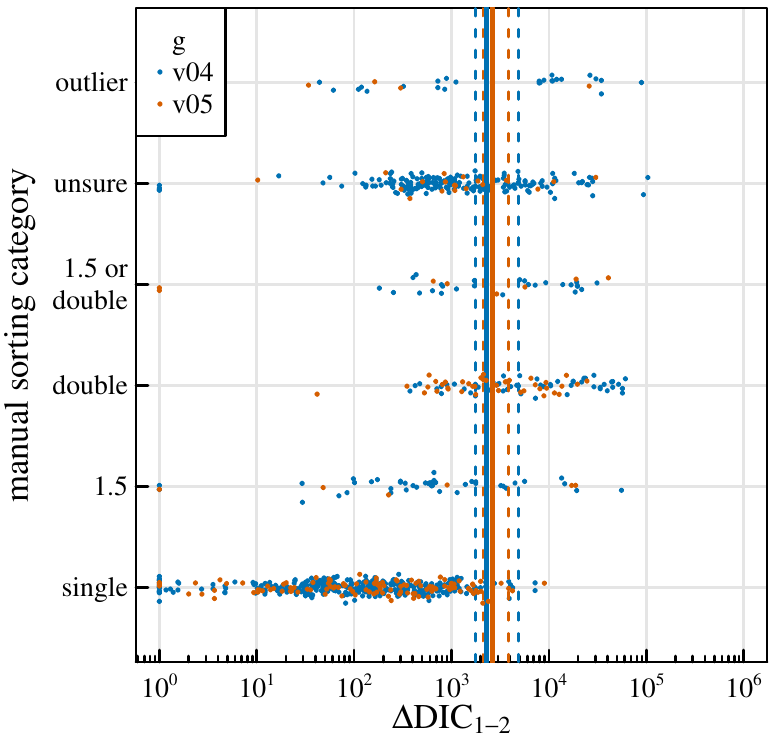}
\includegraphics[width=0.5\textwidth]{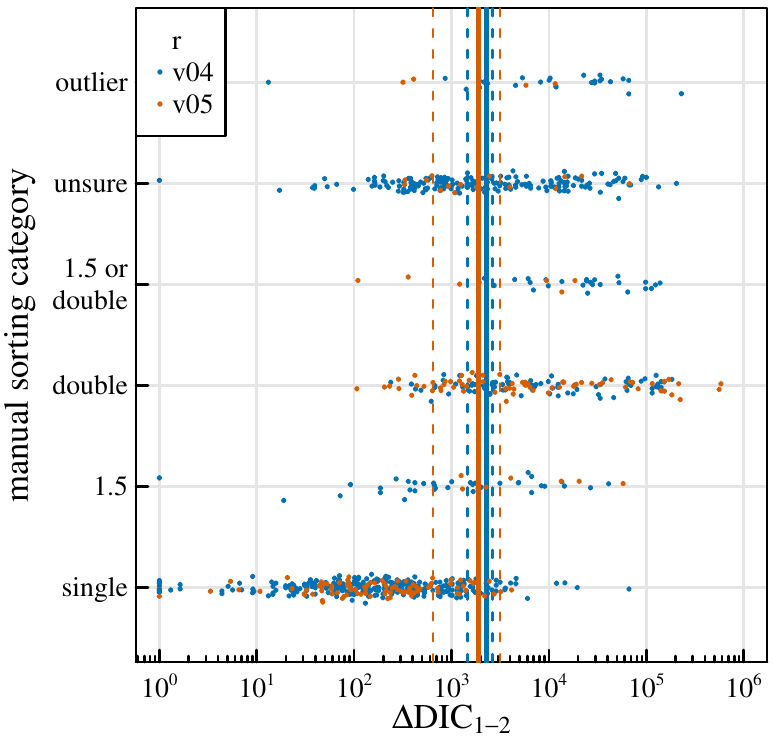}
\includegraphics[width=0.5\textwidth]{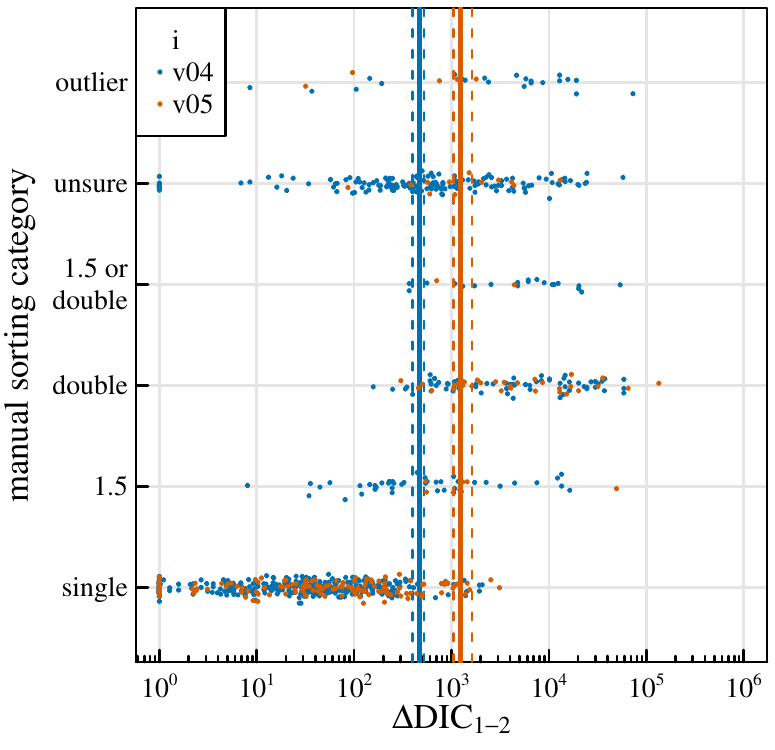}
\includegraphics[width=0.5\textwidth]{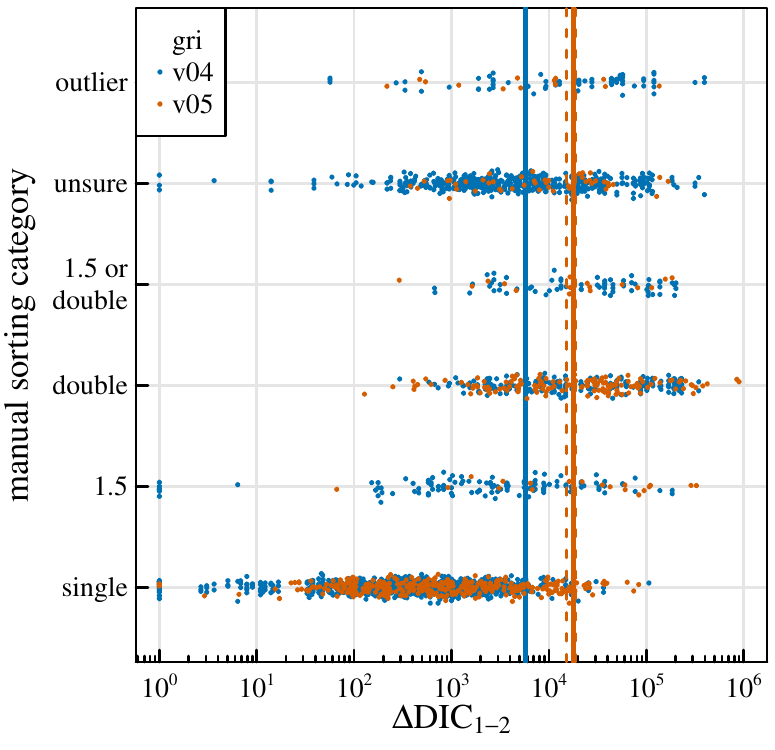}
\caption{The visual classification during manual sorting plotted against $\Delta$DIC$_{1-2}$ for \texttt{v04} (blue points) and \texttt{v05} (orange points) in the $g$ (\textbf{top left}), $r$ (\textbf{top right}) and $i$ (\textbf{bottom left}) bands as well as for the $gri$ joint model selection (\textbf{bottom right}; with DICs summed for all bands). Vertical solid and dashed lines indicate the corresponding $\Delta$DIC$_{1-2}$ cuts for \texttt{v04} (blue) and \texttt{v05} (orange).}
\label{fig:modelselv04v05}
\end{figure}

To investigate such manual calibration differences, Figure~\ref{fig:modelselv04v05} shows the visual classifications in \texttt{v04} (blue) and \texttt{v05} (orange) as a function of the DIC difference between the single and double component models, where we see most confusion in Table~\ref{tab:modelselv04vsv05}. Note that the galaxies shown are not the same and the number of objects is higher for \texttt{v04} ($\sim$\,700 per band; 2000 in the joint selection) than for \texttt{v05} (200 per band, 600 in the joint selection). The corresponding calibrated $\Delta$DIC$_{1-2}$ cuts with their ``unsure" regions (solid and dashed lines) are also indicated. The visual inspection categories that are relevant for calibrating the $\Delta$DIC$_{1-2}$ cut are ``single", ``double" and ``1.5 or double" (cf. Section~\ref{sec:postprocessing}); we show the other categories for completeness only. 

Overall, the distribution of blue and orange points in each panel is similar, indicating that there are no fundamental differences in the visual calibrations for \texttt{v04} and \texttt{v05}. For the $g$ and $r$ bands, the DIC difference cuts are consistent between the two versions, as already mentioned above when comparing Tables~\ref{tab:v04diccuts} and~\ref{tab:v05diccuts}. Hence, the confusion rates in these bands is low with especially the $g$-band coming close to the confusion from fit differences alone (Tables~\ref{tab:modelselv04vsv05} and~\ref{tab:modelselv04vsv05b}). The cuts in the $i$-band and joint selection are not consistent, although they are still relatively close to each other compared to the vast range the DIC differences observed. Therefore, the confusion rates between \texttt{v04} and \texttt{v05} in these bands are still moderate despite the numerical differences in the $\Delta$DIC cuts (Tables~\ref{tab:modelselv04vsv05} as well as~\ref{tab:v04diccuts} and~\ref{tab:v05diccuts}). 

\begin{figure}
\begin{center}
\includegraphics[width=0.8\textwidth]{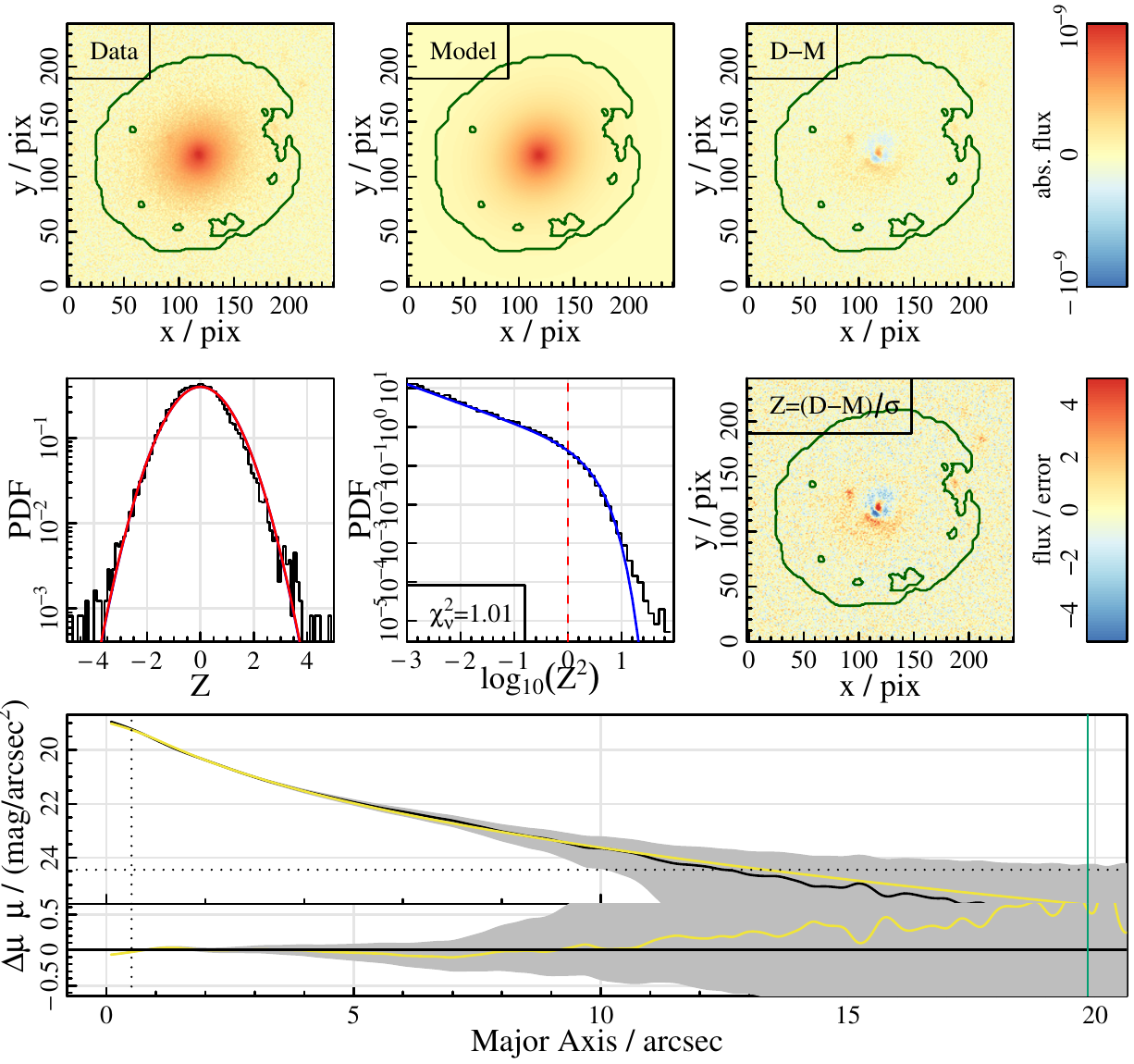}
\caption{The single S\'ersic fit to galaxy 585561 in the $i$-band, which was visually classified as a single S\'ersic object despite its relatively high $\Delta$DIC$_{1-2}$ value of $\sim$\,1500. Panels in the top two rows are the same as those in Figure~\ref{fig:examplefit}, while the bottom row shows the one-dimensional fit only, corresponding to the rightmost panel of the bottom row in Figure~\ref{fig:examplefit}.}
\label{fig:examplefitdicdiff1}
\end{center}
\end{figure}

\begin{figure}[t!]
\begin{center}
\includegraphics[width=0.8\textwidth]{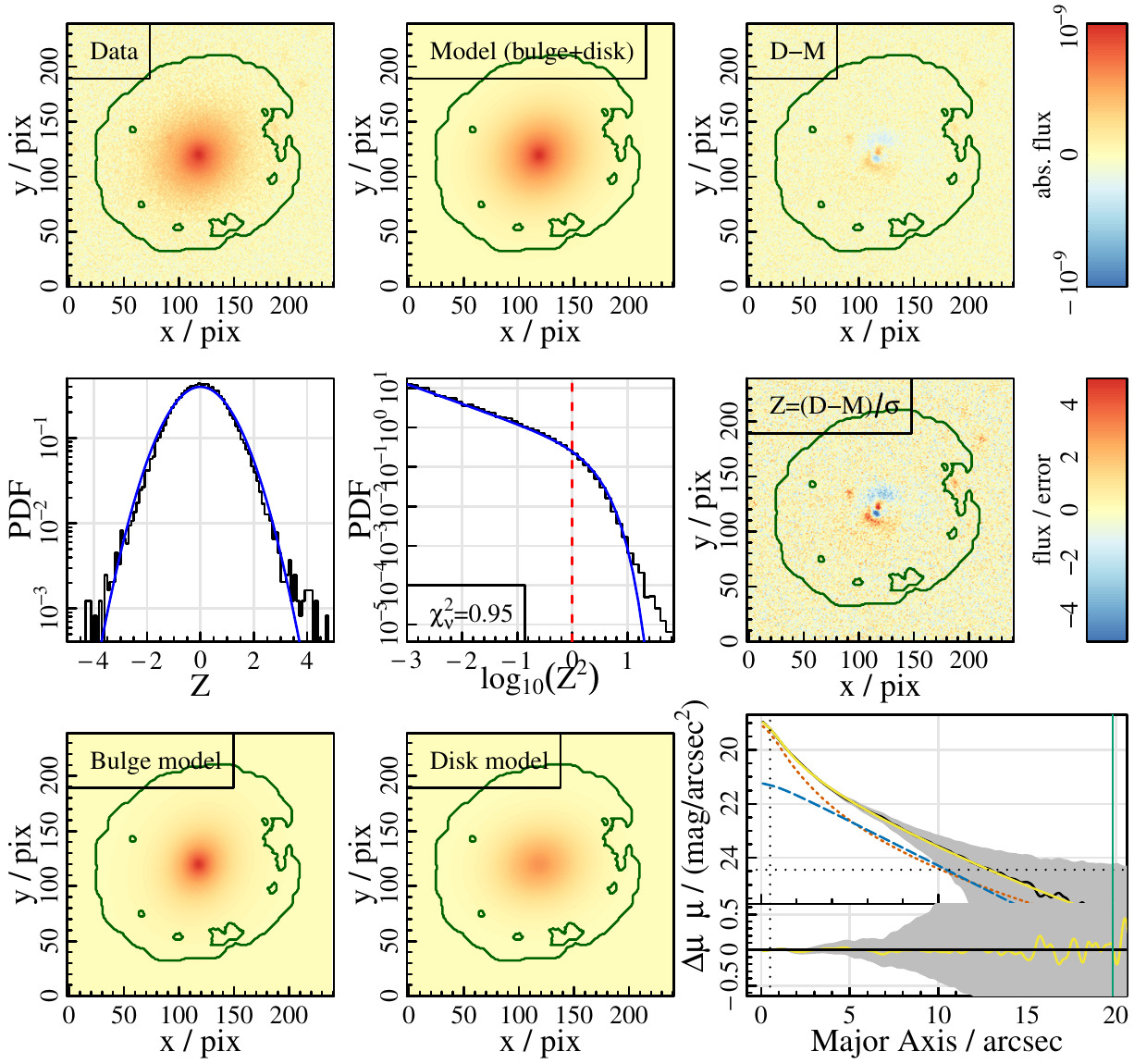}
\caption{The double component fit to galaxy 585561 in the $i$-band, for direct comparison to Figure~\ref{fig:examplefitdicdiff1}. Panels are the same as those in Figure~\ref{fig:examplefit}.}
\label{fig:examplefitdicdiff2}
\end{center}
\end{figure}

The higher cut in \texttt{v05} in the $i$-band appears to be caused by a cluster of orange points around a $\Delta$DIC$_{1-2}$ value of $10^3$, that has no blue counterpart (remember that there are a factor of 3.5 more blue points than orange ones) and is also not observed in the $g$ and $r$ bands. It is possible that this is simply caused by an unfortunate selection of galaxies or random human error and would be downweighted by a larger sample of objects used for calibration. Alternatively, it could be systematic due to the sharpened criteria for the \texttt{v05} calibration. Visual inspection of this cluster of datapoints reveals them all to be borderline cases. Figures~\ref{fig:examplefitdicdiff1} and~\ref{fig:examplefitdicdiff2} show an example, where the double component fit is better than the single S\'ersic fit, but it is debatable whether it is ``significantly" better in the sense that it justifies the increased number of parameters. In the \texttt{v05} calibration we decided that this was not the case, but maybe the decision would have been different during the \texttt{v04} calibration; or even when inspecting the same galaxy at a later or earlier time in the \texttt{v05} calibration, as there is always a certain amount of scatter associated with visual inspection. 

The last point of the model selection that we would like to return to are the high confusion rates against visual inspection for the $g$ and $r$ bands in \texttt{v05} that were apparent from Table~\ref{tab:modelselconfusionv05}. As a reminder, the \texttt{v05} $g$ and $r$ bands and the two versions of the joint model selections have approximately 15\,\%, 17\,\%, 17\,\% and 18\,\% of fits classified wrongly by the automated procedure compared to visual inspection, mostly due to \texttt{NCOMP}\,=\,1 objects that were visually classified as doubles. For all other bands and also for the \texttt{v04} versions of the same bands, the confusion rate is 9\,\% at most. Based on the above analysis of the \texttt{v04} and \texttt{v05} model selection differences, we believe that this is mainly due to the lower number of fits classified as ``unsure" in the \texttt{v05} model selection. 

Going back to Figure~\ref{fig:modelselv04v05}, the objects responsible for the high confusion are orange points in the ``double" manual sorting category but to the left of the orange $\Delta$DIC cut. There is a much higher number of those in the $g$ and $r$ bands than in $i$, partly due to the generally lower number of double component objects in $i$. Most blue points (\texttt{v04} model selection) in that region of $\Delta$DIC$_{1-2}$ instead populate the ``unsure" category, where there is generally a very low number of orange points due to our explicit attempt to classify as few galaxies as possible as ``unsure" in the \texttt{v05} calibration. From this - and the consistency of the \texttt{v04} and \texttt{v05} $\Delta$DIC$_{1-2}$ cuts in the $g$ and $r$ bands - we conclude that the automated model selection itself is not ``worse" in \texttt{v05} than it was in \texttt{v04}. Instead, the higher confusion statistics is simply caused by a large number of ambiguous objects that were more or less randomly assigned to one of the two categories in the \texttt{v05} visual inspection, whereas they were classified as ``unsure" and ignored during \texttt{v04} model selection.

Considering all of the above differences and uncertainties, we conclude that the confusion rates between the model selections in \texttt{v04} and \texttt{v05} as well as those between the \texttt{v05} automated and manual classifications are acceptable. The quality of the \texttt{v05} model selection is comparable to that of \texttt{v04} despite the lower number of visual classifications per band.

\section{Parameter distributions}

After the analysis of the outlier rejection and model selection statistics, we take a closer look at the fitted structural parameters, considering only good fits within the respective model selection category for each band individually. We do not use the joint model selection or take a combined sample across all bands due to the differences in data quality between the bands that result in vastly different relative numbers of galaxies in each category. Section~\ref{sec:v04parameterdistributions} shows the parameter distributions for \texttt{v04} of the catalogue, taken directly from \citet{Casura2022}. Section~\ref{sec:v05parameterdistributions} adds the new \texttt{v05} versions (not included in \citealt{Casura2022} or elsewhere) and compares them to those of \texttt{v04}. Finally, Section~\ref{sec:colours} focuses on galaxy and component colours, again based on \citet{Casura2022}. 

\subsection{\texttt{v04} parameter distributions}
\label{sec:v04parameterdistributions}

\begin{figure}
    \includegraphics[width=\textwidth]{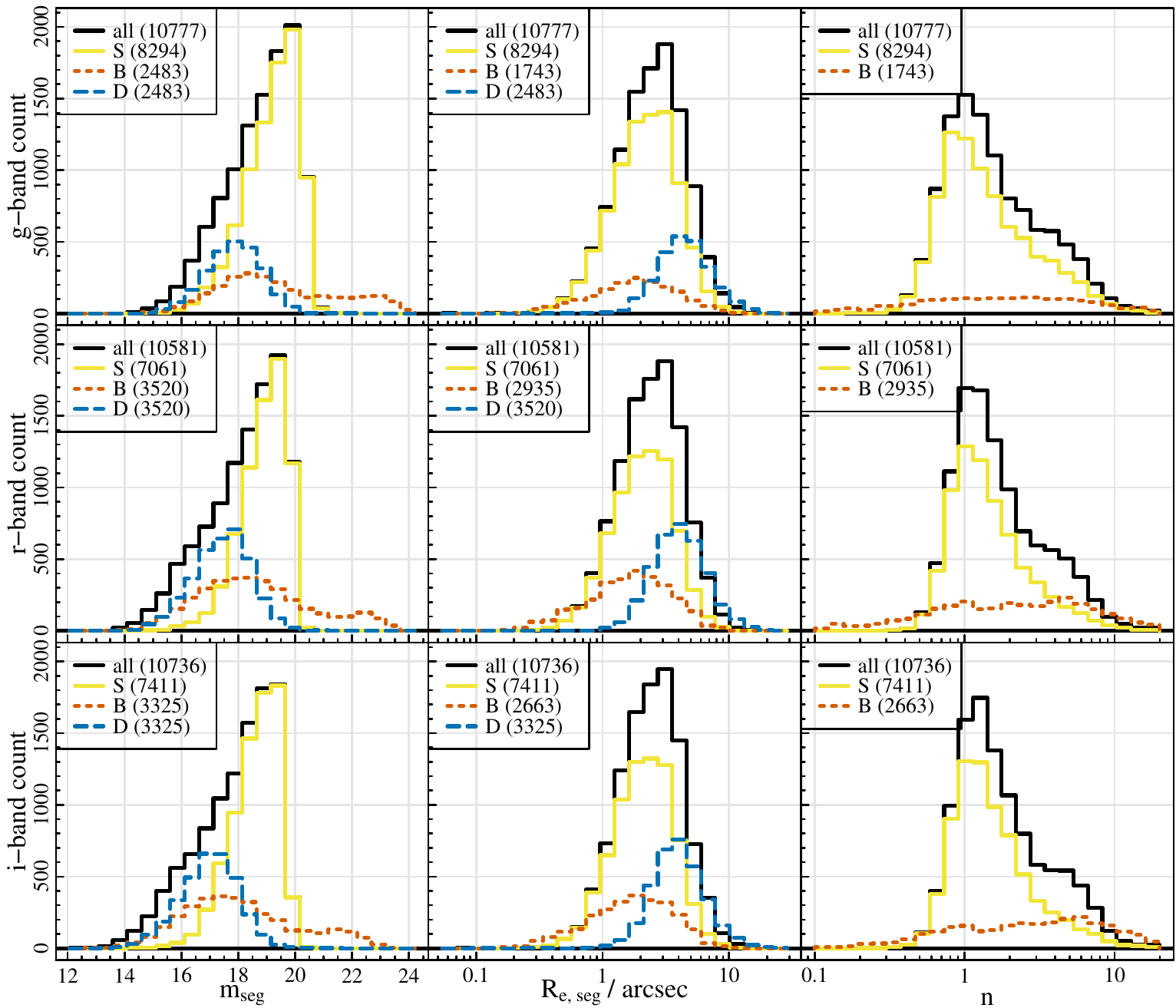}
    \caption{The distribution of the main parameters (limited to segment radii) for all bands and models in \texttt{v04}. Left, middle and right columns show magnitude, effective radius and S\'ersic index while top, middle and bottom rows show the $g$-band, $r$-band and $i$-band respectively. The solid yellow lines are the single S\'ersic (S) values for those galaxies which were classified as single component systems, dotted red and dashed blue lines show bulges (B) and disks (D), respectively, for those objects classified as 1.5- or double component systems. For reference the solid black line shows the single S\'ersic fits for all galaxies with \texttt{NCOMP}\,>\,0 (i.e. including those classified as 1.5- or double component systems). The number of objects in each histogram is given in the legends, where the number of bulges and disks differs for effective radii and S\'ersic indices because these parameters do not exist for 1.5-component fits (point source bulge). We do not show disk S\'ersic indices since they were fixed to 1 (exponential).}	
    \label{fig:resultshists}
\end{figure}

Figure~\ref{fig:resultshists} shows the distribution of the main parameters - magnitude, effective radius and S\'ersic index - for \texttt{v04} of the \texttt{BDDecomp} DMU in all three bands ($g$, $r$ and $i$) for single S\'ersic fits, bulges and disks. The single S\'ersic fit distributions are shown for all galaxies with \texttt{NCOMP}\,>\,0 (i.e. all non-outliers) in black and for those galaxies which were actually classified as single component systems (\texttt{NCOMP}\,=\,1) in yellow. Red dotted and blue dashed lines show bulges and disks, respectively. For disks we show the 1.5-component fits and double component fits combined (i.e. the 1.5-component parameters for objects with \texttt{NCOMP}\,=\,1.5 and double component parameters for those with \texttt{NCOMP}\,=\,2 added into one histogram); the S\'ersic index is not shown since it was fixed to 1. Bulge magnitudes are also shown for 1.5- and double component fits combined; effective radii and S\'ersic indices are only shown for the double component fits since they do not exist in the point source model. The legend indicates the numbers of objects in each histogram, which can also be inferred from Table~\ref{tab:results}. Magnitudes and effective radii are truncated at the segment radii which we found to give more robust results than using the S\'ersic values extrapolated to infinity (see Sections~\ref{sec:postprocessing}, \ref{sec:comparelee} and \ref{sec:systematics}). 

The first thing apparent from Figure~\ref{fig:resultshists} is that the distributions in the three bands are generally very similar, which is reassuring given that the fits were performed independently. Looking at the first column, the single S\'ersic number counts increase up to a sharp drop just before 20\,mag in all bands, which is not surprising given the GAMA survey limit of 19.8\,mag. The faintest of these objects are all classified as single component galaxies (the yellow lines are on top of the black lines), while some of the brighter objects are successfully decomposed into bulges and disks. Disks are generally slightly brighter than bulges. The bulges show a second, smaller peak at very faint magnitudes which we found to be the ones from the 1.5-component fits (unresolved, faint bulges). There is a slight trend for magnitudes to become brighter moving from $g$ to $r$ to $i$ for all components, as expected from typical galaxy colours. We investigate the colours further in Section~\ref{sec:colours}. 

From the middle column it becomes obvious that bulges tend to be smaller than disks by a factor $\sim$\,2, while single S\'ersic fits span a wide range of sizes. Similar to the trend observed in the magnitudes, the smallest objects are classified as single component systems, while some of the larger galaxies can be successfully decomposed. 

\begin{figure}[t!]
\begin{center}
    \includegraphics[width=0.65\textwidth]{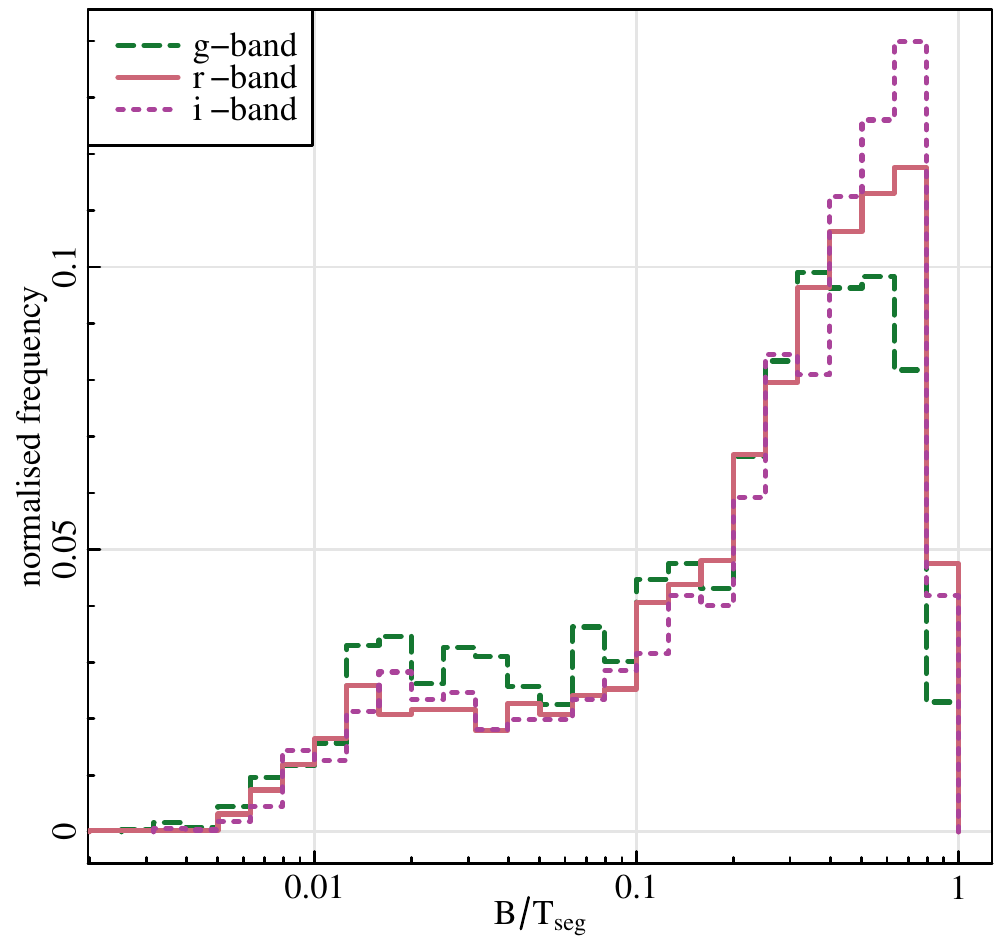}
    \caption{The distribution of the bulge to total flux ratio (limited to segment radii) for the 1.5- and double component fits in all bands in \texttt{v04}. Dashed green, solid light red and dotted dark pink lines refer to the $g$, $r$ and $i$ bands, respectively. The histograms have been normalised by their respective total number of fits (cf. Figure~\ref{fig:resultshists}) to make the bands directly comparable.}	
    \label{fig:BThist}
\end{center}
\end{figure}

\begin{figure}[h!]
\begin{center}
    \includegraphics[width=0.8\textwidth]{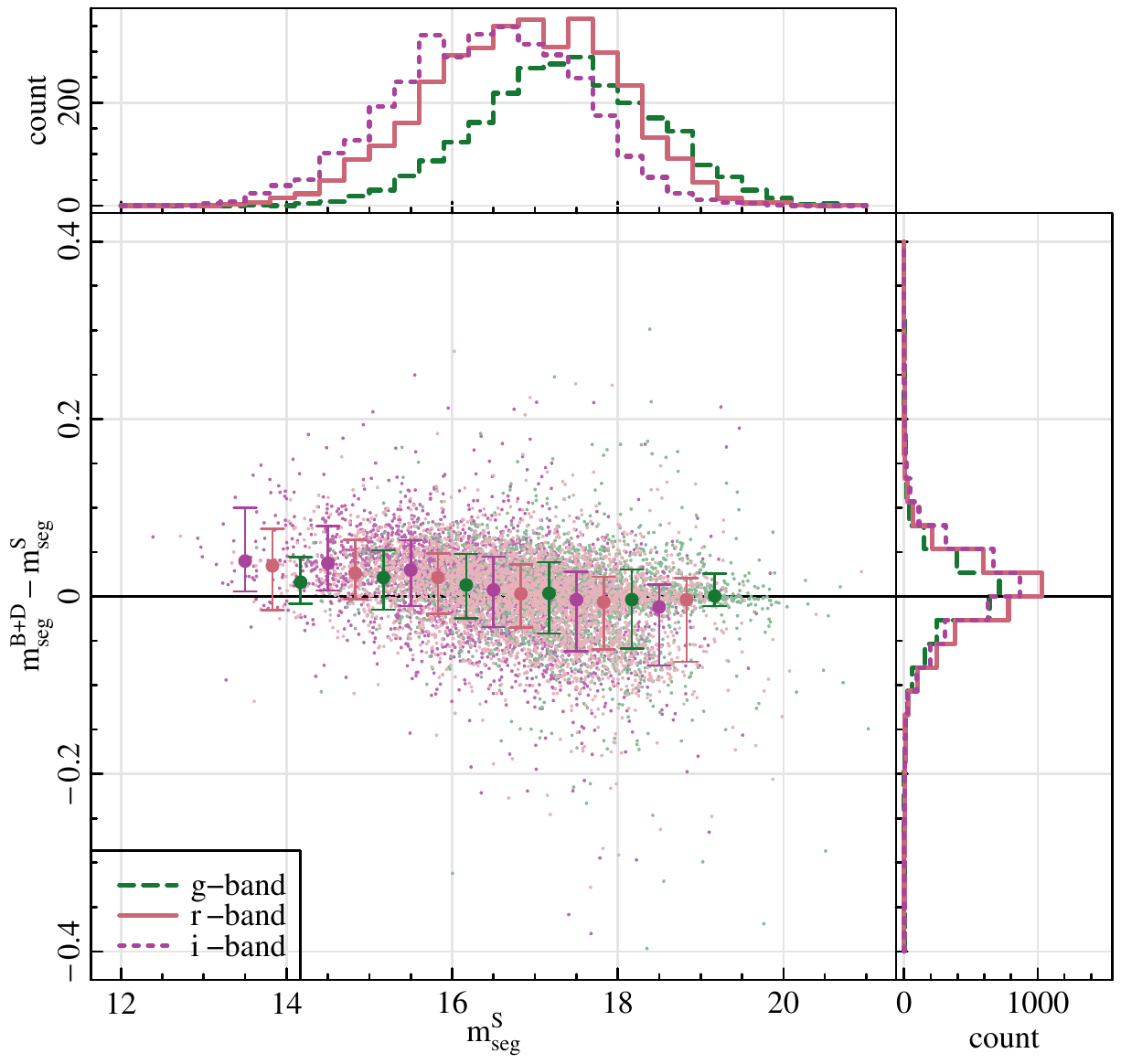}
    \caption{The difference between the single S\'ersic magnitude and the total magnitude derived from the double or 1.5-component fits for those galaxies that were classified as such, for all bands in \texttt{v04} (all magnitudes limited to segment radii). The scatter plot shows the difference between the two magnitudes against the single S\'ersic magnitude for all three bands with the running medians and 1$\sigma$-percentiles overplotted. The top and right panels show the respective marginal distributions. Dashed green, solid light red and dotted dark pink lines refer to the $g$, $r$ and $i$ bands respectively.}	
    \label{fig:magrecovery}
\end{center}
\end{figure}

The S\'ersic indices of single component systems show a clear peak around a value of 1 (exponential), a sharp drop-off at lower values and a longer tail towards higher values. Interestingly, the single S\'ersic distributions showing all systems (black lines) have a secondary ``bump" around a value of 4 or 5 (classical de Vaucouleurs bulge), which is not apparent in those galaxies classified as single component systems (yellow line). Hence most of those high S\'ersic index objects were found to contain bulges and were classified as double component systems. The bulges themselves show a wide range of S\'ersic indices with (at least in $r$ and $i$ bands) a slightly double-peaked nature around values of 1 and 4-6. At this point, we would like to remind the reader that we use the term ``bulge" to refer to all kinds of central components of galaxies, including classical bulges, pseudo-bulges, bars and AGN (cf. Section~\ref{sec:galaxyfitting}). Hence the ``bulge" distribution will include a variety of physical components and their combinations, leading to the wide spread of values. In addition, the S\'ersic index tends to be the parameter with the largest uncertainty, with typical galaxies showing relative errors on their bulge S\'ersic index of 1-10\,\%, adding further scatter to the distribution.

Since the bulge to total flux ratio is a derived parameter that is frequently of interest, we additionally show it in Figure~\ref{fig:BThist} for all three bands; for those galaxies that were classified as a 1.5- or double component fit in the respective band. The majority of systems have intermediate values of B/T with only a few percent at the extreme end above 0.8. The secondary peak at very low B/T values around 0.02 stems from the 1.5-component fits. The B/T ratio generally increases from $g$ to $r$ to $i$, as expected (see Section~\ref{sec:prevcols}). 

Finally, as a first consistency check, we show the difference between the single S\'ersic magnitude and the total magnitude derived from the double or 1.5-component fits, all limited to segment radii, in Figure~\ref{fig:magrecovery}. The distributions for all three bands are highly peaked around zero, with the vast majority of objects having total magnitudes consistent with the single S\'ersic magnitudes within 0.1\,mag (over the entire magnitude range). We only show galaxies that were classified as 1.5- or double component fits here to ensure reliable bulge and disk magnitudes, but note that the distribution is very similar when including objects classified as single S\'ersic fits. This indicates that the total magnitude is well-constrained even in the case when the individual component magnitudes are not (see also Section~\ref{sec:colours}).

\subsection{\texttt{v05} parameter distributions}
\label{sec:v05parameterdistributions}

\begin{figure}
    \includegraphics[width=\textwidth]{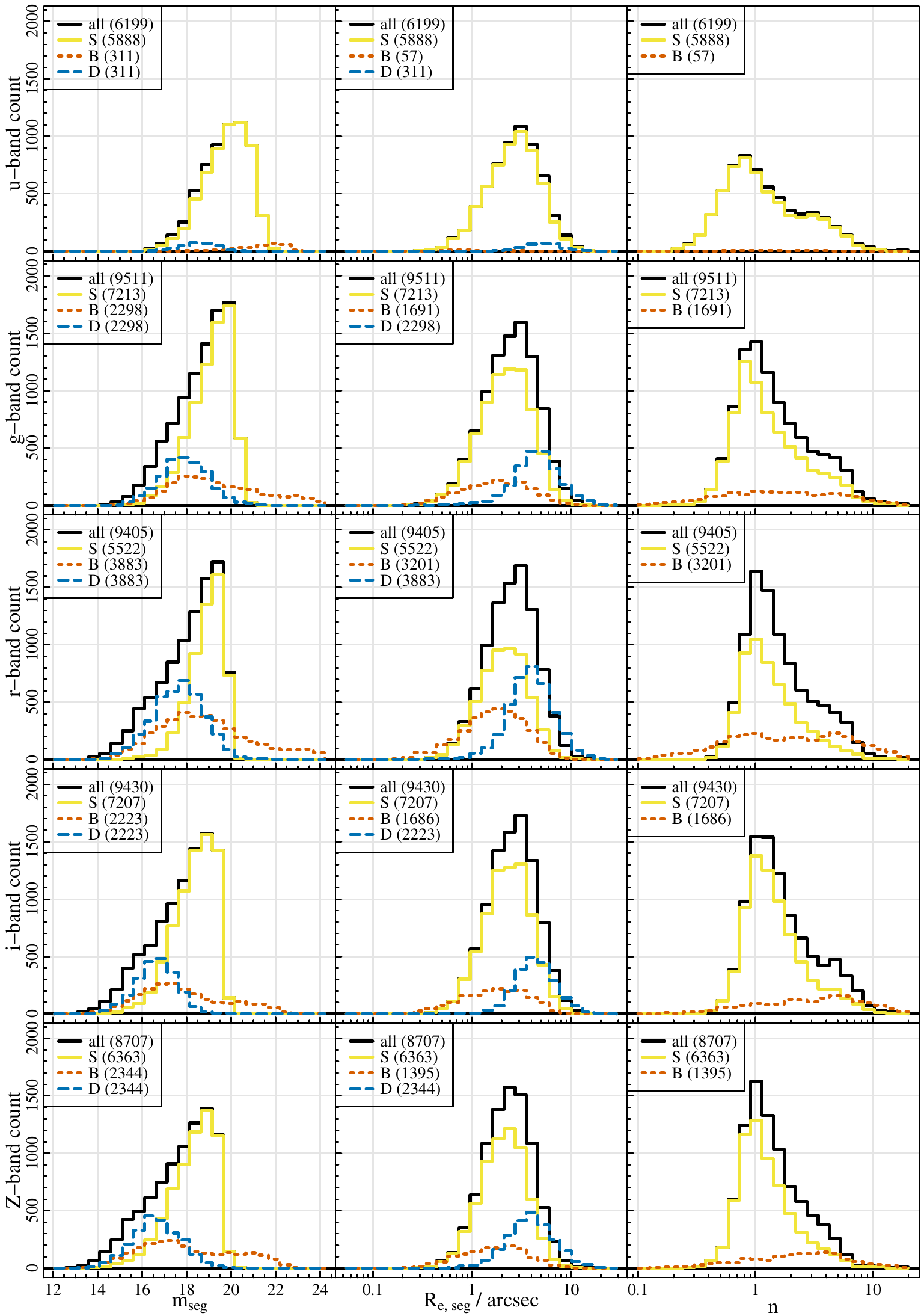}
    \phantomcaption
\end{figure}

\begin{figure}\ContinuedFloat
    \includegraphics[width=\textwidth]{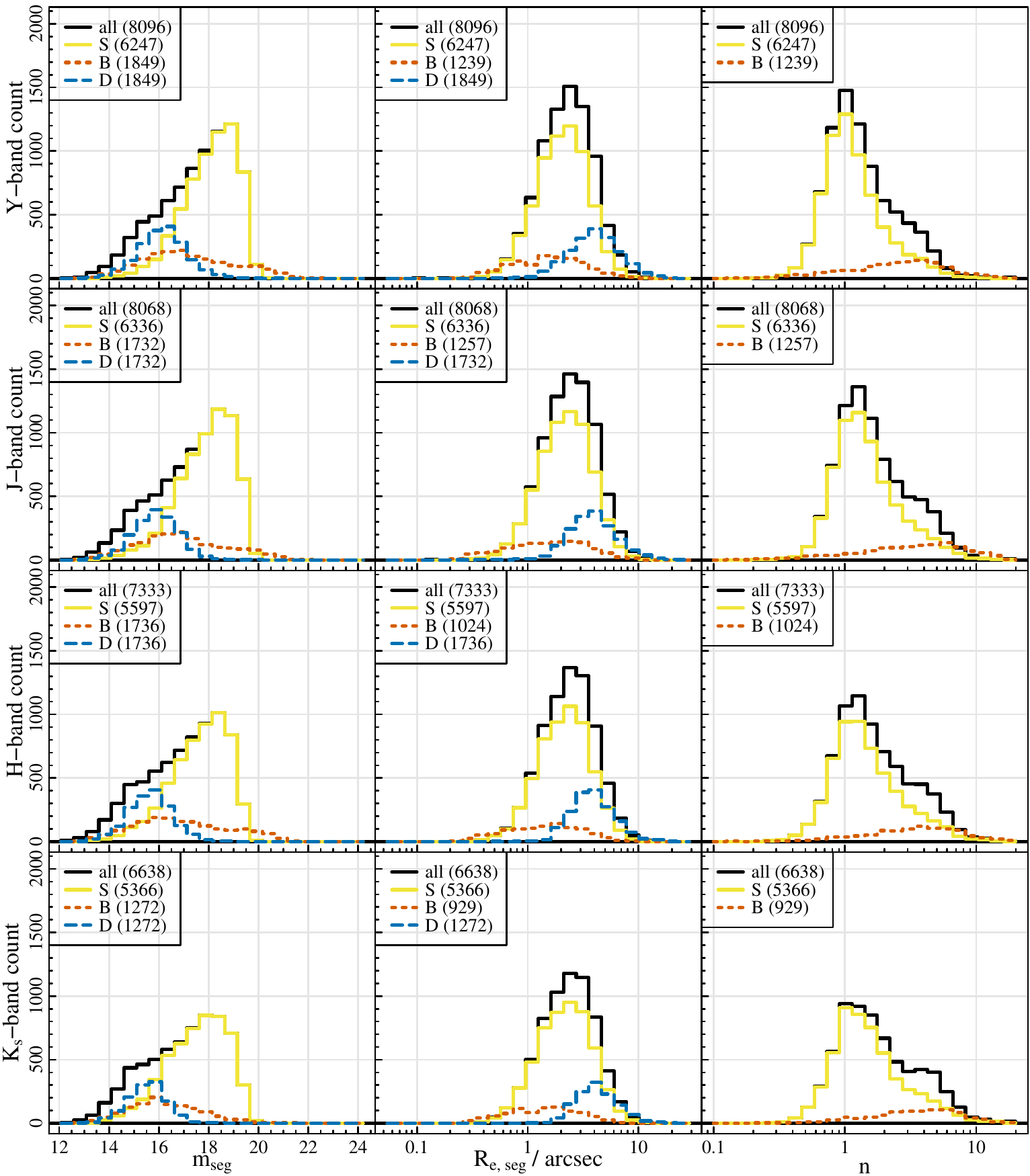}
    \caption{The distribution of the main parameters (limited to segment radii) for all bands and models in \texttt{v05}. Panels are the same as in Figure~\ref{fig:resultshists}, except that we now show all nine bands ($u$, $g$, $r$, $i$, $Z$, $Y$, $J$, $H$, $K_s$ from top to bottom). The axis scales and bin sizes are the same for all bands and also with respect to Figure~\ref{fig:resultshists}.}	
      \label{fig:resultshistsv05}
\end{figure}

Figure~\ref{fig:resultshistsv05} shows the \texttt{v05} equivalent of Figure~\ref{fig:resultshists}: the distribution of the three main S\'ersic parameters in all nine bands for single S\'ersic fits, bulges and disks. The layout of the plot, the colour coding of the lines and the scales of the axes are identical to those in Figure~\ref{fig:resultshists}. A more detailed, direct comparison between the results in \texttt{v04} and \texttt{v05} is shown in Figures~\ref{fig:resultsv04vsv05_R_S} to~\ref{fig:resultsv04vsv05_R_D_SEGRAD} at the end of this section.

Generally, the distributions of the parameters in Figure~\ref{fig:resultshistsv05} are similar in all bands and similar to those in Figure~\ref{fig:resultshists}. The total number of fits in the core bands is lower in \texttt{v05} than in \texttt{v04} due to the joint fitting: in \texttt{v04} we show all fits obtained to individual images of the same galaxy, while in \texttt{v05} there is only one fit per galaxy. For all other bands, the total numbers are lower due to the increased fractions of skipped fits and outliers, cf. Sections~\ref{sec:jointfitting} and~\ref{sec:outlierstats}. The number of objects shown for individual components varies greatly due to the differences in the model selection between bands (Sections~\ref{sec:manualcalibrationchanges} and~\ref{sec:modelseldiffs}), where - as for \texttt{v04} - the brightest and largest objects, often with high single S\'ersic indices, are most frequently decomposed successfully. 

Magnitudes become systematically brighter as a function of wavelength due to typical galaxy colours. Effective radii and S\'ersic indices do not show obvious trends as a function of wavelength, with no indications of inconsistencies in the analysis between different datasets or bands. For all bands, disks tend to be brighter than bulges, with the bulge magnitudes showing a double-peaked nature due to the 1.5- and double component fits. Disks are also typically larger than bulges. The bulge S\'ersic indices again show a large range of values, although with a slight trend towards higher values for longer wavelength bands. 

\begin{figure}[t!]
\begin{center}
    \includegraphics[width=0.8\textwidth]{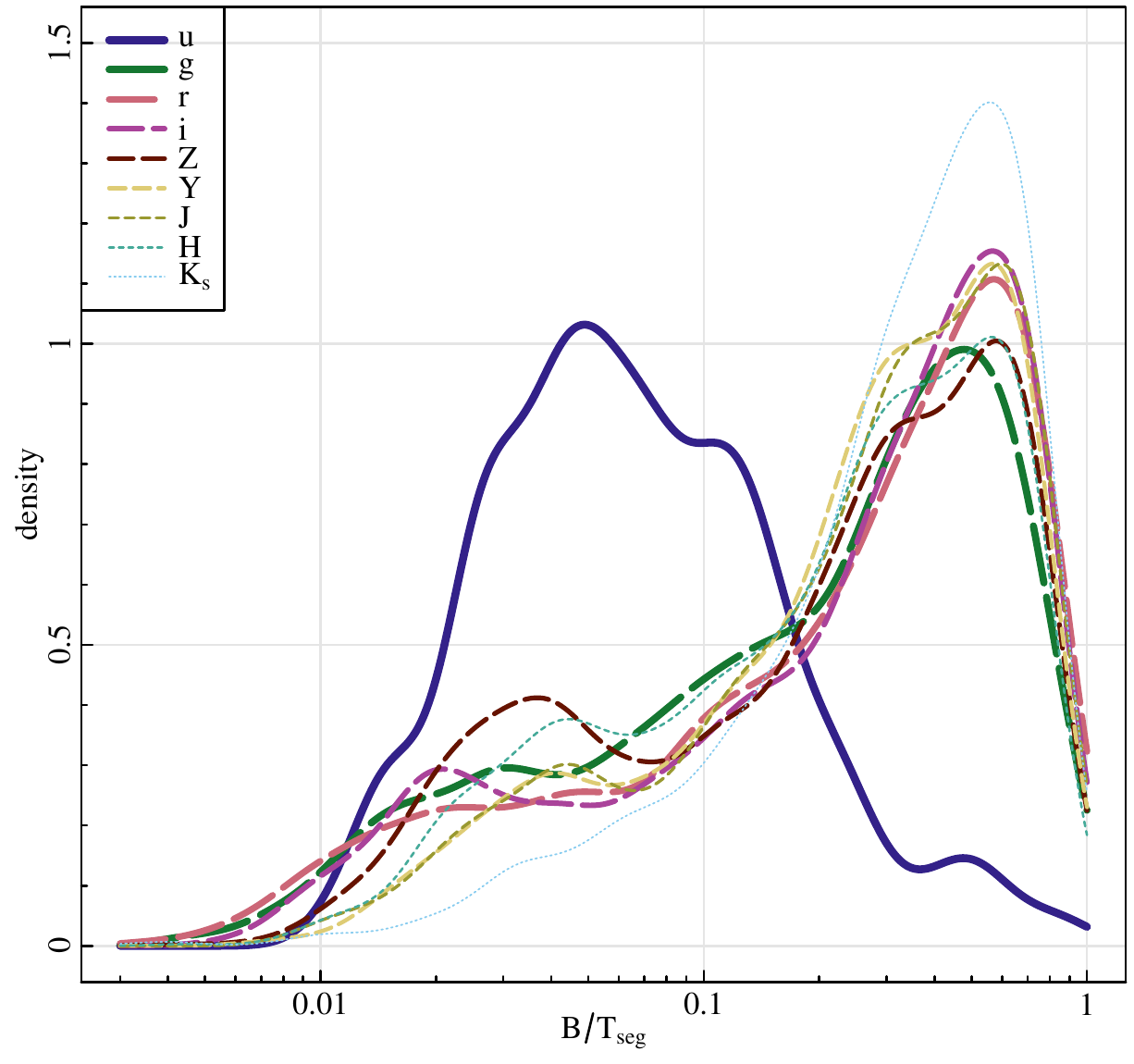}
    \caption{The distribution of the bulge to total flux ratio (limited to segment radii) for the 1.5- and double component fits in all bands in \texttt{v05}. This figure is equivalent to Figure~\ref{fig:BThist}, except that we now show density plots instead of normalised histograms for better visibility due to the higher number of bands.}	
      \label{fig:BThistv05}
\end{center}
\end{figure}

The \texttt{v05} equivalent to Figure~\ref{fig:BThist} is shown in Figure~\ref{fig:BThistv05}. It shows the distribution of the bulge to total flux ratio for all nine bands as a density plot on a logarithmic $x$-axis scale. The general trend observed in Figure~\ref{fig:BThist} persists: the B/T ratio increases as a function of wavelength. The double-peaked nature is again due to the 1.5- and double component fits, where the former populate the region of very low B/T values and the latter the intermediate and high values. The $u$-band is very different from the others since it has a factor of $\sim$\,10 less B/T ratio measurements available than the core bands; and almost all of those are 1.5-component fits (see first row in Figure~\ref{fig:resultshistsv05}). Notable exceptions from the general trend of increased B/T with wavelength are observed for the $Z$ and $H$ bands, both of which have exceptionally high numbers of 1.5-component fits relative to the double component fits (see relative numbers in rows 5 and 8 in Figure~\ref{fig:resultshistsv05}; also visible as larger ``bumps" at low magnitudes). This results in lower B/T values on average for these bands.

\begin{figure}[t!]
\begin{center}
    \includegraphics[width=0.8\textwidth]{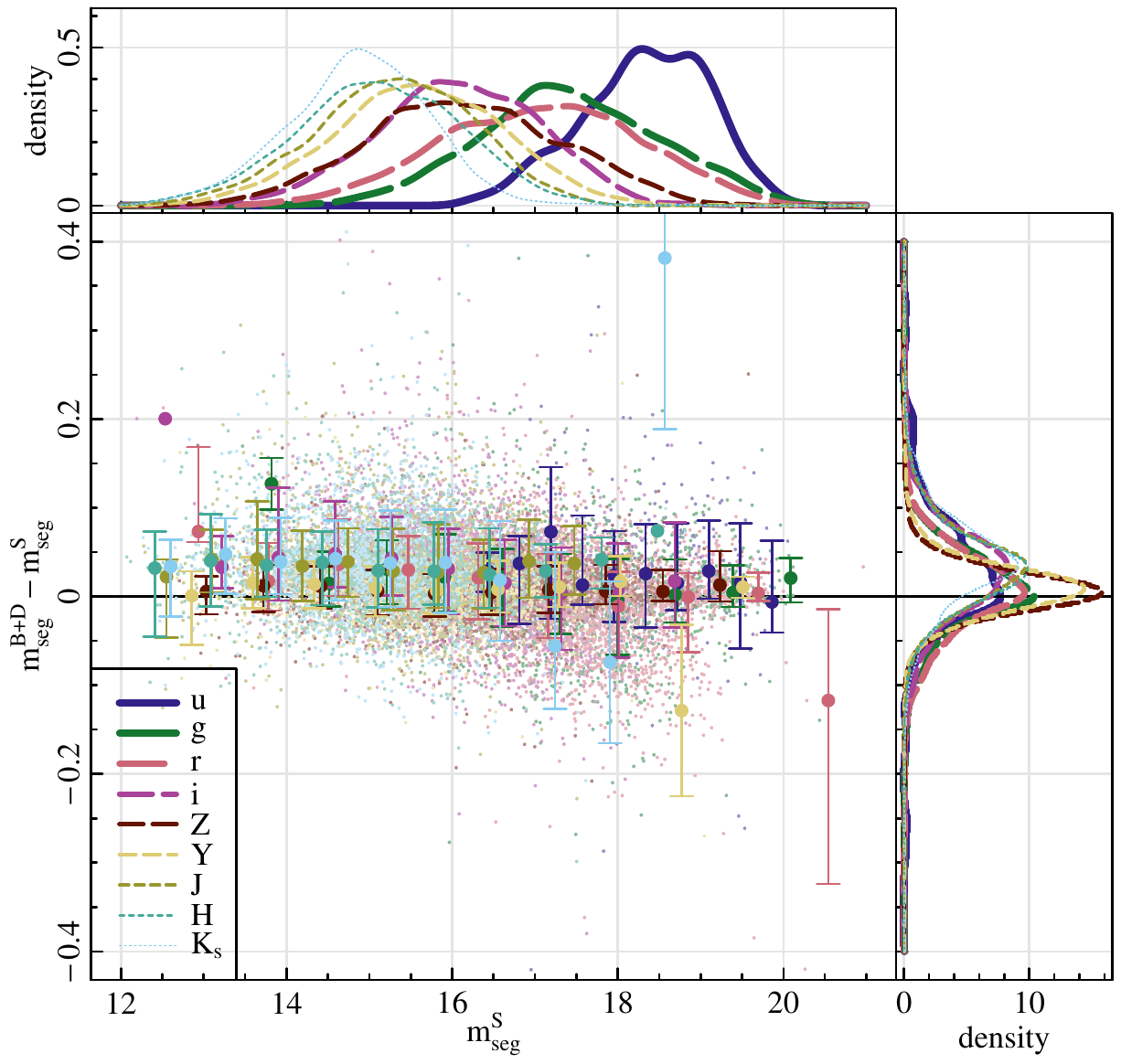}
    \caption{The difference between the single S\'ersic magnitude and the total magnitude derived from the double or 1.5-component fits for those galaxies that were classified as such, for all bands in \texttt{v05} (all magnitudes limited to segment radii). The scatter plot shows the difference between the two magnitudes against the single S\'ersic magnitude for all bands with the running medians and 1$\sigma$-percentiles overplotted. The top and right panels show the respective marginal distributions. This figure is equivalent to Figure~\ref{fig:magrecovery} for \texttt{v04}.}	
      \label{fig:magrecoveryv05}
\end{center}
\end{figure}

Figure~\ref{fig:magrecoveryv05} shows the last of the \texttt{v05} equivalents to Section~\ref{sec:v04parameterdistributions}, namely the difference between the single S\'ersic magnitudes and the total magnitudes derived from the sum of the bulge and disk fluxes (corresponding \texttt{v04} version in Figure~\ref{fig:magrecovery}). For all bands, these two versions of the total magnitude are consistent within 0.1\,mag for the vast majority of galaxies over the entire magnitude range. There is, however, a slight positive bias especially for the $u, i, J, H$ and $K_s$ bands (see the histogram on the right of Figure~\ref{fig:magrecoveryv05}), i.e. the single S\'ersic magnitudes tend to be slightly brighter than the total magnitudes obtained from adding the bulge and disk fluxes. A hint of this is also visible in Figure~\ref{fig:magrecovery}.

\begin{figure}
    \includegraphics[width=1\textwidth]{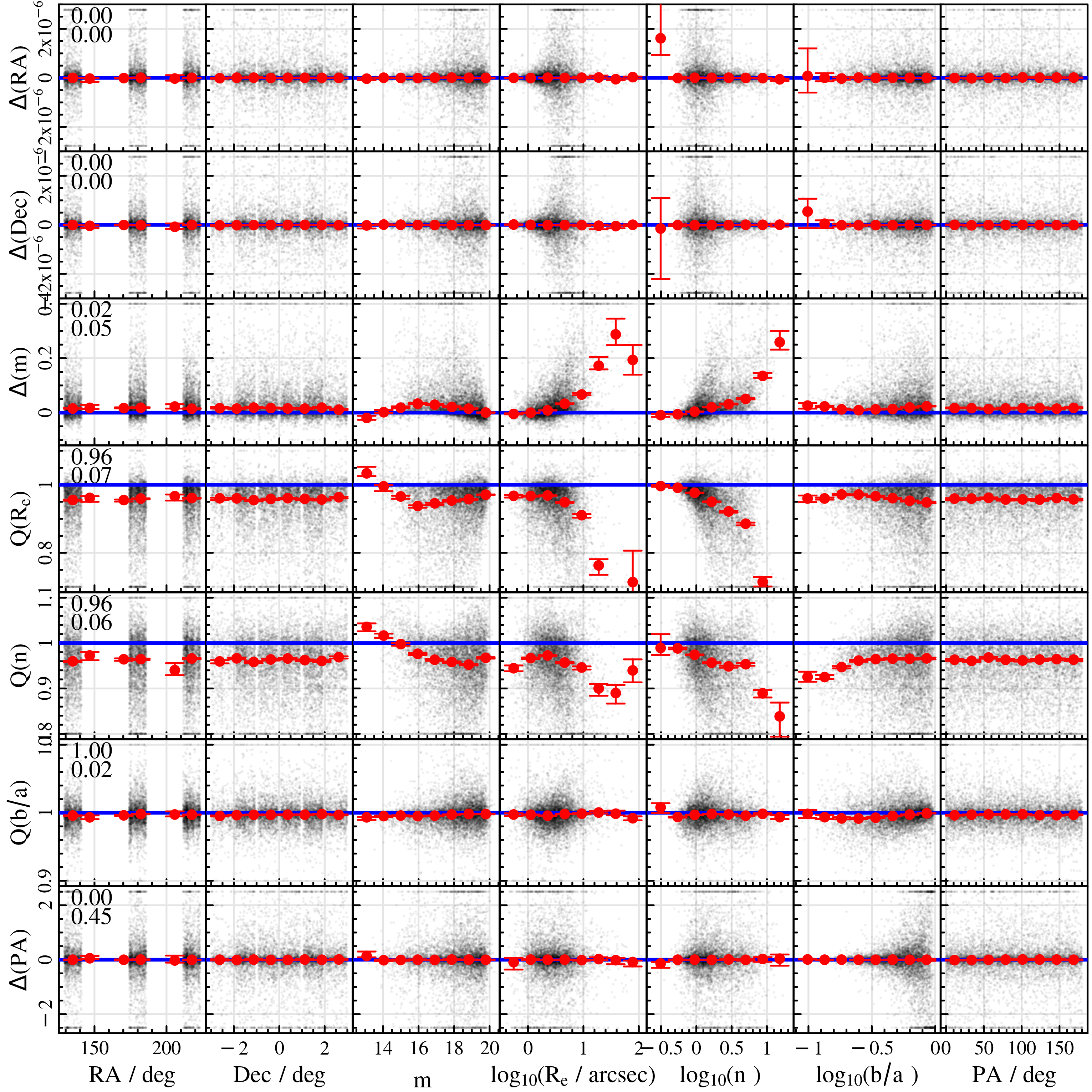}
    \caption{The difference $\Delta$ or quotient Q (for scale parameters) between the \texttt{v05} and \texttt{v04} fits plotted against the \texttt{v04} fits for all single S\'ersic parameters in the $r$-band. Black dots show all fits that had a single data match and were neither skipped nor flagged as outliers in either catalogue, red dots with error bars show the running median and its error in evenly spaced bins and horizontal blue lines indicate no difference between the fits. The numbers in the top left corners of the first row of panels show the median and 1$\sigma$-quantile of the respective distribution in the $y$-direction (which is identical for all panels of a row). Results for the $g$ and $i$ bands are not shown since they are almost indistinguishable from those in the $r$-band.}
\label{fig:resultsv04vsv05_R_S}
\end{figure}

\begin{figure}[t!]
\begin{center}
    \includegraphics[width=0.5\textwidth]{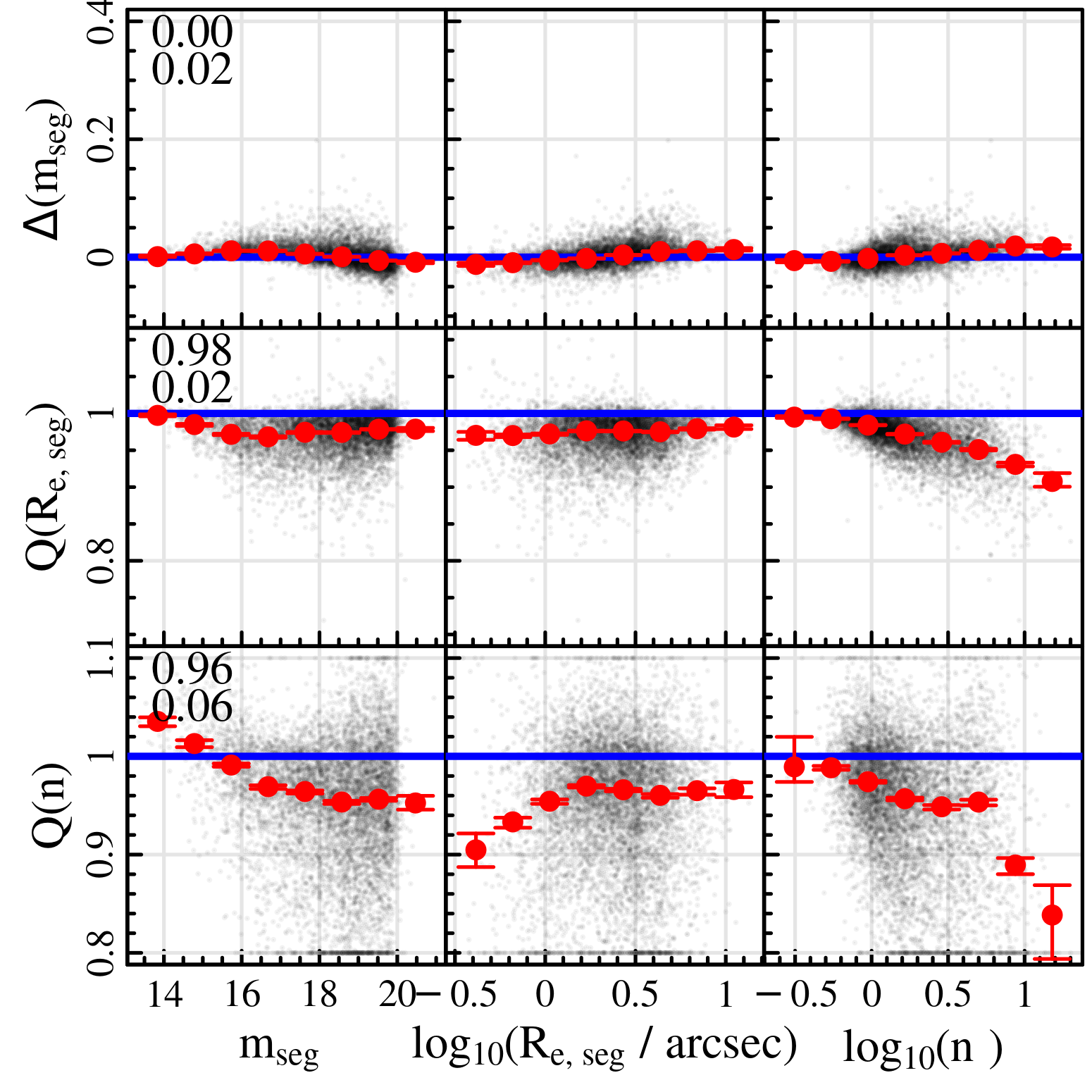}
    \caption{The same as Figure~\ref{fig:resultsv04vsv05_R_S}, but for the three main parameters magnitude, effective radius and S\'ersic index only, truncating the former two to the \texttt{v04} segment radii for both catalogues. The axis scales are the same as in Figure~\ref{fig:resultsv04vsv05_R_S} (central three by three panels).}
\label{fig:resultsv04vsv05_R_S_SEGRAD}
\end{center}
\end{figure}

\begin{figure}[hb!]
\begin{center}
    \includegraphics[width=0.5\textwidth]{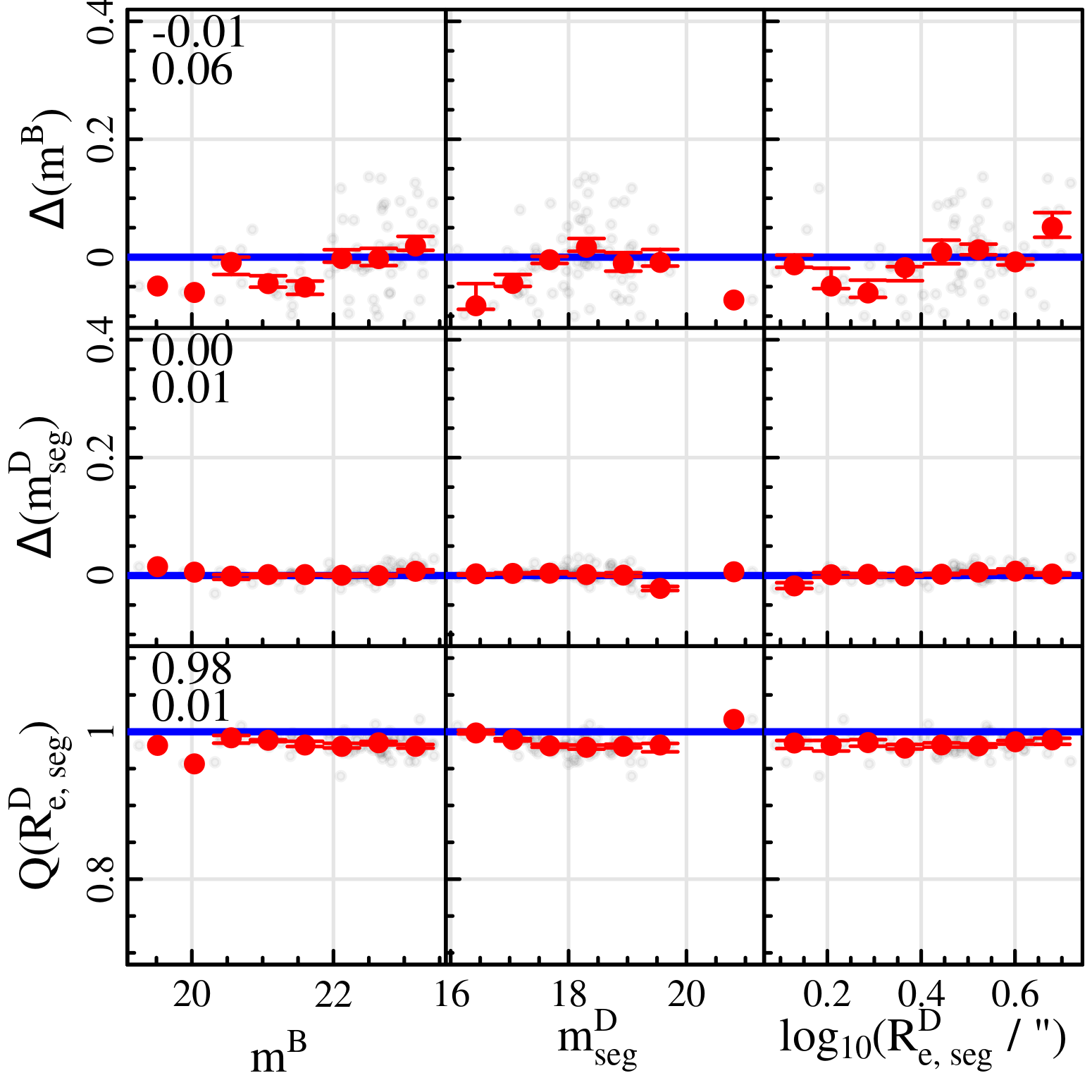}
    \caption{The same as Figure~\ref{fig:resultsv04vsv05_R_S}, but for the main parameters of the 1.5-component fits (bulge and disk magnitudes and disk effective radius); all truncated to the \texttt{v04} segment radii for both catalogues (except for the point source magnitude since the point source flux is entirely contained within the segment by definition). The axis scales are the same as those for the corresponding single S\'ersic parameters in Figures~\ref{fig:resultsv04vsv05_R_S} (central three by three panels) and~\ref{fig:resultsv04vsv05_R_S_SEGRAD}. The sample is limited to objects classified as 1.5-component fits in both catalogues in the $r$-band.}
\label{fig:resultsv04vsv05_R_P_SEGRAD}
\end{center}
\end{figure}

\begin{figure}[ht!]
\begin{center}
    \includegraphics[width=0.8\textwidth]{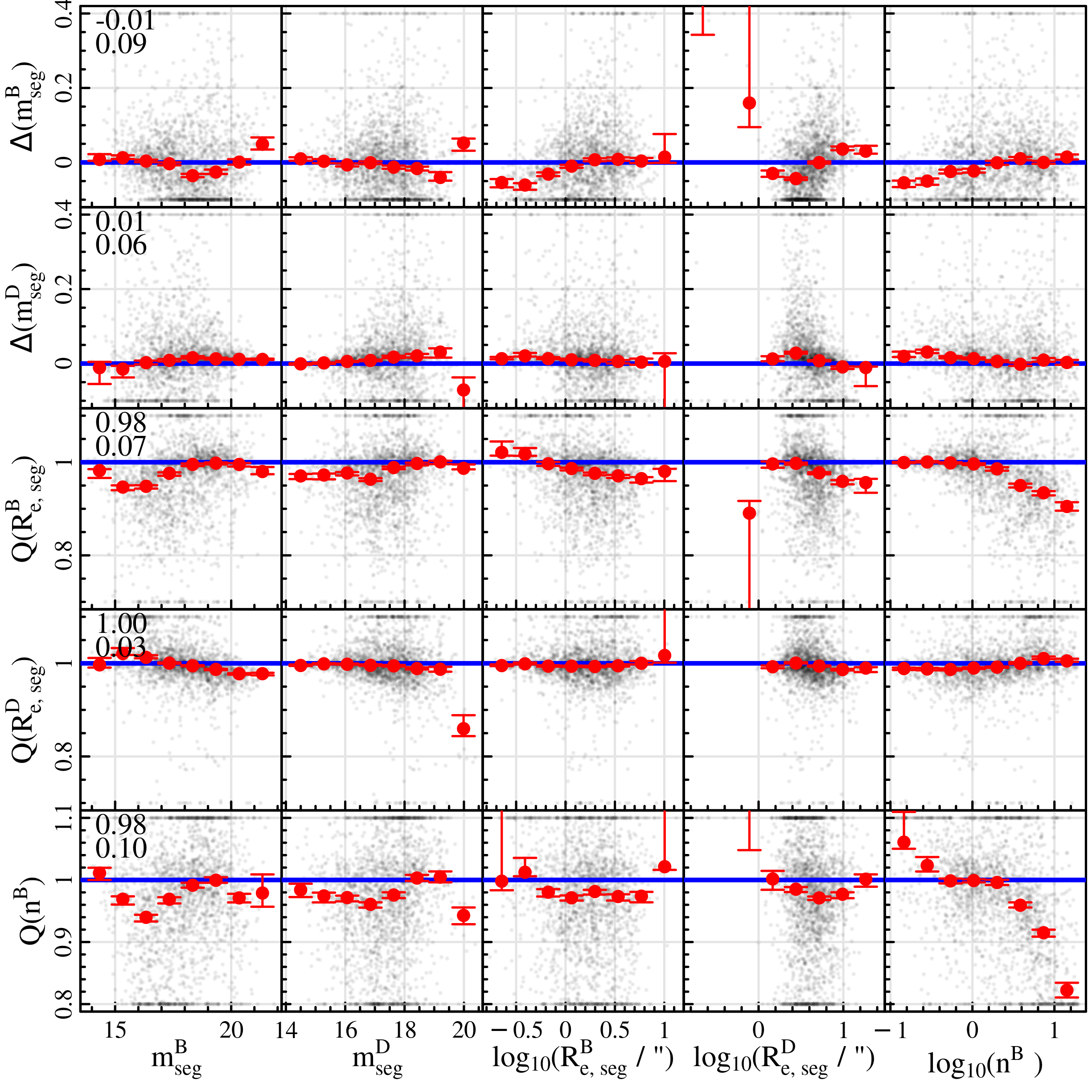}
    \caption{The same as Figure~\ref{fig:resultsv04vsv05_R_S}, but for the main parameters of the double component fits (bulge and disk magnitudes, bulge and disk effective radii and bulge S\'ersic index); with magnitudes and effective radii truncated to the \texttt{v04} segment radii for both catalogues. The axis scales are the same as those for the corresponding single S\'ersic parameters in Figures~\ref{fig:resultsv04vsv05_R_S} (central three by three panels) and~\ref{fig:resultsv04vsv05_R_S_SEGRAD}. The sample is limited to objects classified as double component fits in both catalogues in the $r$-band.}
\label{fig:resultsv04vsv05_R_D_SEGRAD}
\end{center}
\end{figure}

To finish this Section, Figures~\ref{fig:resultsv04vsv05_R_S} to~\ref{fig:resultsv04vsv05_R_D_SEGRAD} provide a direct comparison of the fitted parameters in \texttt{v04} and \texttt{v05} in the $r$-band. For all single S\'ersic parameters, we show the difference between the two fits (\texttt{v05} - \texttt{v04} for position RA and Dec, magnitude $m$, and position angle PA; and \texttt{v05} / \texttt{v04} for the scale parameters effective radius $R_e$, S\'ersic index $n$, and axial ratio $b/a$) against the \texttt{v04} fits (in logarithmic units for scale parameters) in Figure~\ref{fig:resultsv04vsv05_R_S}. Running medians and 1$\sigma$ quantiles are shown as red points with error bars with numerical values indicated in the top left corners of each row. The sample is limited to those objects that had a single data match in the core bands and were neither skipped nor flagged as outlier in either catalogue. The corresponding $g$ and $i$ band plots are nearly indistinguishable with only very slightly increased scatter in the S\'ersic index (1$\sigma$-quantiles of 0.08 for both bands) and position angle (1$\sigma$-quantiles of 0.53 in $g$ and 0.55 in $i$), so we do not show them. 

Galaxy positions in RA and Dec (top two rows of Figure~\ref{fig:resultsv04vsv05_R_S}) are recovered near-perfectly (the $y$-axis range corresponds to 0\farcs01; or 5\,\% of a pixel). The only prominent feature are the three GAMA II equatorial survey regions that are clearly visible in RA. The small gaps in the distribution in Dec at intervals of 1\degr\ (the KiDS tile size) are due to the limitation to objects with a single match, excluding the overlap sample. There are no deviations of RA or Dec as a function of any other parameter (first two rows); nor does any other parameter show systematic trends as a function of RA and Dec (first two columns). Similarly consistent results between \texttt{v04} and \texttt{v05} are obtained for the axial ratio and position angle (bottom two rows; and last two columns), which agree to within $\sim$\,2\,\% and half a degree respectively. 

The three main parameters (central three by three panels in Figure~\ref{fig:resultsv04vsv05_R_S}) show more deviations and also systematic trends: magnitudes are on average 0.01\,mag fainter in \texttt{v05} than in \texttt{v04}; while effective radii and S\'ersic indices are 4\,\% and 3\,\% smaller respectively. Bright and large objects with high S\'ersic indices are particularly strongly affected. These differences are caused by the larger segments in \texttt{v05} as explained in Sections~\ref{sec:postprocessing} and~\ref{sec:segchoices}. We also investigate this effect further in Section~\ref{sec:comparelee}, where we compare the \texttt{v04} results to previous work. 

Truncating the effective radius and magnitude to the \texttt{v04} segment radii (the smaller segments) for both catalogue versions removes the systematic trends in those parameters as shown in Figure~\ref{fig:resultsv04vsv05_R_S_SEGRAD}. The S\'ersic index remains different between the catalogue versions since it cannot be corrected for different segment sizes. For reasons outlined in Section~\ref{sec:v05segmentation}, we still opted for slightly larger segments in \texttt{v05}.

For completeness, Figures~\ref{fig:resultsv04vsv05_R_P_SEGRAD} and~\ref{fig:resultsv04vsv05_R_D_SEGRAD} show the analogous plots for the 1.5- and double component fits, for objects that were classified as such in both catalogues. We show only the three main S\'ersic parameters (for bulges and disks where relevant) and limit magnitudes and effective radii to segment radii. The sample sizes are much smaller (especially in the 1.5-component plot), and the scatter generally increases, in particular for the bulge parameters. The overall agreement between the \texttt{v04} and \texttt{v05} parameters remains high, with systematic trends mostly limited to S\'ersic indices. The average offsets in S\'ersic index decreases to 2\,\% (from 4\,\% for single S\'ersic fits) and the offset in the disk effective radius disappears entirely. This is most likely due to the generally better model fit of the double component model, since the differences only arise when the model cannot represent the data adequately (cf. Section~\ref{sec:postprocessing}).

\subsection{Galaxy and component colours}
\label{sec:colours}

\begin{figure}[t!]
    \includegraphics[width=\textwidth]{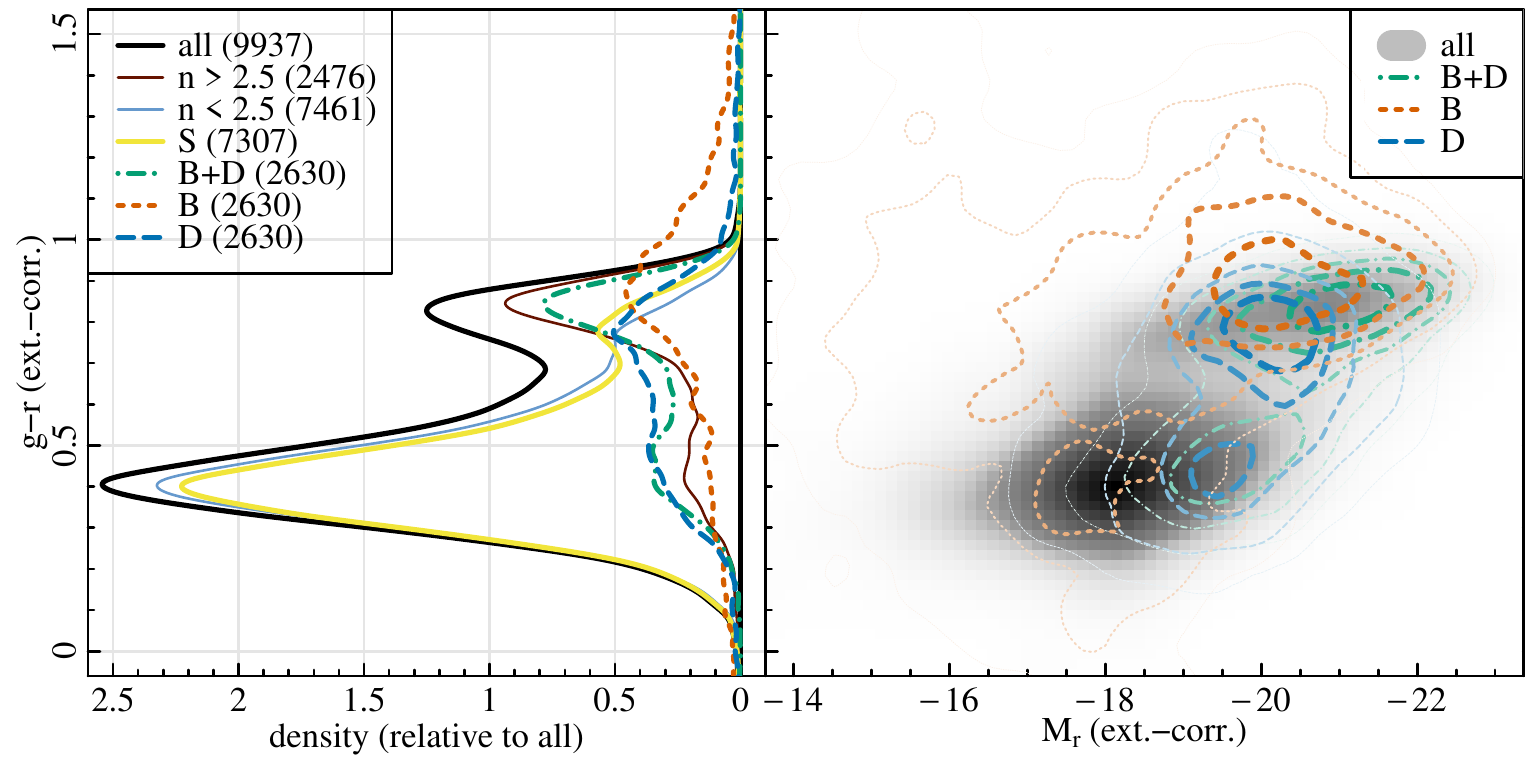}
    \caption{\textbf{Left panel:} The Galactic extinction-corrected $g-r$ colour distributions (limited to segment radii) for galaxies and their components. The colour coding of the lines is the same as for Figures~\ref{fig:resultshists} and~\ref{fig:resultshistsv05}, although with a few additions: The solid black line shows single S\'ersic fits for all galaxies with \texttt{NCOMP}\,>\,0 in the joint model selection; the thinner dark red and light blue solid lines split this sample into those with $n$\,>\,2.5 and $n$\,<\,2.5 in the $r$-band. The solid yellow line gives the single S\'ersic values for those galaxies which were classified as single component systems, dotted red and dashed blue lines show bulges and disks, respectively, for those objects classified as 1.5 or double component systems (always in the joint model selection). The dot-dashed green histogram gives the total galaxy colour (derived from the addition of bulge and disk flux) for 1.5 and double component systems. The number of objects in each histogram is given in the legend.
\textbf{Right panel:} The Galactic extinction-corrected $g-r$ vs. $M_r$ colour-magnitude diagram (limited to segment radii) for galaxies and their components. The colour coding of the lines is the same as for the left panel, although note we do not show the single S\'ersic contours here for clarity. Contours include 10, 25, 50, 75 and 90\,\% of the sample.}	
    \label{fig:colourplots}
\end{figure}

For the study of colours, we focus on the core bands since those are the deepest exposures and most directly comparable in our analysis. Additionally, they benefit from the joint $gri$ model selection, resulting in a relatively high number of objects with reliable bulge and disk magnitudes. To maximise similarity in terms of depth and seeing, we use $g-r$ colours and $r$-band absolute magnitudes, $M_r$. Results for $g-i$ and/or $M_i$ are qualitatively similar, albeit a bit more noisy. Using other bands reduces the number of available components considerably and further increases the scatter (which is already substantial for $g-r$ as we will see below). Following \citet{Casura2022}, we use \texttt{v04} of the \texttt{BDDecomp} DMU. Using \texttt{v05} instead does not change the results. 

The left panel of Figure~\ref{fig:colourplots} shows the distribution of $g-r$ colours for galaxies and their components. The colours are corrected for Galactic extinction, but not for dust attenuation in the emitting galaxy. The Galactic extinction was obtained from \texttt{v03} of the \texttt{GalacticExtinction} catalogue accompanying the equatorial input catalogue on the GAMA database.

The solid black line shows the colour distribution for all single S\'ersic fits that were not classified as outliers in the joint model selection (\texttt{NCOMP}\,>\,0). It is clearly bimodal, with redder colours typically belonging to higher S\'ersic index objects as indicated by the thinner dark red and light blue lines splitting the distribution at $n$\,=\,2.5 (in the $r$-band). Not entirely surprisingly (given the distribution of S\'ersic indices in Figure~\ref{fig:resultshists}), the distribution of single S\'ersic objects actually classified as such (\texttt{NCOMP}\,=\,1, solid yellow line) mostly follows the distribution of low S\'ersic index objects; while the high S\'ersic index objects tend to be classified as double component systems. For the latter, we show total colours with a dash-dotted green line, bulge colours with a dotted red and disk colours with a dashed blue line. As expected, bulges tend to be redder than disks, although the scatter is large. 

The right panel of Figure~\ref{fig:colourplots} shows the corresponding colour-magnitude diagram. Colours and absolute magnitudes are both corrected for Galactic extinction but not for dust at\-tenua\-tion in the emitting galaxy. The absolute magnitude was calculated using the distance modulus provided in v14 of the \texttt{DistancesFrames} catalogue from the GAMA database which we also used to obtain redshifts for the sample selection. 

The grey density plot in the background shows the single S\'ersic fits for all non-outlier (\texttt{NCOMP}\,>\,0) galaxies, corresponding to the black line in the left panel of Figure~\ref{fig:colourplots}. The bimodality of the distribution is even clearer here, with the red sequence and blue cloud being well-separated. The green contours indicate the part of the sample that was classified as 1.5 or double component object\footnote{To be precise, the green contours were derived by adding the respective bulge and disk fluxes of the 1.5 or double component objects (for consistency with the bulge and disk contours), while the grey density plot is based on the single S\'ersic fits (for robustness at low magnitudes). These two versions of the total galaxy magnitude are generally very similar as evidenced by Figure~\ref{fig:magrecovery}.}: as expected, this is concentrated towards the bright end of the galaxy distribution and hence encompasses mostly galaxies located in the red sequence. Correspondingly, bulges and disks are both relatively red, with bulges on average slightly redder than the total galaxies and disks slightly bluer (while both components - obviously - are fainter than the total galaxy). However, both components show a large scatter and overlap with each other: both faint blue bulges exist as well as bright red disks. 

A detailed study of component colours and the different populations in the right panel of Figure~\ref{fig:colourplots} is beyond the scope of this thesis. However, we note that the total galaxy colours show much less scatter; indicating that the scatter results from a different splitting of the light into bulge and disk components in the $g$ and $r$ bands, while the total amount of light is well-constrained (cf. also Figure~\ref{fig:magrecovery}). A further brief investigation into extreme systems (blue bulges with red disks and also excessively red bulges with very blue disks) suggests that they are caused by a variety of remaining uncertainties in our analysis, e.g. swapped components in one of the two bands (Section~\ref{sec:galaxyfitting}), small faint bulges that are barely detected in the $r$-band and missed in $g$, the ``bulge" component dominating both small and large radii in one of the two bands (cf. Section~\ref{sec:modelselectioncaveats}) or failures in the flagging of bad fits, all combined with model selection uncertainties and the necessity of joint model selection to compromise between the bands. While each of these processes by itself only affects a small number of galaxies, in sum across both bands they do reach the 10-20\,\% level. Still, on average our colours do follow the expected trends, as we show in Section~\ref{sec:prevcols} with an overview of similar studies in the literature. We will study the colours of galaxies and their components in more detail in forthcoming work, also including the other bands ($uZYJHK_s$) and taking full account of inclination effects due to dust in the emitting galaxies \citep[see, e.g.][]{Driver2008}. We will then also assess trends in other parameters, such as the component effective radii, with wavelength; and use these to constrain the nature and distribution of dust in galaxy disks.

\section{Catalogue limitations}
\label{sec:cataloguelimitations}

We finish this chapter by pointing out a few limitations of our results that users of the catalogue should be aware of. Most of these have been discussed before and are only summarised here, with references to the relevant sections. 

\subsection{Model limitations}

All of our models are axially symmetric and monotonically decreasing in intensity from the centre. We are unable to capture asymmetries such as spiral arms, offset bulges, tidal tails, mergers, star-forming regions etc.; or disk features such as rings, bumps, truncations or flares. If such features are present in the data, they may bias or skew the model parameters. We also remind the reader that when we talk about ``bulges", what we really mean are the central components. This could be a classical bulge, a pseudo-bulge, an AGN, a bar, or any combination (sometimes resulting in the model trying to fit a mixture between e.g. a bar and a bulge). We make no attempt to distinguish between these cases. 

\subsection{Model selection caveats}
\label{sec:modelselectioncaveats}

Model selection is accurate to >\,90\,\% compared to what could be achieved
by visual classification (Section~\ref{sec:postprocessing}). However, it is important to note that our aim in the model selection is to determine which one of our three models is most appropriate to use for the given data; and not how many physically distinct components an object consists of. The reason for this is that for a given galaxy, the data quality will strongly influence how many fitting parameters can be meaningfully constrained and using more model parameters will inevitably overfit the data and lead to unphysical results. Hence, even in the joint model selection, we base our visual classification on the fit and residuals in individual bands (which is what we fit to), rather than e.g. colour images. Due to the different depths and resolutions of the bands, it is common for the same galaxy to be classified as double component in one band, but single component in another (cf. also discussions in Sections~\ref{sec:manualcalibrationchanges} and~\ref{sec:modelseldiffs}).  

In an attempt to make fitting parameters more directly comparable across bands, we introduced the joint $gri$ model selection (Section~\ref{sec:postprocessing}), yet this is necessarily a compromise between the different bands. For example, we lose bulges that are resolved in the $r$-band but not in $g$ and $i$ due to the larger PSFs; or there may be some ill-constrained $i$-band fitting parameters for an extended low-surface brightness disk that is visible in $r$ and $g$ but not in the shallower $i$-band image. There are also more skipped fits and outliers in the joint model selection than in the band-specific ones because all objects that are skipped or flagged in at least one of the three bands are skipped or flagged in the joint model selection. These problems become significantly worse for the nine-band joint model selection encompassing all of $ugriZYJHK_s$ due to the larger range in depth and seeing for individual bands.

Irrespective of the result of the model selection, we provide all fitted parameters of all models in the catalogue (along with the postage stamps of all fits and a flag indicating the preferred model). This allows users to perform their own selection if desired; but also requires care as not all provided parameters will be meaningful. While single S\'ersic fits to double component objects are mostly reasonable; double component fits to true single component galaxies will have unconstrained and potentially unphysical parameters for at least one of the components.

\begin{figure}
\begin{center}
	\includegraphics[width=0.8\textwidth]{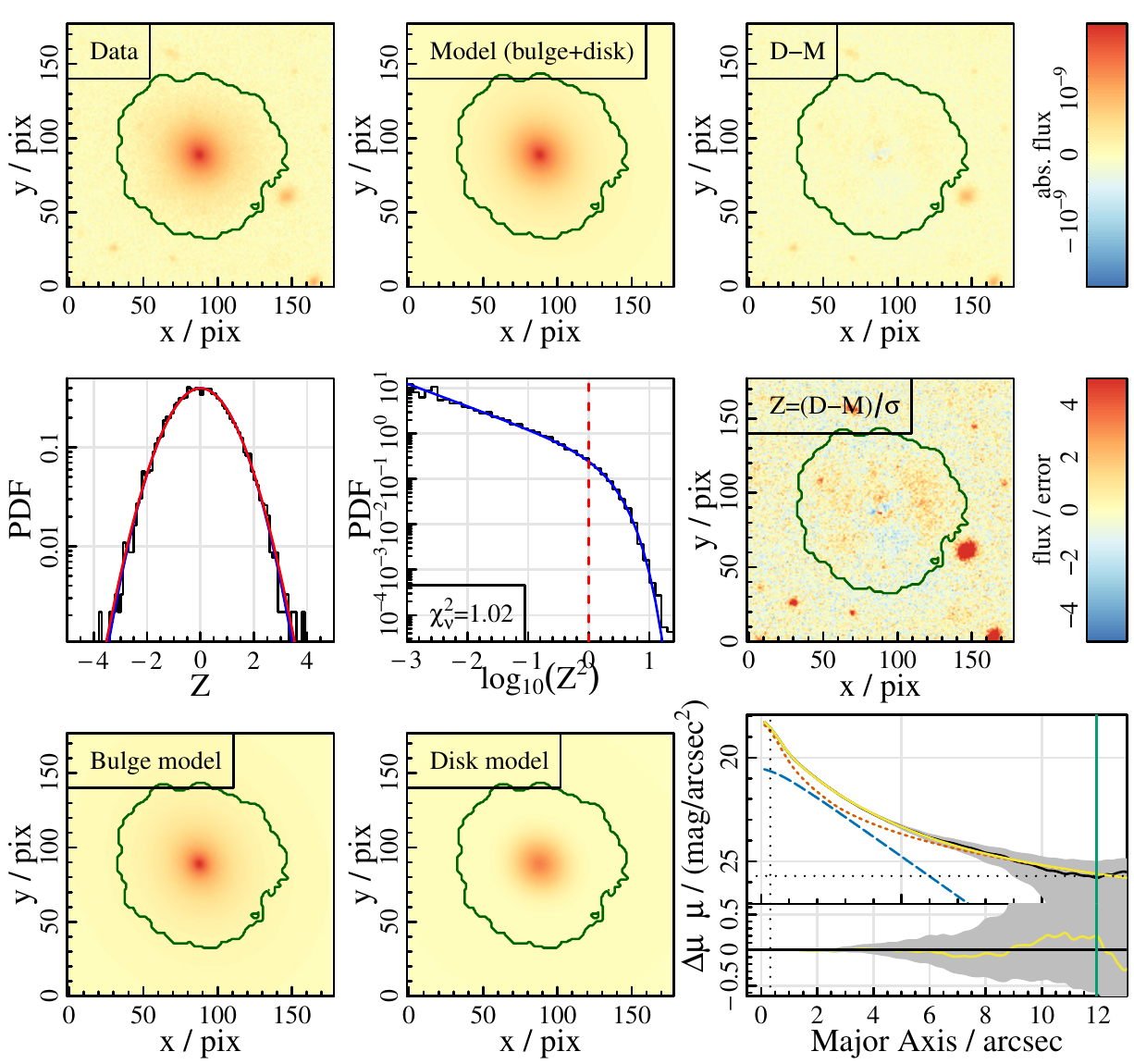}
    \caption{The double component fit to galaxy 549706, classified as double component object but with a very high B/T ratio of 0.71 in the KiDS $r$-band. Panels are the same as those in Figure~\ref{fig:examplefit}.} 
    \label{fig:examplefithighbt}
\end{center}
\end{figure}

We are also aware of a population of objects that are classified as double component fits but have the bulge component dominating both the centre and the outskirts, with the disk only dominating at intermediate radii or even staying ``below" the bulge at all radii. We believe these are essentially single component systems that do not follow a S\'ersic law (e.g. S\'ersic index would be higher at centre than outskirts); and so the freedom of the disk is used to offset this. This population is easily identifyable by the high bulge-to-total ratio (B/T\,$\gtrsim$\,0.6 or 0.7). The single S\'ersic fits may be more appropriate to use in these cases \citep[see also the discussion of this issue in][]{Allen2006}. An example is shown in Figure~\ref{fig:examplefithighbt}.

\subsection{Drawbacks of tight fitting segments}

As detailed in Section~\ref{sec:postprocessing}, we use relatively tight segments around the galaxies for fitting, which results in the best possible fit of the inner regions of the galaxy but can lead to large, unphysical wings. Hence we recommend using only integrated properties, i.e. the summed flux/magnitude within the region that was fitted and the corresponding effective radii and bulge-to-total ratios as given by the corresponding \texttt{*\_SEGRAD} properties in the catalogue. For comparisons to other catalogues using larger fitting segments, their profiles should also be appropriately truncated (see details in Sections~\ref{sec:postprocessing}, ~\ref{sec:segchoices} and \ref{sec:comparelee}). 

\subsection{Sources of systematic uncertainties}

We provide errors for each fitted parameter in the catalogue including our best estimate of systematic uncertainties taken from Table~\ref{tab:errorunderestimate}. However, we do not apply the (small) bias corrections given in the same table, since they are only applicable to large random samples of our galaxies and not to individual objects. In addition, we would like to point out that the systematic errors were estimated from single S\'ersic $r$-band fits. We expect that individual components as well as other bands are affected by similar systematics, but we did not test for this. Also, there are some systematic uncertainties that we do not account for in our simulations, most obviously galaxy features that cannot be captured by our models. For these reasons, the given errors should still be considered as lower limits of the true errors. 

\subsection{GAMA-KiDS RA/Dec offset}

We observed an average offset between the input and output (fitted) positions of galaxies in both RA and Dec of approx. 0.4\,pix (0$\farcs$08). This is due to an offset between the GAMA (SDSS) and KiDS positions; the same offset can
be seen when comparing the KiDS source catalogue with the Gaia catalogue; see also figure~15 in \citet{Kuijken2019}. We correct for this during the outlier rejection, but give the original (uncorrected) fitted values for position otherwise.

\subsection{Completeness limits}

Due to our sample selection (Section~\ref{sec:sampleselection}), our spectroscopic completeness is 100\,\% and even the faintest objects in our sample are well-resolved and bright enough to allow for robust single S\'ersic fits in the core bands. However, this is not the case for the KiDS $u$ and the longest wavelength VIKING bands, as discussed in Section~\ref{sec:outlierstats}: in the shallowest bands ($K_s$ and $u$), we lose approximately 25\,\% of our sample entirely; and robust double component fits can only be obtained for a minority of objects. Also, there is a systemic limit to the component magnitude in that the samples of bulges and disks with magnitudes fainter than the GAMA limit ($r$\,<\,19.8\,mag) are incomplete. For example, a bulge with a magnitude of 22\,mag in the $r$-band will only be contained in our sample if the corresponding disk is bright enough such that the total magnitude is below 19.8\,mag. Hence, the sample of bulges with 22\,mag is incomplete. This applies almost exclusively to the faint bulges from the 1.5-component fits as can be seen in the first column of Figures~\ref{fig:resultshists} and~\ref{fig:resultshistsv05}.

\clearpage
\newpage
\chapter{Quality control}
\label{chap:QC}

After presenting some of the contents of our catalogue of bulge-disk decompositions, we now turn towards demonstrating its robustness. For this, we first compare to previous works in Section~\ref{sec:prevwork}, including work on the same galaxy sample. Section~\ref{sec:simulations} then describes additional internal consistency checks and a detailed study of biases and systematic errors with bespoke simulations. Both of these sections are taken from \citet[their sections~5 and~6]{Casura2022} and are focused on the fitting results of \texttt{v04} of the \texttt{BDDecomp} catalogue. This has the great advantage that duplicate images of the same galaxy were fitted independently and therefore this overlap sample can serve to identify systematic uncertainties. The \texttt{v05} results have been anchored to those of \texttt{v04} in Chapter~\ref{chap:results}, such that the quality control presented here is applicable to that version as well and we do not repeat the analysis for \texttt{v05}. Quality control metrics and diagnostic plots of individual steps of our pipeline, in particular for the preparatory work, have been presented in Section~\ref{sec:pipelinedevelopment}. 

\section{Comparison to previous works}
\label{sec:prevwork}

This section begins the quality control by comparing to previous works. We start with an overview of the catalogue statistics and colours to corresponding literature values; then turn towards a more detailed comparison with the results from \citet{Lange2015} and \citet{Kelvin2012}, who also use GAMA galaxies. In addition, comparisons to \texttt{v04} of our catalogue are provided in \citet[their figure~B2]{Haeussler2022} for the single S\'ersic and double component models and in \citet[their figures~19 and~20]{Robotham2022} for single S\'ersic fits, both finding reasonable agreement.

\subsection{Comparison of catalogue statistics}

As a first check, we compare our model selection statistics to those of other bulge-disk decomposition works, although care must be taken in judging these results since they will depend on the sample selection, data quality and observational band. 

\renewcommand{\arraystretch}{1.5}
\begin{table}[t!]
	\centering
	\caption{Comparison of our \texttt{v04} catalogue statistics to previous works (in the $r$-band unless stated otherwise). All values are given in percent.}
	\label{tab:stats}
	\setlength{\tabcolsep}{3pt}
	\begin{tabu}{p{0.26\textwidth}cccp{.43\textwidth}} 
	    \hline
	    reference & single & double & unsuitable & notes
	    \\ 
	    \hline
	    this work & 47 & 23 & 30 & double including 1.5-comp.; unsuitable including masking (20\,\%)\\
	    \citet{DominguezSanchez2022} & 42 & 55 & 3 & unsuitable corresponding to fit failure, single including galaxies where both single and double fits were acceptable\\
	    \citet{Hashemizadeh2022} & 48 & 45 & 7 & $I$-band; classification by prior visual inspection; difficult objects excluded in sample selection; double S\'ersic fits\\ 
	    \citet{Robotham2022} & 68 & 31 & 1 & stellar mass instead of light; simultaneous nine-band plus SED fit; unsuitable meaning fit failure; lower redshift limit \\
	    \citet{Barsanti2021} & 47 & 28 & 25 & cluster S0 galaxies\\
	    \citet{Dimauro2018} & 27 & 63 & 10 & mostly NIR filters; bright and massive galaxies only ($\log_{10}(M_*/M_\odot)$\,>\,10.3)\\
	    \citet{Lange2016} & 66 & 16 & 18 & selection based on visual morphology; unsuitable counting all flagged galaxies\\
	    \citet{Meert2015} & 44 & 39 & 17 & larger sample up to higher redshift but smaller magnitude range\\
	    \citet{Head2014} & 19 & 35 & 46 & $g$-band; early-type sample; more stringent criteria for ``good" fits\\
	    \citet{Lackner2012} & 35 & 29 & 36 & single corresponding to pure exponential or de Vaucouleurs; unsuitable corresponding to their ``S\'ersic" category\\
	    \citet{Simard2011} & 73 & 26 & 1 & unsuitable corresponding to failure rate of fitting routine; no selection of ``good" fits given\\
	    \citet{Allen2006} & 43 & 34 & 23 & $B$-band; unsuitable galaxies excluded through cuts in redshift, galaxy size and surface brightness \\
	    \hline	   
	\end{tabu}
 \end{table}
 \renewcommand{\arraystretch}{1}

Table~\ref{tab:stats} summarises the corresponding percentages including a few notes on the most important differences of the quoted works to ours (more details on the majority of these studies are given in Section~\ref{sec:prevcols}). In short, for the automated decomposition of large samples of galaxies in the $r$-band, most authors - including ourselves - class roughly half of all galaxies as being well-represented by a single S\'ersic model, with the other half split approximately evenly into double component fits and objects unsuitable for fitting with such simple models. This is also in broad agreement with the morphological classifications obtained by \citet{Driver2022}. 

\subsection{Comparison of component colours to literature}
\label{sec:prevcols}

$g-r$ colours of galaxy components, such as those we present in Figure~\ref{fig:colourplots} and Section~\ref{sec:colours}, are not found frequently in the literature, although a number of authors have presented bulge-disk decompositions in several bands. For example, \citet{Simard2011} perform bulge-disk decompositions for a large sample of galaxies in the SDSS $g$ and $r$ bands but only present colour-magnitude diagrams for the total galaxies (their figures 9 and 10). These are visually comparable to our total galaxy colours as indicated by the dot-dashed green contours in the right panel of Figure~\ref{fig:colourplots}. \citet{Mendel2014} add the SDSS $u$, $i$ and $z$ bands to the analysis of \citet{Simard2011} and present component masses in $ugriz$ but also do not study component colours. 

Similarly, \citet{Meert2015} present a large $r$-band catalogue which is extended to include the $g$ and $i$ bands in \citet{Meert2016}. Colour-magnitude diagrams, however, are again only presented for total galaxies, with the authors noting that component colours can be calculated from their catalogue but should be used with care since they are subject to large uncertainites. 

More recently, \citet{Dimauro2018} provide (UVJ) component colours in their catalogue but defer their study to future work; while \citet{Bottrell2019} present $ugriz$ colour-magnitude diagrams for total galaxies (colour-coded by B/T); but again not for individual components. \citet{DominguezSanchez2022} decompose a sample of $\sim$\,10\,000 galaxies in the $g, r$ and $i$ bands, but do not study the colours. 

Among the first to show component colours for a large sample of galaxies were \citet{Lackner2012} in their study of $\sim$\,70\,000 $z$\,<\,0.05 SDSS galaxies in the $g$, $r$ and $i$ bands. However, in contrast to our fits, their $g$ and $i$ band fits are not independent. Instead, in order to decrease the noisyness of the colours, the structural parameters are taken from the $r$-band and only the magnitude is adjusted. Additionally, \citet{Lackner2012} (along with e.g. \citealt{Mendel2014}) fix the S\'ersic index of the bulge to either 1 or 4 for their double component fits to limit the number of free parameters since the data is insufficient to constrain the bulge light profile. Keeping these differences in mind, their figure~32 showing the $g-r$ vs. $M_r$ colour-magnitude diagram for bulges and disks as contours superimposed on the greyscale background for all galaxies can be compared to the right panel of our Figure~\ref{fig:colourplots} (for a more detailed description of Figure~\ref{fig:colourplots}, see Section~\ref{sec:colours}). In general, both plots are very similar\footnote{For reference, the \citet{Lackner2012} cyan contours represent 6684 galaxies with a bulge S\'ersic index of 1 and the magenta contours show 14042 objects with a bulge S\'ersic index of 4. Also note that their $x$-axis is reversed with respect to ours.}: the grey background shows a large blue cloud and a well-separated red sequence. The double component fits populate the red sequence, green valley and the brighter part of the blue cloud. The bulges tend to be slightly redder than the red sequence but with a large scatter especially at the faint end. Disks spread from the red sequence towards the green valley with a smaller population also in the blue cloud. \citet{Lackner2012} also note the large scatter in colour for bulges in particular, despite their fitting constraints and lower reshift limit. Hence it is not surprising that even with the higher quality KiDS data and our new fitting routines, we get a large scatter in component colours, especially since we leave the bulge S\'ersic index free and perform independent fits in both bands. The latter can lead to very extreme colours since it is not guaranteed that the ``bulge" and ``disk" models actually fit the same features in both images (in particular when there are additional features present that are not fully captured by the models; see also Section~\ref{sec:colours}). 

\citet{Kim2016} found similar difficulties when performing $g$-band and $r$-band decompositions on $\sim$\,10\,000 large bright and approximately face-on SDSS galaxies. While they leave the S\'ersic index free as we do, the $g$-band structural parameters are again taken from the $r$-band fits with only the magnitudes adjusted. Despite this, they find it necessary to remove almost 40\,\% of their sample after fitting because they show excessively red bulge colours (combined with low B/T values in the $r$-band). After this cut, the $g-r$ vs. $M_r$ colour-magnitude diagram for bulges shown in their figure 7 is slightly less noisy than ours (Figure~\ref{fig:colourplots}), although still comparable. \citet{Kim2016} did not study the properties of the disks in their sample. 

One of the most direct comparisons to make is with \citet{Kennedy2016} who study GAMA galaxies in the G09 region (a subset of our sample) in the $ugrizYJHK$ bands from the SDSS \citep{York2000} and the United Kingdom Infra-red Telescope Infrared Deep Sky Survey (UKIDSS, \citealt{Lawrence2007}). They use the \texttt{MEGAMORPH} multi-band fitting method with \texttt{GALAPAGOS-2} and \texttt{GALFITM} \citep{Haeussler2013, Vika2013} to perform simultaneous S\'ersic plus exponential fits across all 9 bands. The structural parameters are constrained to be the same in all bands, with only the component magnitudes allowed to vary freely, therefore providing robust colours. While the paper focuses on studying $u-r$ colours, the corresponding catalogue on the GAMA database (\texttt{MegaMorph:MegaMorphCatv01}) contains the information for the fits in all 9 bands such that $g-r$ colours can easily be derived. 

\begin{figure}[t!]
\begin{center}
    \includegraphics[width=0.8\textwidth]{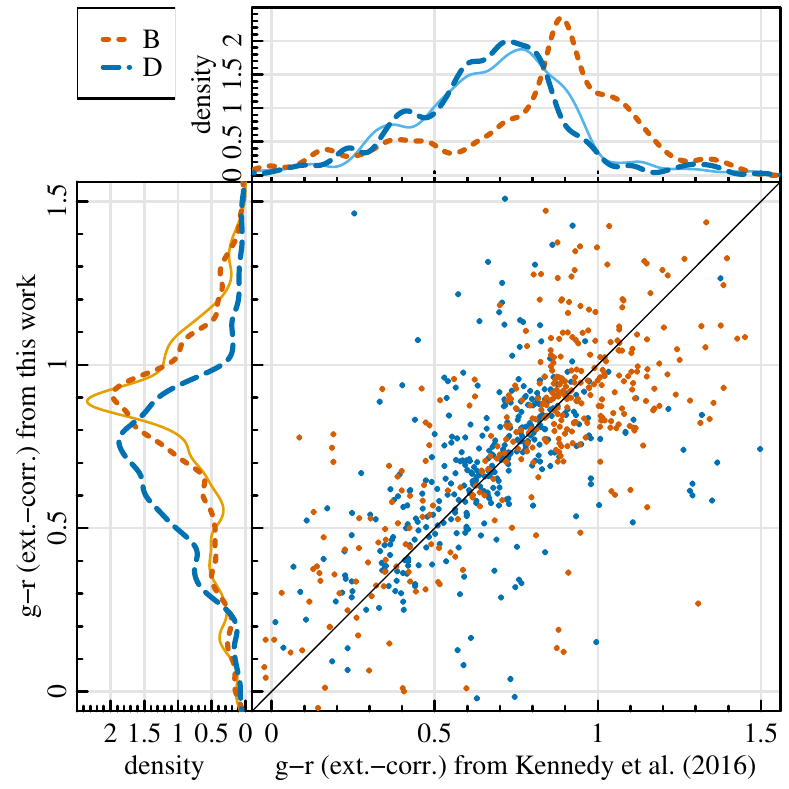}
    \caption{Our Galactic extinction-corrected $g-r$ component colours (limited to segment radii) compared against those from \citet{Kennedy2016} for a subsample of 390 objects that appear in both catalogues and were classified as double component fits in the joint \texttt{v04} model selection. The scatter plot shows the direct comparison, while the density plots show the respective distributions in both catalogues (ours on the left, \citealt{Kennedy2016} on the top). Bulges are again shown in red with dotted lines and disks in blue with dashed lines. To aid the direct comparison of the distributions, the lighter solid lines also show the \citet{Kennedy2016} bulge distribution on the left and our disk distribution on the top.}	
    \label{fig:colourvskennedy}
\end{center}
\end{figure}

This comparison is shown in Figure~\ref{fig:colourvskennedy} for those galaxies that were present in both catalogues and classified as double component fits (\texttt{NCOMP}\,=\,2) in the joint model selection of our fits (\citealt{Kennedy2016} perform neither model selection nor outlier rejection). In addition to the scatter plot directly comparing the component colours, we show the corresponding distributions in the left (this work) and top \citep{Kennedy2016} panels of Figure~\ref{fig:colourvskennedy}. As always, bulges are shown in red (points and dotted lines) and disks in blue (points and dashed lines). To aid the direct comparison of the distributions, we additionally show the \citet{Kennedy2016} bulge colour distribution in the left panel as a solid orange line and the disk distribution from this work in the top panel as a light blue solid line. Component colours are all corrected for Galactic extinction and limited to segment radii for our fits. 

Despite the large scatter, it can be seen that our component colours are generally in agreement with those from \citet{Kennedy2016} with no systematic differences. The scatter in both catalogues is also comparable, although \citet{Kennedy2016} perform multi-band simultaneous fits with fixed structural parameters that should lead to more robust component colours. This advantage of their work seems to be balanced by advantages of our work, such as the improved data quality of KiDS, the robustness of the fitting procedure with \texttt{ProFound} and \texttt{ProFit} and our post-processing steps (in particular outlier rejection and model selection). 

In addition to the large $g-r$ component colour studies discussed above, there are a number of publications focusing on the $g-i$ colours of bulges and disks for samples ranging between $\sim$\,100 and $\sim$\,1000 objects (i.e. roughly a factor of 10 smaller than ours), namely \citet{Gadotti2009, Head2014, Vika2014, Fernandez-Lorenzo2014, Cook2019, Barsanti2021}. We briefly compare our work to their results here, noting that all above authors have more stringent constraints on their fits than we do and also report problems in deriving bulge colours. For example, \citet{Fernandez-Lorenzo2014}, although they fit the galaxies in both bands, use fixed aperture photometry to derive more stable bulge colours. \citet{Vika2014}, while performing \emph{simultaneous} multi-band fits, do not allow for any variation of structural parameters (except magnitudes) with wavelength. \citet{Head2014}, in addition to varying magnitudes, allow for a trend in disk sizes with wavelength in approximately 30\,\% of their sample, noting that this leads to increased scatter. \citet{Cook2019}, who use \texttt{ProFit} like this work, allow disks to deviate slightly from the exponential profile but fix all bulges to be exactly round (axial ratio of 1), again only allow magnitudes and disk sizes to vary between bands and employ a sophisticated, visually-guided re-fitting procedure to obtain physically meaningful fits for ``difficult" objects. \citet{Barsanti2021}, also employing \texttt{ProFit}, additionally allow for differing bulge sizes and S\'ersic indices in the different bands (but fixing bulge and disk axial ratios and position angles and performing model selection in the $r$-band only), but class approximately half of their double component fits as ``unreliable". \citet{Gadotti2009}, fitting bulges, bars and disks to a sample of face-on, visually-selected ``well-behaved" galaxies refrain from automated fitting and instead treat each galaxy individually. 

After these notes on the inherent difficulties associated with deriving component colours, we can now turn to the corresponding results: \citet{Head2014}, in their study of early-type red-sequence galaxies in the Coma cluster, measure an average $g-i$ difference between bulges and disks of 0.09\,$\pm$\,0.01\,mag. Similarly, \citet{Barsanti2021} find a bulge-disk $g-i$ difference of 0.11\,$\pm$\,0.02\,mag for their sample of S0 cluster galaxies. \citet{Fernandez-Lorenzo2014}, on the other hand, have a sample of mostly late-type spirals (with B/T\,<\,0.1 for $\sim$\,66\,\% of their objects) and find a difference of 0.29\,mag in the median $g-i$ bulge and disk colours, i.e. a factor of $\sim$\,3 larger. In line with this, \citet{Vika2014} report that the bulge and disk colours are similar for early-type galaxies but differ significantly for late-types. The $g-i$ differences for the different morphological classes given in their table 2 range from 0.03\,$\pm$\,0.04\,mag for ellipticals to 0.28\,$\pm$\,0.06\,mag for late-type spirals; with the overall average (comprising approximately two thirds late-types) being 0.19\,$\pm$\,0.04\,mag. Similarly, the average $g-i$ colour difference of the \citet{Gadotti2009} sample of varying galaxy types amounts to 0.18\,$\pm$\,0.04\,mag (from the online-version of their table~2). 

Our results are perfectly in line with this: the median bulge-disk $g-i$ colour difference for our 1.5- or double component fits is 0.17\,$\pm$\,0.01\,mag, consistent with the \citet{Vika2014} and \citet{Gadotti2009} results. Limiting to objects with a total $g-i$\,>\,1 (red-sequence galaxies) reduces the value to 0.14\,$\pm$\,0.01\,mag; while focusing on 2-component fits only (excluding 1.5-component fits) yields 0.10\,$\pm$\,0.02\,mag, in agreement with \citet{Head2014} and \citet{Barsanti2021}. This is because our double component galaxies lie predominantly on the red sequence, as can be seen in Figure~\ref{fig:colourplots}. 1.5-component fits on the other hand, have very small (namely unresolved) bulges by definition and hence belong to the class of late-type spirals. In fact, 87\,\% of the 1.5-component objects have a ($g$-band) B/T ratio less than 0.1, with the median value as low as 0.02. Computing the bulge-disk $g-i$ colour difference for this sample of objects yields a value of 0.46\,$\pm$\,0.02\,mag, suggesting that the trend described in \citet{Vika2014} continues at very low B/T. 

From all of these comparisons we conclude that our component colours - although noisy - are in line with previous work. In order to increase the colour robustness while preserving the ability to capture physical trends with wavelength (i.e. not fixing the structural parameters to be the same in all bands), a simultaneous fit in all bands is needed. This has many advantages as shown by the \texttt{MEGAMORPH} project team using \texttt{GALAPAGOS} and \texttt{GALFITM} \citep{Haeussler2013, Vika2013, Haeussler2022}, especially for automated analyses, since it naturally ensures smooth wavelength trends while preserving physical variation and additionally allows more robust fits to fainter magnitudes. With \texttt{ProFit} v2.0.0, released in February 2021, now supporting a multi-band fitting mode - and the newly developed package \texttt{ProFuse} \citep{Robotham2022} even combining this with a spectral analysis - this is certainly an interesting avenue to explore in future work and could provide a valuable alternative. It would also solve some of the other challenges we faced during the individual fits, as we discuss in Sections~\ref{sec:jointfitting}, ~\ref{sec:manualcalibrationchanges} and~\ref{sec:outlierstats}.

\subsection{Comparison to size-stellar mass relations of \citet{Lange2015}}
\label{sec:sizemass}

\begin{figure}
    \includegraphics[width=0.5\textwidth]{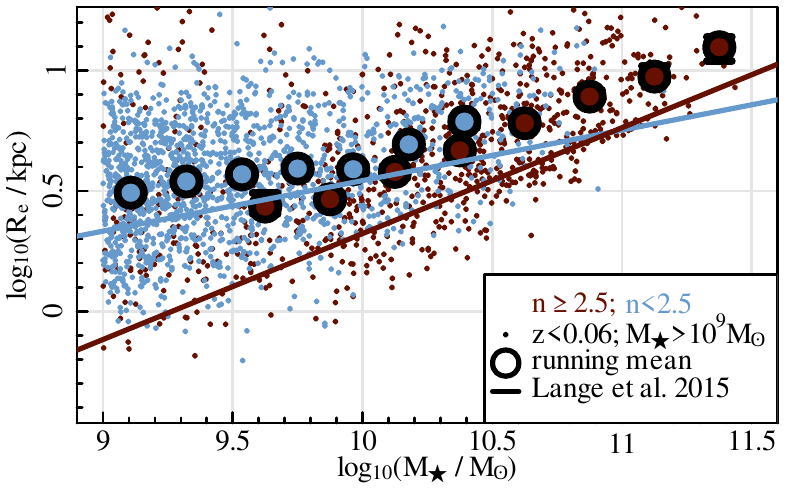}
    \includegraphics[width=0.5\textwidth]{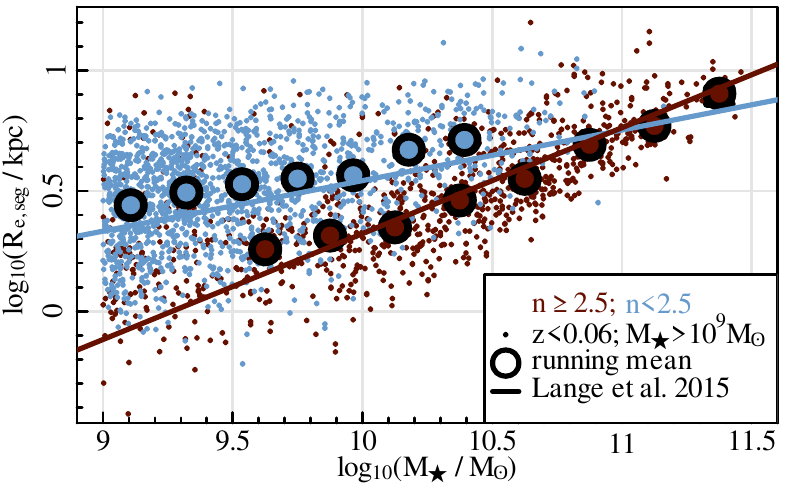}
    \caption{The size-stellar mass relation for our $r$-band fits (dots) compared to the \citet{Lange2015} fits (lines). The sizes are obtained from our single S\'ersic effective radii (\textbf{left panel}: ex\-tra\-po\-la\-ted to infinity; \textbf{right panel}: limited to segment radii) and the distance moduli provided in the \texttt{DistancesFrames} catalogue originally described by \citet{Baldry2012}. The stellar masses are taken from the most recent version of the \texttt{StellarMasses} catalogue initially presented in \citet{Taylor2011}. The sample is limited to the redshift range 0.0001\,<\,$z$\,<\,0.06 (redshifts also from \texttt{DistancesFrames}) and the stellar mass range $M_*$\,>\,$10^9$\,$M_\odot$. Large circles with error bars indicate the running median with its error (usually smaller than the data point). Solid lines show the single exponential $M_*-R_e$ relation fits obtained by \citet{Lange2015} for their single component $r$-band sample, split by a S\'ersic index cut at $n$\,=\,2.5 (taken from their tables~2 and 3).}	
    \label{fig:sizemass}
\end{figure}

Figure \ref{fig:sizemass} shows the size-stellar mass relation obtained from our $r$-band single S\'ersic fits in combination with the redshifts and distance moduli of v14 of the \texttt{DistancesFrames} catalogue \citep{Baldry2012} and v19 of the \texttt{StellarMasses} catalogue \citep{Taylor2011}; both from the GAMA database. The aperture-derived stellar masses have been scaled to match the S\'ersic total flux using the \texttt{fluxscale} keyword provided in the \texttt{StellarMasses} catalogue. The $g$-band and $i$-band results are very similar to those from the $r$-band so we do not show them. 

The sample is limited to objects which were not flagged during our outlier rejection (Section~\ref{sec:postprocessing}) and split into early- and late-type galaxies according to our fitted S\'ersic index ($n\lessgtr2.5$; analogous to \citealt{Lange2015}). We also limit the redshift range to 0.0001\,<\,$z$\,<\,0.06 and the stellar mass range to $M_*$\,>\,$10^9$\,$M_\odot$, thus avoiding the need for volume corrections. For comparison, we show the $M_*-R_e$ relations obtained by \citet{Lange2015} by fitting a single power law to the single component $r$-band fits of \citet{Kelvin2012} (pre-release of \texttt{SersicPhotometry:SersicCatSDSSv09}) combined with an earlier version of the stellar masses catalogue of \citet{Taylor2011} (\texttt{StellarMasses:StellarMassesv16}). We note that the stellar masses did not change much between v16 and v19: the mean and standard deviation of $\Delta\log_{10}(M_*/M_\odot)$ are 0.006 and 0.07 respectively for our sample. The two panels show the results obtained with effective radii taken directly from the S\'ersic fits (left; extrapolated to infinity by definition) or limited to the segment radius within which they were fitted (right; see Section~\ref{sec:postprocessing}). To guide the eye, we also show the running median and its error for our data; where the error is taken as the 1$\sigma$-quantile divided by the square-root of data points within that bin (usually smaller than the size of the data points). 

In both cases, the slope of the mass-size relation obtained from our data agrees well with the \citet{Lange2015} fit results.\footnote{For reference, the slopes of the plotted lines are 0.21 and 0.44 for the late- and early-types respectively; taken from tables~2 and~3 of \citet{Lange2015}).} There is an offset in the absolute sizes, but those will inherently depend on the exact definition of the size measurement at hand as well as the (depth of the) data used. Already calculating effective radii within the segments within which we fitted for them (right panel) brings our results much closer to those of \citet{Lange2015}; although the measurements are then not directly comparable to their fits anymore since they use S\'ersic values extrapolated to infinity (which will, in turn, depend on the segment size used for fitting). We now discuss these issues further by directly comparing our fits to those of \citet{Kelvin2012}, which the \citet{Lange2015} results were based on.

\subsection{Comparison to single S\'ersic fits of \citet{Kelvin2012}}
\label{sec:comparelee} 

\begin{figure}
    \includegraphics[width=0.5\textwidth]{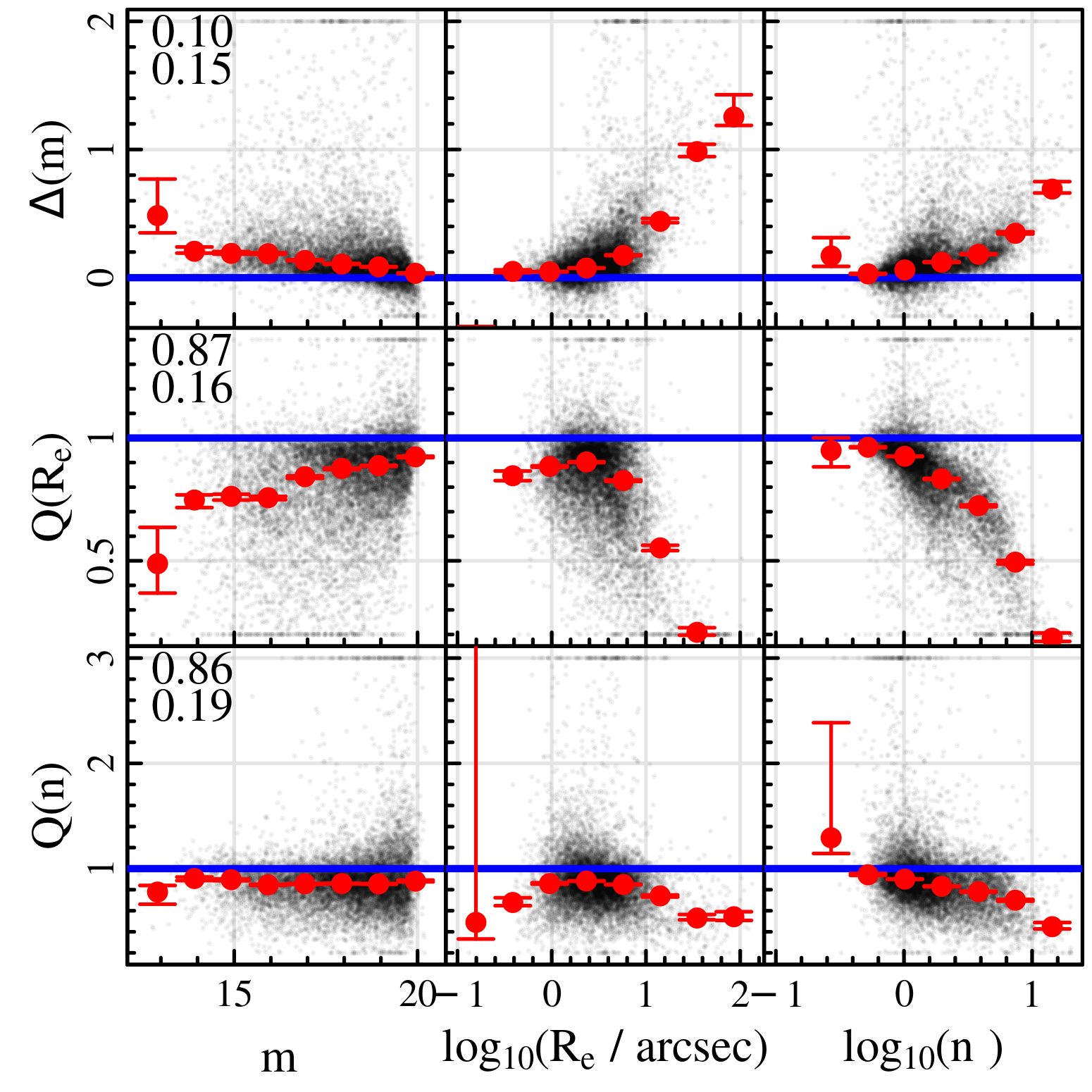}
	\includegraphics[width=0.5\textwidth]{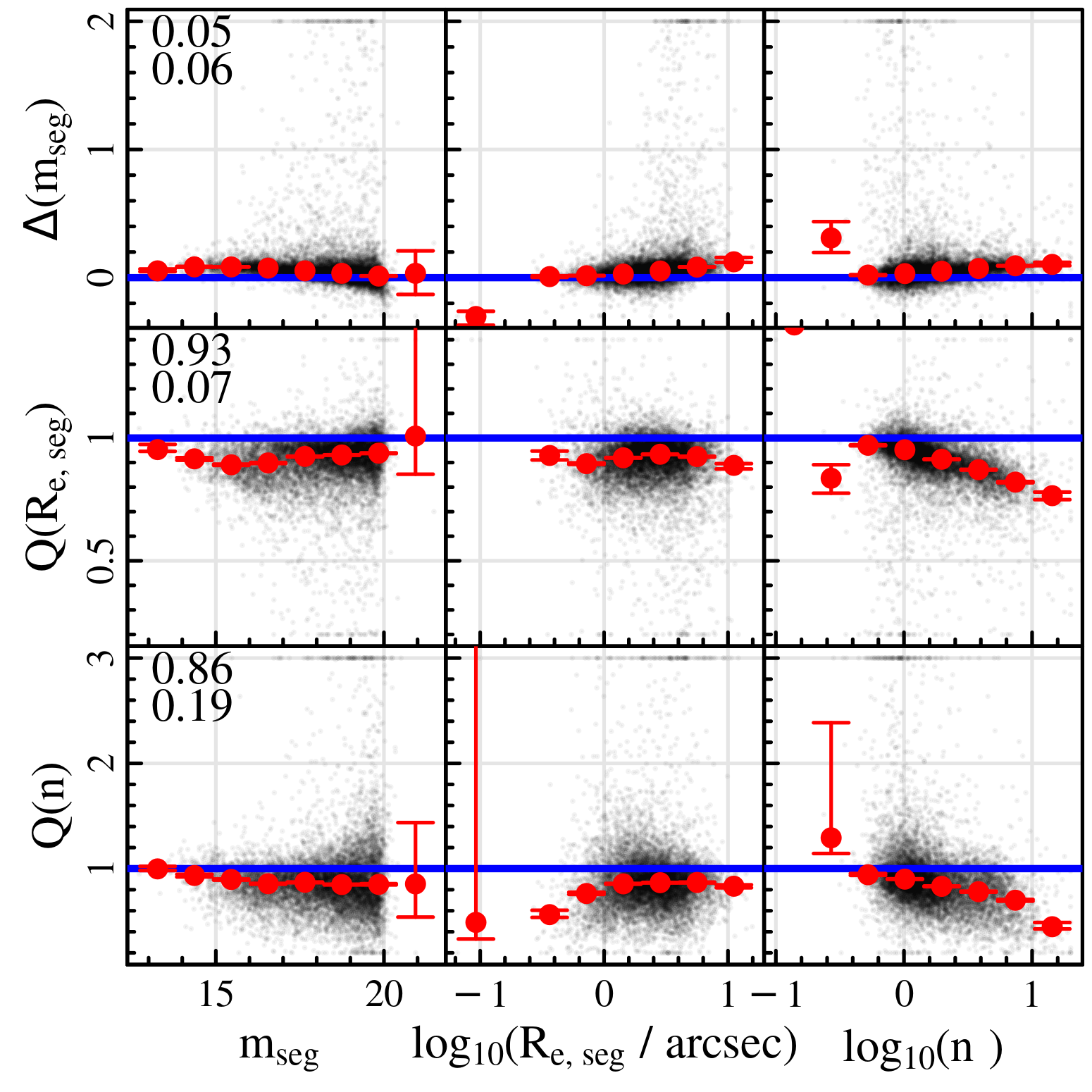}
    \caption{The difference $\Delta$ or quotient Q (for scale parameters) between the \citet{Kelvin2012} fits and our fits plotted against our fits for the three most important single-S\'ersic parameters magnitude, effective radius and S\'ersic index in the $r$-band. The top panels show the S\'ersic parameters extrapolated to infinity, while for the bottom panels we calculated the magnitude and radius within the segment radius for both our and the \citet{Kelvin2012} fits. Outliers are clipped to the plotting interval; which is the same in both cases. Black dots show all fits, red dots with error bars show the running median and its error in evenly spaced bins and horizontal blue lines indicate no difference between the fits. The numbers in the top left corners of the first row of panels show the median and 1$\sigma$-quantile of the respective distribution in the $y$-direction (which is identical for all panels of a row). The sample is limited to fits that are available in the \citet{Kelvin2012} catalogue and were classified as single component in our model selection.}	
    \label{fig:comparelee}
\end{figure}

To further investigate the size offset observed in Section~\ref{sec:sizemass}, we directly compared our fits to those of \citet{Kelvin2012} (\texttt{SersicPhotometry:SersicCatSDSSv09} on the GAMA database). Since \citet{Kelvin2012} do not provide double component fits, the analysis is limited to single S\'ersic fits. We again use the $r$-band as an example for the discussion, but note that results are very similar for the $g$ and $i$ bands. 

The \citet{Kelvin2012} fits are based on the Structural Investigation of Galaxies via Model Analysis (\texttt{SIGMA}) code applied to data from SDSS DR7. \texttt{SIGMA} is a wrapper around \texttt{Source Extractor} \citep{Bertin1996}, \texttt{PSF Extractor} \citep{Bertin2011} and \texttt{GALFIT 3} \citep{Peng2010} performing similar steps to what we do in our pipeline (Section~\ref{sec:pipelineoverview}), i.e. source identification, background subtraction, PSF estimation and 2D model fits to the surface brightness profile of the galaxies. The differences lie in the data and code used, where we upgrade SDSS to KiDS, \texttt{Source Extractor} to \texttt{ProFound}, \texttt{PSF Extractor} to a combination of \texttt{ProFound} and \texttt{ProFit} and \texttt{GALFIT} to \texttt{ProFit}; with all the advantages described in Sections~\ref{sec:kids}, \ref{sec:profit} and \ref{sec:profound}. In addition, we also perform multi-component fits and model selection. For the comparison to the \citet{Kelvin2012} results, we focus on the three most important single S\'ersic fit parameters: magnitude $m$, S\'ersic index $n$ and effective radius $R_e$, which tend to be the least ``well-behaved" (position, axial ratio and angle are generally more easily constrained and uncorrelated, see e.g. Figure~\ref{fig:resultsv04vsv05_R_S} in Section~\ref{sec:v05parameterdistributions}). 

Figure~\ref{fig:comparelee} shows the difference between our fits and the \citet{Kelvin2012} fits for these three parameters. In line with Figures~\ref{fig:resultsv04vsv05_R_S} to~\ref{fig:resultsv04vsv05_R_D_SEGRAD}, we show the difference in magnitudes (\citet{Kelvin2012} fits - our fits), while for the effective radius and S\'ersic index (scale parameters) we show the quotient (\citet{Kelvin2012} fits / our fits) on the $y$-axis; always plotted against our fitted values on the $x$-axis (in logarithmic space for scale parameters). Again, we show the results for S\'ersic parameters extrapolated to infinity (left panels) and for the magnitude and effective radii calculated within the segment radius (right panels), where we limit both our fits and those of \citet{Kelvin2012} to our fitting segment (which are generally smaller than the fitting regions used in \citet{Kelvin2012}) to obtain directly comparable results. 

For the S\'ersic parameters extrapolated to infinity (left panels), large differences can be seen in all fitted parameters, including systematic trends across the parameter space. Note the larger axis range relative to Figures~\ref{fig:resultsv04vsv05_R_S} to~\ref{fig:resultsv04vsv05_R_D_SEGRAD}. This shows once again that fitted S\'ersic parameters are not directly comparable given the differences in the data, code and processing steps with a wealth of potentially different systematic uncertainties (Section~\ref{sec:postprocessing}). However, when we limit the analysis to our segment sizes (bottom panels), the fits become much more comparable. On average, now, our fits are $\sim$\,0.03\,mag brighter and approximately 7\,\% larger than the \citet{Kelvin2012} fits to the same galaxies, which is not surprising given the increased depth and resolution of KiDS compared to SDSS and the numerous sources of different systematic uncertainties (e.g., differing sky subtraction and PSF estimation). Also, there are fewer trends across the parameter space, indicating that systematic differences arise mainly from the extrapolation to infinity. 

The exception to this is the S\'ersic index, which still shows some trends. The reason is that the S\'ersic index, unlike the magnitude and effective radius, cannot be corrected for different fitting regions. The \citet{Kelvin2012} fits, which were performed within larger fitting regions than our fits, will inevitably have to compromise more between the inner and outer regions of the galaxy to be fitted (unless the light profile truly follows a single S\'ersic profile with no deviations out to very large radii, which is rarely the case). Our tight fitting segments, on the other hand, will result in better fits to the inner regions of the galaxy at the expense of producing unphysical wings when extrapolated beyond the fitting segment (Section~\ref{sec:postprocessing}). Thus, fitted S\'ersic indices are always a weighted average (or compromise) across a range of radii and their absolute values will never be directly comparable between catalogues unless the fitting regions are exactly the same (or the galaxies studied follow perfect S\'ersic profiles).

\section{Systematic uncertainties and biases}
\label{sec:simulations}

The MCMC chain errors returned by the fitting procedure do not include systematic uncertainties which arise due to galaxy features not accounted for in the models, nearby other objects, imperfect PSF estimation, background subtraction inaccuracies and similar effects. For an individual galaxy, the presence of such ``features" will systematically shift the fitted parameters away from the true values, thus introducing a bias. For a statistically large enough sample of galaxies, however, most of these effects are expected to cancel out on average since they are random from one galaxy to the next (e.g. nearby other sources shifting the fitted positions). These ``random systematics" can - for statistical samples - be accounted for by simply increasing the given parameter errors such that in most cases, the true values are included in the credible intervals again. Such systematics can be studied using overlap sample galaxies, i.e. those that appeared in more than one KiDS tile (cf. Section~\ref{sec:sampleselection}). In addition, there can be ``one-sided effects" that lead to an overall bias across the sample, e.g. due to excess flux from nearby objects. These can only be detected using simulations. In the following, we study both of these effects using our bespoke simulations, the overlap sample of real galaxies and the overlap sample of simulated galaxies; where we refer to the random systematics as ``error underestimates" and to the one-sided effects as ``biases". 

The final corrections for both of these effects are listed in Table~\ref{tab:errorunderestimate}. In short, biases are very small ($\lesssim$\,1\,\%), while systematic errors are a factor of 2-3 larger than the random MCMC errors alone. The error underestimate corrections are also applied to the released catalogues (for \texttt{v04} and \texttt{v05}), while the bias corrections are not since they are only valid for a large random subset of our galaxies and not for individual objects. Note that the systematic error studies were carried out on results from \texttt{v03} of the \texttt{BDDecomp} DMU, while the remainder of this chapter refers to \texttt{v04}. However, since \texttt{v04} is statistically identical to \texttt{v03} (see Section~\ref{sec:v04}), the results can directly be transferred. We would also like to point out that we focus on single S\'ersic $r$-band fits in this section. We expect individual components in the 1.5- and double component fits as well as the other bands to be affected by similar systematics. Effects are likely to become worse for fainter and/or less well-resolved objects (i.e. bulges in particular; and objects in the KiDS $u$ and the longer wavelength VIKING bands).

\subsection{Overlap sample comparison}
\label{sec:overlapstudies}

\begin{figure}
    \includegraphics[width=0.5\textwidth]{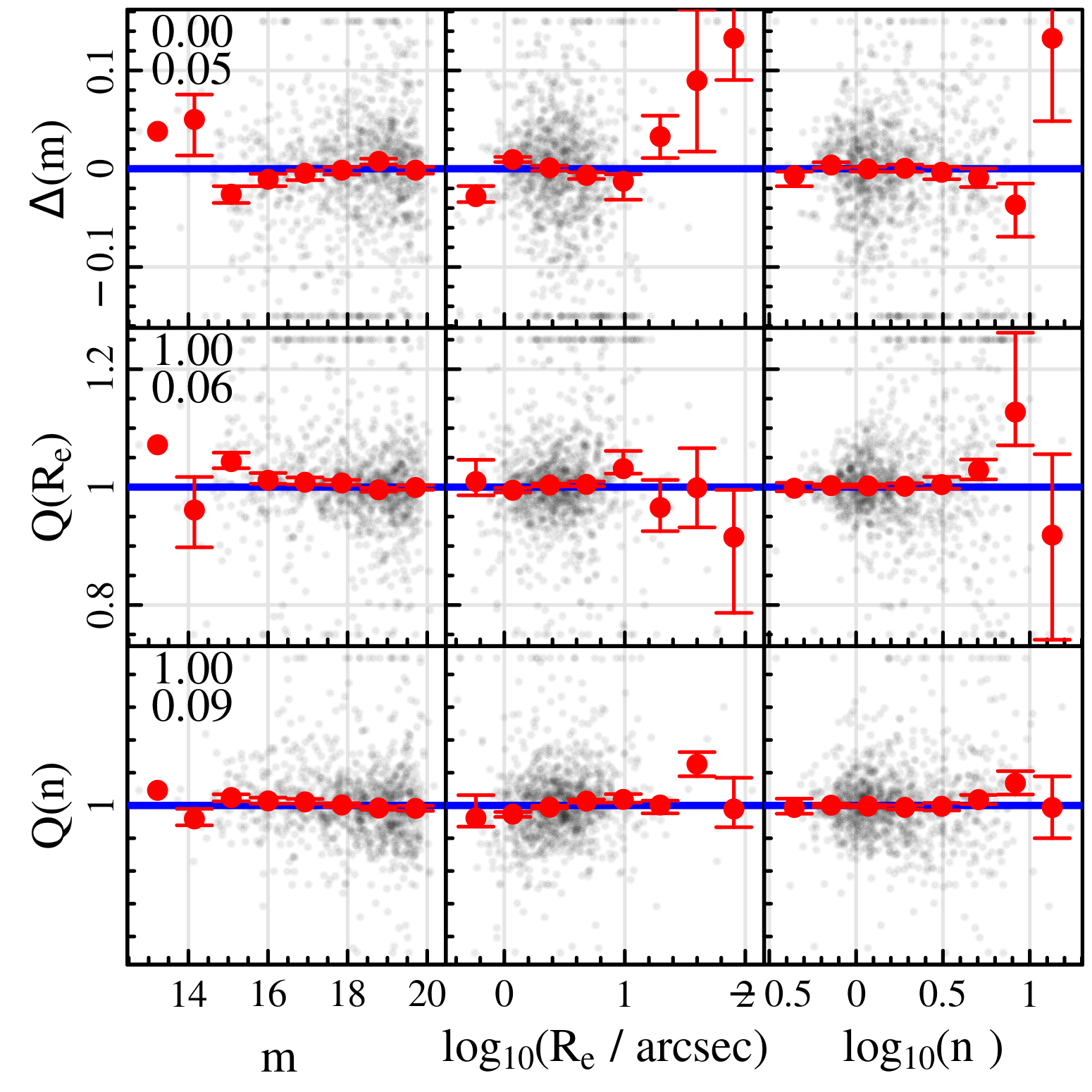}
    \includegraphics[width=0.5\textwidth]{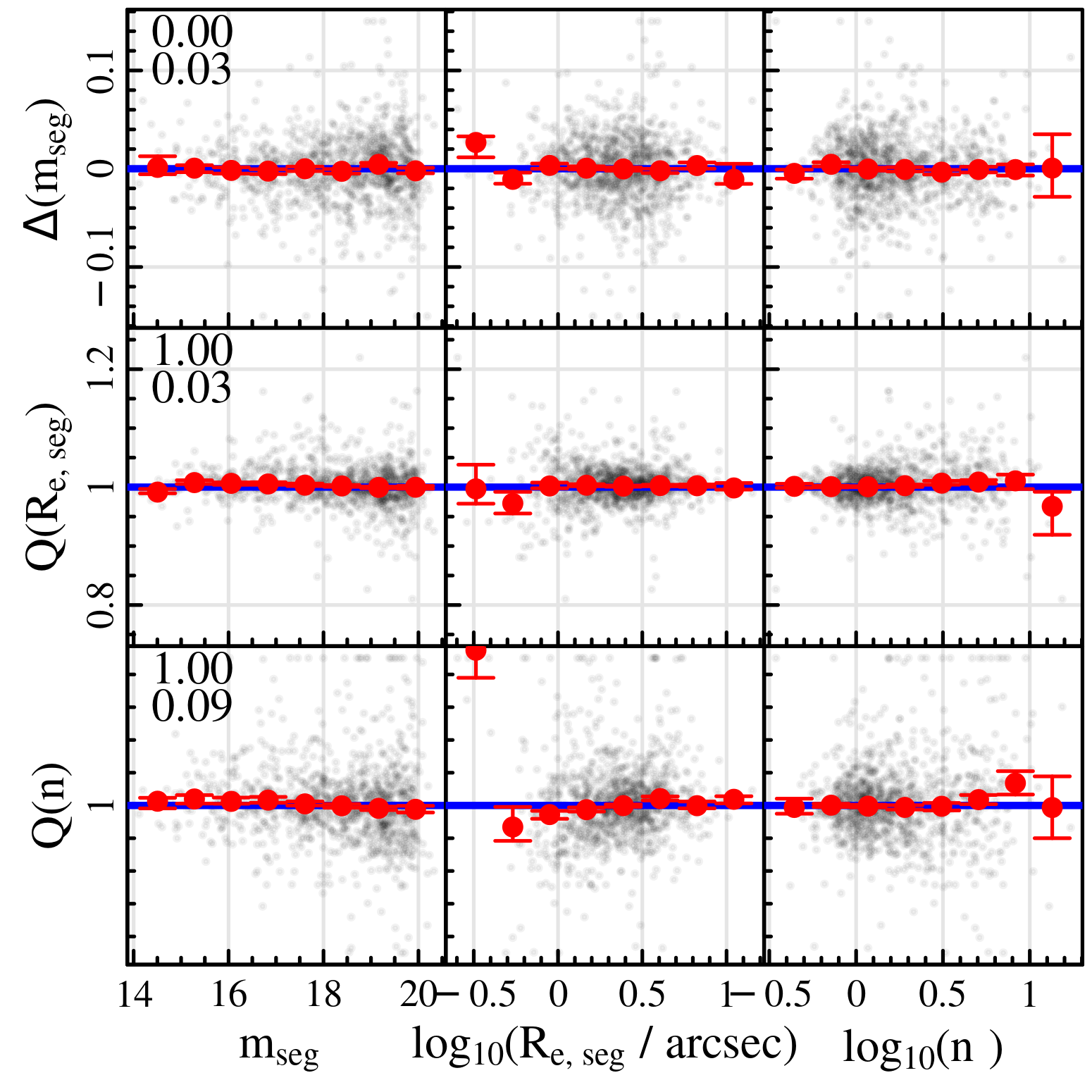}
    \caption{Similar to Figure~\ref{fig:comparelee} but now showing an internal consistency check of our catalogue using galaxies that were imaged (and successfully fitted) at least twice in the KiDS $r$-band. We pass these duplicate observations of the same physical objects through our pipeline independently and then compare the fit to the shallower image with the fit to the deeper (higher signal-to-noise ratio) image. Note the different plotting ranges relative to Figure~\ref{fig:comparelee} (especially on the $y$-axis).}	
    \label{fig:compareoverlap}
\end{figure}
As an internal consistency check, we compared the fit results obtained from multiple observations of the same physical object (Section~\ref{sec:sampleselection}) in Figure~\ref{fig:compareoverlap}. The plots are very similar to the ones in Figure~\ref{fig:comparelee} (see description in Section~\ref{sec:comparelee}; but note the different $y$-axis scale), except that we now show the differences between two of our own fits to different KiDS images of the same galaxy (in the overlap region between the KiDS tiles). Hence all fits shown in Figure~\ref{fig:compareoverlap} are based on KiDS data and use the exact same pipeline for analysis, though the different observations are treated entirely independently. We always use the deeper image as the reference image (the image depth at the edge of KiDS tiles can vary greatly depending on the number of dithers - between 1 and 5 - that cover the area). For the right set of panels, we evaluate the magnitude and effective radius within whichever of the two segment radii is smaller to avoid extrapolation and obtain consistent results. 

For both versions (S\'ersic parameters extrapolated to infinity on the left and truncated to segment radii on the right), there are very little differences between the two fits to the same galaxy; and there are no systematic trends across the parameter space. The running median is consistent with 0 (or 1, for scale parameters) in almost all bins, which shows that there are no inherent systematic differences in our fits related to image depth. This holds true despite the segments being systematically larger for the deeper images. The difference in segment size is too small on average to detect systematic trends across the entire sample: the median difference between the two segment radii is only 6\,\%. However, there are a number of outliers visible as rows of points at the top and bottom of the panels since they were clipped to the plotting intervals (for plotting only). These correspond to the fits where the segment sizes differ substantially. Evaluating both fits within the same region before comparison (right panels) removes those outliers and reduces the overall scatter.

\subsection{Simulations: parameter recovery}
\label{sec:parameterrecovery}

As a final test of our pipeline, we ran simulations where we insert single S\'ersic model galaxies convolved with an appropriate PSF at random locations in the KiDS data. To obtain a realistic distribution of parameters for the model galaxies (including correlations), we use the fitted single S\'ersic parameters of a random sample of 1000 $r$-band galaxies that were not classified as outliers. The PSF to convolve with is taken as the model PSF that was fitted to the nearest real galaxy (at the position where the model galaxy is inserted), which is close to the real PSF at that image location. We then simply add the PSF-convolved galaxy to the KiDS data and run the resulting image through our entire pipeline (segmentation, sky subtraction, PSF estimation, galaxy fitting, outlier flagging, model selection). 

In this way we are able to check for intrinsic biases in our entire pipleline, with 3 exceptions: issues due to galaxy features not represented by our models (bars, spiral arms, disk breaks, mergers, etc.), problems in the data processing performed by the KiDS team (if any), and deviations of the true PSF from a Moffat function.

\begin{figure}
    \includegraphics[width=0.5\textwidth]{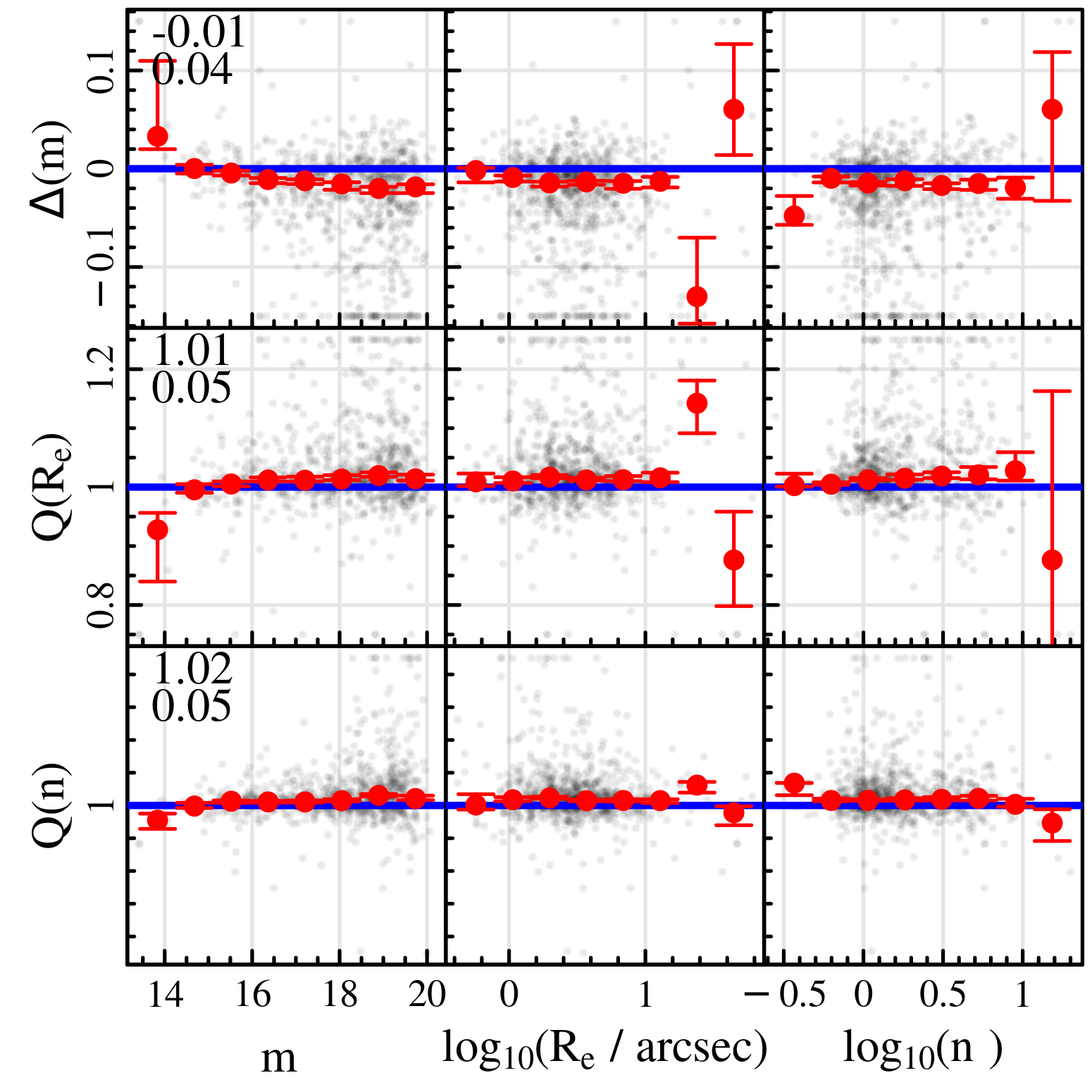}
    \includegraphics[width=0.5\textwidth]{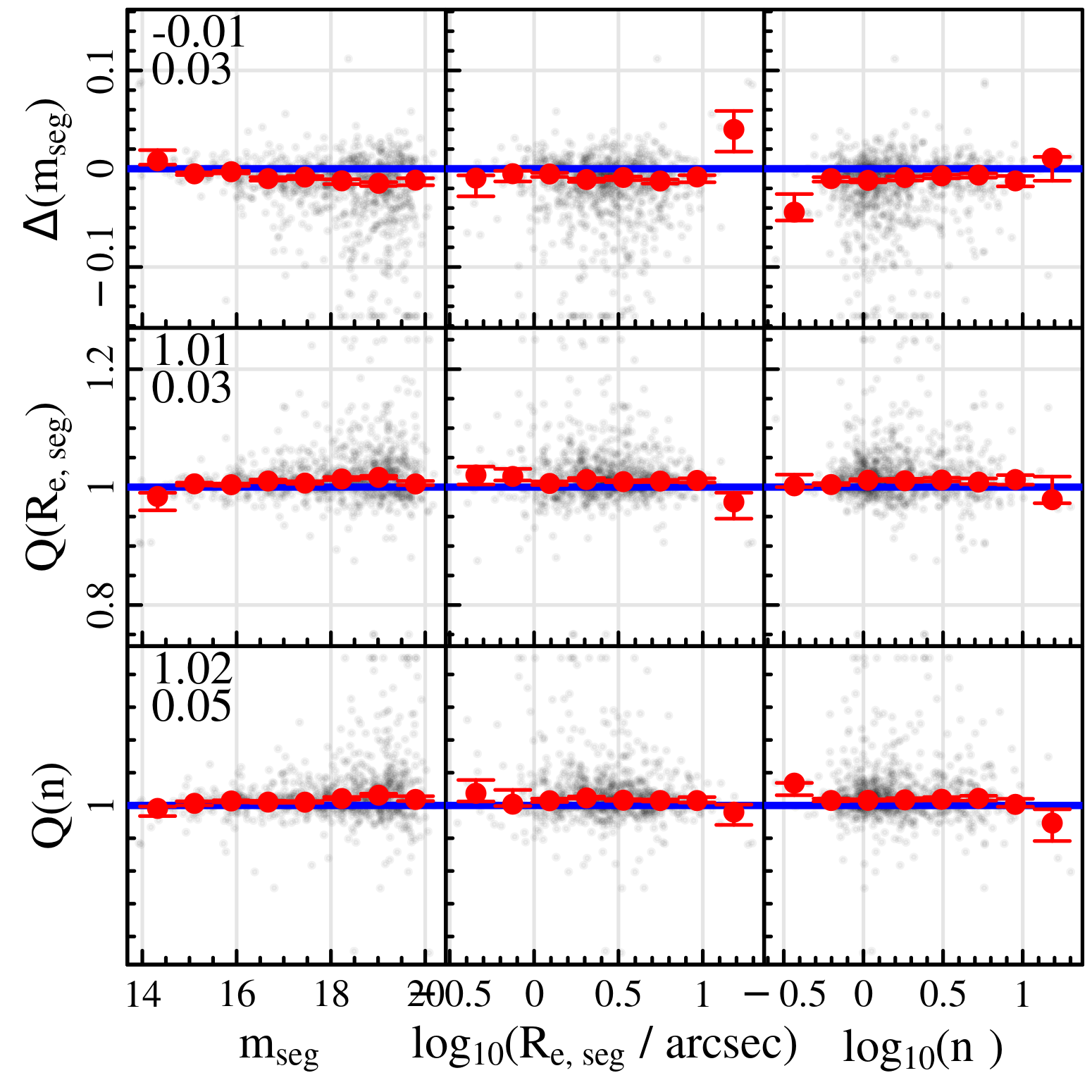}
    \caption{Similar to Figure~\ref{fig:compareoverlap} but comparing the fitted parameters of simulated images to the true (input) parameters; both limited to segment radii for the right set of panels.}	
    \label{fig:comparesimulations}
\end{figure}

In Figure~\ref{fig:comparesimulations} we show the corresponding plots to Figures~\ref{fig:comparelee}~and~\ref{fig:compareoverlap}; where on the $x$-axis we now have the true (input) parameters of our simulated galaxies and on the $y$-axis the difference between the fitted and the true values; both limited to segment radii for the right set of panels. 

\begin{figure}
\begin{center}
    \includegraphics[width=0.8\textwidth]{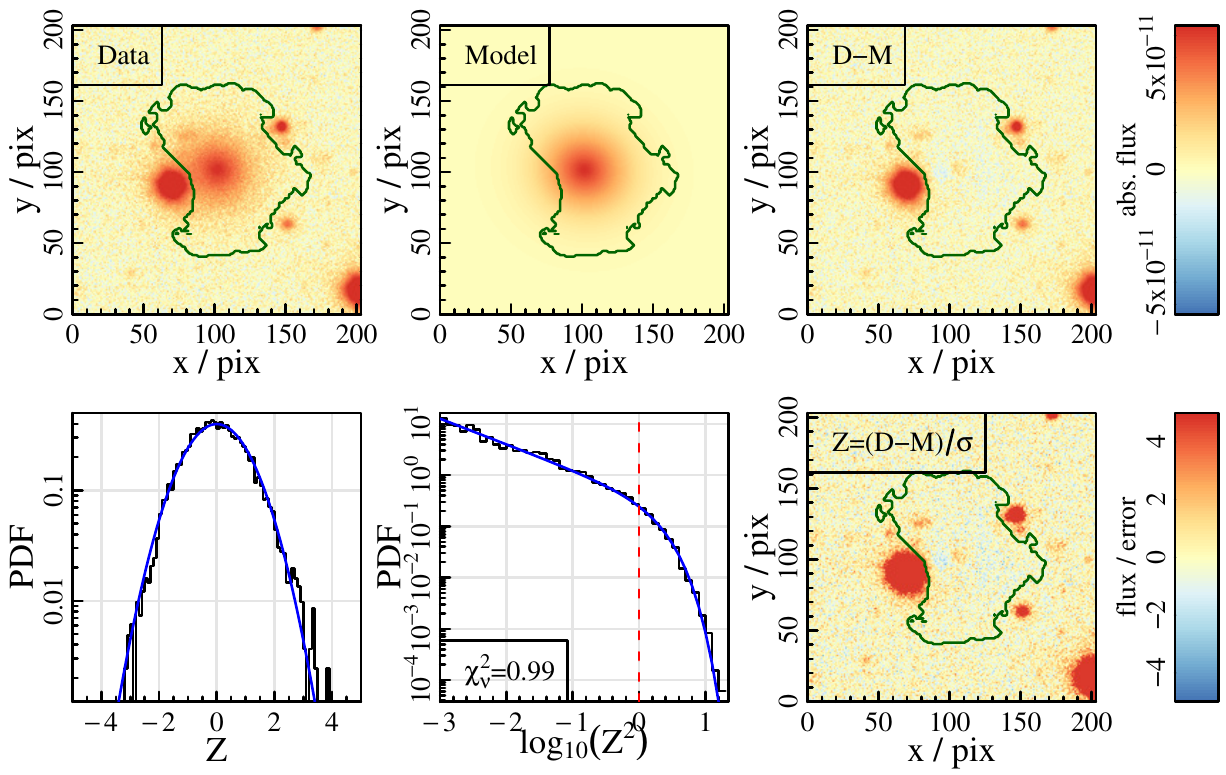}
    \caption{An example fit to a simulated galaxy where the difference between the true and the fitted magnitude is large due to the wings of a nearby bright object and a small faint object included in the segmentation map. Panels are the same as the top two rows in Figure~\ref{fig:examplefit}.}	
    \label{fig:examplebadsim}
\end{center}
\end{figure}

Generally, all parameters are recovered well, although for both versions of the plot the magnitudes show a slight offset of $\sim$\,0.01\,mag (with corresponding trends in effective radius and S\'ersic index since these parameters are correlated); worsening for faint objects. This offset is driven by a number of galaxies scattering to very low values, i.e. where the fit attributes significantly more flux to the galaxy than what we put into the simulation. Visual inspection of these simulated objects revealed that all of them have additional flux from other objects included in the segmentation maps. 
Figure~\ref{fig:examplebadsim} shows an example, where the difference between the fitted and the true magnitude is -0.17 for the values extrapolated to infinity and -0.14 for the segment truncated values. Truncating to segment radii only leads to limited improvement of the agreement since the cause of the offsets observed in Figure~\ref{fig:comparesimulations} is flux from nearby objects; and all simulated galaxies are intrinsically represented well by a single S\'ersic model as they were created as such.

Nearby objects affect approximately 5-10\,\% of our simulated fits in this way. Since it is a one-sided effect (there are no sources with negative flux), it results in a slight overall bias across the sample. This is expected to occur at a similar level also in the fits to real galaxies and could only be improved by simultaneously fitting nearby sources (see also the discussion of this issue in \citealt{Haeussler2007}). However, for this work we decided against this option as explained in Section~\ref{sec:galaxyfitting}. We may revisit this decision in future work. 

\subsection{Simulations: model selection accuracy}
\label{sec:simulationsmodelselection}

Since we know that all of our input galaxies were perfect single S\'ersic systems, the model selection and outlier rejection statistics can be used to judge the failure rate of these routines (cf. also Sections~\ref{sec:postprocessing}~and~\ref{sec:statistics}). We simulated 1000 galaxies at random locations; which resulted in 1126 objects to be fit (due to the overlap regions between KiDS tiles). Of these, 262 (23\,\%) were skipped; which is similar to the fraction of skipped fits for real galaxies, as expected since the main reason for this are the KiDS masks. Of the remaining objects, 94\,\% are classified as single component fits, 3\,\% are 1.5- or double component fits and 3\,\% are flagged as outliers. 

\begin{figure}
\begin{center}
    \includegraphics[width=0.8\textwidth]{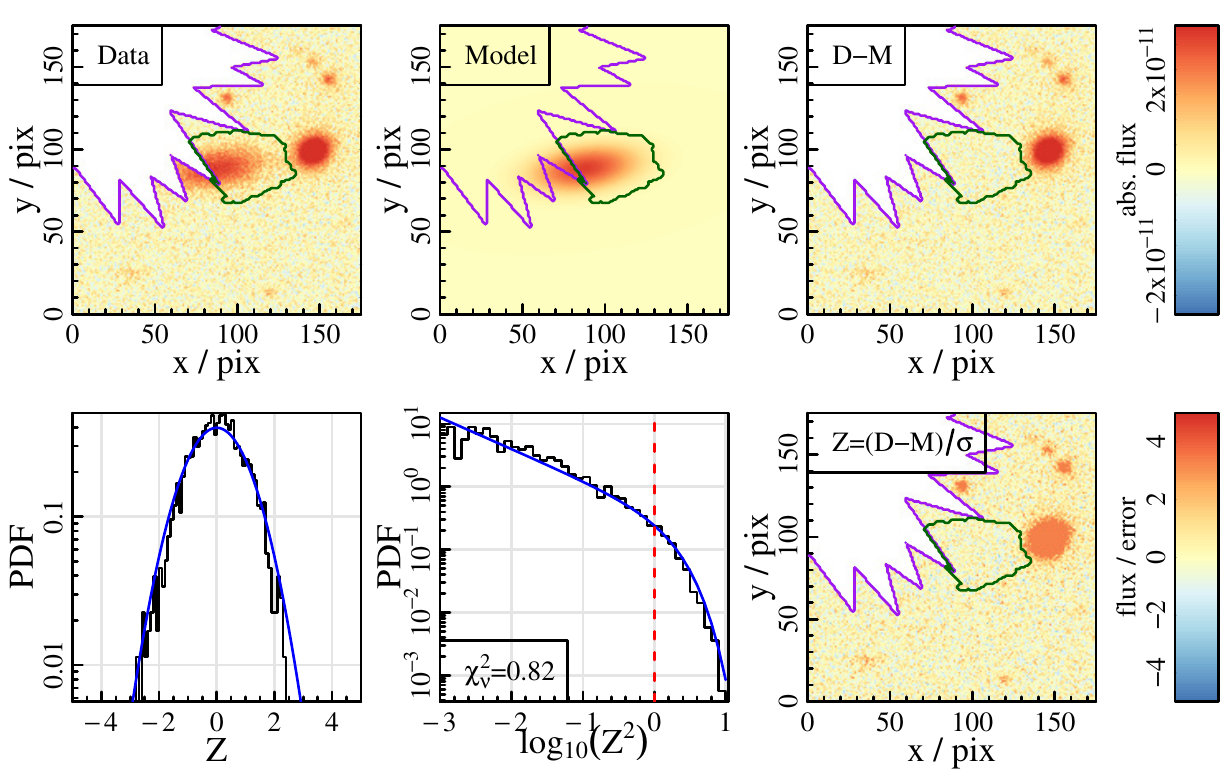}
    \caption{An example fit to a simulated galaxy which was flagged as a bad fit due to a nearby masked area from a bright star chopping up the segmentation map. Panels are the same as the top two rows in Figure~\ref{fig:examplefit}.}
    \label{fig:exampleoutliersim}
\end{center}
\end{figure}

The number of outliers is significantly less than the 11\,\% of real $r$-band galaxies flagged (Section \ref{sec:postprocessing}) because in the simulations all galaxies are intrinsically ``well-behaved". Figure~\ref{fig:exampleoutliersim} shows an example of the most commonly occuring reason for being flagged as an outlier (in the simulations), namely the mask of a nearby bright star chopping up the segmentation map. 

\begin{figure}
\begin{center}
    \includegraphics[width=0.8\textwidth]{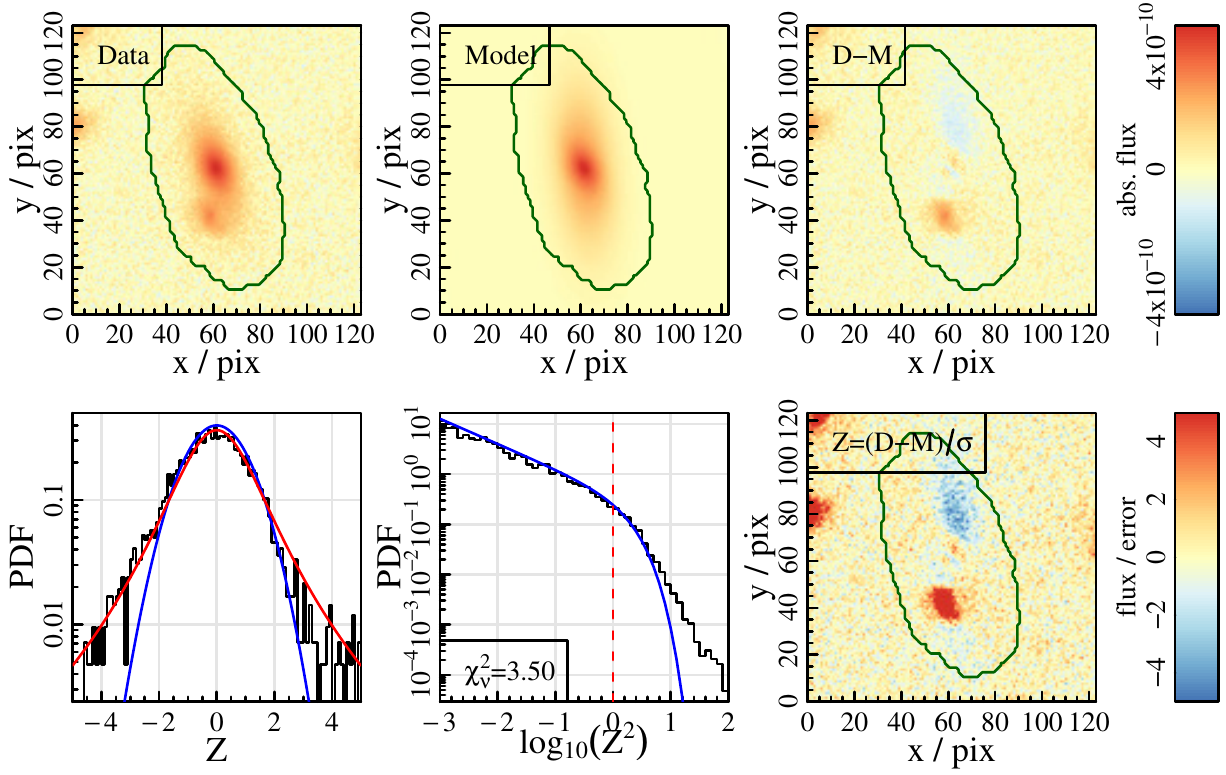}
    \includegraphics[width=0.8\textwidth]{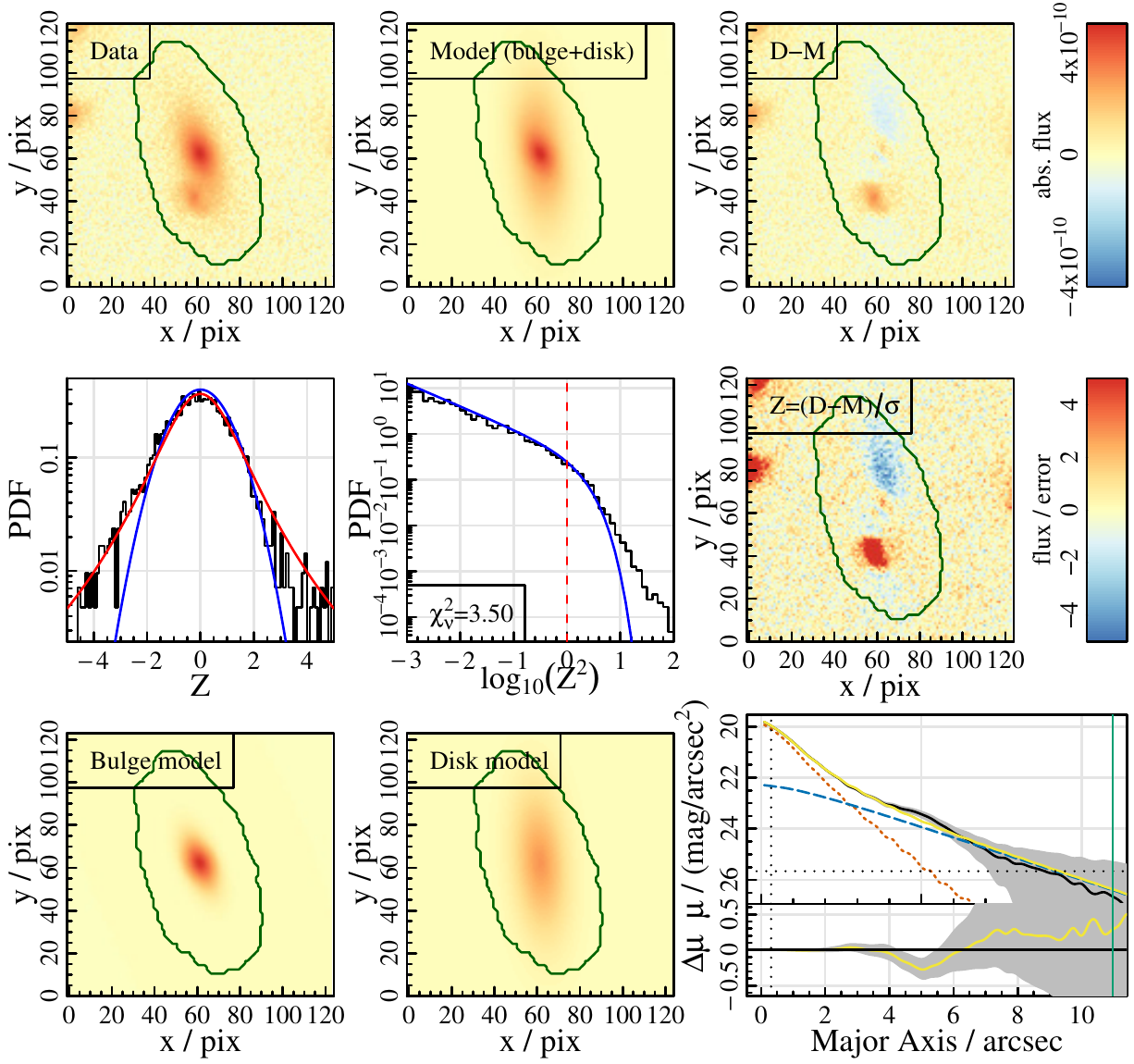}
    \caption{An example fit to a simulated galaxy which was classified as a double component fit. \textbf{First two rows:} the single S\'ersic fit. \textbf{Last three rows:} the double component fit. Panels are the same as in Figure~\ref{fig:examplefit}.} 
    \label{fig:exampledoublesim}
\end{center}
\end{figure}

The fact that 97\,\% of non-outlier simulated galaxies are correctly classified as single component fits confirms that model selection is accurate provided the galaxy can be unambiguously assigned to the single S\'ersic model (cf. Section~\ref{sec:postprocessing}). We visually inspected the 3\,\% 1.5- and double component fits and found nearby interfering objects in all of them. Figure~\ref{fig:exampledoublesim} shows an example, where the fit attempts to capture the additional ``features" with the freedom of a second component. Note that since we only simulated single S\'ersic objects, we cannot comment on the accuracy of the model selection procedure for double component objects here. However, our model selection procedure was optimised on all types of real galaxies (not just single S\'ersic objects); and our comparison to visual inspection in Section~\ref{sec:postprocessing} also indicates a high accuracy for double component systems. 

\subsection{Systematic uncertainties}
\label{sec:systematics}

\begin{figure}
    \includegraphics[width=\textwidth]{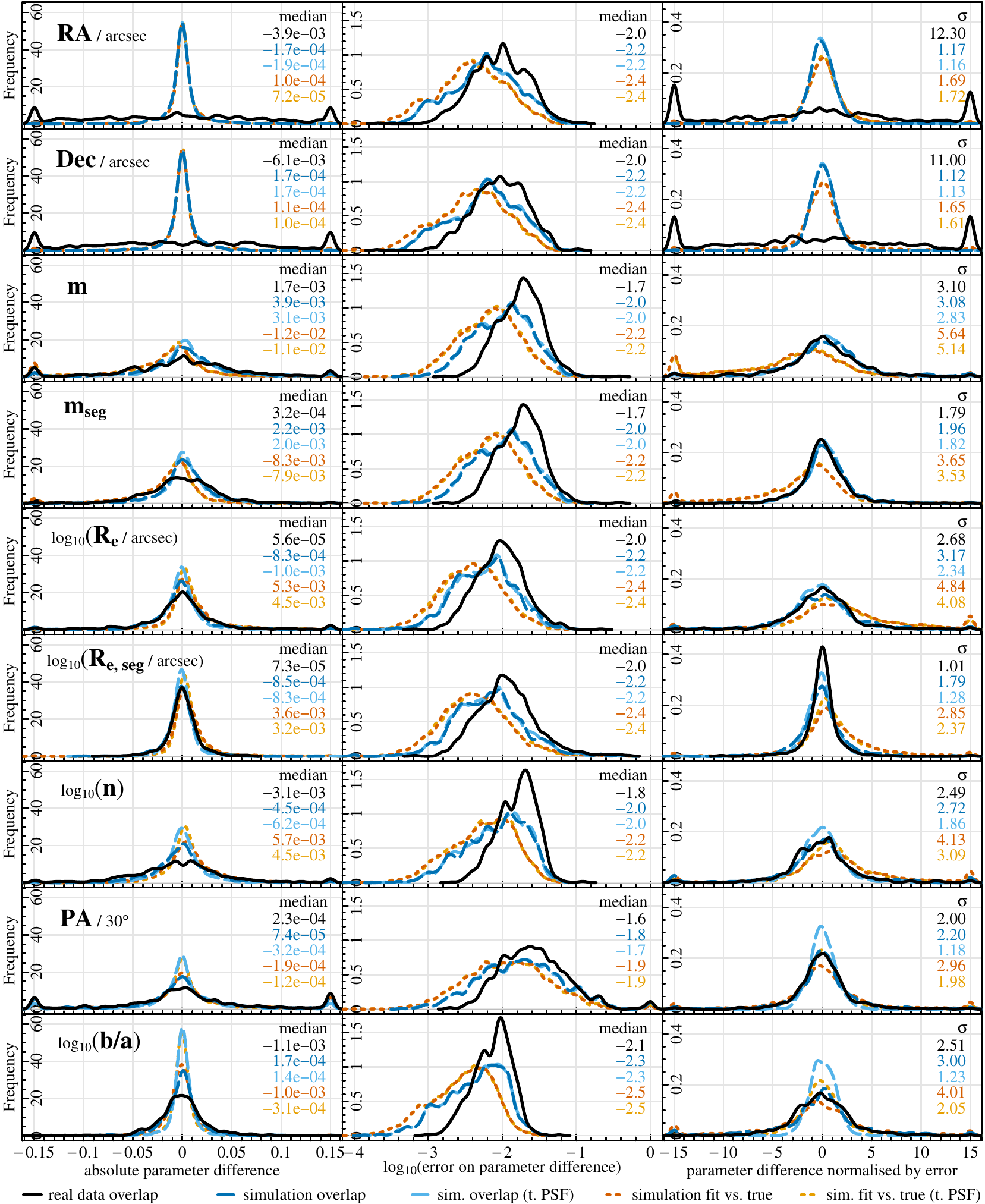}
    \caption{For all single S\'ersic parameters as labelled top to bottom: \textbf{Left column:} The distribution of the absolute difference between the fitted and true values for simulated galaxies; or between the fitted values to two versions of the same (simulated or real) galaxy in the overlap sample. The legend at the bottom indicates which difference is shown; scale parameters are treated in logarithmic space throughout. See text for details. \textbf{Middle column:} The error on the parameter difference shown in the left column. \textbf{Right column:} The parameter difference normalised by its error (i.e. left column divided by middle column).}
    \label{fig:differrnorm}
\end{figure}

Figure~\ref{fig:differrnorm} shows the results of our systematic error study, which we will now discuss in detail. Going from left to right we show three different plot types as labelled on the $x$-axis and described in more detail in the caption and below. Going from top to bottom, each plot type is shown for each single S\'ersic parameter as labelled in the left panels; and the colours of the lines in the plot indicate which sample was used according to the legend at the bottom of the figure. The left panels show the distribution of absolute differences between the values fitted to both versions of a galaxy (in the indicated units); the middle panels show the corresponding error distribution (errors added in quadrature for the two fits; in the same units as the fit); and the right panels show the distribution of the absolute difference divided by its error (unitless). The solid black lines labelled ``real data overlap" are using the overlap sample. The dashed dark blue lines labelled ``simulation overlap" show the same for the overlap sample in simulated galaxies, where we have run more simulations specifically to boost the number of simulated overlap galaxies to a similar value as we have for the real overlap sample ($\sim$\,700). The dashed light blue line labelled ``sim. overlap (t. PSF)" is the same as the dashed dark blue line, just that when fitting the galaxy, instead of the estimated PSF we passed in the true PSF (i.e. the one we used to convolve the model galaxy with originally). The dotted dark orange line labelled ``simulation fit vs. true" shows the difference between the fitted value and the true value (instead of between the two fitted values in the overlap region) for the same sample of galaxies. Note the errors here now are just the errors of the fit since the true values do not have errors. The dotted light orange line (``sim. fit vs. true (t. PSF)") is the same for the run that used the true PSFs.
 
All values are clipped to the plotting intervals (for plotting only). For scale parameters, all distributions are shown in logarithmic space (which the parameters were also fitted in). To make the scales comparable to the other parameters, the angle is shown in units of 30\degr\ (which it was also fitted in to make the MCMC step size comparable to the other parameters). For the magnitude and effective radius, we show both the fitted S\'ersic values and the segment truncated values. Comparing rows 3 and 4 (S\'ersic and segment magnitudes) or rows 5 and 6 (S\'ersic and segment effective radii) against each other, it becomes clear immediately that the distributions for the segment values are narrower, i.e. limiting to segment sizes increases the stability and reduces the scatter in those parameters as already observed in many previous sections. Note, though, that our simulated galaxies do follow perfect single S\'ersic profiles, so the differences between the segment and S\'ersic values are generally expected to be much smaller in the simulations than in real data (see Section~\ref{sec:postprocessing} for details). 

For each distribution, we also give the median of the absolute difference and the errors; and the 1$\sigma$-quantile (half of the range containing the central 68\,\% of data) of the normalised difference. These values (with uncertainties) are also given in Table~\ref{tab:errorunderestimate}.

\subsubsection{Overlap sample: real vs. simulated}

Focusing on the real data and simulation overlap samples (solid black and dashed dark blue lines), which are most directly comparable, it can be seen that the distributions of the pa\-ra\-me\-ter differences (left column) are broader for the real galaxies than for the simulated galaxies for all parameters; i.e. for two versions of the same galaxy in the overlap sample, the fitted values are on average closer to each other in the simulation than in the real data. 
This could be due to two reasons: either irregular galaxy features in combination with noise (i.e. perfect single S\'ersic objects are just more easily constrained/less easily influenced by noise fluctuations); or differences in the KiDS data processing between tiles (e.g. inaccuracies in their background subtraction procedure) that affect the real galaxies but not the simulated ones since those were added later. In reality, it is probably a combination of these two effects (with the first one presumably dominating). All steps of our own analysis affect both the simulated and the real galaxies and on average will have the same effect on both. 

The errors (second column) do reflect this additional uncertainty in real galaxies in that they are larger by 0.2-0.3\,dex for all parameters. In fact, the errors on the simulated galaxies seem to be more severly underestimated than those on the real galaxies, which becomes clear when looking at the parameter differences normalised by the respective errors (third column). In an ideal world, these would all be Gaussians centred on zero with a standard deviation of 1. As there will always be a few outliers due to interfering objects or image artifacts, instead of the mean and standard deviation we will consider their more robust equivalents, the median and 1$\sigma$-quantile (shown in plots). All overlap sample distributions (simulated and real, i.e. black, dark blue and light blue lines) are centred on zero, as already expected from the results shown in Figure~\ref{fig:compareoverlap}. However, it can be seen that for all parameters except segment effective radius, the 1$\sigma$-quantile is larger than 1 both for the simulated and the real sample: values generally range between 2 and 3 with simulated galaxies performing slightly worse due to the underestimated errors (which are most likely caused by the PSFs, see discussion below). 

The exception to this is position (RA and Dec, top two rows), for which the normalised distribution is a factor of approximately 10 broader for real galaxies than for simulated ones. In fact, a considerable fraction of these distributions fall outside of the plotting range, such that the clipping to these intervals results in prominent peaks at the plot edges (top right two panels of Figure~\ref{fig:differrnorm}). We believe this to be mainly due to the accuracy of the astrometric solution of the KiDS data, which shows a scatter of approximately $0\farcs$04 in DR4.0 in both RA and Dec (\citealt{Kuijken2019}; and we also confirmed this using the KiDS $r$-band source catalogues). This is a factor of 4 larger than the median MCMC error on position (top two panels in the middle row of Figure~\ref{fig:differrnorm}). Accounting for this additional source of scatter between the tiles (which only affects real objects but not the simulations since those were inserted after the astrometric calibration) would bring the normalised error distribution for the real data overlap sample into much closer agreement with the simulated version. The remaining factor of $\sim$\,2-3 difference could also be due to the astrometry, considering that the overlap sample by definition sits at tile edges, where the astrometric solution is the most uncertain; or - as for all other parameters - due to irregular galaxy features in combination with noise (see discussion above). As a last point we would like to note that the absolute differences in position are usually still within 1 pixel ($0\farcs$2), i.e. although it stands out from the plot, this is a sub-pixel effect.

\subsubsection{Simulated overlap: imperfect vs. true PSFs}

These distributions of the simulated overlap sample can be compared to their equivalent distributions using the true PSFs (dashed light blue lines in Figure~\ref{fig:differrnorm}). This allows us to determine which parameters are affected by imperfect PSF estimates. However, we note that we can only make qualitative and relative statements here since we do not know how close to the truth our estimated PSFs for the real galaxies are. When simulating our galaxies, we convolve it with the model PSF fitted to the nearest real galaxy, i.e. this is the true PSF (cf. Section~\ref{sec:parameterrecovery}). When processing the simulated galaxy through our pipeline, the estimated PSF is then obtained by fitting nearby stars in the usual way. The nearest galaxy (which the true PSF is based on) is typically around 200\arcsec\ away, with the distribution ranging between $\sim$\,0 and $\sim$\,500\arcsec. This is close enough to provide a realistic PSF for the position of interest since KiDS tiles are much larger than this ($\sim$\,1\,deg$^2$) and the PSF varies only slowly across the tiles. However, it is further away than the stars used to obtain the estimated PSF, which are typically within $\sim$\,100-200\arcsec\ and can at the very maximum be $\sqrt{2}$\,$\times$\,200\arcsec\ away since the large cutouts used for PSF fitting are 400\arcsec\ on each side. This is expected since the density of stars is much higher than that of GAMA galaxies in KiDS data. However, it implies that the deviations of the estimated to the true PSFs in the simulations will on average be larger than for real data, leading to the errors on simulated galaxies being more severely underestimated as noted above. 

Comparing the simulations with true PSFs to the real data, it can be seen that the simulations now perform better than the real data for all parameters except segment effective radius (1$\sigma$-quantiles between 1 and 2.8). In addition, comparing the simulations with the true and the false PSFs against each other, we can assess which parameters are most affected (relatively speaking): position angle and axial ratio are most severely influenced; followed by S\'ersic index, effective radius and magnitude, while the position is nearly unaffected. This makes sense: the axial ratio and position angle are very sensitive to mistakes in the ellipticity and orientation of the PSF; while the fitted S\'ersic index, effective radius and magnitude depend on the concentration and FWHM of the PSF. The position is only very weakly affected since the PSF is always centred and symmetric. 

Note that all 1$\sigma$-quantiles are still larger than 1 (ranging from 1.1 to 2.8) even for the simulations with true PSFs. This indicates that the error underestimates of these parameters are not exclusively caused by the effects studied so far (galaxy and/or image processing features not accounted for in the simulations and PSF uncertainties) but there is an additional contribution from features that are also present in the simulations such as nearby objects, noise fluctuations, background subtraction inaccuracies or image artifacts.

\subsubsection{Simulated sample: fitted vs. true values}

Finally, for the simulated samples we can compare the fitted values to the true values (instead of the overlap sample comparison), which is shown as dotted orange lines in Figure~\ref{fig:differrnorm}. This allows us to detect biases, but is less directly comparable to the sample of real galaxies where the true value is unknown. Note the errors are generally slightly smaller compared to the overlap studies since for those, the errors for both fits were added in quadrature while the true values now do not have errors. Correspondingly, the normalised distributions are slightly broader even though the absolute differences between parameters are comparable. Most notably, however, the median of the distribution is now shifted away from zero for magnitude, effective radius and S\'ersic index (see also Table~\ref{tab:errorunderestimate}).
This is due to the bias caused by nearby objects already described in Section~\ref{sec:parameterrecovery}: magnitudes are too bright by $\sim$\,0.01\,mag; effective radii and S\'ersic indices too large by approximately 1\,\% (always a bit better for segment values and/or simulations using the true PSFs). All other parameters still have their distributions centred on zero, i.e. do not show any bias - at least not one that we can test with our simulations. This makes sense since position, axial ratio and angle will be influenced by nearby objects (and other effects) as well, but without any preferred direction and so on average this leads to an error underestimate rather than an overall bias. Using the true PSFs (dotted light orange lines) narrows all distributions slightly as expected; but again there is only an error underestimate rather than an overall bias introduced by the wrong PSFs since they are ``randomly wrong". 

One source of potential bias that we cannot test with the simulations are galaxy features not accounted for in the models. If, for example, there is a large population of galaxies that have bars; and these bars lead to the bulge axial ratios being systematically underestimated, this is again a one-sided effect that could lead to an overall bias. For this reason, we explicitly use the term ``bulge" in its broadest sense, including all kinds and possible combinations of central galaxy components. In addition, such features could further increase the error underestimate because they will tend to influence both fits to a galaxy in the overlap sample in the same way and hence are difficult to detect in the above analysis. If there are systematic one-sided deviations of the true PSFs from Moffat functions, these could lead to one-sided systematically wrong PSF estimates which could in turn also introduce an additional bias that cannot be tested by the simulations which use Moffat model PSFs. However, based on the PSF quality control, we do not believe that our PSF estimates are systematically wrong, see discussion in Section~\ref{sec:psfdetails}.

\subsubsection{Corrections for systematics and their validity}

\begin{table}
	\centering
	\caption{Biases and error underestimates for all single S\'ersic parameters derived from our systematic error studies (Section~\ref{sec:systematics}). The bias is additive (indicated with $\pm$) for those parameters that were treated in linear space and multiplicative (indicated with $\divideontimes$) for those treated in logarithmic space. Error underestimates are always multiplicative. The column ``bias/$\sigma$" gives the significance of each bias.}
	\label{tab:errorunderestimate}
	\begin{tabular}{llrr} 
		\hline
		\multirow{2}{*}{param.} & \multicolumn{1}{c}{bias} & \multirow{2}{*}{bias/$\sigma$} & \multicolumn{1}{c}{error}\\
		 & \multicolumn{1}{c}{(using true PSFs)} & & \multicolumn{1}{c}{underest.}\\
		\hline
RA & $\pm\,\, (7 \pm 18) \times 10^{-5}$\,arcsec &   0.39 & $ 12.27 \pm 0.71 $ \\ 
 Dec & $\pm\,\, (10 \pm 19) \times 10^{-5}$\,arcsec &   0.56 & $ 10.98 \pm 0.53 $ \\ 
 $m$ & $\pm\,\, (-11.4 \pm 0.8) \times 10^{-3}$  & -13.83 & $  3.10 \pm 0.16 $ \\ 
 $m_{seg}$ & $\pm\,\, (-8 \pm 0.5) \times 10^{-3}$  & -14.92 & $  1.79 \pm 0.09 $ \\ 
 $R_{e}$ & $\divideontimes\,\, (1.0105 \pm 0.0009)$  &  11.61 & $  2.68 \pm 0.12 $ \\ 
 $R_{e, seg}$ & $\divideontimes\,\, (1.0074 \pm 0.0005)$  &  14.39 & $  1.01 \pm 0.05 $ \\ 
 $n$ & $\divideontimes\,\, (1.010 \pm 0.001)$  &  10.93 & $  2.49 \pm 0.10 $ \\ 
 PA & $\pm\,\, (-4 \pm 19) \times 10^{-3}$\,deg &  -0.19 & $  2.00 \pm 0.10 $ \\ 
 b/a & $\divideontimes\,\, (0.9993 \pm 0.0004)$  &  -2.07 & $  2.51 \pm 0.09 $ \\ 
        \hline
	\end{tabular}
\end{table}

Table~\ref{tab:errorunderestimate} summarises the results of the systematic error studies: for all single S\'ersic parameters (plus the segment magnitude and segment effective radius), we give the average bias and error underestimates with uncertainties. The bias is estimated from the median of the offset between fitted and true values in the simulation using the true PSFs (light orange numbers in the first column of Figure~\ref{fig:differrnorm}). The errors on the median are taken as the 1$\sigma$-quantiles of these distributions divided by the square-root of the number of data points ($\sim$\,2000). Loga\-rith\-mic parameters are converted back into linear space to simplify bias correction in the catalogue. Nonetheless, scale parameters have multiplicative correction factors while location parameters have additive corrections (in given units). In other words: to correct for the bias, subtract the values in Table~\ref{tab:errorunderestimate} from the catalogue values for position, magnitude and position angle; and divide by the given values for effective radius, S\'ersic index and axial ratio. Users should note, however, that due to the way these biases were estimated, they do not include all sources of potential bias (e.g. galaxy features such as bars and spiral arms; or systematically wrong PSFs). Also, we recommend to apply the bias correction only to statistically large and random samples; they are average values not applicable to individual galaxies as evident from Figure~\ref{fig:comparesimulations}. 

The next column in Table~\ref{tab:errorunderestimate} gives the significance of each bias, which is the deviation of the median from 0 (or 1 for scale parameters) divided by its error. It can be seen that position, position angle and axial ratio are not biased (consistent with 0/1 within 2$\sigma$), while magnitude, effective radius and S\'ersic index are biased (deviation from 0/1 of >\,5$\sigma$); as found and discussed before. 

Finally, the last column in Table~\ref{tab:errorunderestimate} gives the error underestimates estimated from the width of the distribution of the normalised difference between fits to two versions of the same galaxy in the overlap sample (black numbers in the last column of Figure~\ref{fig:differrnorm}). The uncertainties in this case were estimated by bootstrapping the distributions 1000 times each to get an estimate of the variation of the distribution width. Since these distributions were normalised (by the respective errors), there is no need to convert between linear and logarithmic space and the given values can directly be used as correction factors for the MCMC errors. We have applied the relevant correction to all quoted errors in the catalogue, but also give the original (purely random) errors for completeness. Also note that since these values are now based on real data, they do include PSF uncertainties (in contrast to the biases). Also, since the overlap sample is in many ways the worst in terms of data quality (sitting at tile edges), these are likely upper limits. However, they still do not include error underestimates caused by (galaxy) features not accounted for in the models such as bars, rings, spiral arms or similar, as well as nearby objects. Since these are physical (rather than related to the data taking or image processing), they will be present in both versions of the overlap sample galaxy and influence both fits in similar ways - leading to a (random) bias on individual galaxy fits; and hence an error underestimate for a large enough sample. These issues should be kept in mind when using the catalogue.

\clearpage
\newpage
\chapter{Summary, conclusion and outlook}
\label{chap:conclusion}

In this thesis we presented our pipeline for the single S\'ersic fits and bulge-disk decompositions of 13096 galaxies at redshifts $z$\,<\,0.08 in the GAMA II equatorial survey regions in the KiDS $u$, $g$, $r$, $i$ and the VIKING $Z, Y, J, H, K_s$ bands. The galaxy modelling is done using \texttt{ProFit}, the Bayesian two-dimensional surface profile fitting code of \citet{Robotham2017}, fitting three models to each galaxy:
\begin{enumerate} 
\item{a single S\'ersic component,}
\item{a two-component model consisting of a S\'ersic bulge plus exponential disk and }
\item{a two-component model consisting of a point source bulge plus exponential disk (for unresolved bulges).}
\end{enumerate} 
The preparatory work (image segmentation, background subtraction and obtaining initial parameter guesses) is carried out using the sister package \texttt{ProFound} \citep{Robotham2018}; with the PSF estimated by fitting nearby stars using a combination of \texttt{ProFound} and \texttt{ProFit}. Segmentation maps are defined on joint $gri$-images, while the remaining analysis is performed individually in each band except for the model selection, for which we offer both a per-band and a joint version. The analysis is fully automated and self-contained with no dependency on additional tools. 

In addition to the galaxy fitting, we performed a number of post-processing steps including the flagging of bad fits and model selection. An overview of the number of galaxies successfully fitted in each band as well as the number classified in each category is given in Tables~\ref{tab:results} and~\ref{tab:resultsv05} as well as Figure~\ref{fig:ncompstats}. For our planned applications of the catalogue, which involves the statistical study of dust attenuation effects, we need fits that are most directly comparable to each other. Hence, we choose to model a maximum of two components for each galaxy even if more features may be present; and focus on achieving good fits in the high signal-to-noise regions of the galaxies by using relatively small segments for fitting. Consequently, we recommend using truncated magnitudes and effective radii for all analyses instead of the S\'ersic values which are extrapolated to infinity. The quality of the fits was ensured by visual inspection, comparing to previous works \citep{Kelvin2012, Lange2015}, studying independent fits of galaxies in the overlap regions of KiDS tiles and bespoke simulations. The latter two were also used for a detailed analysis of how systematic uncertainties affect our fits. 

We found that the combination of \texttt{ProFound} and \texttt{ProFit} is well-suited to the automated analysis of large datasets. The fully Bayesian MCMC treatment enabled by \texttt{ProFit} is able to overcome the main shortcomings of traditionally used downhill-gradient based optimisers, namely their susceptibility to initial guesses and their inability to easily derive realistic error estimates. The watershed deblending algorithm used by \texttt{ProFound} is less prone to catastrophic segmentation failures and allows us to extract more complex object shapes than other commonly used algorithms based on elliptical apertures; while still preserving the total flux well. With its wealth of utility functions, it not only facilitates the robust segmentation of large sets of images but also provides sky background estimates and reasonable initial guesses for the MCMC fitting. 

These characteristics, in combination with our own routines for quality assurance, led to results that are robust across a variety of galaxy types and image qualities and in reasonable agreement with previous studies given the different data, code and focus of the analysis. The outlier rejection routine efficiently identifies objects for which none of our models is appropriate such as irregular galaxies or those compromised by masked areas. Model selection is based on a $\Delta$DIC cut and accurate to >\,90\,\% compared to what could be achieved by visual inspection. There is a minimal bias in the fitted magnitude, effective radius and S\'ersic index of approximately 0.01\,mag, 1\,\% and 1\,\% respectively (on average across the full sample) caused by excess flux from nearby other objects. The errors obtained from the MCMC chains are underestimated with respect to the true errors by factors of typically between 2 and 3 (see Table~\ref{tab:errorunderestimate}) and can easily be corrected for statistically large samples of galaxies.

All results are integrated into the GAMA database as part of the \texttt{BDDecomp} DMU. The DMU consists of a number of catalogues giving the results of the preparatory work, the 2D surface brightness distribution fits and the post-processing of all 13096 galaxies in our full sample ($z$\,<\,0.08 in the GAMA II equatorial survey regions) in all bands; with additional diagnostic plots and all fit inputs available on the GAMA file server (see Section~\ref{sec:bddecompdmu} for details). So far, four DMU versions have been released, while \texttt{v05} will be made available alongside the publication of this thesis. The full DMU is currently available to GAMA team members with a version restricted to SAMI galaxies available to the SAMI team. It will be made publicly available in one of the forthcoming GAMA data releases. Readers interested in using (parts of) the catalogue before it is publicly released are encouraged to contact the authors to explore the possibilities for a collaboration\footnote{\url{http://www.gama-survey.org/collaborate/}}.

Both the GAMA and the SAMI teams have actively made use of the catalogue versions already released, with many more studies in progress and planned (see Chapter~\ref{chap:intro}). We therefore detailed not only the final version, but also different stages during pipeline development, including quality control steps of the preparatory work, in Chapter~\ref{chap:pipeline}. Since \texttt{v05} of the catalogue has not been previously published, a particular focus is placed on ensuring the reliability of those results by comparing it to \texttt{v04} (Chapter~\ref{chap:results}). The latter has benefitted from an extensive quality control detailed in Chapter~\ref{chap:QC}. 

In addition to the projects of collaborators mentioned above, we have many own plans and ideas for scientific analyses on the basis of our bulge-disk decomposition results. These include studying bulge and disk colours and trends of structural parameters with wavelength as well as deriving component stellar masses with the aims of investigating the stellar mass functions, stellar mass-to-light ratios and scaling relations such as the size-mass relation for individual components as a function of wavelength. By comparison with dust radiative transfer models, we will then also be able to constrain the nature and distribution of dust in galaxy disks, thereby contributing to the understanding of systematic uncertainties affecting studies of galaxy structure, formation and evolution, which in turn are vital for our understanding of the universe as a whole (see Section~\ref{sec:scienceaims}). 

On the technical side, further development of the pipeline foresees a number of minor improvements in the near future, for example considering the uncertainty of the astrometric solution for the joint fit, simultaneously fitting nearby sources and improving our treatment of the VIKING data by accounting for the slight differences in pixel size between frames, using the provided confidence maps and generating bright star masks. In the longer term, we will exploit the relatively new multi-band fitting functionality of \texttt{ProFit} - or better yet the new package \texttt{ProFuse} - to achieve simultaneous fits across all wavelength bands with physically motivated variations of structural parameters. Improvements to the preparatory work pipeline foresee the creation of a joint segmentation map including all bands, eliminating all manually calibrated tuning parameters and simultaneously fitting all suitable stars for more robust PSF estimates. In the post-processing, the main focus of development is on minimising manual intervention - especially the re-calibration of model selection - to allow easy scaling of the code to larger samples of galaxies from potentially different bands and datasets.

To summarise, we obtained a catalogue of robust structural parameters for the components of a sample of 13096 nearby GAMA galaxies while at the same time contributing to the advancement of image analysis, surface brightness fitting and post-processing routines for quality assurance in the context of automated large-scale bulge-disk decomposition studies. The further development of such methods and new approaches is vital to fully exploit the data of future sky surveys that will provide multi-wavelength imaging for millions of galaxies at unprecedented depth and resolution. The resulting measured parameters in turn are crucial to test and improve theoretical models and simulations and ultimately understand the formation, structure, composition and evolution of our universe.

\newpage
\thispagestyle{plain}
\section*{Acknowledgements}

GAMA is a joint European-Australasian project based around a spectroscopic campaign using the Anglo-Australian Telescope. The GAMA input catalogue is based on data taken from the Sloan Digital Sky Survey and the UKIRT Infrared Deep Sky Survey. Complementary imaging of the GAMA regions is being obtained by a number of independent survey programmes including GALEX MIS, VST KiDS, VISTA VIKING, WISE, Herschel-ATLAS, GMRT and ASKAP providing UV to radio coverage. GAMA is funded by the STFC (UK), the ARC (Australia), the AAO, and the participating institutions. The GAMA website is \url{http://www.gama-survey.org/}.

Based on observations made with ESO Telescopes at the La Silla Paranal Observatory under programme IDs 177.A-3016, 177.A-3017, 177.A-3018 and 179.A-2004, and on data products produced by the KiDS consortium. The KiDS production team acknowledges support from: Deutsche Forschungsgemeinschaft, ERC, NOVA and NWO-M grants; Target; the University of Padova, and the University Federico II (Naples).

This publication has made use of data from the VIKING survey from VISTA at the ESO Paranal Observatory, programme ID 179.A-2004. Data processing has been contributed by the VISTA Data Flow System at CASU, Cambridge and WFAU, Edinburgh.

The total computing time for all runs and test runs performed in the context of this thesis sums to approximately 676\,700 hours on CPU. With an electricity usage of 12\,W per CPU, our total energy consumption for the project is 8120\,kWh. Assuming an average CO$_2$ equivalent of 500\,g\,/\,kWh (corresponding to the average of the German energy mix during the period of the project\footnote{\url{https://www.umweltbundesamt.de/presse/pressemitteilungen/bilanz-2019-co2-emissionen-pro-kilowattstunde-strom}}), results in 4\,tons of CO$_2$ emitted by this PhD thesis. This is comparable to the worldwide average of CO$_2$ emissions per person per year.\footnote{\url{https://de.statista.com/statistik/daten/studie/167877/umfrage/co-emissionen-nach-laendern-je-einwohner/}} It is still a lower limit since it does not count any non-automated calculations nor failed runs; and completely ignores other activities related to the project, such as travelling.  

\newpage
\thispagestyle{plain}
\section*{Thanks}

First of all, I would like to thank my supervisor Jochen Liske for going through this project with me start to end. Jochen is a very skilled and enthusiastic scientist who is never short of ideas and was a great source of inspiration for me. On the more technical side, much support came from Aaron Robotham. I am very grateful to him for always being available and helping me out countless times during the course of my PhD. Furthermore, let me thank Robi Banerjee, Peter Hauschildt and Dieter Horns for being part of my defense committee. 

It was great to be welcomed into the GAMA team, a very friendly and supportive group of scientists. Working with the SAMI and KiDS teams was also a pleasure, with members of both collaborations always happy to help out. My stays at ICRAR will remain in good memory due to the hospitality and friendlyness of everybody I met there. While it is impossible to name everyone who contributed to the success of this project, a few people who come to my mind - apart from Jochen and Aaron of course - are Dan Taranu, Angus Wright, Jarkko Laine, Robin Cook, Hosein Hashemizadeh, Stefania Barsanti, Ben Henderson, Sree Oh, Simon Driver, Boris H{\"au{\ss}ler, Benne Holwerda, Ned Taylor, Amanda Moffett, Konrad Kuijken, Luke Davies, Michelle Cluver and Christos Georgiou. 

A special thanks goes to Denis Wittor and Volker Heesen for proof-reading this thesis and suggesting many improvements. Janis Kummer was so kind to provide his thesis to use as a template and J\"org Knoche offered much needed emergency IT-support on several occassions. Together with all other current and former members of the Hamburger Sternwarte - many of whom I count to my friends - they made my PhD time a lot of fun and one of the best periods of my life. 

Last but not least, my thanks goes to my family and friends - especially my parents, but also my sister, grandparents, some of the wider family and many close friends - all of whom always gave me the feeling that I could achieve anything if only I wanted to. I am extremely lucky to have Lars, the best husband that I could have wished for and an amazing dad to our son Tim. Without his unlimited support - and the happy nature of Tim - I could have never finished this project.

\clearpage
\newpage

\pagenumbering{roman}
\clearpage
\newpage


\bibliographystyle{thesis}

\addcontentsline{toc}{chapter}{Bibliography}
\bibliography{/home/sarah/Desktop/papers/zzzreferences}

\clearpage

\end{document}